\documentclass{SciPost}

\hypersetup{
    colorlinks,
    linkcolor={red!50!black},
    citecolor={blue!50!black},
    urlcolor={blue!80!black}
}

\usepackage[bitstream-charter]{mathdesign}
\usepackage{bm} 
\usepackage{textcomp}  

\graphicspath{{/}{images/}}

\DeclareSymbolFont{usualmathcal}{OMS}{cmsy}{m}{n}
\DeclareSymbolFontAlphabet{\mathcal}{usualmathcal}

\fancypagestyle{SPstyle}{
\fancyhf{}
\lhead{\colorbox{scipostblue}{\bf \color{white} ~SciPost Physics Codebases }}
\rhead{{\bf \color{scipostdeepblue} ~Submission }}

\fancyfoot[C]{\textbf{\thepage}}
}

\newcommand{\kdotpy}{\texttt{kdotpy}}
\newcommand{\kdotp}{\ensuremath{\vec{k}\cdot\vec{p}}}
\newcommand{\HOME}{\raisebox{0.5ex}{\texttildelow}}
\newcommand{\uscore}{\char`_}

\newcommand {\ii}      {\mathrm{i}}
\newcommand {\ee}      {\mathrm{e}}
\newcommand {\nm}      {\,\mathrm{nm}}
\newcommand {\textmicrom}  {\ensuremath{\text{\textmu}\mathrm{m}}}
\newcommand {\microm}  {\,\textmicrom}
\newcommand {\meV}     {\,\mathrm{meV}}
\newcommand {\lB}      {l_B}
\newcommand {\sigmaH}  {\sigma_\mathrm{H}}

\newcommand {\abs}[1]  {\lvert#1\rvert}

\newcommand {\ddx}[2]  {\frac{\mathrm{d}#1}{\mathrm{d}#2}}

\renewcommand {\vec}   {\mathbf}
\newcommand {\vecsigma} {\bm{\sigma}}

\newcommand {\ket}[1]      {\lvert#1\rangle}
\newcommand {\ketbf}[1]    {\lvert\mathbf{#1}\rangle}
\newcommand {\bra}[1]      {\langle#1\rvert}

\newcommand {\braket}[2]    {\langle#1\vert#2\rangle}
\newcommand {\bramidket}[3] {\langle#1\vert#2\vert#3\rangle}

\newcommand {\avg}[1]   {\langle#1\rangle}
\newcommand {\half}       {\tfrac{1}{2}}
\newcommand {\threehalf}  {\tfrac{3}{2}}
\newcommand {\spinup}  {\mathord\uparrow}
\newcommand {\spindn}  {\mathord\downarrow}

\newcommand {\muB}     {\mu_B}
\newcommand {\kB}      {k_B}

\newcommand {\Ec}      {E_\mathrm{c}}
\newcommand {\Ev}      {E_\mathrm{v}}
\newcommand {\Eg}      {E_\mathrm{g}}
\newcommand {\Deltaso} {\Delta_\mathrm{SO}}
\newcommand {\VH}      {V_\mathrm{H}}

\renewcommand{\Im}{\mathop{\mathrm{Im}}}

\DeclareMathOperator{\diag}{diag}
\DeclareMathOperator{\tr}{tr}

\begin{document}

\pagestyle{SPstyle}

\begin{center}{\Large \textbf{\color{scipostdeepblue}{
kdotpy: \ensuremath{\mathbf{k}\cdot\mathbf{p}} theory on a lattice for simulating semiconductor band structures\\
}}}\end{center}

\begin{center}\textbf{
Wouter Beugeling\textsuperscript{1,2$\star\dagger$},
Florian Bayer\textsuperscript{1,2},
Christian Berger\textsuperscript{1,2},
Jan B\"ottcher\textsuperscript{3},
Leonid Bovkun\textsuperscript{1,2},
Christopher Fuchs\textsuperscript{1,2},
Maximilian Hofer\textsuperscript{1,2},
Saquib Shamim\textsuperscript{1,2},
Moritz Siebert\textsuperscript{1,2},
Li-Xian Wang\textsuperscript{1,2},
Ewelina M. Hankiewicz\textsuperscript{3},
Tobias Kie\ss{}ling\textsuperscript{1,2},
Hartmut Buhmann\textsuperscript{1,2} and
Laurens W. Molenkamp\textsuperscript{1,2}
}\end{center}

\begin{center}
{\bf 1} Physikalisches Institut (EP3), Universit\"{a}t W\"{u}rzburg, Am Hubland, 97074 W\"{u}rzburg, Germany
\\
{\bf 2} Institute for Topological Insulators, Am Hubland, 97074 W\"{u}rzburg, Germany
\\
{\bf 3} Institut f\"ur Theoretische Physik und Astrophysik (TP4), Universit\"{a}t W\"{u}rzburg, Am Hubland, 97074 W\"{u}rzburg, Germany
\\[\baselineskip]
$\star$ \href{mailto:kdotpy@uni-wuerzburg.de}{\small kdotpy@uni-wuerzburg.de}\,,\quad
$\dagger$ \href{mailto:wouter.beugeling@physik.uni-wuerzburg.de}{\small wouter.beugeling@physik.uni-wuerzburg.de}
\end{center}

\section*{\color{scipostdeepblue}{Abstract}}
\boldmath\textbf{%
The software project \kdotpy{} provides a Python application for simulating electronic band structures of semiconductor devices with \kdotp{} theory on a lattice. The application implements  the widely used Kane model, capable of reliable predictions of transport and optical properties for a large variety of topological and non-topological materials with a zincblende crystal structure.
The application automates the tedious steps of simulating band structures. The user inputs the relevant physical parameters on the command line, for example materials and dimensions of the device, magnetic field, and temperature. The program constructs the appropriate matrix Hamiltonian on a discretized lattice of spatial coordinates and diagonalizes it. The physical observables are extracted from the eigenvalues and eigenvectors and saved as output. The program is highly customizable with a large set of configuration options and material parameters.%
}

\boldmath\textbf{%
The project is released as free open source software under the GNU General Public License, version 3. The code that accompanies this article is available from our Gitlab repository at \url{https://git.physik.uni-wuerzburg.de/kdotpy/kdotpy/-/tags/v1.0.0}.%
}

\vspace{\baselineskip}




\vspace{10pt}
\noindent\rule{\textwidth}{1pt}
\tableofcontents
\noindent\rule{\textwidth}{1pt}
\vspace{10pt}


\clearpage
\section{Introduction}

In the field of solid-state physics, particularly in the branch of semiconductor
physics, the theory of electronic band structures plays a central role.
Semiconductor materials are inherently very complex systems due to the typically
large number of atomic orbitals of their constituent atoms. A reduction to the
essential degrees of freedom (namely, the electronic bands close to the
Fermi level at charge neutrality) can be achieved by effective models, such as
the perturbative method known as \kdotp{} theory. 
The Kane model \cite{Kane1957}, a \kdotp{} theory that was originally
proposed for describing the electronic band structure of indium antimonide (InSb),
has since become an established model for the large category of binary
semiconductor compounds with zincblende crystal structure, and in particular
for narrow gap semiconductor compounds. The Kane model has been widely used in
the semiconductor community, because the resulting band structures are well
suited for predicting transport phenomena as well as optical (spectroscopic)
properties. It is therefore not only a valuable asset upon analyzing data, but
also a very powerful tool for designing device functionality.

The minimal number of the degrees of freedom in the Kane model is still too
large to permit exact analytic solutions of the Schr\"odinger equation, that defines the band
structure. The problem can be approached numerically, but setting up numerical
simulations and extracting physical observables can be a tedious task, for example
due to the quantity and complexity of the matrix elements in the Hamiltonian.
These issues mandate a software
solution where the tedious tasks are automated, i.e., an application which takes
physical parameters like geometry and materials of a device as input, and which
produces physical observables as its output.

Here, we present the Python application \kdotpy{} as a means to deal with these
challenges.
The motivation for developing \kdotpy{} has been the need to analyze the
experimental results of the research group of L.~W.~Molenkamp in W\"urzburg.
In many regards the situation that triggered the development of \kdotpy{} was archetypical for an experimental group engaged in the study of sophisticated semiconductor heterostructures. The complexity of the collected data called for a description beyond simplified effective models. It was well understood that \kdotp{} theory provided a suitable framework for a realistic description of the experimental findings and a Fortran program set up by Pfeuffer-Jeschke and Novik~\cite{PfeufferJeschke2000_thesis,NovikEA2005} was already used for band structure simulations for many years. However, the modelling used in this program proved to be
inadequate for simulating edge channels and surface states. 
Moreover, it lacks an intuitive user interface, and the program structure and the programming style have proven to be major obstacles towards maintainability and towards future extension. In particular, the fact that the initial developers had moved on to other positions and a general lack of detail in the documentation of the code constitute major entrance thresholds for inclusion of novel functionalities.

In the \kdotpy{} project, we attempt to steer clear of these problems
by applying modern concepts regarding project design and development workflow.
The development strategy is guided by the philosophy that \kdotpy{}'s primary
goal is to support current (experimental and theoretical) research, for which
feedback from the community is an important aspect.
While \kdotpy{} has started out with the specific task of analyzing transport
phenomena in Mn-doped HgTe quantum wells \cite{ShamimEA2020SciAdv,ShamimEA2021NatCommun,ShamimEA2022NatCommun},
it has since been extended to tackle a much broader variety of problems~\cite{MahlerEA2021NanoLett,MullerEA2021NanoLett,ThenapparambilEA2023NanoLett,WangEA2024AdvSci}. As \kdotp{} theory is much more widely applicable than the study of
topological insulators or HgTe in particular, we have equipped \kdotpy{} with
the infrastructure for simulating any material whose band structure is described
by the Kane model. This includes other II-VI materials as well as III-V semiconductors
like GaAs and InSb.
We expect that \kdotpy{} will thus be beneficial for a broad community
spanning various disciplines in semiconductor physics.
We intend to create a diverse and international user base and encourage active
discussion among users and developers, in the spirit of open science and open
software development. The community driven nature facilitates
sustainable maintenance and future project development.

In line with modern standards in scientific research, we strive towards transparency
by publishing this project as open source software. We believe that the choice of
Python as a widely known programming language facilitates inspection of our
code and thus fits the ideas of open science.
We aim for a high level of reproducibility and for compliance with the FAIR
principles for research data \cite{WilkinsonEA2016_FAIR} by providing rich
metadata in the output files and by avoiding the use of proprietary file formats.

The purpose of this article is to provide a comprehensive overview of all
important components of \kdotpy{} together with relevant physical background.
We emphasize that reading the complete manuscript is not required for being able
to use the application: We encourage the reader to focus on the parts of personal
interest and/or those needed to address a specific physical problem.
The article is structured as follows:
In Sec.~\ref{sec_physics_background}, we provide an in-depth review of the
physical concepts and theories essential for understanding the implementation.
In Sec.~\ref{sec_implementation}, we discuss all important components of the
implementation, making the connection between the physics and the project's
source code. Installation instructions are
provided in Sec.~\ref{sec_installation}. We illustrate the usage of \kdotpy{}
with several detailed examples in Sec.~\ref{sec_usage}.
The Appendices contain technical details of the implementation as well as a
comprehensive reference of configuration options, material parameters,
command-line arguments, and several other aspects.

\clearpage
\section{Physics background}
\label{sec_physics_background}

The Kane model \cite{Kane1957} implemented in \kdotpy{} was originally
proposed as a theory for the electronic band structure of InSb. It has proven to
be a suitable model for a broad class of binary semiconductor compounds
with a zincblende crystal structure, to which InSb also belongs. The Kane model
defines an $8\times 8$ Hamiltonian depending on the momentum coordinate $\vec{k}$.
The basis of the  Hilbert space is formed by two $s$ orbitals and six $p$ orbitals.
These states are treated exactly at the $\Gamma$ point ($\vec{k}=0$), while
couplings to other (so-called \emph{remote}) bands are treated perturbatively in
the framework of \kdotp{} theory.

Importantly, the Kane model takes into account spin-orbit interaction, which is
essential for explaining the energetic positions of the electronic bands at the
$\Gamma$ point \cite{Kane1957}. In compounds of heavy elements, spin-orbit
coupling and other relativistic effects become strong enough to result in an
inversion of the band ordering. The latter is the key ingredient for the
formation of topological phases, which constitute a major research branch at
the Department of Physics in W\"urzburg since the first experimental
realizations of such materials in mercury telluride (HgTe) quantum wells
\cite{KonigEA2007}. \kdotp{} theory is well suited towards accessing topological
properties and accordingly this was a main focus upon setting up the \kdotpy{} project.

The numerical approach also admits analysis of more complicated configurations
than bulk materials. In particular, experiments are typically performed with
devices with several layers of zincblende-type materials stacked on top of a
substrate. The latter constellation is described by taking the bulk Hamiltonian
and substituting the momentum coordinate $k_z$ by
its spatial representation $-\ii\hbar\partial_z$, a derivative with respect to the
spatial coordinate $z$. For simulating the band structure, the $z$ coordinate is
discretized to a finite set $\{z_j\}$, so that the Hilbert space dimension
(and the size of the Hamiltonian matrix) becomes $8n_z$. The different layers in
the system are modelled by making the coefficients in the Kane model $z$-dependent.
This approach goes back to Burt \cite{Burt1992,Burt1999}
and has since been used by many others, see, e.g.,
Refs.~\cite{PfeufferJeschke2000_thesis,NovikEA2005,Winkler2003_book}.
This discretization method distinguishes \kdotpy{} from the aforementioned Fortran
program \cite{PfeufferJeschke2000_thesis,NovikEA2005}, which uses an expansion
into Legendre polynomials in order to convert the Hamiltonian to matrix form. The
present approach has the benefits that the modelling is more intuitive and that
it can more reliably resolve essential features like edge channels \cite{ShamimEA2020SciAdv}
and surface states \cite{MahlerEA2021NanoLett}.

The bulk Kane model and the reduction to lower dimensional geometries are discussed
Secs.~\ref{sec_kane} and \ref{sec_lower_dim_geometries}, respectively. The
remainder of Sec.~\ref{sec_physics_background} is dedicated to the effects of
magnetic fields, bulk inversion asymmetry, and strain, as well as to topology.

\subsection{\kdotp{} theory and the Kane model in a nutshell}
\label{sec_kane}

\subsubsection{\kdotp{} theory}

The \kdotp{} method aims at finding an effective model for the band structure
near a high-symmetry point. The method is based on perturbation theory on
the Bloch wave functions, where eigenenergies and eigenstates 
at $\vec{k}=0$ are treated as exact and the contributions for $\vec{k}\not=0$
as perturbation.
The Schr\"odinger equation
$[\hat{\vec{p}}^2/2m + V(\vec{r})]\psi_{n,\vec{k}}(\vec{r}) = E_{n}(\vec{k})\psi_{n,\vec{k}}(\vec{r})$
acting on the Bloch wave functions
$\psi_{n,\vec{k}}(\vec{r}) = \ee^{\ii \vec{k}\cdot \vec{r}}u_{n,\vec{k}}(\vec{r})$
can be written in terms of the Bloch functions $u_{n,\vec{k}}(\vec{r})$ as
\begin{equation}\label{eqn_schrodinger_bloch}
  \left[
   \frac{\hbar^2 \vec{k}^2}{2m}
   + \frac{2\hbar \vec{k}\cdot\hat{\vec{p}}}{2m}
   + \frac{\hat{\vec{p}}^2}{2m}
   + V (\vec{r})
  \right] u_{n,\vec{k}}(\vec{r})
  =  E_{n}(\vec{k}) u_{n,\vec{k}}(\vec{r}),
\end{equation}
where $\hat{\vec{p}}=-\ii\hbar\nabla$ is the momentum operator, $\vec{k}$ is
a vector of real values, and the index $n$ labels the bands.
The Hamiltonian $H_0=\hat{\vec{p}}/2m + V(\vec{r})$
at $\vec{k}=0$ is treated as exact, whereas
$H'_\mathbf{k} = \hbar^2 \vec{k}^2/2m + \hbar \vec{k}\cdot\hat{\vec{p}}/m$
is treated as perturbation. The second term in $H'_\mathbf{k}$ gives
\kdotp{} theory its name. The unperturbed eigenenergies $E_{n}(0)$ and
eigenfunctions $\ket{u_{n,0}}$, that solve $H_0 \ket{u_{n,0}} = E_{n}(0)\ket{u_{n,0}}$,
are assumed to be known. The first order perturbation to the energy is
\begin{equation}\label{eqn_kdotp_pert_e1}
 E_{n}^{(1)}(\vec{k})
 = \bramidket{u_{n,0}}{H'_\mathbf{k}}{u_{n,0}}
 = \frac{\hbar^2 \vec{k}^2}{2m}.
\end{equation}
An additional term proportional to $\bramidket{u_{n,0}}{\vec{k}\cdot\hat{\vec{p}}}{u_{n,0}}$
vanishes if we assume that the dispersions attain a maximum or
minimum at $\vec{k}=0$. The first order perturbation to the eigenfunctions is
\begin{equation}\label{eqn_kdotp_pert_u1}
  \ket{u_{n,\vec{k}}^{(1)}}
  = \frac{\hbar}{m}\sum_{n':E_{n'}\not= E_{n}}
      \frac{\vec{k}\cdot\vec{P}_{n'n}}{E_{n}(0)-E_{n'}(0)}\ket{u_{n',0}},
\end{equation}
where $\vec{P}_{n'n} \equiv \bramidket{u_{n',0}}{\hat{\vec{p}}}{u_{n,0}}$ and the
sum runs over all bands $n'$ for which $E_{n'}(0)\not=E_n(0)$. From
Eq.~\eqref{eqn_kdotp_pert_u1}, we obtain the second order perturbation of the
energy as
\begin{equation}\label{eqn_kdotp_pert_e2}
 E_{n}^{(2)}(\vec{k})
 = \bramidket{u_{n,0}}{H'_\mathbf{k}}{u^{(1)}_{n,\vec{k}}}
 = \frac{\hbar^2}{m^2}\sum_{n':E_{n'}\not= E_{n}}
     \frac{\abs{\vec{k} \cdot \vec{P}_{n'n}}^2}{E_{n}(0)-E_{n'}(0)}
   +\mathcal{O}(\vec{k}^3),
\end{equation}
noting that $\vec{P}_{nn'} = \vec{P}_{n'n}^*$ and discarding terms of cubic
order in momentum. This observation completes the perturbation theory up to
lowest nontrivial order in momentum, as
\begin{align}
  E_n(\vec{k}) &=
  E_n(0) + \frac{\hbar^2 \vec{k}^2}{2m}
  + \frac{\hbar^2}{m^2}\sum_{n':E_{n'}\not= E_{n}}
     \frac{\abs{\vec{k} \cdot \vec{P}_{n'n}}^2}{E_{n}(0)-E_{n'}(0)}
  + \mathcal{O}(\vec{k}^3),\label{eqn_kdotp_pert_e_all}\\
  \ket{u_{n,\vec{k}}^{(1)}}
  &= \frac{\hbar}{m}\sum_{n':E_{n'}\not= E_{n}}
      \frac{\vec{k}\cdot\vec{P}_{n'n}}{E_{n}(0)-E_{n'}(0)}\ket{u_{n,0}}
  + \mathcal{O}(\vec{k}^2)\label{eqn_kdotp_pert_u_all}.
\end{align}
Importantly, it is possible to consider only those bands $n$ with energies
$E_{n}(0)$ sufficiently close to the charge neutrality point, or in other words,
to reduce the number of degrees of freedom by choosing a restricted selection of
the $\ket{u_{n,0}}$ as basis. However, the perturbations involve a sum over all
bands $n'$, also those outside of the basis. These bands are known as
\emph{remote bands} in the context of \kdotp{} theory.

\subsubsection{The Kane model}

In the seminal paper by Kane \cite{Kane1957}, \kdotp{} theory was used to
calculate the band structure of InSb. As a matter of fact, Kane's model applies
to many more semiconductor materials with a zincblende lattice structure.
The important extra ingredient in Kane's analysis is spin-orbit coupling, one of
the relativistic corrections that affect the atomic orbitals.
The Hamiltonian $H_0$ (cf.\ Eq.~\eqref{eqn_schrodinger_bloch}) is modified by
adding the spin-orbit term \cite{Kane1957}
\begin{equation}\label{eqn_hso}
 H_\mathrm{SO} = \frac{\hbar}{4m^2c^2}(\nabla V \times \hat{\vec{p}})\cdot\vecsigma
\end{equation}
where $\vecsigma = (\sigma_x,\sigma_y,\sigma_z)$ is the vector of Pauli matrices
acting on the spin degree of freedom and $V$ is the potential near the atomic
core.

In InSb, the relevant atomic orbitals are the $5s$ orbitals of In and the $5p$
orbitals of Sb. Together with the spin degree of freedom ($S=\frac{1}{2}$), this
yields the eight component basis,
\begin{equation}
  \ket{S,\spinup},-\tfrac{1}{\sqrt{2}}\ket{X+\ii Y,\spindn},\ket{Z,\spinup},\tfrac{1}{\sqrt{2}}\ket{X-\ii Y,\spindn},
  \ket{S,\spindn},+\tfrac{1}{\sqrt{2}}\ket{X-\ii Y,\spinup},\ket{Z,\spindn},\tfrac{1}{\sqrt{2}}\ket{X+\ii Y,\spinup},
\end{equation}
where $\ket{S}$ labels the $s$ orbital and $\ket{X},\ket{Y},\ket{Z}$ the $p$ orbitals.
In this basis, the Hamiltonian at $\vec{k}=0$ is represented by the diagonal matrix $H_0=\diag(E_s,E_p,E_p,E_p,E_s,E_p,E_p,E_p)$, and $H_\mathrm{SO}$
can be written as
\begin{equation}
 H_\mathrm{SO}=\begin{pmatrix}h_\mathrm{SO}&0\\0&h_\mathrm{SO}\end{pmatrix}
 \qquad\text{with}
 \qquad
 h_\mathrm{SO}=
 \begin{pmatrix}
 0&0&0&0\\
 0&-\frac{1}{3}\Deltaso&\frac{\sqrt{2}}{3}\Deltaso&0\\
 0&\frac{\sqrt{2}}{3}\Deltaso&0&0\\
 0&0&0&\frac{1}{3}\Deltaso
 \end{pmatrix}
\end{equation}
where
$\Deltaso=(3\ii\hbar/4m^2c^2)\bramidket{X}{(\partial_x V\,\hat{p}_y-\partial_y V\,\hat{p}_x)}{Y}$ is the spin-orbit splitting; the eigenvalues of $h_\mathrm{SO}$ are
$0,\frac{1}{3}\Deltaso,\frac{1}{3}\Deltaso,-\frac{2}{3}\Deltaso$.
Thus, $h_\mathrm{SO}$ partially lifts the degeneracy between
the $p$ orbital states (six states at energy $E_p$) to four states at $E_p+\frac{1}{3}\Deltaso$
and two states at $E_p-\frac{2}{3}\Deltaso$.

The eigenstates of $H_0+H_\mathrm{SO}$ are formed by the basis
\begin{align}
  \ketbf{1}&=\ket{\Gamma_6,+\half}=\ket{S,\spinup},\nonumber\\
  \ketbf{2}&=\ket{\Gamma_6,-\half}=\ket{S,\spindn},\nonumber\\
  \ketbf{3}&=\ket{\Gamma_8,+\threehalf}=\frac{1}{\sqrt{2}}\ket{X+\ii Y,\spinup},\nonumber\\
  \ketbf{4}&=\ket{\Gamma_8,+\half}=\frac{1}{\sqrt{6}}\left[\ket{X+\ii Y,\spindn}-2\ket{Z,\spinup}\right],\nonumber\\
  \ketbf{5}&=\ket{\Gamma_8,-\half}=-\frac{1}{\sqrt{6}}\left[\ket{X-\ii Y,\spinup}+2\ket{Z,\spindn}\right],\label{eqn_orbital_basis}\\
  \ketbf{6}&=\ket{\Gamma_8,-\threehalf}=-\frac{1}{\sqrt{2}}\ket{X-\ii Y,\spindn},\nonumber\\
  \ketbf{7}&=\ket{\Gamma_7,+\half}=\frac{1}{\sqrt{3}}\left[\ket{X+\ii Y,\spindn}+\ket{Z,\spinup}\right],\nonumber\\
  \ketbf{8}&=\ket{\Gamma_7,-\half}=\frac{1}{\sqrt{3}}\left[\ket{X-\ii Y,\spinup}-\ket{Z,\spindn}\right],\nonumber
\end{align}
where the notation $\ket{\Gamma_r,m_j}$ refers to the irreducible representation
$\Gamma_r$ ($r=6,7,8$) of the point group $T_d$ under which the states transform
as well as the total angular momentum quantum number ($J_z$ eigenvalue) $m_j$.
In the basis of Eq.~\eqref{eqn_orbital_basis}, the Hamiltonian $H_0+H_\mathrm{SO}$
can be diagonalized as
\begin{equation}\label{eqn_h0_hso}
  H_0 + H_\mathrm{SO} = \begin{pmatrix}
        \Ec&0&\\
        0 & \Ec\\
        &&\Ev&0&0&0\\
        &&0&\Ev&0&0\\
        &&0&0&\Ev&0\\
        &&0&0&0&\Ev\\
        &&&&&&\Ev-\Deltaso&0\\
        &&&&&&0&\Ev-\Deltaso
        \end{pmatrix},
\end{equation}
with $\Ec\equiv E_s$, $\Ev\equiv E_p+\frac{1}{3}\Delta$, and
$\Ev-\Deltaso \equiv E_p-\frac{2}{3}\Delta$ referring to the
`conduction band', `valence band' and split-off band energies at $\Gamma$, respectively.
Note that in the notation we use the labels for `conduction' and `valence' band
for the $\Gamma_6$ and $\Gamma_8$ bands, respectively, without regard to the
actual band ordering. For inverted materials, $\Ec$ lies in the valence band and
$\Ev$ in the conduction band.

We note that between references
(cf.\ Refs.~\cite{Kane1957,Weiler1981bookchapter,PfeufferJeschke2000_thesis,NovikEA2005},
for example),
the basis may differ slightly by complex phases and order. Here, we group the
multiplets belonging to each representation ($\Gamma_6$, $\Gamma_8$, $\Gamma_7$)
together. The Hamiltonian thus has the block structure
\begin{equation}\label{eqn_ham_blocks}
  H = \left(\begin{array}{c|c|c}
        H_{66}&H_{68}&H_{67}\\ \hline
        H_{86}&H_{88}&H_{87}\\ \hline
        H_{76}&H_{78}&H_{77}
      \end{array}\right).
\end{equation}
Where necessary due to space restriction, we will write terms of the full
$8\times 8$ Hamiltonian in terms of these smaller blocks.

The $\vec{k}\cdot\hat{\vec{p}}$ term in the Hamiltonian introduces matrix elements
proportional to $k_x\bramidket{S}{p_x}{X}$, $k_y\bramidket{S}{p_y}{Y}$
and $k_z\bramidket{S}{p_z}{Z}$.
To write these terms, one defines the \emph{Kane parameter}
\begin{equation}\label{eqn_p_kane}
  P= \frac{\hbar}{m}\abs{\bramidket{S}{\hat{p}_x}{X}}
   = \frac{\hbar}{m}\abs{\bramidket{S}{\hat{p}_y}{Y}}
   = \frac{\hbar}{m}\abs{\bramidket{S}{\hat{p}_z}{Z}}.
\end{equation}
The ambiguity in the definitions of $P$ in the literature
(again, cf.\ Refs.~\cite{Kane1957,Weiler1981bookchapter,PfeufferJeschke2000_thesis,NovikEA2005})
is resolved by choosing a real and positive value. The value of the Kane
parameter is of similar magnitude for many materials due to the similarity of
the orbital wave functions. 

The next step towards the formulation of the full Kane model is to identify
all possible inversion symmetric terms up to quadratic order. (We will discuss
inversion asymmetric terms in Sec.~\ref{sec_bia}.) Using the
conventions of Ref.~\cite{NovikEA2005}, we write
\begin{equation}\label{eqn_hkane}
 H_\mathrm{Kane} = H_0 + H_\mathrm{SO}+ H_\vec{k}
\end{equation}
in terms of the constant part $H_0 + H_\mathrm{SO}$ [Eq.~\eqref{eqn_h0_hso}]
and the momentum-dependent part
\begin{equation}\label{eqn_ham_kp}
  H_\vec{k} =
  \begin{pmatrix}
        T_\vec{k} & 0 & -\sqrt{\frac{1}{2}}Pk_+ & \sqrt{\frac{2}{3}}Pk_z & \sqrt{\frac{1}{6}}Pk_- & 0 & -\sqrt{\frac{1}{3}}Pk_z & -\sqrt{\frac{1}{3}}Pk_-\\
        0 & T_\vec{k} & 0 & -\sqrt{\frac{1}{6}}Pk_+ & \sqrt{\frac{2}{3}}Pk_z & \sqrt{\frac{1}{2}}Pk_-& -\sqrt{\frac{1}{3}}Pk_+& \sqrt{\frac{1}{3}}Pk_z\\
        -\sqrt{\frac{1}{2}}Pk_-&0&U_\vec{k}+V_\vec{k}&-S^-_\vec{k}&R_\vec{k}&0&\frac{1}{\sqrt{2}}S^-_\vec{k}&-\sqrt{2}R_\vec{k}\\
        \sqrt{\frac{2}{3}}Pk_z & -\sqrt{\frac{1}{6}}Pk_-&-S^{-\dagger}_\vec{k} &U_\vec{k}-V_\vec{k}&C_\vec{k}&R_\vec{k}&\sqrt{2}V_\vec{k}&-\sqrt{\frac{3}{2}}\tilde{S}^-_\vec{k}\\
        \sqrt{\frac{1}{6}}Pk_+ & \sqrt{\frac{2}{3}}Pk_z&R^\dagger_\vec{k}&C^\dagger_\vec{k}&U_\vec{k}-V_\vec{k}&S^{+\dagger}_\vec{k}&-\sqrt{\frac{3}{2}}\tilde{S}^+_\vec{k}&-\sqrt{2}V_\vec{k}\\
        0 & \sqrt{\frac{1}{2}}Pk_+ & 0 &R_\vec{k}^\dagger&S^{+}_\vec{k} & U_\vec{k}+V_\vec{k} & \sqrt{2}R^\dagger_\vec{k}&\frac{1}{\sqrt{2}}S^+_\vec{k}\\
        -\sqrt{\frac{1}{3}}Pk_z & -\sqrt{\frac{1}{3}}Pk_- & \frac{1}{\sqrt{2}}S^{-\dagger}_\vec{k} & \sqrt{2}V_\vec{k} & -\sqrt{\frac{3}{2}}\tilde{S}^{+\dagger}_\vec{k} & \sqrt{2}R_\vec{k} & U_\vec{k} & C_\vec{k}\\
        -\sqrt{\frac{1}{3}}Pk_+ & \sqrt{\frac{1}{3}}Pk_z & -\sqrt{2}R_\vec{k}^\dagger & -\sqrt{\frac{3}{2}}\tilde{S}^{-\dagger}_\vec{k} & -\sqrt{2}V_\vec{k}&\frac{1}{\sqrt{2}}S^{+\dagger}_\vec{k} & C_\vec{k}^\dagger & U_\vec{k}
  \end{pmatrix}
\end{equation}
with
\begin{align}
  T_\vec{k} &= \frac{\hbar^2}{2m}(2F+1) |\mathbf{k}|^2\nonumber\\
  U_\vec{k} &= -\frac{\hbar^2}{2m}\gamma_1 |\mathbf{k}|^2&
  V_\vec{k} &= -\frac{\hbar^2}{2m}\gamma_2\left(k_x^2+k_y^2-2k_z^2\right)\nonumber\\
  R_\vec{k} &= \frac{\hbar^2}{2m}\sqrt{3}\left(\gamma_2\left(k_x^2-k_y^2\right) - 2\ii\gamma_3 k_xk_y\right)\label{eqn_ham_kp_matrix_elements}\\
  S^\pm_\vec{k} &= -\frac{\hbar^2}{2m}\sqrt{3}\left(k_\pm\{\gamma_3, k_z\}+ k_\pm[\kappa,k_z]\right)&
  \tilde{S}^\pm_\vec{k} &= -\frac{\hbar^2}{2m}\sqrt{3}\left(k_\pm\{\gamma_3, k_z\}-\frac{1}{3} k_\pm[\kappa,k_z]\right)\nonumber\\
  C_\vec{k}&=\frac{\hbar^2}{2m}2k_-[\kappa,k_z].\nonumber
\end{align}
where $|\mathbf{k}|^2 = k_x^2 + k_y^2 + k_z^2$ and $k_\pm = k_x \pm \ii k_y$.
The band energies $\Ec$, $\Ev$ and $\Ev-\Deltaso$, the Kane parameter $P$, and
the band structure parameters $F$, $\gamma_{1,2,3}$, $\kappa$ are material properties,
that contain contributions from couplings with remote bands.

\subsubsection{Axial symmetry}

Most of the matrix elements of Eq.~\eqref{eqn_ham_kp_matrix_elements} preserve
axial symmetry, i.e., they are invariant under a rotation around the $z$ axis.
The only exception is $R_\vec{k}$, which can be expanded into an axial and
nonaxial part, $R_\vec{k} = R_\vec{k}^\mathrm{ax} + R_\vec{k}^\mathrm{nonax}$ with
\begin{equation}\label{eqn_r_ax_r_noax}
  R_\vec{k}^\mathrm{ax}   = \frac{\hbar^2}{2m}\frac{\sqrt{3}}{2}(\gamma_2+\gamma_3)k_-^2,\qquad
  R_\vec{k}^\mathrm{nonax}= \frac{\hbar^2}{2m}\frac{\sqrt{3}}{2}(\gamma_2-\gamma_3)k_+^2.
\end{equation}
In some calculations, we use the \emph{axial approximation}, where we
approximate $R_\vec{k} \approx R_\vec{k}^\mathrm{ax}$ and neglect the non-axial
part $R_\vec{k}^\mathrm{nonax}$. We note that the non-axial contributions are
significant in many cases, so that the axial approximation should be used with
care. This issue will be discussed in more detail in the context of Landau levels
in Sec.~\ref{sec_ll_formalism}.

\subsubsection{Band structure parameters}

\begin{table*}
 \begin{tabular}{l|rrr|r|rrrrr}
  & \multicolumn{1}{c}{$\Ec$} & \multicolumn{1}{c}{$\Ev$} &  \multicolumn{1}{c|}{$\Ev-\Deltaso$} & \multicolumn{1}{c|}{$P$} & \multicolumn{1}{c}{$2F+1$}
  & \multicolumn{1}{c}{$\gamma_1$} & \multicolumn{1}{c}{$\gamma_2$} & \multicolumn{1}{c}{$\gamma_3$} & \multicolumn{1}{c}{$\kappa$} \\
  & \multicolumn{1}{c}{$\mathrm{meV}$} & \multicolumn{1}{c}{$\mathrm{meV}$} & \multicolumn{1}{c|}{$\mathrm{meV}$} & \multicolumn{1}{c|}{$\mathrm{meV}\,\mathrm{nm}$} & &
  & & &\\
  \hline
  HgTe    &  $-303$ &    $0$ & $-1080$ &   $846$ &     $1$ &
   $4.10$ &  $0.50$ & $1.30$ & $-0.40$ \\
  CdTe    &  $1036$ & $-570$ & $-1480$ &   $846$ &  $0.82$ & 
   $1.47$ & $-0.28$ & $0.03$ & $-1.31$
 \end{tabular}
 \caption{Coefficients of the inversion-symmetric \kdotp{} Hamiltonian $H_k$ at zero
 temperature. The values are adapted from Refs.~\cite{PfeufferJeschke2000_thesis,NovikEA2005}.
 We have taken into account the valence band offset $E_\mathrm{VBO}=-570\meV$
 for CdTe, taking $\Ev=0\meV$ for HgTe as the reference energy. These are also the
 default parameters implemented in $\kdotpy$.}
 \label{table_kdotpcoeff}
\end{table*}

For the simulations in \kdotpy, we use the material-dependent band structure parameters
established by Refs.~\cite{PfeufferJeschke2000_thesis,NovikEA2005}, summarized
in Table~\ref{table_kdotpcoeff}. Since \kdotpy{} has started out as a simulation
program for HgTe and CdTe, we use these materials as an example here, noting that
the modelling can be applied for a much wider variety of materials. We define
the band energy
of the $\Gamma_8$ orbitals of unstrained HgTe as the reference energy $E=0$.
Thus, $\Ev=0\meV$ for HgTe by definition. The valence band energy for CdTe is
determined by the valence band offset $E_\mathrm{VBO}=-570\meV$ of CdTe with
respect to HgTe. (All band energies, including the valence band offset, are
treated as a material parameters in this model.)

All parameters come with substantial error bars. Especially for the Luttinger
parameters $\gamma_{1,2,3}$ one finds substantial variations among various
literature sources \cite{Weiler1981bookchapter,FriedrichEA1994}.
In order to allow for adjustments of the material parameters,
we have equipped \kdotpy{} with an interface where the values can be changed in
the form of a configuration file (see Sec.~\ref{sec_matparam} and
Appendix~\ref{app_matparam_reference}). We have supplied the default
parameters as listed in Table~\ref{table_kdotpcoeff}.

For ternary compounds that are formed by alloying two binary compounds for
example Hg$_{1-x}$Cd$_x$Te from HgTe and CdTe, we determine the parameters by
suitable interpolation. These interpolations are linear to first approximation,
but higher-order corrections are typically also taken into consideration.
For example, the gap energy $\Eg = \Ec - \Ev$ for Hg$_{1-x}$Cd$_x$Te is
\begin{equation}
 \Eg(\text{Hg$_{1-x}$Cd$_x$Te}) = -303 (1 - x) + 1606 x - 132 x (1 - x)
\end{equation}
in units of meV, at zero temperature \cite{LaurentiEA1990,BeckerEA2000}. This
expression is a quadratic perturbation to the linear interpolation, with
the coefficient of the latter term being known as the \emph{bowing parameter}. The
Luttinger parameters $\gamma_{1,2,3}$ and $\kappa$ are approximated by cubic
polynomials, approximating the result of the interpolation scheme used in
Ref.~\cite{PfeufferJeschke2000_thesis}. In general, material parameters like
$\Eg$ are temperature-dependent \cite{LaurentiEA1990,BeckerEA2000}.

\subsubsection{Kane models with different number of orbitals}

The Kane model can also be formulated with a different number of orbitals. For
example, a simpler $6\times 6$ version can be used, where the $\Gamma_7$ orbitals
are omitted. Another version contains $14$ orbitals, which adds a quadruplet of
$\Gamma_8$ states and a doublet of $\Gamma_7$ states in the conduction band. We
will neither consider the $14$ orbital in this work nor have implemented it into
\kdotpy, because the six additional
orbitals add complexity that is not necessary for most purposes for which we use
\kdotpy. The so-called Luttinger model with only the four $\Gamma_8$
orbitals \cite{Luttinger1956} is also left out of consideration because it
cannot capture the essential physics of topological insulators, where $\Gamma_6$
and $\Gamma_8$ are inverted.

If one changes perspective between the $6$ and $8$ orbital Kane model, one needs
to take into account renormalizations of the coefficients. The reason for this
renormalization is that in the $6$ orbital model, the $\Gamma_7$ orbitals become
remote bands, and thus one must consider their additional perturbative contributions
$\vec{P}_{n'n}$ (cf. Eqs.~\eqref{eqn_kdotp_pert_e_all} and \eqref{eqn_kdotp_pert_u_all}).
This perturbation contributes to (for example) the
band masses of the $\Gamma_6$ and $\Gamma_8$ orbitals. In the $8$ orbital model,
the $\Gamma_7$ orbitals are treated explicitly and thus its perturbative
contribution on the other bands is absent. To ascertain that both models yield
the same dispersion, the band masses of the $\Gamma_6$ and $\Gamma_8$ orbitals
need to be renormalized. Similar renormalizations apply to other coefficients
and between different pairs of models. Details are provided, for example, in the
tables of Winkler \cite{Winkler2003_book}.

\subsection{Lower dimensional geometries}
\label{sec_lower_dim_geometries}

So far, we have considered the Hamiltonian in terms of the momentum coordinates
$(k_x, k_y, k_z)$. These momenta are good quantum numbers by virtue of Bloch's
theorem. For Bloch's theorem to be valid, the system must have (discrete)
translational symmetry in three dimensions. Thus, for a bulk crystal, this
description is appropriate.

However, experimentally relevant systems are of finite size; thus the
translational symmetry is broken. Nevertheless, if the dimensions of the system
are sufficiently large, the system may still be treated as approximately infinite.
The decisive criterion is the size of the system compared to the de Broglie
wavelength of the particles, or equivalently, the (quantum mechanical) confinement
energy compared to other energy scales in the system. 

Typically, we have a `hybrid' situation, where the system is infinite in some and
finite in other directions. 
For semiconductor devices simulated by \kdotpy, we typically consider `2D' and `1D'
geometries, where $n$D refers to the number $n$ of dimensions in which the system
has translational symmetry, or, in other words, the number of momentum coordinates.

A prominent example of a 2D geometry is a quantum well system.
In the 2D geometry, translational symmetry is broken in $z$ direction whereas it
is preserved in the $x$ and $y$ direction. The appropriate coordinates in this
geometry are thus $(k_x, k_y, z)$. This system may be thought of as an
infinite slab of material. More generically, any stack of layers of different
materials, called a `layer stack', is described as a 2D geometry. In this case,
the material
parameters (like $\Ec$, $\Ev$, $\gamma_{1,2,3}$, etc.) are treated as a function of
$z$ \footnote{We treat the Kane parameter $P$ on equal footing as the other
material parameters. In some works (see, e.g., Ref.~\cite{PfeufferJeschke2000_thesis}),
it is argued that $P$ must be equal in all layers, as a result of how it is defined
in terms of the atomic orbitals, cf.\ Eq.~\eqref{eqn_p_kane}. Since the model
yields physical result also if we relax this restriction, we allow $P$ to have
different values between layers.}.
The material parameters are effectively constant in the bulk of each layer
and transition smoothly between them at the interfaces between layers.
Many important topological aspects, like surface states, can exist only
at the interface between a topological and a trivial material and thus need to be
simulated in the 2D geometry.

The 1D geometry is equivalent to an infinite wire, with translational symmetry only
in $x$ direction and a finite size both in $y$ and $z$ direction, with coordinates
$(k_x, y, z)$. The extent in $y$ and $z$ direction must not necessarily be equal
or even similar in size. Typically, in order to simulate the edge states of the
quantum spin Hall effect, one considers a `ribbon' or `strip' of material, where
the thickness is $\sim10\nm$ and the width is $\sim500\nm$ or larger
\cite{ShamimEA2020SciAdv}.

\subsubsection{Dimensional reduction}
\label{sec_dimensional_reduction}

The reduction of the bulk Hamiltonian (3D) to a lower dimensionality is a two
step process for each of the confined dimensions.
Firstly, the momentum coordinate in the Hamiltonian is substituted by its
representation in spatial coordinates. For example, $k_z$ is substituted by the
derivative $-\ii \partial_z$ (where $\partial_z = \partial/\partial z$).
Likewise, $k_z^2$ is substituted by the second-order derivative $-\partial_z^2$.
These substitutions essentially encode an inverse Fourier transform.
Secondly, we choose a finite basis for each spatial direction. This step is
necessary, because the computational Hilbert space must be of finite dimension.
Here, we specifically choose the basis defined by a finite set of points at
which the wave functions are evaluated. We consider uniformly spaced grids of
the form $\{z_j\}_{j=j_\mathrm{min}, \ldots, j_\mathrm{max}}$,
with $z_j=j \Delta z$, where $\Delta z$ is the grid resolution and $j$ is an
integer index taken from the finite range $[j_\mathrm{min}, j_\mathrm{max}]$.
We note that our choice differs from the Fortran code of Refs.~\cite{PfeufferJeschke2000_thesis,NovikEA2005}, which uses a set of
envelope functions based on Legendre polynomials.

In the chosen basis, the discretization of the first derivatives is given by
\begin{equation}
  \partial_z \psi(z) = \lim_{dz\to0}\frac{\psi(z+dz)-\psi(z-dz)}{2 dz}
    \approx \frac{\psi(z+\Delta z) - \psi(z-\Delta z)}{2\Delta z}
    \to \partial_z \psi_j = \frac{\psi_{j+1} - \psi_{j-1}}{2\Delta z},
\end{equation}
where in the final step, we substitute $z=z_j$ and write $\psi_j=\psi(z_j)$ and
$\psi_{j\pm1} = \psi(z_{j\pm1}) = \psi(z_j \pm \Delta z)$. The second derivative
is obtained by applying the same principle twice, but substituting $dz \to \Delta z / 2$,
\begin{align}
  \partial_z^2 \psi(z) &= \lim_{dz\to0}\frac{\psi(z+2dz)-2\psi(z)+\psi(z-2dz)}{4 dz^2}\nonumber\\
    &\approx \frac{\psi(z+\Delta z) - 2\psi(z) + \psi(z-\Delta z)}{\Delta z^2}
    \to  \partial_z^2 \psi_j =
    \frac{\psi_{j+1} - 2\psi_j + \psi_{j-1}}{\Delta z^2}.
\end{align}
The inverse-Fourier and discretization steps combined lead to the substitution rules
\begin{align}
  k_z \psi &\to \hat{k}_z\psi_j =\frac{-\ii}{2\Delta z}(\psi_{j+1} - \psi_{j-1}),\\
  k_z^2 \psi &\to \hat{k}_z^2 \psi_j = \frac{-1}{\Delta z^2}(\psi_{j+1} - 2\psi_j + \psi_{j-1}),
\end{align}
where we use the notation $\hat{k}_z$ to emphasize that this object is an operator.
The substitution rules for $k_y$ and $k_y^2$ are analogous to those for $k_z$
and $k_z^2$, respectively.

\subsubsection{Hermitian discretization}

In a layered system (2D geometry), where the material parameters are functions
of $z$, special care needs to be taken that these functions generally do not commute with
$\hat{k}_z=-\ii\partial_z$. For this reason, we find anticommutators and commutators
of the form $\{Q,k_z\}$ and $[Q, k_z]$, respectively, in the off-diagonal matrix
elements of the Hamiltonian, where $Q$ denotes a generic $z$-dependent material
parameter.

The operator $\hat{k}_z$ being a derivative, this leads to contributions
involving $\partial_z Q$. To clarify this statement, it is useful to expand the
matrix element $\bramidket{\phi}{\{Q, \hat{k}_z\}}{\psi}$ in terms of its spatial
representation as an integral over $z$,
\begin{align}
  \bramidket{\phi}{\{Q, \hat{k}_z\}}{\psi}
  &= -\ii\int dz\, \phi^*(z)\,(Q(z)\partial_z+\partial_z Q(z))\,\psi(z)\nonumber\\
  &= -\ii\int dz\, (2\phi^*(z)Q(z)\psi'(z) + \psi^*(z)Q'(z)\psi(z)),\label{eqn_anticomm_matrix_element}
\end{align}
where a prime denotes the $z$ derivative. In the same representation, invoking
integration by parts and assuming that the states vanish at the boundaries
of the integration domain, one can also write
\begin{equation}\label{eqn_anticomm_matrix_element2}
  \bramidket{\phi}{\{Q, \hat{k}_z\}}{\psi}
  = -\ii\int dz\, \bigl(\phi^*(z)Q(z)\partial_z\psi(z)+ \phi^*(z) \partial_z (Q(z)\psi(z))\bigr)
  = -\ii\bigl(\bramidket{\phi}{Q}{\partial_z\psi} - \bramidket{\partial_z\phi}{Q}{\psi}\bigr).
\end{equation}
For the matrix element $\bramidket{\phi}{[Q, \hat{k}_z]}{\psi}$ involving a
commutator, we find that
\begin{equation}\label{eqn_comm_matrix_element}
  \bramidket{\phi}{[Q, \hat{k}_z]}{\psi}
  = -\ii\int dz\, \phi^*(z)\,(Q(z)\partial_z-\partial_z Q(z))\,\psi(z)
  = -\ii\int dz\, \psi^*(z)Q'(z)\psi(z)
  = \bramidket{\phi}{Q'}{\psi}.
\end{equation}
In other words, the commutator $[Q, \hat{k}_z]$ only contributes where
$Q'(z)\equiv \partial_z Q(z)$ is nonzero,
which is only near the interfaces between layers, as the material parameters are 
constant in the bulk of each layer. The commutator terms can thus be interpreted
as interface terms.

On the diagonal of the Hamiltonian, the matrix elements $T_\mathbf{k}$ and
$U_\mathbf{k}$ contain effective-mass terms, quadratic in the momentum $k_z$.
The correct substitution to operator form is given by a symmetrized triple
product, in general
\begin{equation}\label{eqn_symmetrized_triple_product1}
  qk_ik_j \to \{k_i q k_j\}_\mathrm{S}
  \equiv \tfrac{1}{2}(\hat{k}_i \hat{q} \hat{k}_j + \hat{k}_j \hat{q} \hat{k}_i),
\end{equation}
where $i,j=x,y,z$ and $\hat{q}$ is a hermitian operator
\cite{MorrowBrownstein1984,EinevollEA1990,PfeufferJeschke2000_thesis}.
For the relevant diagonal matrix elements with effective mass terms of the form
$Qk_z^2$, the substitution rule reads $Qk_z^2 \to \hat{k}_z Q(z) \hat{k}_z$.
In the spatial representation, these terms can be written as
\begin{equation}\label{eqn_symmetrized_triple_product2}
  \bramidket{\phi}{\hat{k}_z Q \hat{k}_z}{\psi}
  = -\int dz\,\phi^*(z)\partial_z (Q(z)\partial_z\psi(z))
  = \int dz\,\phi^{\prime*}(z)Q(z)\psi'(z)
  = \bramidket{\phi'}{Q}{\psi'},
\end{equation}
where $\psi'(z)\equiv \partial_z\psi(z)$.

The first-order terms are discretized as follows. We take the form of Eq.~\eqref{eqn_anticomm_matrix_element2} and substitute
$\psi(z)\to(\psi(z+\frac{1}{2}\Delta z)+\psi(z-\frac{1}{2}\Delta z))/2$ and
$\partial_z\psi(z)\to(\psi(z+\frac{1}{2}\Delta z)-\psi(z-\frac{1}{2}\Delta z))/\Delta z$ and analogous rules for $\phi^*(z)$ and $\partial_z\phi^*(z)$.
We thus obtain
\begin{align}
 &\frac{-\ii}{2\Delta z}\sum_z\left[
  \left(\phi^*(z+\tfrac{1}{2}\Delta z)+\phi^*(z-\tfrac{1}{2}\Delta z)\right) Q(z)
  \left(\psi(z+\tfrac{1}{2}\Delta z)-\psi(z-\tfrac{1}{2}\Delta z)\right)\right.\nonumber\\
  &\qquad\qquad-\left.\left(\phi^*(z+\tfrac{1}{2}\Delta z)-\phi^*(z-\tfrac{1}{2}\Delta z)\right) Q(z)
  \left(\psi(z+\tfrac{1}{2}\Delta z)+\psi(z-\tfrac{1}{2}\Delta z)\right)\right]
\end{align}
By expanding and erasing the terms that cancel, we find
\begin{equation}
 \frac{-\ii}{\Delta z}\sum_z\left(
  \phi^*(z-\tfrac{1}{2}\Delta z) Q(z) \psi(z+\tfrac{1}{2}\Delta z)-
  \phi^*(z+\tfrac{1}{2}\Delta z) Q(z) \psi(z-\tfrac{1}{2}\Delta z)
  \right)
\end{equation}
In our computational basis, the Hilbert space is defined as the wave functions on
the discrete coordinates $z_j$, so we must align $z\pm\frac{1}{2}\Delta z$ with
these values. This implies that $Q(z)$ is evaluated at intermediate points
$z_j \pm \frac{1}{2}\Delta z$, which is not a problem since $Q$ is a function
which we can evaluate anywhere. The resulting expression can be written in three
equivalent forms, related by shifts of the `dummy variable' $j$,
\begin{align}
 \bramidket{\phi}{\{Q,\hat{k}_z\}}{\psi}
 &\to\frac{-\ii}{\Delta z}\sum_j\left(
   \phi^*_{j}\,Q(z_j+\tfrac{1}{2}\Delta z)\,\psi_{j+1}-
   \phi^*_{j}\,Q(z_j-\tfrac{1}{2}\Delta z)\,\psi_{j-1}
   \right)\label{eqn_discrete_first_derivative1}\\
 &=\frac{-\ii}{\Delta z}\sum_j Q(z_j-\tfrac{1}{2}\Delta z)\left(
   \phi^*_{j-1}\,\psi_{j}-\phi^*_{j}\,\psi_{j-1}
   \right)\label{eqn_discrete_first_derivative2}\\
 &=\frac{-\ii}{\Delta z}\sum_jQ(z_j+\tfrac{1}{2}\Delta z)\left(
   \phi^*_{j}\,\psi_{j+1}-\phi^*_{j+1}\,\psi_{j}
   \right)\label{eqn_discrete_first_derivative3}.
\end{align}
We use the first form [Eq.~\eqref{eqn_discrete_first_derivative1}] to extract
the action of the Hamiltonian matrix, $(H\psi)_i = \sum_j H_{ij} \psi_j$.
From the second and third forms [Eqs.~\eqref{eqn_discrete_first_derivative2} and \eqref{eqn_discrete_first_derivative3}], we find that the resulting Hamiltonian
matrix is hermitian.

For the matrix elements of commutator form $[Q, \hat{k}_z]$, we simply
substitute the derivative $Q'$ into Eq.~\eqref{eqn_comm_matrix_element},
\begin{equation}
  \bramidket{\phi}{[Q, \hat{k}_z]}{\psi}
  \to \sum_j\phi^*_j Q'(z_j) \psi_j,
\end{equation}
where $Q'(z_j)$ is the derivative of $Q$ evaluated at the grid point $z_j$. In
$\kdotpy$, we use the discrete derivative
$Q'(z_j) = (Q'(z_j+\Delta z) - Q'(z_j-\Delta z))/2\Delta z$.

The discretization of the quadratic terms follows from
Eq.~\eqref{eqn_symmetrized_triple_product2}, where we substitute
$\psi'(z)=\partial_z\psi(z)\to(\psi(z+\frac{1}{2}\Delta z) - \psi(z-\frac{1}{2}\Delta z))/\Delta z$. We thus obtain
\begin{equation}
 -\frac{1}{\Delta z^2}\sum_z 
  \left(\phi^*(z+\tfrac{1}{2}\Delta z)-\phi^*(z-\tfrac{1}{2}\Delta z)\right) Q(z)
  \left(\psi(z+\tfrac{1}{2}\Delta z)-\psi(z-\tfrac{1}{2}\Delta z)\right).
\end{equation}
We expand and align $z\pm\tfrac{1}{2}\Delta z$ with the discrete coordinates $z_j$,
and obtain
\begin{align}
\bramidket{\phi}{\hat{k}_z Q\hat{k}_z}{\psi}
 &\to\frac{1}{\Delta z^2}\sum_j\left(
   \phi^*_{j}\,Q(z_j+\tfrac{1}{2}\Delta z)\,(\psi_{j+1}-\psi_{j})+
   \phi^*_{j}\,Q(z_j-\tfrac{1}{2}\Delta z)\,(\psi_{j-1}-\psi_{j})
   \right)\label{eqn_discrete_second_derivative1}\\
 &=-\frac{1}{\Delta z^2}\sum_j
   (\phi^*_{j}-\phi^*_{j-1})\,Q(z_j-\tfrac{1}{2}\Delta z)\,(\psi_{j}-\psi_{j-1})
   \label{eqn_discrete_second_derivative2}\\
 &=-\frac{1}{\Delta z^2}\sum_j
    (\phi^*_{j+1}-\phi^*_{j})\,Q(z_j+\tfrac{1}{2}\Delta z)\,(\psi_{j+1}-\psi_{j}).
   \label{eqn_discrete_second_derivative3}
\end{align}
The first form [Eq.~\eqref{eqn_discrete_second_derivative1}] again defines
the action of the Hamiltonian matrix $(H\psi)_i = \sum_j H_{ij} \psi_j$ and
the second and third forms [Eqs.~\eqref{eqn_discrete_second_derivative2}
and \eqref{eqn_discrete_second_derivative3}] show that the expression is
hermitian if $\psi=\phi$.

\subsection{Magnetic fields}

\subsubsection{Peierls substitution}
\label{sec_peierls_subst}

In (classical) Hamiltonian mechanics, the motion of a charged particle with
dispersion $p^2/2m$ and charge $-e$ in a magnetic field $\vec{B}$ is given by
\begin{equation}
 H = \frac{1}{2m}\left(\vec{p} + e\vec{A}\right)^2,
\end{equation}
in terms of the magnetic gauge field $\vec{A}$ that satisfies
$\vec{B} = \nabla\times \vec{A}$. In this expression, $\vec{p}$ acts as the
canonical momentum, while $\vec{\Pi} = \vec{p} + e\vec{A}$ is the kinetic
momentum, equal to mass times velocity. In order to obtain the Hamiltonian for
the motion of a particle in a magnetic field from a generic zero-field
Hamiltonian, one applies the Peierls substitution
\begin{equation}\label{eqn_peierls_substitution}
  \vec{p} \to \vec{\Pi} = \vec{p} + e \vec{A},
\end{equation}
or equivalently, $\vec{k} \to \vec{k} + (e/\hbar) \vec{A}$.

The electromagnetic gauge field $\vec{A}$ is subject to gauge invariance by the
transformation $\vec{A}\to\vec{A}+\nabla \Lambda$ where $\Lambda$ is a function
depending on the spatial coordinates. This gauge transformation does not alter
the relation $\vec{B} = \nabla \times \vec{A}$. (Here, and in the remainder of
the work, we assume that $\vec{B}$ and $\vec{A}$ are not time-dependent.)
This leaves us with a freedom to choose the gauge conveniently. For example,
for a perpendicular magnetic field $\vec{B} = (0,0,B_z)$, two common gauge
choices are the symmetric gauge $\vec{A}=B_z(-y/2,x/2,0)$ and the Lorentz gauge
$\vec{A}=B_z(-y,0,0)$. For the numerical simulations, one should notice that
the symmetric gauge breaks translational invariance in $x$ and $y$ direction,
while the Lorentz gauge breaks it in $y$ direction only. Thus, for simulations
of a device in a magnetic field, \kdotpy{} uses the Lorentz gauge so that the
translational symmetry in $x$ direction is left intact. This simulation is done
in the strip geometry, i.e., the 1D geometry with momentum coordinates $k_x$ and
spatial coordinates $y$ and $z$.

For in-plane fields $\vec{B}=(B_x,B_y,0)$, on similar grounds we choose a gauge
that depends only on the $z$ coordinate, leaving translational symmetries in $x$
and $y$ directions intact (if they are not yet broken by some other means). The
gauge that satisfies this property is $\vec{A}=(B_y z, -B_x z, 0)$. For generic
magnetic fields $\vec{B}=(B_x,B_y,B_z)$ we simply take the sum of in-plane and
out-of-plane, and use the gauge
\begin{equation}\label{eqn_gauge}
  \vec{A} = (B_y z - B_z y, -B_x z, 0).
\end{equation}
The appropriate geometry for the simulations depends on the out-of-plane
component: If $B_z\not=0$, a 1D geometry is needed, while for a purely in-plane
field, the 2D geometry is typically the appropriate one.

In quantum mechanics, the momentum and gauge fields discussed above must be
replaced by the appropriate operators. Importantly, the momentum $\hat{k}$ and
gauge field $\hat{A}$ operators (which is a function of spatial coordinates) do
not commute in general. We quantize the quadratic terms in a symmetric way
\begin{align}
  \left(k_i+\frac{e}{\hbar}A_i\right)\left(k_j+\frac{e}{\hbar}A_j\right)
  &\to \hat{k}_i\hat{k}_j + \frac{e}{\hbar}(\hat{k}_i A_j + \hat{k}_j A_i)
   + \left(\frac{e}{\hbar}\right)^2 A_i A_j\\
  &=-\partial_i\partial_j
    -\ii\frac{e}{\hbar}\left(\partial_i A_j + \partial_j A_i + A_i \partial_j + A_j \partial_i\right)
    +\left(\frac{e}{\hbar}\right)^2 A_i A_j
\end{align}
with $i,j = x,y,z$, because this form assures that the Hamiltonian is hermitian
\cite{Sakurai1994_book}. The Hamiltonian is gauge invariant under the gauge
transformation $\vec{A} \to \vec{A}+\nabla \Lambda$ and
$\ket{\psi} \to \exp(-\ii e \Lambda / \hbar)\ket{\psi}$, where $\Lambda$ is a function of
the spatial coordinates \cite{Sakurai1994_book}.

Importantly, the kinetic momentum operators do not commute,
\begin{equation}
  \left[k_i+\frac{e}{\hbar}A_i, k_j+\frac{e}{\hbar}A_j\right]
  = -\ii\frac{e}{\hbar}(\partial_i A_j - \partial_j A_i)
  = -\ii\frac{e}{\hbar} \sum_k\epsilon_{ijk} B_k,
\end{equation}
which raises the question how terms of the form $(k_i+(e/\hbar)A_i)(k_j+(e/\hbar)A_j)$
should be quantized. Obviously, if $i=j$, there is no ambiguity. For the
off-diagonal terms, we assume the following quantization rules
\begin{align}
  k_\pm^2
  &\to (\hat{k}'_x)^2 - (\hat{k}'_y)^2 \pm \ii (\hat{k}'_x\hat{k}'_y+\hat{k}'_y\hat{k}'_x)\nonumber\\
  & = -\partial_x^2 +\partial_y^2\mp2\ii\partial_x\partial_y
      +2\frac{e}{\hbar}(A_x \pm \ii A_y)(-\ii\partial_x \pm\partial_y)
      \mp\frac{e}{\hbar}B_z+\left(\frac{e}{\hbar}\right)^2(A_x\pm \ii A_y)^2
      \label{eqn_kpm2_quantized}\\
  k_\pm k_z
  &\to \tfrac{1}{2}\{\hat{k}'_x \pm \ii \hat{k}'_y,\hat{k}'_z\}\nonumber\\
  &= -\ii (\partial_x \pm \ii\partial_y)\partial_z
    -\ii\frac{e}{\hbar}(A_x \pm \ii A_y)\partial_z \mp \frac{e}{2\hbar}(B_x\pm \ii B_y)
    \label{eqn_kpmkz_quantized}
\end{align}
where we have defined $\hat{k}'_i = \hat{k} + (e/\hbar)A_i$ and we have assumed
the gauge defined by Eq.~\eqref{eqn_gauge}. The terms $eB_z/\hbar$ and $e(B_x\pm \ii B_y)/2\hbar$ in Eqs.~\eqref{eqn_kpm2_quantized} and \eqref{eqn_kpmkz_quantized}, respectively, can be interpreted as originating from the non-commutative character of the kinetic momentum operators. For the sake of doing the numerics, we further substitute
$A_x\pm\ii A_y= (B_y\mp\ii B_x)z - B_z y = \mp\ii (B_x\pm \ii B_y)z - B_z y$.

\subsubsection{Out-of-plane magnetic field in strip geometry}

The case of a pure out-of-plane field, $\vec{B} = (0,0,B_z)$, can be treated in
the strip geometry, i.e., the 1D case with coordinates $(k_x, y, z)$. We choose
the gauge $\vec{A}=(A_x,0,0)$ with $A_x = -B_z (y-y_0)$, where the
\emph{gauge origin} $y_0$ can be freely chosen.

In \kdotpy{}, we choose $y_0=0$ to be the centre of the strip, so that $A_x$ is
antisymmetric under reflection in $y$. The momentum coordinate is simply shifted
according to the Peierls substitution, $k_x \to k_x - (e/\hbar)B_z y$. The
operators $\hat{k}_\pm^2$ are calculated along similar lines as
Eq.~\eqref{eqn_kpm2_quantized}, but with $\hat{k}'_x=k_x - (eB_z/\hbar) y$,
\begin{equation}
 k_\pm^2 \to \left(k_x - \frac{eB_z}{\hbar} y\right)^2
   \pm 2\ii\left(k_x - \frac{eB_z}{\hbar} y\right)(-\ii\partial_y) + \partial_y^2 \mp \frac{eB_z}{\hbar}.
\end{equation}
We describe its action in the Hilbert space defined by the discrete coordinates
$(y_i,z_j)$ on a grid with resolutions $(\Delta y, \Delta z)$ as the matrix element
\begin{align}
\bramidket{\phi}{\hat{k}_\pm^2}{\psi}
  = &\sum_{i,j}\phi^*_{i,j}\biggl[\left(k_x - \frac{eB_z}{\hbar} y\right)^2 \psi_{i,j}
   \pm 2\left(k_x - \frac{eB_z}{\hbar} y\right)
      \left(\frac{\psi_{i+1,j}-\psi_{i-1,j}}{2\Delta y}\right)\biggr.\nonumber\\
   &\hspace{10em}{}\biggl.-\frac{\psi_{i+1,j}-2\psi_{i,j}+\psi_{i-1,j}}{(\Delta y)^2}
   \mp \frac{eB_z}{2\hbar}\left(\psi_{i+1,j}+\psi_{i-1,j}\right)\biggr]\label{eqn_kpm2_discrete1}\\
  = &\sum_{i,j}\biggl[\phi^*_{i,j}\left(k_x - \frac{eB_z}{\hbar} y\right)^2 
   \mp 2\frac{\phi^*_{i+1,j}-\phi^*_{i-1,j}}{2\Delta y}\left(k_x - \frac{eB_z}{\hbar} y\right)\biggr.
      \nonumber\\
   &\hspace{10em}{}\biggl.-\frac{\phi^*_{i+1,j}-2\phi^*_{i,j}+\phi^*_{i-1,j}}{(\Delta y)^2}
   \pm \frac{eB_z}{2\hbar}\left(\phi^*_{i+1,j}+\phi^*_{i-1,j}\right)\biggr]\psi_{i,j}\label{eqn_kpm2_discrete2}
\end{align}
where $\psi_{i,j}\equiv \psi(y_i,z_j)$. Equation~\eqref{eqn_kpm2_discrete1} is
implemented as matrix element in \kdotpy. Equation~\eqref{eqn_kpm2_discrete2} is
the conjugate form obtained from Eq.~\eqref{eqn_kpm2_discrete1} by shifting the
dummy variables $i\to i\pm 1$ and $y\to y\pm\Delta y$. Note that the variable
shift yields a contribution
$(eB_z/\hbar)(\phi^*_{i+1,j}+\phi^*_{i-1,j})\psi_{i,j}$ from the second term,
which has the same structure as the final term. For this reason, we use the
non-diagonal form $\mp(eB_z/2\hbar)\phi^*_{i,j}(\psi_{i+1,j}+\psi_{i-1,j})$ for
the matrix element of the final term, and not the diagonal form
$\mp(eB_z/\hbar)\phi^*_{i,j}\psi_{i,j}$ as one could have expected naively.

\subsubsection{In-plane magnetic field in slab and strip geometries}

Let us consider a pure in-plane field, $\vec{B}=(B_x,B_y,0)$. The gauge choice
is $\vec{A}=(A_x,A_y,0)$ with $A_x = B_y (z-z_0)$ and $A_y = -B_x (z-z_0)$.
For \kdotpy{} we choose $z_0=0$, i.e., the centre of the layer stack. 

In the slab (2D) geometry, $k_\pm^2$ simply evaluates as
$(k'_x)^2 - (k'_y)^2 \pm 2\ii k'_x k'_y$ with $k'_x = k_x +(eB_y/\hbar) z$
and $k'_y = k_y - (eB_x/\hbar) z$. With $k'_\pm = k'_x\pm \ii k'_y$ and
$B_\pm = B_x \pm \ii B_y$, we find $k'_\pm = k_\pm \mp \ii (e B_\pm/\hbar) z$.
For the $S^\pm_\vec{k}$ matrix elements in the Hamiltonian,
we consider $k'_\pm \{\gamma_3, \hat{k}_z\}$. Combining Eqs.~\eqref{eqn_discrete_first_derivative1} and \eqref{eqn_kpmkz_quantized}, we
find the matrix element
\begin{align}
 \bramidket{\phi}{k'_\pm \{\gamma_3, \hat{k}_z\}}{\psi}
  = &\sum_{j}\phi^*_{j}\biggl[
    -\frac{\ii}{\Delta z}\left(k_\pm\mp\ii\frac{eB_\pm}{\hbar}z\right)
    \left(\gamma_3(z+\tfrac{1}{2}\Delta z)\psi_{j+1}-\gamma_3(z-\tfrac{1}{2}\Delta z)\psi_{j-1}\right)
    \biggr.\nonumber\\
    &\hspace{10em}{}\biggl.\mp \frac{eB_\pm}{2\hbar}
    \left(\gamma_3(z+\tfrac{1}{2}\Delta z)\psi_{j+1}+\gamma_3(z-\tfrac{1}{2}\Delta z)\psi_{j-1}\right)\biggr]\label{eqn_kpmkz_discrete1}\\
  = &\sum_{j}\biggl[
    \frac{\ii}{\Delta z}\left(k_\pm\mp\ii\frac{eB_\pm}{\hbar}z\right)
    \left(\phi^*_{j+1}\gamma_3(z+\tfrac{1}{2}\Delta z)-\phi^*_{j-1}\gamma_3(z-\tfrac{1}{2}\Delta z)\right)
    \biggr.\nonumber\\
    &\hspace{10em}{}\biggl.\pm \frac{eB_\pm}{2\hbar}
    \phi^*_{j+1}\left(\gamma_3(z+\tfrac{1}{2}\Delta z)+\phi^*_{j-1}\gamma_3(z-\tfrac{1}{2}\Delta z)\right)\biggr]\psi_j\label{eqn_kpmkz_discrete2}
\end{align}
where $k'_\pm = k_\pm \mp \ii (eB_\pm/\hbar) z$ and $\psi_j\equiv\psi(z_j)$.
Again, these two expressions are related by a shift in variables $j\to j\pm 1$
and $z\to z\pm\Delta z$.

In the strip geometry, we obtain $\bramidket{\phi}{k'_\pm \{\gamma_3, \hat{k}_z\}}{\psi}$
by simply substituting $k_y \to \hat{k}_y$, where
$\hat{k}_y\psi_{i,j} = -\ii(\psi_{i+1,j}-\psi_{i-1,j})/\Delta y$. The bracketed
parts in Eqs.~\eqref{eqn_kpmkz_discrete1} and \eqref{eqn_kpmkz_discrete2} do not
contain $y$-dependent terms other than the wave functions $\phi$ and $\psi$;
$\hat{k}_y$ commutes with the other factors.

\subsubsection{Generic magnetic fields}

For a generic magnetic field $\vec{B} = (B_x,B_y,B_z)$, we can straightforwardly
adapt the equations above. For the $k_\pm^2$ terms, we simply make the substitutions
$\hat{k}'_x=k_x-(eB_z/\hbar)y+(eB_y/\hbar)z$ and $\hat{k}'_y = -\ii\partial_y - (eB_x/\hbar)z$,
adding the in-plane component of the field to the gauge compared to
Eq.~\eqref{eqn_kpm2_quantized}. The derivative $\partial_y$ commutes with the
terms proportional with $z$, hence no extra terms of the form $eB_x/\hbar$ or
$eB_y/\hbar$ (constant in space) appear.
Likewise, for $k'_\pm \{\gamma_3,\hat{k}_z\}$, the out-of-plane component
appears as an extra term $(eB_z/\hbar)y$ in $k_x+(e/\hbar)A_x$ compared to
Eqs.~\eqref{eqn_kpmkz_discrete1} and \eqref{eqn_kpmkz_discrete2}. This term
commutes with $\partial_z$.

\subsection{Landau level formalism}
\label{sec_ll_formalism}

\subsubsection{Out-of-plane field in the axial approximation}

If the magnetic field is purely out-of-plane, we have the commutator relation
$[\hat{k}'_-,\hat{k}'_+] = 2eB_z/\hbar$ while $\hat{k}'_z$ commutes with both
$\hat{k}'_x$ and $\hat{k}'_y$. This structure is formally equivalent to the
ladder operators of the harmonic oscillator with commutation relation
$[a,a^\dagger] = 1$, if we define
\begin{equation}\label{eqn_ll_ladder_operators}
  a = \sqrt{\frac{\hbar}{2eB_z}}\hat{k}'_- = \frac{1}{\sqrt{2}}\lB \hat{k}'_-,
  \qquad
  a^\dagger = \sqrt{\frac{\hbar}{2eB_z}}k'_+ = \frac{1}{\sqrt{2}}\lB \hat{k}'_+,
\end{equation}
where $\lB=\sqrt{\hbar/eB_z}$ is the magnetic length. (Here, we have tacitly
assumed $B_z>0$.) The eigenstates of the number operator $a^\dagger a$ are
denoted $\ket{n}$, where $n$ is the eigenvalue, i.e., $a^\dagger a\ket{n} = n\ket{n}$.
The raising and lowering operator act as $a^\dagger \ket{n} = \sqrt{n+1}\ket{n+1}$
and $a\ket{n}=\sqrt{n-1}\ket{n-1}$, respectively.

In this context, these eigenstates are called Landau level states (see, e.g.,
Refs. \cite{PfeufferJeschke2000_thesis,NovikEA2005}). The Hamiltonian can be
reformulated in terms of ladder operators by substituting $k_+$ and $k_-$ by
$a^\dagger$ and $a$, respectively. The combination
$k_x^2+k_y^2 = \tfrac{1}{2}(k_+k_-+k_-k_+)$ is substituted by a term proportional
to $a^\dagger a + a a^\dagger = 2n+1$. 

It can be shown \cite{PfeufferJeschke2000_thesis,NovikEA2005} that in the axial
approximation, the eigenstates of the Hamiltonian can be written in the form
\begin{equation}\label{eqn_ll_eigenstate}
  \ket{\Psi^{(n)}} = \begin{pmatrix}
    f^{(n)}_1(z)\ket{n}\\
    f^{(n)}_2(z)\ket{n+1}\\
    f^{(n)}_3(z)\ket{n-1}\\
    f^{(n)}_4(z)\ket{n}\\
    f^{(n)}_5(z)\ket{n+1}\\
    f^{(n)}_6(z)\ket{n+2}\\
    f^{(n)}_7(z)\ket{n}\\
    f^{(n)}_8(z)\ket{n+1}
  \end{pmatrix},
\end{equation}
where $n=-2,-1,0,1,\ldots$ is called the Landau level index (LL index). For
$n=-2,-1,0$, some of the indices $n'$ in $\ket{n'}$ on the right-hand side are
negative; these components are understood to be zero.
For $n\geq 1$, the dimension of the subspace spanned by
$\ket{\Psi^{(n)}}$ is $8 n_z$, where $n_z$ is the number of degrees of freedom
in the $z$ direction. (For discrete coordinates $z_j$, as in the \kdotpy{} calculation,
$n_z$ is equal to the number of grid points.) For $n=-2,-1,0$, the dimensionality
of the subspace is reduced to $1 n_z$, $4 n_z$, and $7 n_z$, respectively, where
$1$, $4$, and $7$ refer to the number of nonzero components in
Eq.~\eqref{eqn_ll_eigenstate}. This observation is important for the
implementation, as we will discuss in detail later.

The Landau level index $n$ is a conserved quantity in axial approximation
[where we neglect $R^\mathrm{nonax}_\mathbf{k}$, Eq.~\eqref{eqn_r_ax_r_noax}],
meaning that the Hamiltonian is block-diagonal in the basis
$\{\ket{\Psi^{(n)}}\}_{n=-2,-1,0,1,\ldots}$, i.e.,\linebreak[4]
$\bramidket{\Phi^{(n')}}{H^\mathrm{ax}}{\Psi^{(n)}} = 0$ if $n'\ne n$. This
property allows to calculate Landau level spectra separately for each Landau
level $n=-2,-1,0,\ldots,n_\mathrm{max}$, where $n_\mathrm{max}$ is the desired
maximal Landau level index. The result is
a set of magnetic-field dependent energy eigenvalues $E^{(n)}_j(B_z)$ for each
Landau index $n$, where $j$ runs over all eigenstates within each Landau level.
Due to the fact that the basis is finite, by choosing a Landau level cutoff
$n_\mathrm{max}$, the spectrum is inherently incomplete. This is typically a
problem only at small magnetic fields, where the energy spacing between Landau
levels is small. This energy spacing is equivalent to the cyclotron energy
$\hbar\omega_c=\hbar e B_z/m$ known from the theory of the quantum Hall effect.

\subsubsection{Landau level formalism with axial symmetry breaking}

If the nonaxial term $R_\vec{k}^\mathrm{nonax}$ [Eq.~\eqref{eqn_r_ax_r_noax}]
is added, the Landau level index is no longer a conserved quantum number.
In order to demonstrate this, let us define
\begin{equation}
  H^\mathrm{nonax} =
   R_\vec{k}^\mathrm{nonax}
    (\ket{\mathbf{3}}\bra{\mathbf{5}} + \ket{\mathbf{4}}\bra{\mathbf{6}}) +
   R_\vec{k}^{\mathrm{nonax}\dagger}
    (\ket{\mathbf{5}}\bra{\mathbf{3}} + \ket{\mathbf{6}}\bra{\mathbf{4}})
\end{equation}
to be the nonaxial part of the Hamiltonian, expressed in the orbital basis of
Eq.~\eqref{eqn_orbital_basis} (with bold-face numbers indicating the orbitals).
In the ladder operator formalism, 
$R_\vec{k}^\mathrm{nonax} \sim a^\dagger a^\dagger$, so that
\begin{equation}\label{eqn_ll_eigenstate_noax}
 H^\mathrm{nonax}\ket{\Psi^{(n)}} \sim
  [(\ket{\mathbf{3}}\bra{\mathbf{5}} + \ket{\mathbf{4}}\bra{\mathbf{6}})a^\dagger a^\dagger +
  (\ket{\mathbf{5}}\bra{\mathbf{3}} + \ket{\mathbf{6}}\bra{\mathbf{4}})a a]\ket{\Psi^{(n)}}
  \sim \begin{pmatrix}
    0\\
    0\\
    f^{(n)}_5(z)\sqrt{n+2}\sqrt{n+3}\ket{n+3}\\
    f^{(n)}_6(z)\sqrt{n+2}\sqrt{n+3}\ket{n+4}\\
    f^{(n)}_3(z)\sqrt{n-1}\sqrt{n-2}\ket{n-3}\\
    f^{(n)}_4(z)\sqrt{n}\sqrt{n-1}\ket{n-2}\\
    0\\
    0
  \end{pmatrix},
\end{equation}
where $\sim$ indicates that we have suppressed the prefactors in the notation.
Examining Eqs.~\eqref{eqn_ll_eigenstate} and \eqref{eqn_ll_eigenstate_noax},
we observe that $\bramidket{\Phi^{(n')}}{H^\mathrm{nonax}}{\Psi^{(n)}}$
is generally nonzero if $n'=n\pm 4$ and zero otherwise. Thus, the nonaxial terms
couple Landau levels with indices differing by $4$.
In the basis of $\ket{\Psi^{(n)}}$, the total Hamiltonian (axial and nonaxial
terms) has matrix elements between the $n$ and $n'$ blocks
for $n' -n = -4, 0, 4$. We note that the Landau index modulo $4$ remains conserved,
because $\bramidket{\Phi^{(n')}}{H^\mathrm{nonax}}{\Psi^{(n)}} = 0$
if $n'-n$ is not divisable by $4$.

The block off-diagonal ($n'\ne n$) nonaxial terms are
typically much weaker than dominant axial terms on the block diagonal ($n'=n$),
so that the former can be viewed as perturbation to the latter. For this reason,
basis of Landau level states is still useful even if the Landau index is not
conserved. In typical Landau level spectra, the Landau index is often almost conserved,
unless two levels of indices $n$, $n'$ with $n'-n=\pm4$ come close in energy: In
the latter case, hybridization between the states occurs and the spectrum
exhibits an anticrossing.

In \kdotpy, we always use the Landau level basis in the Landau level mode
(\texttt{kdotpy ll}). In the axial approximation where the Hamiltonian matrix
is block diagonal, the diagonalization can thus be performed block by block,
which gives a performance bonus in view of the smaller matrix size.
If nonaxial terms are considered, the Hamiltonian is written as one large matrix
with all $\ket{\Psi^{(n)}}$ with $n=-2,-1,0,\ldots,n_\mathrm{max}$. This
Hamiltonian is diagonalized as a whole. In \kdotpy, we do not perform perturbation
theory explicitly, but always diagonalize the full matrix. Due to the upper
limit $n_\mathrm{max}$, the levels with $n=n_\mathrm{max} - 3,n_\mathrm{max} - 2,
n_\mathrm{max} - 1, n_\mathrm{max}$ are not coupled to their counterpart with $n'=n+4$;
as a result, there is a slight inaccuracy in the energies of the highest levels.
This numerical error can often be estimated by raising $n_\mathrm{max}$ and
analyzing how well the energies have converged to definite value.

\subsection{Other magnetic couplings}

\subsubsection{Zeeman effect}

The electrons are subject to additional magnetic coupling, for example the
Zeeman effect $H_\mathrm{Z}=g\muB \vec{B}\cdot\vec{S}$, where $g$ is the
gyromagnetic ratio or ``$g$-factor'' ($g\approx 2$ in vacuum), $\muB$ is the Bohr magneton,
and $\vec{S}$ is the vector of spin operators $(\hat{S}_x,\hat{S}_y,\hat{S}_z)$.
The nonzero blocks (cf.\ Eq.~\eqref{eqn_ham_blocks}) of the Zeeman Hamiltonian
$H_Z$ are
\begin{align}
   H_\mathrm{Z,66} &= g_\mathrm{e}\muB \begin{pmatrix}
        \tfrac{1}{2}B_z&\tfrac{1}{2}B_-\\
        \tfrac{1}{2}B_+&-\tfrac{1}{2}B_z 
   \end{pmatrix},
   &H_\mathrm{Z,88} &= 2\kappa\muB \begin{pmatrix}
       -\tfrac{3}{2}B_z & -\tfrac{1}{2}\sqrt{3} B_- & 0 & 0 \\
        -\tfrac{1}{2}\sqrt{3} B_+ & -\tfrac{1}{2}B_z & -B_- & 0\\
       0 & -B_+ & \tfrac{1}{2}B_z & -\tfrac{1}{2}\sqrt{3} B_-\\
       0 & 0 & -\tfrac{1}{2}\sqrt{3} B_+ & \tfrac{3}{2}B_z
   \end{pmatrix}\nonumber\\
   H_\mathrm{Z,77} &= 2(\kappa+\tfrac{1}{2})\muB \begin{pmatrix}
        -\tfrac{1}{2}B_z&-\tfrac{1}{2}B_-\\
        -\tfrac{1}{2}B_+&\tfrac{1}{2}B_z 
   \end{pmatrix},
   &H_\mathrm{Z,87} &= 2(\kappa+1)\muB \begin{pmatrix}
       \sqrt{3/8}B_- & 0\\
       -\sqrt{1/2}B_z & \sqrt{1/8}B_-\\
       -\sqrt{1/8}B_+ & -\sqrt{1/2}B_z\\
       0 & -\sqrt{3/8}B_+
   \end{pmatrix}\label{eqn_hzeeman}
\end{align}
where $g_e$ is the $g$-factor for the $\Gamma_6$ block and $-2\kappa$ acts as
the $g$-factor of the combined $\Gamma_8,\Gamma_7$ block \cite{PfeufferJeschke2000_thesis,NovikEA2005,Winkler2003_book}.

\subsubsection{Exchange coupling in Mn-doped materials}
\label{sec_exchange_mn}

\begin{table*}
 \begin{tabular}{l|rr|rrrr}
  & \multicolumn{1}{c}{$g_e$} & \multicolumn{1}{c|}{$\kappa$} & \multicolumn{1}{c}{$N_0 \alpha$} & \multicolumn{1}{c}{$N_0 \beta$} & \multicolumn{1}{c}{$g_\mathrm{Mn}$} & \multicolumn{1}{c}{$T_0$} \\
  & & & \multicolumn{1}{c}{$\mathrm{meV}$} & \multicolumn{1}{c}{$\mathrm{meV}$} 
  & & \multicolumn{1}{c}{$\mathrm{K}$}\\
  \hline
  HgTe               &  $2$ & $-0.4\phantom{0}$ & -- & -- & -- & -- \\
  CdTe               &  $2$ & $-1.31$           & -- & -- & -- & -- \\
  Hg$_{1-y}$Mn$_y$Te &  $2$ & $-0.4\phantom{0}$ & $400$ & $-600$ & $2$ & $2.6$ 
 \end{tabular}
 \caption{Coefficients of the magnetic couplings (Zeeman and paramagnetic exchange).
 The values of $\kappa$ also appear in Table~\ref{table_kdotpcoeff}.
 The values relevant for the paramagnetic exchange in Hg$_{1-y}$Mn$_y$Te are
 based on Ref.~\cite{NovikEA2005}. The value for $T_0$ is appropriate only
 for $y\sim 0.02$. (However, \kdotpy{} implements it as this constant value, to
 remain consistent with past models.)
 }
 \label{table_zeeman_exchange_coeff}
\end{table*}

In the dilute magnetic semiconductor Hg$_{1-y}$Mn$_{y}$Te, the Mn atoms carry a
finite magnetic moment, such that the material behaves paramagnetically if the
Mn content $y$ does not exceed a few percent \cite{Furdyna1988}.
The coupling between the Mn magnetic moments and the carrier spins has a similar
matrix structure as the Zeeman effect [Eq.~\eqref{eqn_hzeeman}], but the
response to the external magnetic field $\vec{B}$ is nonlinear. This coupling
is modelled by the (paramagnetic) exchange
Hamiltonian \cite{Furdyna1988,NovikEA2005,ShamimEA2020SciAdv},
\begin{equation}
 H_\mathrm{ex} = \sum_l C^{(l)} \langle\vec{m}\rangle \cdot \vec{S}^{(l)},
\end{equation}
where $l$ labels the blocks, (the $\Gamma_6$ block and the $\Gamma_8,\Gamma_7$
block), $\langle\vec{m}\rangle$ is the average Mn
spin and $\vec{S}^{(l)}$ the
spin operators of the respective block. The coupling constants $C^{(l)}$
(cf. $g_e$ and $2\kappa$  in Eq.~\eqref{eqn_hzeeman}) are 
phenomenologically determined material parameters. For Hg$_{1-y}$Mn$_{y}$Te,
we assume that they are proportional to the Mn content $y$,
\begin{equation}
  C^{\Gamma_6} = -y N_0 \alpha
  \qquad\text{and}\qquad
  C^{\Gamma_8,\Gamma_7} = -y N_0 \beta,
\end{equation}
with $N_0 \alpha=400\meV$ and $N_0 \beta=-600\meV$ \cite{Furdyna1988,NovikEA2005}.

The paramagnetic response of the Mn magnetic moments to the external field is
modelled by the empirical law \cite{Furdyna1988,LiuEA2008PRL101,NovikEA2005}
\begin{equation}\label{eqn_magnetization_mn}
  \langle\vec{m}\rangle = -S_0 \frac{\vec{B}}{\lvert \vec{B}\rvert}B_{5/2}\left(
    \frac{\tfrac{5}{2}g_\mathrm{Mn}\muB \lvert \vec{B}\rvert}{\kB (T+T_0)}
  \right),
\end{equation}
where $B_{5/2}$ is the Brillouin function for spin $J=\tfrac{5}{2}$,
\begin{equation}\label{eqn_brillouin_def}
 B_J(x) = \frac{2J+1}{2J}\coth\left(\frac{2J+1}{2J}x\right)
  - \frac{1}{2J}\coth\left(\frac{1}{2J}x\right).
\end{equation}
The effective total spin $S_0=-\frac{5}{2}$ and the temperature offset $T_0$ are
material parameters. For historical reasons, we have used the value
$T_0=2.6\,\mathrm{K}$, but recent experiments have proved that $T_0$ depends
on the Mn content $y$ and is typically smaller than this value, especially for
smaller $y$ \cite{MandalEA2024}.
In \kdotpy, the values $g_\mathrm{Mn}$ and $T_0$ are treated as
material parameter and need not be constant in the Mn content, though the default
material definitions for Hg$_{1-y}$Mn$_{y}$Te contain the $y$-independent values
for historical reasons.

\subsection{Bulk inversion asymmetry}
\label{sec_bia}

The terms in Hamiltonian $H_\mathrm{Kane}$ [Eq.~\eqref{eqn_hkane}] are symmetric under
spatial inversion, i.e., the transformation given by
$(x,y,z)\to(-x,-y,-z)$ and $(k_x,k_y,k_z)\to(-k_x,-k_y,-k_z)$.
For $H_\mathrm{Kane}$, the relevant point group is $O_h$.
However, the zincblende crystal structure is not symmetric under inversion; it
has point group $T_d$. This means that terms breaking spatial inversion symmetry
are allowed to appear in the Hamiltonian. Indeed, \emph{bulk-inversion asymmetry}
(BIA) is known to affect the band structure, although the inversion symmetric
terms remain dominant. In other words, bulk-inversion asymmetry can be treated
as a perturbation to the inversion symmetric Hamiltonian $H_\mathrm{Kane}$.

From representation theory of the point group $T_d$, it can be derived which
BIA terms are permitted. Like before, we consider only those terms up to quadratic
order in momentum. There is a single independent term linear in momentum, given by
the Hamiltonian blocks \cite{Winkler2003_book}
\begin{equation}
   H_{\mathrm{BIA},88} = C \begin{pmatrix}
        0 & -\frac{1}{2} k_+ & k_z & -\frac{1}{2}\sqrt{3} k_-\\
        -\frac{1}{2} k_- & 0 & \frac{1}{2}\sqrt{3} k_+ & -k_z\\
        k_z & \frac{1}{2}\sqrt{3} k_- & 0 & -\frac{1}{2} k_+\\
        -\frac{1}{2}\sqrt{3} k_+ & -k_z & -\frac{1}{2} k_- & 0
   \end{pmatrix},\qquad
   H_{\mathrm{BIA},87} = \frac{1}{2\sqrt{2}} C \begin{pmatrix}
        k_+ & 2 k_z \\
        0 & -\sqrt{3} k_+\\
        \sqrt{3} k_- & 0\\
        2 k_z & -k_-
   \end{pmatrix},
   \label{eqn_hbia_linear}
\end{equation}
and $H_\mathrm{BIA,78}=H_\mathrm{BIA,87}^\dagger$.
The coefficient $C$ is the material parameter for the strength of linear BIA.
There are three independent quadratic terms, given by
\begin{align}
   H_{\mathrm{BIA},68} &= \frac{1}{\sqrt{6}}B_{8+} \begin{pmatrix}
        \sqrt{3}k_-k_z& 2 \ii k_x k_y & k_+ k_z & 0\\
        0 & k_- k_z & 2 \ii k_x k_y & \sqrt{3} k_+ k_z
   \end{pmatrix} +
   \frac{1}{3\sqrt{2}}B_{8-} \begin{pmatrix}
        0 & \sqrt{3} K_4 & 0 & K_5\\
        -K_5 & 0 & -\sqrt{3} K_4 & 0
   \end{pmatrix}\nonumber\\
      H_{\mathrm{BIA},67} &= \frac{1}{\sqrt{3}}B_7 \begin{pmatrix}
        -\ii k_xk_y & -k_+ k_z\\
        k_- k_z & \ii k_xk_y 
   \end{pmatrix},
   \label{eqn_hbia_quadratic}
\end{align}
where $K_4\equiv k_x^2 - k_y^2$ and $K_5 \equiv k_x^2 + k_y^2 - 2 k_z^2$, and
the coefficients $B_{8+}$, $B_{8-}$, and $B_7$ are material parameters. The blocks
$H_{\mathrm{BIA},76}$ and $H_{\mathrm{BIA},86}$ are given by the respective hermitian
conjugates. The blocks $H_{66}$ and $H_{77}$ do not contain BIA terms up to
quadratic order.
\begin{table*}
 \begin{tabular}{l|r|rrr}
  & \multicolumn{1}{c|}{$C$} & \multicolumn{1}{c}{$B_{8+}$} &  \multicolumn{1}{c}{$B_{8-}$} & \multicolumn{1}{c}{$B_7$} \\
  & \multicolumn{1}{c|}{$\mathrm{meV}\,\mathrm{nm}$} & \multicolumn{1}{c}{$\mathrm{meV}\,\mathrm{nm}^2$} & \multicolumn{1}{c}{$\mathrm{meV}\,\mathrm{nm}^2$} & \multicolumn{1}{c}{$\mathrm{meV}\,\mathrm{nm}^2$} \\
  \hline
  HgTe    &  $-7.4\phantom{0}$ & $-106.46$ & $-13.77\phantom{0}$ & $-100\phantom{.0}$ \\
  CdTe    &  $-2.34$ & $-224.1\phantom{0}$ & $-6.347$ &   $-204.7$ 
 \end{tabular}
 \caption{Coefficients of the BIA Hamiltonian $H_\mathrm{BIA}$ up to quadratic
 order in momentum. The values for CdTe are taken from Ref.~\cite{Winkler2003_book},
 which in turn cites Ref.~\cite{CardonaEA1988} for $C$.
 The values for $C$, $B_{8+}$ and $B_{8-}$ for HgTe are based on the result of
 a calculation by Di~Sante and Sangiovanni based on density-functional theory (DFT)
 \cite{DiSanteSanGiovanni_DFT}. The value
 for $B_7$ for HgTe is an educated guess approximately equal to $B_{7,\mathrm{CdTe}}B_{8+,\mathrm{HgTe}}/B_{8+,\mathrm{CdTe}}$. The values in this table are included
 as material parameters in \kdotpy.
 }
 \label{table_bia_coeff}
\end{table*}
Representative values of the material parameters for CdTe
\cite{Winkler2003_book,CardonaEA1988} and HgTe \cite{DiSanteSanGiovanni_DFT}
are listed in Table~\ref{table_bia_coeff}. These values are also included as
material parameters in \kdotpy.

For nonzero magnetic fields, the BIA terms also lead to extra contributions
from the Peierls substitution. For a strip geometry, this leads to extra
contributions involving the gauge field $\vec{A}$ (see Sec.~\ref{sec_peierls_subst}).
In the Landau level formalism (see Sec.~\ref{sec_ll_formalism}), additional
ladder operators appear. For growth direction (001), this leads to terms
coupling Landau levels with indices $n,n'$ with $n'-n=\pm 2$. Whereas for
the inversion symmetric LL Hamiltonian $n\mod 4$ is a conserved quantum number,
the BIA lowers this symmetry such that only $n\mod 2$ remains conserved.

\subsection{Strain}
\label{sec_strain}

\subsubsection{Strain and stress}
\label{sec_strain_stress}

The materials used in a heterostructure typically have slight
differences in their equilibrium lattice constant. Two adjacent 
epitaxially grown layers can be strained as to match their in-plane
lattice constants, if the mismatch between the equilibrium lattice constants is
sufficiently small. This is known as pseudomorphic growth
(see Ref.~\cite{Leubner2016_thesis} and references therein).
In practice, the in-plane lattice constant is generally
determined by that of the substrate, which is usually much thicker than the
epitaxial layers on top. If the epitaxial layers are too thick (more than a few 
hundred nm for HgTe and CdTe), relaxation occurs towards the equilibrium
lattice constant.

The physical quantity \emph{strain} is the deviation of the lattice
constant of the strained layer $a_\mathrm{s}$ relative to the equilibrium lattice constant
$a_0$; in one dimension, strain is the dimensionless quantity 
\begin{equation}\label{eqn_strain_1d}
  \epsilon = \frac{a_s - a_0}{a_0}.
\end{equation}
In a three-dimensional crystal, strain leads to a displacement
of each of the three
lattice vectors $\vec{a}_i$ ($i=x,y,z$ in a crystal with cubic symmetry). 
The displacement can be written in terms of a tensor $A_{ij}$, which satisfies
\cite{DeCaroTapfer1995part1}
\begin{equation}
  \vec{a}_{i,s} = \sum_{j=1}^3 (\delta_{ij} + A_{ij})\vec{a}_{j,0},
\end{equation}
where $\delta_{ij}$ is the Kronecker delta. The non-rotational (i.e., symmetrical)
part of $A_{ij}$ is the strain tensor
\begin{equation}
  \epsilon_{ij} = \frac{1}{2}(A_{ij} + A_{ji}). 
\end{equation}
The strain tensor $\epsilon_{ij}$ is symmetric by definition. The diagonal terms
$\epsilon_{ii}$ represent \emph{linear strain} and the off-diagonal terms
$\epsilon_{xy}, \epsilon_{yz}, \epsilon_{zx}$ represent \emph{shear strain}.

A strained crystal is not in equilibrium and thus experiences forces that pushes
the crystal back to its equilibrium lattice constant, or conversely, external
forces are needed to bring a crystal into a strained state. Here, we consider
the linear response regime, where Hooke's law is valid, i.e., the force $F$
and the displacement $u$ are linearly proportional, $F=k u$, with the stiffness
constant $k$. The generalization of Hooke's law to a crystal in three dimensions
relates the \emph{stress} tensor $\sigma_{ij}$ to the strain tensor $\epsilon_{kl}$
as
\begin{equation}\label{eqn_stiffness_tensor}
  \sigma_{ij} = \sum_{k,l=1}^3 S_{ijkl} \epsilon{kl},
\end{equation}
in terms of the stiffness tensor $C_{ijkl}$. The stiffness tensor is a rank-4
tensor with $81=3\times3\times3\times3$ components (also known as elasticity
modules). The stress tensor and
stiffness tensor (elasticity modules) both carry
the units of pressure, typically $\mathrm{GPa}$ in this context.

For a crystal with cubic symmetry, the stiffness tensor has only $3$ independent
components \cite{DeCaroTapfer1995part1}. If the strain and stress tensors
are vectorized as
$\tilde{\epsilon} = (\epsilon_{xx},\epsilon_{yy},\epsilon_{zz},2\epsilon_{yz},2\epsilon_{zx},2\epsilon_{xy})$ and\linebreak[4]
$\tilde{\sigma} = (\sigma_{xx},\sigma_{yy},\sigma_{zz},\sigma_{yz},\sigma_{zx},\sigma_{xy})$,
respectively, Eq.~\eqref{eqn_stiffness_tensor} can be written as
$\tilde{\sigma} = \tilde{C}\tilde{\epsilon}$
in terms of the elasticity matrix
\begin{equation}
 \tilde{C} = \begin{pmatrix}
   C_{11} & C_{12} & C_{12} & 0 & 0 & 0 \\
   C_{12} & C_{11} & C_{12} & 0 & 0 & 0 \\
   C_{12} & C_{12} & C_{11} & 0 & 0 & 0 \\
   0 & 0 & 0 & C_{44} & 0 & 0 \\
   0 & 0 & 0 & 0 & C_{44} & 0 \\
   0 & 0 & 0 & 0 & 0 & C_{44}
  \end{pmatrix}.
\end{equation}
The three independent components $C_{11}$, $C_{12}$, and $C_{44}$ are material
parameters that can be found in the literature.

\subsubsection{Strain Hamiltonian}

\begin{table*}
 \begin{tabular}{l|r|rrr}
  & \multicolumn{1}{c|}{$C_1$} & \multicolumn{1}{c}{$D_d$} &  \multicolumn{1}{c}{$D_u$} & \multicolumn{1}{c}{$D_u'$} \\
  & \multicolumn{1}{c|}{$10^3\,\mathrm{meV}$} & \multicolumn{1}{c}{$10^3\,\mathrm{meV}$} & \multicolumn{1}{c}{$10^3\,\mathrm{meV}$} & \multicolumn{1}{c}{$10^3\,\mathrm{meV}$} \\
  \hline
  HgTe    &  $-3.83$ & $0\phantom{.0}$ & $2.25\phantom{0}$ & $\tfrac{1}{2}\sqrt{3}\times 2.08$\\
  CdTe    &  $-4.06$ & $-0.7$ & $1.755$ & $\tfrac{1}{2}\sqrt{3}\times 3.2\phantom{0}$
 \end{tabular}
 \caption{Deformation potentials for HgTe and CdTe, see
 Refs.~\cite{PfeufferJeschke2000_thesis} and references therein.
 Note that some references use the alternative notation
 $C = C_1$, $a=D_d$, $b=-\frac{2}{3}D_u$ and $d=-\frac{2}{\sqrt{3}}D_u'$
 \cite{PfeufferJeschke2000_thesis,BirPikus1974_book}. 
 The values in this table are included as material parameters in \kdotpy.
 }
 \label{table_strain_coeff}
\end{table*}

The effect of strain on the band structure is formalized in the Bir-Pikus
formalism \cite{BirPikus1974_book}: The strain Hamiltonian is obtained from
the \kdotp{} Hamiltonian by substitution of all terms quadratic in momentum as
\begin{equation}\label{eqn_bir_pikus_substitution}
  k_i k_j \to \epsilon_{ij}
\end{equation}
and changing the band mass parameters $F$, $\gamma_{1,2,3}$ to the
\emph{deformation potentials} (i.e., strain coefficients) $C_1$, $D_d$, $D_u$, and $D_u'$
\cite{PfeufferJeschke2000_thesis,NovikEA2005,Winkler2003_book}.
Taking Eq.~\eqref{eqn_ham_kp} as a starting point, we find the strain Hamiltonian
\begin{equation}\label{eqn_ham_strain1}
  H_\mathrm{s} = \begin{pmatrix}
        T_\mathrm{s} & 0 & 0 & 0 & 0 & 0 & 0 & 0\\
        0 & T_\mathrm{s} & 0 & 0 & 0 & 0 & 0 & 0\\
        0 & 0 & U_\mathrm{s}+V_\mathrm{s}&S_\mathrm{s}&R_\mathrm{s}&0&-\frac{1}{\sqrt{2}}S_\mathrm{s}&-\sqrt{2}R_\mathrm{s}\\
        0 & 0 & S^*_\mathrm{s} & U_\mathrm{s}-V_\mathrm{s}& 0 &R_\mathrm{s}&\sqrt{2}V_\mathrm{s}&\sqrt{\frac{3}{2}}S_\mathrm{s}\\
        0 & 0 & R^*_\mathrm{s} & 0 &U_\mathrm{s}-V_\mathrm{s}&-S_\mathrm{s}&\sqrt{\frac{3}{2}}S^*_\mathrm{s}&-\sqrt{2}V_\mathrm{s}\\
        0 & 0 & 0 &R_\mathrm{s}^* & -S^*_\mathrm{s} & U_\mathrm{s}+V_\mathrm{s} & \sqrt{2}R^*_\mathrm{s}&-\frac{1}{\sqrt{2}}S^*_\mathrm{s}\\
        0 & 0 & -\frac{1}{\sqrt{2}}S^*_\mathrm{s} & \sqrt{2}V_\mathrm{s} & \sqrt{\frac{3}{2}}S_\mathrm{s} & \sqrt{2}R_\mathrm{s} & U_\mathrm{s} & 0\\
        0 & 0 & -\sqrt{2}R_\mathrm{s}^* & \sqrt{\frac{3}{2}}S^*_\mathrm{s} & -\sqrt{2}V_\mathrm{s}&-\frac{1}{\sqrt{2}}S_\mathrm{s} & 0 & U_\mathrm{s}
        \end{pmatrix}
\end{equation}
with
\begin{align}
  T_\mathrm{s} &= C_1 \tr \epsilon\nonumber\\
  U_\mathrm{s} &= D_d \tr \epsilon,&
  R_\mathrm{s} &= \frac{1}{\sqrt{3}}D_u (\epsilon_{xx} - \epsilon_{yy})
                    -\frac{2}{\sqrt{3}}D_u'\ii \epsilon_{xy},
  \label{eqn_ham_strain2}\\
  V_\mathrm{s} &= -\frac{1}{3}D_u (\epsilon_{xx} +\epsilon_{yy} -2 \epsilon_{zz}),&
  S_\mathrm{s} &= \frac{2}{\sqrt{3}}D_u'(\epsilon_{xz} - \ii\epsilon_{yz}),\nonumber
\end{align}
where $\tr \epsilon = \epsilon_{xx} + \epsilon_{yy} + \epsilon_{zz}$.
The values of the deformation potentials $C_1$, $D_d$, $D_u$, and $D_u'$ for
HgTe and CdTe that have been provided with \kdotpy{}, are listed in
Table~\ref{table_strain_coeff} and have been taken from
Ref.~\cite{PfeufferJeschke2000_thesis}. We note that this and other literature
sources (e.g., Ref.~\cite{BirPikus1974_book}) use the alternative set of deformation
potentials $C = C_1$, $a=D_d$, $b=-\frac{2}{3}D_u$ and $d=-\frac{2}{\sqrt{3}}D_u'$.

Symmetry allows further terms in the strain Hamiltonian. For example, application
of the Bir-Pikus substitution to cubic momentum terms leads to strain terms 
linear in strain and momentum, i.e., involving linear combinations of
$\epsilon_{ij}k_l$. (See the tables in Ref.~\cite{Winkler2003_book} for 
examples of these terms.) Here, we neglect these `higher-order' terms, as they
are negligible for these materials \cite{PfeufferJeschke2000_thesis}. For a
systematic study of these higher-order strain terms, we refer to
Ref.~\cite{KetkarEA2024_inprep}.

\subsection{Topology}

\subsubsection{Berry connection, Berry curvature, Chern number}
\label{sec_berry}

The topological character of a material is typically understood theoretically in
terms of the Chern number of the bands. In a nutshell,  band inversion can cause
the Chern numbers to become non-trivial, which leads to measurable signatures in
the Hall conductance.

A common picture to understand the concept of a Chern number is the analogy with
a winding number of the spin of a spin-$\frac{1}{2}$ particle \cite{BernevigHughes2013book}.
The spin eigenstate $\ket{\psi(\vec{k})}$ can be represented by a momentum
dependent vector $\vec{d}(\vec{k})$ on the Bloch sphere. Then one can consider
the area swept out by $\vec{d}(\vec{k})$ where the momenta $\vec{k}$ span the
complete Brillouin zone. The signed area is $4\pi C$, i.e., the area $4\pi$
of the unit sphere times the Chern number $C$. If the Brillouin zone is finite,
the Chern number is integer, whereas for the continuum limit of infinitely large
Brillouin zones, it may also be half-integer.

The above picture can be formalized in terms of concepts from differential geometry.
If we consider an eigenstate $\ket{\psi(\vec{k})}$ that is differentiable in the
Brillouin zone, we can define the Berry connection \cite{BernevigHughes2013book}
\begin{equation}\label{eqn_berryconnection}
  \vec{A}^{(\psi)}(\vec{k}) = \ii\bramidket{\psi(\vec{k})}{\nabla_{\vec{k}}}{\psi(\vec{k})}
\end{equation}
at each point $\vec{k}$. This quantity is analogous
to the magnetic vector potential, hence it also goes by the name Berry vector
potential \cite{BernevigHughes2013book}. The Berry phase is an integral of the
Berry curvature over the closed contour $\mathcal{C}$ that surrounds the
Brillouin zone,
\begin{equation}\label{eqn_berryphase1}
 \gamma^{(\psi)} = \int_\mathcal{C}d\vec{k}\cdot \vec{A}^{(\psi)}(\vec{k}).
\end{equation}
The Berry phase is an integer number times $2\pi$, which defines the
Chern number $C^{(\psi)}$ as $\gamma^{(\psi)} =2\pi C^{(\psi)}$ \cite{BernevigHughes2013book}.

The integral of Eq.~\eqref{eqn_berryphase1} can be rewritten by virtue of
Stokes' theorem as an integral over the interior $\mathcal{S}$ of the Brillouin zone,
\begin{equation}\label{eqn_berryphase2}
 \gamma^{(\psi)} = \int_\mathcal{S} d\vec{S} \cdot \nabla_{\vec{k}} \times \vec{A}^{(\psi)}(\vec{k})
 \equiv \int_\mathcal{S} d\vec{S} \cdot \Omega^{(\psi)}(\vec{k}),
\end{equation}
where $d\vec{S}$ is a surface element and $\Omega^{(\psi)}(\vec{k})$ is the
Berry curvature, defined by
$\Omega^{(\psi)}(\vec{k}) = \nabla_{\vec{k}} \times \vec{A}^{(\psi)}$, or in
component notation ($i, l, m = 1,2,3$),
\begin{equation}\label{eqn_berrycurvature1}
  \Omega^{(\psi)}_i(\vec{k}) = \sum_{l,m}\epsilon_{ilm}F^{(\psi)}_{lm}(\vec{k})\\
\end{equation}
with
\begin{equation}\label{eqn_berrycurvature2}
  F^{(\psi)}_{lm}(\vec{k})= \partial_{k_l} \vec{A}^{(\psi)}_m(\vec{k}) - \partial_{k_m} \vec{A}^{(\psi)}_l(\vec{k})
  = \ii\braket{\partial_{k_l}\psi(\vec{k})}{\partial_{k_m}\psi(\vec{k})} - \ii\braket{\partial_{k_m}\psi(\vec{k})}{\partial_{k_l}\psi(\vec{k})}.
\end{equation}
The analogue of the Berry curvature in electrodynamics is the magnetic field $\vec{B}=\nabla\times\vec{A}$.

In practice, the Berry connection is challenging to calculate numerically: The
derivative is discretized, where we evaluate the eigenstates at $\vec{k}$ and
$\vec{k}+d\vec{k}$. However, the result of the numerical evaluation has an
indefinite overall phase factor ($U(1)$ gauge factor) which can vary randomly
even between two points $\vec{k}$ and $\vec{k}+d\vec{k}$ close in momentum space.
In Ref.~\cite{BernevigHughes2013book}, the Berry curvature $\Omega^{(\psi)}_i(\vec{k})$
[Eqs.~\eqref{eqn_berrycurvature1} and \eqref{eqn_berrycurvature2}] is rewritten
by `insertion of one', $\sum_{\psi'} \ket{\psi'}\bra{\psi'}$. This manipulation
eventually yields a gauge invariant form of the Berry curvature
\begin{equation}\label{eqn_berrycurvature3}
  \Omega^{(\psi)}_i(\vec{k}) 
  = -\Im \sum_{\psi'\not=\psi} \sum_{l,m}
    \frac{
     \epsilon_{ilm}\bramidket{\psi'(\vec{k})}{(\partial_{k_l}H)}{\psi(\vec{k})} \bramidket{\psi(\vec{k})}{(\partial_{k_m}H)}{\psi'(\vec{k})}
    }{
     (E_{\psi'}-E_\psi)^2
    },
\end{equation}
where the summation is over all states in the spectrum other than $\ket{\psi}$
itself, and $E_\psi$ and $E_{\psi'}$ are the energy eigenvalues. This expression
is manifestly $U(1)$ invariant, as all phase factors from
the eigenstates come in conjugate pairs. The momentum derivatives of the
Hamiltonian $\partial_{k_l}H$ ($l=1,2,3$) can be evaluated analytically or
numerically; even for the numerical derivative, there are no indefinite phase
factors to be taken care of.

\subsubsection{Hall conductance}
\label{sec_chern_hall}

The bulk-boundary theorem (see, e.g., Ref.~\cite{BernevigHughes2013book})
connects the Chern numbers to the Hall
conductance $\sigmaH = \sigma_{xy}$ of a device in a magnetic field, as
\begin{equation}\label{eqn_bulk_boundary}
  \sigmaH = \frac{e^2}{h}\sum_{\text{occupied states }\psi} C^{(\psi)}.
\end{equation}
That is, the sum of the Chern numbers of the occupied bands yields the Hall
conductance in units of the conductance quantum $e^2/h$. This result can be used
to find the Hall conductance $\sigmaH$ in a Landau fan: For each additional
occupied Landau level, the Hall conductance is increased by $C^{(\psi)}e^2/h$.
Typically, $C^{(\psi)} = 1$ for all Landau levels, so that finding the Hall
conductance simplifies to a mere counting problem of occupied states.
In \kdotpy{}, the Hall conductance can be determined either with Chern numbers
calculated with Eq.~\eqref{eqn_berrycurvature3} or with simulated Chern numbers
based on the assumption that $C^{(\psi)} = 1$. The implementation is discussed
in more detail in Sec.~\ref{sec_berrychern_implementation}.

\clearpage
\section{Implementation}
\label{sec_implementation}

\subsection{Overview of the program}

\subsubsection{General remarks}

Like its name suggests, \kdotpy{} implements \kdotp{} theory in a Python program.
The program is designed as a command-line interface (CLI): The operation of the
program is determined primarily by a sequence of command line arguments entered
by the user on the shell (e.g., bash). Moreover, the behaviour and output can be
adjusted with configuration settings.

\subsubsection{Package structure}
\label{sec_package_structure}

The \kdotpy{} application is actually a collection of several subprograms, that
do the actual work. Most of these subprograms are Python scripts themselves.
The subprograms are referred to by the first two arguments on the command line,
for example \texttt{kdotpy 2d}. The command \texttt{kdotpy} invokes the main script \texttt{main.py} (provided that the \texttt{kdotpy} module has been installed
successfully with PIP, see Section~\ref{sec_installation}). In this example,
\texttt{main.py} imports \texttt{kdotpy-2d.py} and runs its \texttt{main()}
function. The latter contains the `recipe' of this subprogram, which we will
discuss in greater detail in Sec.~\ref{sec_workflow}.

The \kdotpy{} module defines five subprograms that we classify as \emph{calculation
subprograms}, because they do the actual computational work:
constructing a Hamiltonian, diagonalizing it, and processing the eigenvalues and
eigenstates.

\begin{itemize}
\item \texttt{kdotpy 2d}:    Calculates a dispersion of a structure with two
                  translational degrees of freedom. The $z$ direction is kept
                  spatial and is appropriately discretized.
                  A typical configuration for this case is a quantum well.

\item \texttt{kdotpy 1d}:    Calculates a dispersion of a structure with one
                  translational degree of freedom, for example a Hall bar. The
                  $y$ and $z$ direction are kept spatial and are discretized. Due
                  to the very large size of the problem, it usually makes
                  sense to run this subprogram on a cluster that supports `jobs'
                  with large memory requirements.

\item \texttt{kdotpy ll}:    Calculates a Landau level spectrum for a configuration
                  similar to the `2D' mode, using the Landau level formalism
                  of Sec.~\ref{sec_ll_formalism}.

\item \texttt{kdotpy bulk}:  Calculates the dispersion for the `bulk', i.e., with
                  translational degrees of freedom in all $3$ directions.

\item \texttt{kdotpy bulk-ll}: Calculates a Landau level spectrum for a geometry
 with translational symmetry in the $z$ direction.

\end{itemize}
                       
The package also contains several auxiliary subprograms for several other tasks
other than doing the actual calculations:
\begin{itemize}
\item \texttt{kdotpy merge}:
This subprogram is used for re-plotting data from earlier runs of \kdotpy.
It takes as input one or more XML files. The name \texttt{kdotpy merge} derives
from the fact that it can be used to merge several data files, and show all
results in a single plot.

\item \texttt{kdotpy compare}:
Similar to \texttt{kdotpy merge}, but shows data
files (or sets of data files) using different colours/symbols in a single plot.

\item \texttt{kdotpy batch}:
A simple tool for running batch calculations with \kdotpy{}, for example
iterating a calculation while varying one of the input parameters. This
subprogram serves a similar purpose as a shell script, but has a
convenient monitor that shows the estimated time at which
the calculations will finish.

\item \texttt{kdotpy test}: Test suite. This tool runs all standardized tests
or a selection of them, as to test whether \kdotpy{} runs properly.
                    
\item \texttt{kdotpy config}: For viewing and modifying the configuration settings.

\item \texttt{kdotpy help}: Show the built-in help file (with command line and
configuration option reference) in a terminal viewer. With \texttt{kdotpy help \emph{item}},
can search for a specific \texttt{\emph{item}} in the help file. 

\item \texttt{kdotpy doc}: With \texttt{kdotpy doc \emph{object}}, one gets
developer information on the given \texttt{\emph{object}}, e.g., a function or class.
This tool shows the relevant docstring on the terminal.

\item \texttt{kdotpy version}: Show the version of \texttt{kdotpy}.
                    
\end{itemize}
The final three are not separate Python scripts, but functions called directly
from the main program. For detailed instructions on how to use \kdotpy{} from
the command line, we refer to Sec.~\ref{sec_usage}.

The Python code is structured as follows. The root directory of
contains the metadata, such as the \texttt{pyproject.toml} file, the
\texttt{README.md} file and the license text. The code itself is found in the 
subdirectory \texttt{src/kdotpy}. This directory contains the main module
(in \texttt{main.py}), the scripts for the subprograms, and several other
source files that are imported by the subprograms. Inside \texttt{src/kdotpy},
the following subdirectories collect several larger components of the program:
\begin{itemize}
 \item \texttt{bandalign}: For band alignment, see Sec.~\ref{sec_bandalign}.
 \item \texttt{cmdargs}: Functions that parse the command line arguments, see
 Sec.~\ref{sec_cmdargs}.
 \item \texttt{density}: Functions for calculating carrier density, density of
 states, etc., see Sec.~\ref{sec_dos}.
 \item \texttt{diagonalization}: The infrastructure for diagonalization of the
 Hamiltonians, see Sec.~\ref{sec_diagonalization}.
 \item \texttt{hamiltonian}: For construction of the Hamiltonians, see Sec.~\ref{sec_hamiltonian_implementation}.
 \item \texttt{materials}: For parsing the files that contain the material
 parameters, see Sec.~\ref{sec_matparam}.
 \item \texttt{ploto}: For plot output, see Sec.~\ref{sec_plots}.
 \item \texttt{tableo}: For table output, see Sec.~\ref{sec_tables}.
 \item \texttt{xmlio}: For XML input and output, see Sec.~\ref{sec_xml}.
\end{itemize}
The \texttt{materials} subdirectory also contains the default materials file.
Finally, the built-in helpfile is located at \texttt{docs/helpfile.txt} inside
\texttt{src/kdotpy}. The components will be discussed below in the remainder of
Sec.~\ref{sec_implementation}, roughly in the order in which they are executed
in a \kdotpy{} calculation.

\subsubsection{Calculation subprogram workflow}
\label{sec_workflow}

\begin{figure}
\includegraphics[width=130mm]{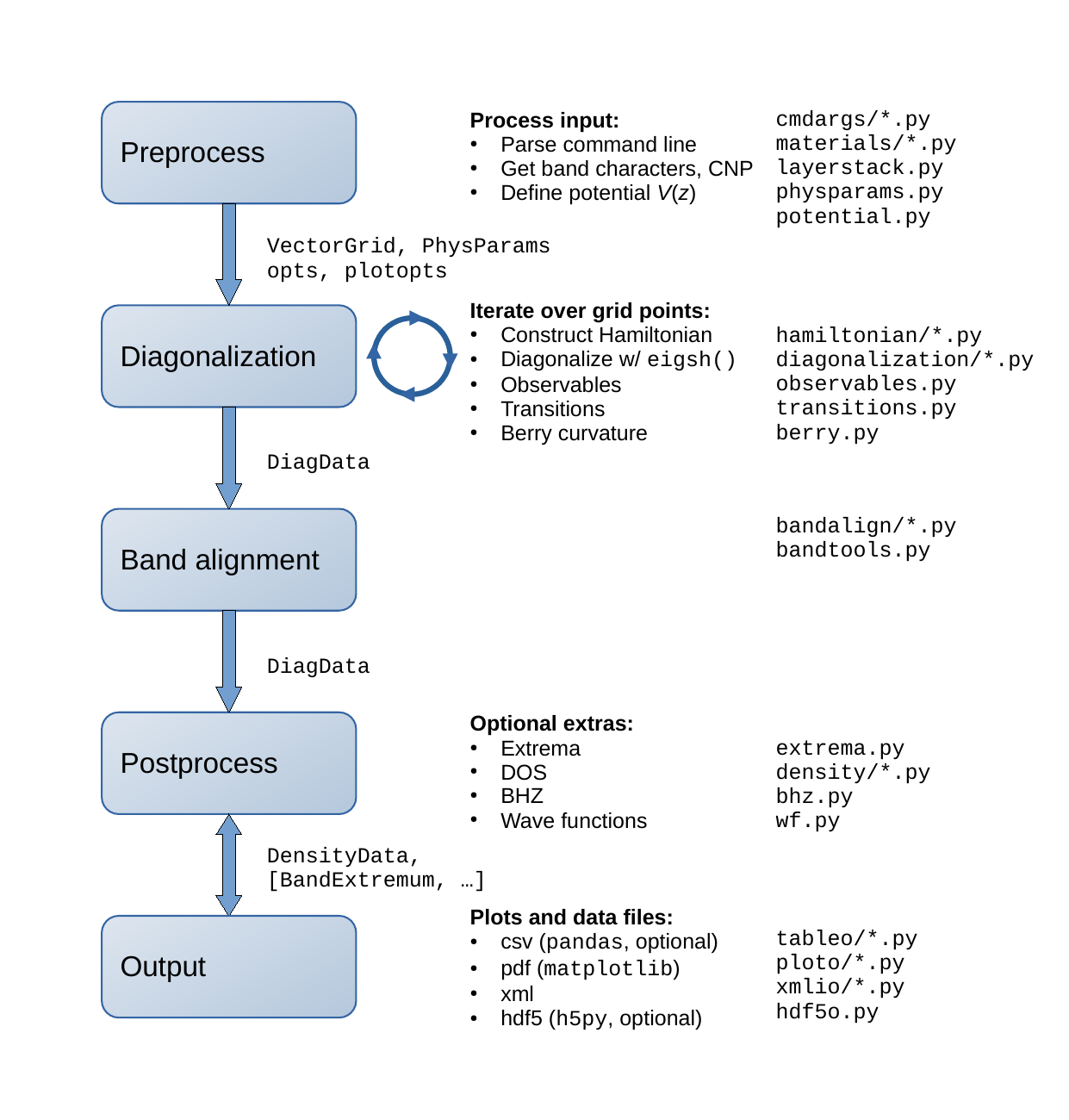}
\caption{Flow diagram of \kdotpy, that illustrates the workflow for the calculation
subprograms \texttt{kdotpy 1d}, \texttt{kdotpy 2d}, \texttt{kdotpy bulk},
\texttt{kdotpy ll}, and \texttt{kdotpy bulk-ll}. The diagram shows the most important
components of the program and some relevant data classes and source files (not a
complete list).}
\label{fig_flowdiagram}
\end{figure}

The five calculation subprograms differ substantially in the details, but they
all follow a similar workflow, going through the same sequence of stages. 
Figure~\ref{fig_flowdiagram} illustrates a schematic flow diagram with the
important stages. The sequence can be summarized as follows.

\begin{itemize}
 \item Preprocessing:
  \begin{itemize}
   \item Parse command line arguments, read configuration values and material parameters
   \item Define electrostatic potential (read from file and/or calculated with the self-consistent Hartree method)
   \item Prepare for diagonalization (determine band characters, charge neutrality point (CNP))
  \end{itemize}
  
  \item Diagonalization (iteration over $k$ or $B$ points):
  \begin{itemize}
   \item Construct Hamiltonian
   \item Diagonalize using \emph{diagsolver} (e.g., \texttt{eigsh})
   \item Calculate observables, transitions, Berry curvature
  \end{itemize}

 \item Band alignment

 \item Postprocessing and output:
  \begin{itemize}
   \item Optional extras: Extrema, DOS, BHZ, wave functions
   \item Output to files: \texttt{csv}, \texttt{pdf}, \texttt{xml}, \texttt{hdf5}, etc.
  \end{itemize}
\end{itemize}

\subsubsection{Data structures and data flow}

The different components of the program `communicate' via classes which contain
the important data. In the following list, we provide the most important ones:

\begin{itemize}
  \item \texttt{VectorGrid}: The class \texttt{VectorGrid} contains the $k$ and/or
  $B$ values on which the diagonalization is performed. The important class attributes
  are the vector type (\texttt{vtype}) and one, two or three arrays of values.
  The class represents a sequence of vector values (if the dimensionality is one) or a
  cartesian product of two or three sequences of values.
  
  \item \texttt{PhysParams}: The class \texttt{PhysParams} contains the physical
  parameters and coefficients related to the simulated structure, such as its size,
  the resolution of the spatial coordinates, material parameters, and strain values.

\item \texttt{DiagDataPoint}: Each instance of \texttt{DiagDataPoint} contains
the eigenvalues and (optionally) the eigenvectors at a single value $(k,B)$.
The eigenvectors are calculated and stored temporarily, and usually deleted after
the observables have been calculated in order to reduce the memory requirement.
The observables are stored in the \texttt{DiagDataPoint} instance, together with
other eigenstate data such as Landau level index, band index, and subband character.
The \texttt{DiagDataPoint} class defines several member functions for retrieving
and manipulating the eigenstate data.

\item \texttt{DiagData}: The class \texttt{DiagData} contains the result of the
diagonalization as an array of\linebreak[4] \texttt{DiagDataPoint} instances. Optionally, it
also contains a \texttt{VectorGrid} instance to give a multidimensional view of
the data. (The data is stored as a flat list of \texttt{DiagDataPoint} instances.)
The \texttt{DiagData} class contains several member functions for retrieving
eigenstate data for the complete data set, for example for the output functions.
Importantly, after successful band alignment, these member functions can access
the eigenstates `by band', e.g., the eigenvalues of all eigenstates with the
same band index. This is essential for making the plots and for some of the
postprocessing functions.

\item \texttt{DensityData}: The result of the calculation of the density of states
is stored in a \texttt{DensityData} instance. The key quantities inside this class
are the integrated density of states (IDOS) and the corresponding energy values.
The class can store an IDOS integrated over all $k$ as well as separate values
per $k$ or $B$. In the latter case, also a grid of $k$ or $B$ values is stored.
The member functions allow retrieval of the IDOS and the ordinary density of
states (as energy derivative of the IDOS), where the values can be scaled to the
desired set of units (inverse units of length, in terms of nm, cm, or m).
Broadening can also be applied.

\end{itemize}

Option values taken from command-line input are stored in the \texttt{dict}
instances \texttt{opts}, \texttt{plotopts}, \texttt{modelopts}, etc.

\subsection{Input and preprocessing}
\label{sec_input}

\subsubsection{Command line parsing}
\label{sec_cmdargs}

The \kdotpy{} program is designed as a non-interactive command-line interface (CLI).
The behaviour of the program is determined by the string of arguments provided
by the user in the shell command that starts \kdotpy. The first argument after
the program name \texttt{kdotpy} determines the subprogram that is used, for
example
\begin{verbatim}
kdotpy 2d 8o noax msubst CdZnTe 4% mlayer HgCdTe 68% HgTe HgCdTe 68%
llayer 10 7 10 zres 0.25 k -0.6 0.6 / 120 kphi 45 split 0.01
\end{verbatim}
runs the script \texttt{kdotpy-2d.py} with further arguments being passed to
this script. (For a detailed explanation of all arguments in this example, please
refer to the Usage Example of Sec.~\ref{sec_tutorial_basic}.) Unlike many other
programs, the option arguments are not preceded by hyphen or double-hyphen, for
ease of input. The argument list \texttt{sys.argv} passed to the script is
wrapped in the instance \texttt{sysargv} of the \texttt{CmdArgs} class, which
keeps track of all arguments that have been parsed. This allows \kdotpy{} to
track if there are any unparsed arguments; if that is the case, a warning message
is shown at the end.

The source file \texttt{cmdargs/cmdargs.py} provides several wrapper functions
for extracting information from the command line arguments. For example, the
functions \texttt{cmdargs.options()},\linebreak[4] \verb+cmdargs.plot_options()+, and
\texttt{cmdargs.bandalign()} return \texttt{dict} instances with option values
used by several other functions in \kdotpy. Some features are enabled by
simply putting the appropriate argument in the command line, e.g., \texttt{extrema}
in the above example for doing the extrema analysis. This is simply tested by
\verb+"extrema" in sysargv+; the \texttt{CmdArgs} class automatically marks
the argument as parsed.

We have provided a comprehensive list of all command line options in
Appendix~\ref{app_commands}, sorted thematically. For up-to-date information on
commands (for versions other than the present one, v1.0.0), we recommend to
consult the wiki~\cite{kdotpy_wiki} and/or the built-in help.

\subsubsection{Configuration options}

For user preferences that are unlikely to change between different calculations,
\kdotpy{} uses a system of configuration settings. A comprehensive reference for
the configuration options is provided as Appendix~\ref{app_config}. The user may
also consult the wiki~\cite{kdotpy_wiki} and/or built-in help for up-to-date
information.

The configuration settings are stored in the file \texttt{\HOME/.kdotpy/kdotpyrc},
which is read and parsed at the start of the
program. This file contains \texttt{\emph{key}=\emph{value}} pairs for all known
configuration options defined in \texttt{config.py}. If any configuration option
is missing from \texttt{\HOME/.kdotpy/kdotpyrc}, the \texttt{\emph{key}=\emph{value}}
pair is added as a comment (i.e., preceded by \texttt{\#}) with the default value.
Using this system, the configuration file remains usable if new features are
added to \kdotpy{} and the new configuration options will become visible to the
user.

It is also possible to use a custom configuration file by using
\texttt{config \emph{filename}} as a command line argument. Multiple files may
be loaded in sequence, where the files are processed from left to right, and
configuration values loaded later override those loaded earlier. For flexibility,
it is also possible to adjust configuration values on the command line by
providing \texttt{config} followed by semicolon-separated \texttt{\emph{key}=\emph{value}}
pairs, like
\begin{verbatim}
kdotpy 2d ... config 'key1=value1;key2=value2' ...
\end{verbatim}
where the single quotes are required by the shell. Both variants of the
\texttt{config} argument may be combined in any order.

The configuration values are typically parsed \emph{ad hoc}: they are evaluated
at the moment they are needed. The configuration values are not checked for
validity at the start of the program. Thus, error messages related to invalid
configuration values may occur at any point during runtime.

An auxiliary subprogram \texttt{kdotpy config} is provided in order to read and
manipulate settings easily. It also gives access to the help file which contains
detailed information on every configuration option. The command-line syntax for
the configuration tool \texttt{kdotpy config} is summarized
in Sec~\ref{sec_kdotpy_config}.

\subsubsection{Vector grids}
\label{sec_vector_grids}

The array of momenta $\vec{k}$ or magnetic fields $\vec{B}$ where the
Hamiltonian is evaluated is called the `grid'. In general, the command line parser
yields a \texttt{ZippedKB} object, which contains the values of $\vec{k}$ and $\vec{B}$.
This object may represent a dispersion ($\vec{k}$ is a \texttt{VectorGrid}
instance, $\vec{B}$ is single valued), a magnetic-field dependence ($\vec{k}$ is
single valued, $\vec{B}$ is a \texttt{VectorGrid} instance), or a single point
(when both are single valued). A combination of $\vec{k}$- and $\vec{B}$-dependence
is not permitted.

The \texttt{VectorGrid} class stores a multi-dimensional array of $\vec{k}$ or $\vec{B}$
values as a cartesian product of values. It may have any one of the following
coordinate types:
\begin{itemize}
  \item \texttt{x}, \texttt{y}, \texttt{z}: Values along one of the cartesian
  coordinate axes.
  
  \item \texttt{xy}: Cartesian coordinates $(x,y)$ in the $z=0$ plane.
  
  \item \texttt{xyz}: Cartesian coordinates $(x,y,z)$.
  
  \item \texttt{pol}: Polar coordinates $(r,\phi)$, that convert to cartesian
  coordinates as $(x,y) = r(\cos\phi,\sin\phi)$ with $z=0$.
  
  \item \texttt{cyl}: Cylindrical coordinates $(r,\phi,z)$, that convert to
  cartesian coordinates as\linebreak[4] $(x,y,z)=(r\cos\phi,r\sin\phi,z)$.
  
  \item \texttt{sph}: Spherical coordinates $(r,\theta,\phi)$, where $\theta$
  is the polar angle in $[0,\pi] = [0,180^{\circ}]$ and $\phi$ the azimuthal angle.
  The relation to cartesian coordinates is given by\linebreak[4]
  $(x, y, z) = (r \sin \theta \cos \phi, r\ \sin \theta \sin \phi, r \cos \theta)$.

\end{itemize}
In \kdotpy, the radial coordinate $r$ of the angular coordinate systems may 
be chosen negative, in contrast to the usual condition that $r\geq 0$.

The input of the grid on the command line is done by combination of the vector
components on the command line, followed by a single value or a range, for example
\begin{verbatim}
kdotpy 2d ... k 0 0.2 / 20 kphi 45 bz 0.1 ...
\end{verbatim}
yields the momentum values given by the polar coordinates $(k_i, 45^\circ)$ with
$k_i = \{0, 0.01, \ldots, 0.2\}\,\mathrm{nm}^{-1}$ in a magnetic field
$\vec{B}=(0,0,B_z)$ with $B_z=0.1\,\mathrm{T}$. The valid component arguments
for momenta, are \texttt{k}, \texttt{kx}, \texttt{ky}, \texttt{kz}, \texttt{kphi}, and
\texttt{ktheta}; for magnetic fields, they are \texttt{b}, \texttt{bx}, \texttt{by},
\texttt{bz}, \texttt{bphi}, and \texttt{btheta}. The arguments \texttt{k} and \texttt{b}
without explicit component refer to the radial direction if used in combination
with angular coordinates. If they are used by themselves, \texttt{k} is equivalent
to \texttt{kx} and \texttt{b} is equivalent to \texttt{bz}, i.e., the `natural'
directions for momentum and magnetic field, respectively. The values are in units
of $\mathrm{nm}^{-1}$ and $T$ for momentum and magnetic field, respectively, and
degrees for the angular components.

Values and ranges are input as follows:
\begin{itemize}
\item \texttt{k \emph{a}}: Single value $a$.

\item \texttt{k \emph{a} * \emph{b}}: Single value $a\,b$, where $a$ is an integer
  and $b$ is a floating-point value.

\item \texttt{k \emph{a} \emph{b} / \emph{c}}: Values $a$ and $b$ are the minimum
 and maximum of the range, respectively. If $c$ is a floating point number, it
 denotes the step size $\Delta = c$. If $c$ is an integer, the step size is
 $\Delta = (b-a)/c$. The resulting values are
 $\{a,a+\Delta, a+2\Delta, \ldots, b\}$. Note that if $c$ is an integer, the
 number of values in this range is $c+1$.
 
\item \texttt{k \emph{b} / \emph{c}}: Equivalent to \texttt{k 0 \emph{b} / \emph{c}}.

\item \texttt{k \emph{a} \emph{b} \emph{c} / \emph{d}}: The single momentum value
  $a+ (b-a) c / d$, i.e., the $c$'th element from the range given by
  \texttt{k \emph{a} \emph{b} / \emph{d}}. The values $c$ and $d$ must be integers.
  
\end{itemize}
Here, \texttt{k} can be replaced by any of the momentum and magnetic field
components listed above. Also, a \texttt{//} (double-slash) operator can be used
instead of \texttt{/}. In this case, quadratic stepping is used, which is useful
in particular for magnetic fields: \texttt{bz \emph{a} \emph{b} // \emph{c}}
(where $c$ is an integer) yields
the values $B_z=\{a, a+\Delta, a+4\Delta, a+9\Delta, \ldots, b\}\,\mathrm{T}$ where
$\Delta = (b-a)/c^2$. The number of values is $c+1$ like with the single slash \texttt{/}.
The typical use case of \texttt{//} is with the perpendicular magnetic field
in Landau-level mode, with $a=0$.

The dimensionality of the grid is determined automatically by \kdotpy{} based on
the input of single values and ranges. For momenta, it is always smaller than or
equal to the geometric dimension (i.e., the number of momentum components).
For magnetic fields, multi-dimensional grids are not supported; the maximum
dimension of the grid is 1.

\subsubsection{Material parameters}
\label{sec_matparam}

Many of the coefficients of the Hamiltonian are material dependent. The values
of these parameters are defined in separate input files. The program provides
a default materials file (\texttt{materials/default}) with a set of parameters
for HgTe, CdTe, (Hg,Cd)Te, (Hg,Mn)Te, and (Cd,Zn)Te. The user may override
these materials and define new ones by providing custom material definitions in
their own files in the directory \texttt{\HOME/.kdotpy/materials}.

The materials files are formatted similar to the \texttt{.ini} format and are
parsed by Python's\linebreak[4] \texttt{configparser} module. The `sections' in the file serve
as the material labels; for example the definitions for HgTe are preceded
by the section head \texttt{[HgTe]}. The section contains several
\texttt{\emph{param} = \emph{value}} pairs. The parser simply reads this data
and puts them as keys and values of a \texttt{dict} instance. Comments (everything
that follows \verb+#+) are ignored. In addition, the user may also adjust
material parameters on the command line with the argument \texttt{matparam}
followed by several key-value pairs.

The `left-hand size' \texttt{\emph{param}} may be any of the recognized material
parameters, as listed in Appendix~\ref{app_matparam_reference}, or a user defined
`auxiliary' parameter which is used in the value of some other parameter.
The `right-hand size' \texttt{\emph{value}} must represent a number or a valid
Python expression where a restricted set of functions and operators may be used:
\begin{itemize}
  \item The arithmetic operators \texttt{+}, \texttt{-}, \texttt{*}, \texttt{/},
  \texttt{**}. Exponentiation is represented by \texttt{**}, not \verb+^+.
  \item The \kdotpy{} defined polynomial functions \texttt{linint},
  \texttt{linearpoly}, \texttt{quadrpoly}, \texttt{cubicpoly}, and \texttt{poly}.
  These represent the functions $x\mapsto a(1-x)+bx$, $x\mapsto c_0+c_1x$,
  $x\mapsto c_0 + c_1x + c_2x^2$, $x\mapsto c_0+c_1x+c_2x^2+c_3x^3$, and
  $x\mapsto c_0 +c_1 x + \ldots + c_n x^n$, respectively.
  \item The comparison functions \texttt{geq}, \texttt{leq}, \texttt{gtr}, and
  \texttt{less}, which implement the binary comparison operators $\geq$, $\leq$,
  $>$, and $<$, respectively.
  \item Functions of the Python \texttt{math} module, like \texttt{sqrt} and \texttt{log}.
  \item The mathematical constants \texttt{pi} and \texttt{e}
  \item The physical constants defined in \texttt{physconst.py}, see
  Appendix~\ref{app_units_reference}.
\end{itemize}
The value may also contain any other predefined or auxiliary material parameter.
If the expression is not valid Python syntax or if
any `forbidden' keyword is used (e.g., \texttt{import}), \kdotpy{} will raise an
exception. The restricted function and variable labels also cannot be used as
\texttt{\emph{param}} on the left-hand side of \texttt{=}. (The following are
also not permitted as \texttt{\emph{param}}: All Python keywords except \texttt{as},
the numerical constants \texttt{inf} and \texttt{nan}, and the reserved variables
\texttt{x}, \texttt{y}, \texttt{z}, and \texttt{T}.) The following examples are
an excerpts from the default definitions for HgTe and Hg$_{1-x}$Cd$_x$Te,
\begin{verbatim}
## HgTe, mercury telluride
[HgTe]
compound    = HgTe
composition = 1, 1     
P           = sqrt(18800. * hbarm0)
Ev          = 0.0
Ec          = -303.0 + 0.495 * T ** 2 / (11.0 + T)  # Eg of HgCdTe for x = 0
gamma1      = 4.1
gamma2      = 0.5
gamma3      = 1.3
\end{verbatim}
\begin{verbatim}
## HgCdTe (Hg_{1-x} Cd_x Te), mercury cadmium telluride
[HgCdTe]
compound    = HgCdTe
linearmix   = HgTe,CdTe,x 
composition = 1 - x, x, 1  # is also set automatically by linearmix 
P           = sqrt(18800. * hbarm0)
Eg          = -303 * (1 - x) + 1606 * x - 132. * x * (1 - x) + \
              (0.495 * (1 - x) - 0.325 * x - 0.393 * x * (1 - x)) * T ** 2 / \
              (11.0 * (1 - x) + 78.7 * x + T)  # identical to Ref. [HgCdTe2]
Eg0         = -303.0 + 0.495 * T ** 2 / (11.0 + T)
Evoff       = -570. * (Eg - Eg0) / (1606 - -303)
Ev          = Evoff
Ec          = Evoff + Eg
gamma1      = poly( 4.1, -2.8801,  0.3159, -0.0658, x)
gamma2      = poly( 0.5, -0.7175, -0.0790,  0.0165, x) 
gamma3      = poly( 1.3, -1.3325,  0.0790, -0.0165, x)
\end{verbatim}
The second example for Hg$_{1-x}$Cd$_x$Te also illustrates the `special parameter'
\texttt{linearmix}, which creates a third material as linear combination of
two other materials. Here, \texttt{linearmix = HgTe,CdTe,x} indicates that all
material parameters are interpolated as
$p_\mathrm{HgCdTe} = (1-x)\,p_\mathrm{HgTe} + x\,p_\mathrm{CdTe}$. The subsequent
definitions (like \texttt{gamma1} in this example) override the linearly
interpolated ones.

The expressions on the right-hand side are parsed by an abstract syntax tree (AST)
parser based on Python's
\texttt{ast} module, so that it accepts the `whitelisted' functions and operators
only. If an expression value evaluates to a finite value (not $\pm\infty$, not
\texttt{nan}), the parser substitutes that value, otherwise it leaves the expression
unevaluated. The latter is common when a material parameter depends on the
composition (e.g., $x$ in Hg$_{1-x}$Cd$_x$Te). This is acceptable until the layer
stack is built; at that point, all material parameters must evaluate to numerical
values.

Using the built-in and custom defined materials is done by using the material
label (the section label in the materials file, e.g., \texttt{[HgTe]}, without
the brackets), possibly followed by up to three
numerical values. For example, \texttt{HgTe} simply yields the material \texttt{HgTe}
and \texttt{HgCdTe 68\%} yields the material \texttt{HgCdTe} with the substitution
$x\to0.68$. As the material label can be any alphanumerical string starting with a
letter (numbers, hyphens and underscores are also allowed at following positions),
it is not needed to use molecular formulas as labels. For example, the label
\verb+mercury_telluride+ would be acceptable too. This also gives the freedom
to define multiple sets of parameters for one material, e.g., using labels
\verb+CdTe_Novik+ and \verb+CdTe_Weiler+ for CdTe parameters based on
the different literature sources (Refs.~\cite{NovikEA2005} and
\cite{Weiler1981bookchapter}, respectively).

\subsubsection{Building the layer stack}

The structure of the sample as function of the growth direction $z$ is known as
the \emph{layer stack} in the language of \kdotpy. The relevant properties of
the materials in the device are stored in a \texttt{LayerStack} instance. This
includes the material and thickness of each layer, the $z$ resolution, the coordinates
of the interfaces, the width of the interface smoothening, and the `layer names'
(like `well', `barrier', etc.).
The \texttt{LayerStack} object is itself an attribute of the \texttt{PhysParams}
instance that is used to construct the Hamiltonian.

Importantly, the \texttt{LayerStack} class provides a function that creates
a cache of all material parameters, which is stored within \texttt{PhysParams}.
The $z$ dependence of each material parameter $Q$ is calculated in the following
manner. We first construct a set of layer weights functions $w_l(z)$ for each layer
$l$. Without interface smoothening, $w_l(z) = 1$ for $z$ inside the layer $l$
and $0$ elsewhere. For a finite interface smoothening width $\delta_\mathrm{if}$
(by default $\delta_\mathrm{if}=0.075\nm$), the weight is
\begin{equation}\label{eqn_layer_weights}
 w_l(z) = \frac{1}{2}\left[ 
   \tanh\left(\frac{z - z_{\mathrm{min},l}}{\delta_\mathrm{if}}\right) -
   \tanh\left(\frac{z - z_{\mathrm{max},l}}{\delta_\mathrm{if}}\right)
 \right],
\end{equation}
where the layer $l$ spans the interval $[z_{\mathrm{min},l}, z_{\mathrm{max},l}]$.
The weights are normalized as\linebreak[4] $\tilde{w}_l(z) = w_l(z)/\sum_{l'}w_{l'}(z)$,
so that $\sum_l\tilde{w}_l(z)=1$ everywhere inside the layer stack. If we write
the value of the material parameter $Q$ in layer $l$ as $Q_l$, the function
$Q(z)$ is constructed as
\begin{equation}\label{eqn_qz}
  Q(z) = \sum_l \tilde{w}_l(z)Q_l.
\end{equation}
This function has the desired property that it interpolates smoothly between
$Q_{l}$ and $Q_{l+1}$ if $\delta_\mathrm{if}$ is finite, while it approaches
the values $Q_l$ away from the interfaces.

In view of the discrete representation of the $z$ coordinates, we need to
evaluate $Q(z)$ only in a discrete array of $z$ values. If the $z$ resolution is
$\Delta z$, we require evaluation at integer and half-integer multiples of $\Delta z$,
because the action of the first and second degree derivatives in $z$, given by
Eqs.~\eqref{eqn_discrete_first_derivative1} and \eqref{eqn_discrete_second_derivative1}, respectively, involves $Q(z_j\pm\frac{1}{2}\Delta z)$.
The functions $Q(z)$ are thus evaluated on a discretized set of coordinates
\begin{equation}
  \{
  z_\mathrm{min}-\tfrac{1}{2}\Delta z, z_\mathrm{min}, z_\mathrm{min}+\tfrac{1}{2}\Delta z,
  \ldots,
  z_\mathrm{max}-\tfrac{1}{2}\Delta z, z_\mathrm{max}, z_\mathrm{max}+\tfrac{1}{2}\Delta z
  \}
\end{equation}
where $z_\mathrm{min}$ and $z_\mathrm{max}$ refer to the bottom and top of the
complete layer stack. In \texttt{physparams.py}, this is achieved by setting
\begin{verbatim}
self.cache_z = -0.5 + 0.5 * np.arange(2 * self.nz + 1)
\end{verbatim}
where coordinates are interpreted in units of $\Delta z$ with $0$ located at
$z_\mathrm{min}$ at the bottom of the layer stack.

\subsubsection{Handling of strain}

In \kdotpy, we use a simplified strain model for the diagonal components of the
strain tensor.  (The shear components are assumed to be zero.)
By default, we strain each layer to match the in-plane lattice constants to
that of the substrate. This defines
\begin{equation}\label{eqn_kdotpy_inplane_strain}
 \epsilon_\parallel = \epsilon_{xx}=\epsilon_{yy}
   =\frac{a_\mathrm{s} - a_0}{a_0}
\end{equation}
in terms of the target (`strained') lattice constant $a_\mathrm{s}$ and the 
equilibrium lattice constant $a_0$ of each layer material, cf.\ Sec.~\ref{sec_strain}.
The target lattice constant $a_\mathrm{s}$ is determined by one of three
command-line arguments
\begin{itemize}
 \item \texttt{msubst} followed by a material: The target lattice constant
 $a_\mathrm{s}$ is taken as the equilibrium lattice constant of the substrate.
 The latter is defined as a material parameter \texttt{a} (see
 Appendix~\ref{app_matparam_reference}).
 \item \texttt{alattice} followed by a value: The target lattice constant
 $a_\mathrm{s}$ is taken to be the input value in $\mathrm{nm}$.
 \item \texttt{strain} followed by a value: 
 Equation~\eqref{eqn_kdotpy_inplane_strain}
 is bypassed and $\epsilon_\parallel$ is taken directly from the input value.
\end{itemize}
From minimization of the strain-energy tensor \cite{DeCaroTapfer1993,DeCaroTapfer1995part1},
under the assumptions that the substrate is thick (so that its lattice constant stays
at the equilibrium value) and that shear strain is absent, we find that the
out-of-plane strain of the epitaxial layers equals
\begin{equation}\label{eqn_oop_strain_relation}
 \epsilon_\perp = \epsilon_{zz} = -\frac{2C_{12}}{C_{11}} \epsilon_\parallel
\end{equation}
For both HgTe and CdTe, $C_{12}/C_{11}\approx0.69$ \cite{PfeufferJeschke2000_thesis},
which is presently hardcoded as a constant. (In a future \kdotpy{} release, strain
handling will be improved, by treating the elasticity modules
$C_{11}$, $C_{12}$, and $C_{44}$ as material parameters.)

The \text{strain} command-line argument also allows for several other strain
configurations. In short, the values following \text{strain} are interpreted
as $\epsilon_{xx}$, $\epsilon_{yy}$, and $\epsilon_{zz}$, in order. If any of
these values is not specified (omitted or input as \texttt{-}), it is determined
from the other value(s), following these rules:
\begin{itemize}
 \item If $\epsilon_{xx}$ is specified but $\epsilon_{yy}$ is not, then set
 $\epsilon_{yy}=\epsilon_{xx}$, and vice versa.
 \item If $\epsilon_{zz}$ is specified but $\epsilon_{xx}$ and $\epsilon_{yy}$
 are not, then set
 \begin{equation}\label{eqn_strain_rule1}
  \epsilon_{xx} = \epsilon_{yy} = -\frac{C_{12}/C_{11}}{1+C_{12}/C_{11}} \epsilon_{zz}.
 \end{equation}
 This relation minimizes the strain-energy tensor with $\epsilon_{xx}$ and $\epsilon_{yy}$
 as free variables \cite{DeCaroTapfer1995part1}.
 \item If $\epsilon_{xx}$ and $\epsilon_{yy}$ are specified but $\epsilon_{zz}$
 is not, then set
 \begin{equation}\label{eqn_strain_rule2}
   \epsilon_{zz} = -(C_{12}/C_{11}) (\epsilon_{xx} + \epsilon_{yy})
 \end{equation}
 which generalizes Eq.~\eqref{eqn_oop_strain_relation} to the case where
 $\epsilon_{xx}\not=\epsilon_{yy}$. (Presently, only $\epsilon_{xx}=\epsilon_{yy}$
 is supported, but this will change in a future version.)
\end{itemize}
The relations \eqref{eqn_strain_rule1} and \eqref{eqn_strain_rule2} apply to
growth direction (001). If strain is combined with a nontrivial orientation
(command-line argument \texttt{orient}), a correct result is not guaranteed.

\subsubsection{Electrostatic potentials}
\label{sec_potentials}

In addition to the intrinsic potentials defined by the band edges, \kdotpy{} can
also model electrostatic potentials $V(z)$. The potential is added to the
Hamiltonian as a diagonal matrix, adding $V(z_j)$ to each point $z_j$ in the
$z$-coordinate basis. The potential can be read from a file or calculated
from a carrier density profile with certain boundary conditions. In \texttt{kdotpy 2d}
and \texttt{kdotpy ll}, the potential $V(z)$ is initialized in the following order:

\begin{itemize}
\item Read the potential from a file, if the command line argument
\texttt{potential \emph{filename}} is provided. The input file must be a CSV file
with two columns, labelled \texttt{z} and \texttt{potential} in the first row.
The values that follow represent the coordinates $z_i$ and the values $V(z_i)$.
The input coordinate values need not align with the $z$ coordinates of the lattice:
Interpolation and extrapolation is used to calculate $V(z)$ on all $z$ coordinates
of the lattice. (Thus, certain coordinate values may be omitted, e.g., when $V(z)=0$
for all $z \in[z_1,z_2]$, one can specify the potential at $z_1$ and $z_2$ and omit
all intermediate values.) A numerical multiplier can be added in order to scale
the potential, for example \texttt{potential v.csv 10} adds $10 V(z)$ to the
Hamiltonian, where $V(z)$ is the potential defined in \texttt{v.csv}.
Multiple arguments and numerical multipliers may be added
in order to input linear combinations, for example with
\texttt{potential v1.csv 10 potential v2.csv -5} the potential is
$10V_1(z)-5V_2(z)$. The parsing of potential files is done by the function
\verb+read_potential()+ in \texttt{potential.py}.

\item Calculate a Hartree potential self-consistently, if the command line argument
\texttt{selfcon} is given. This method considers the electric charge density profile
due to the occupied eigenstates, solves $V(z)$ from the Poisson equation, and then
feeds this back into the Hamiltonian, after which diagonalization yields a new set
of eigenstates. If this iterative process converges, it yields a self-consistent
solution of the Schr\"odinger and the Poisson equation. We discuss this method in
greater detail in Sec.~\ref{sec_selfcon}.

\item Define a potential from boundary conditions, specified on the command line.
This is a `static' calculation, not done self-consistently. The following commands
are the most common ones to define a potential:
  \begin{itemize}
   \item \texttt{vinner \emph{v0}}, where \texttt{\emph{v0}} is a numerical value,
        representing an energy $V_0$ in meV.
        The potential $V(z)$ is fixed at $V(z_0) = 0$ in the centre of the `well'
        layer and a potential difference $V_0$ is applied between the bottom and
        top interface of this layer. (This is equivalent to setting the potential
        to $-\frac{1}{2}V_0$ and $\frac{1}{2}V_0$ at the bottom and top
        interface of the well layer.) Extrapolation to the other layers is done
        by assuming a constant electric field.
   \item \texttt{vouter \emph{v0}}, where \texttt{\emph{v0}} is a numerical value,
        representing an energy $V_0$ in meV. This command is similar to
        \texttt{vinner}, with the difference that a potential difference $V_0$
        is applied between the bottom and top of the full layer stack.
   \item \texttt{vsurf \emph{v0} \emph{w}}, where \texttt{\emph{v0}} and
        \texttt{\emph{w}} represent an energy $V_0$ in meV and a width $w$ in nm.
        The resulting potential depends on the minimal distance $d_\mathrm{if}(z)$
        between $z$ and each interface $z_i$. The potential is
        \begin{equation}
         V(z) = V_0 (1 - d_\mathrm{if}(z) / w)
        \end{equation}
        if $d_\mathrm{if}(z)\leq w$, otherwise $0$. If the command is followed
        by \texttt{q}, this is changed to the quadratic dependence
        $V(z) = V_0 (1 - d_\mathrm{if}(z) / w)^2$
   \end{itemize}
The parsing of the potential options is done by \verb+gate_potential_from_opts()+,
whereas the solution of $V(z)$ given the boundary conditions is done by
\verb+solve_potential()+, both in\linebreak[4] \texttt{potential.py}. We provide more details
on solving $V(z)$ in Appendix~\ref{app_solution_poisson}.

\end{itemize}

These options may be combined, e.g., if one uses a potential input file together
with the self-consistent Hartree method, then the input potential serves as the
initial potential for the iterative solution process. For \texttt{kdotpy 1d},
the self-consistent Hartree method is not available. The extra option
\texttt{potentialy \emph{filename}} provides the capability of defining a
potential $V(y)$. For \texttt{kdotpy bulk} and \texttt{kdotpy bulk-ll}, none of
the potential options is available.

\subsection{Construction of Hamiltonian}
\label{sec_hamiltonian_implementation}

\subsubsection{Bulk geometry}

The \kdotp{} Hamiltonian is an $8\times 8$ matrix $H_{pq}(k_x,k_y,k_z)$ with
orbital indices $p,q$ and depending on the momentum coordinates $(k_x,k_y,k_z)$.
For bulk materials, the band structure can simply be obtained by substituting
values $(k_x,k_y,k_z)$ and \emph{diagonalizing} the $8\times 8$ matrix as to
obtain $8$ eigenvalues and $8$ eigenvectors.

\subsubsection{Two-dimensional geometry}

As discussed in Sec.~\ref{sec_lower_dim_geometries}, layered structures break
translational symmetry in the $z$ direction. The momentum $k_z$ is substituted by
the operator $-\ii\partial_z$ and the coordinates are discretized to $z=\{z_j\}$.
This set has to be finite for the sake of calculation. With $n_z$
coordinates in this set, the dimension of the Hilbert space is $8n_z$.
The Hamiltonian is thus represented as an $8n_z\times8n_z$ matrix
$H_{i,p;j,q}(k_x,k_y)$ where $i,j$ represent the $z$ coordinates and $p,q$ the
orbitals.

In \kdotpy, the matrix $H_{i,p;j,q}(k_x,k_y)$ is constructed as a
sparse matrix from $8\times 8$ blocks. (The blocks are $6\times 6$ is the
six-orbital Kane model is used. For the sake of clarity, we assume the eight-orbital
model in the following discussion.) The $8\times 8$ blocks that constitute
several terms of the Hamiltonian are defined in the submodule
\verb+hamiltonian.blocks+. They are summed up in a function from
\verb+hamiltonian.full+. For the 2D geometry, the appropriate function is
\verb+hz()+, defined with the argument signature
\begin{verbatim}
hz(z, dz, k, b, params, **kwds)
\end{verbatim}
This function returns an $8\times 8$ matrix for each combination of arguments.
The indices $i,j$ for the $z$ coordinates encoded as \texttt{z} and \texttt{z + dz},
respectively. The arguments \texttt{k} and \texttt{b} are momentum and magnetic
field. The argument \texttt{params} is the \texttt{PhysParams} instance needed
to construct the matrices and \texttt{**kwds} denotes several additional options.

In line with the algorithm discussed in Sec.~\ref{sec_lower_dim_geometries}, 
\texttt{hz} returns a nonzero result only for \texttt{dz} being $0$ or $\pm 1$.
For \texttt{dz = 0}, the results are the $8\times 8$ submatrices $H_{i,p;i,q}$
appearing on the block diagonal of the full Hamiltonian matrix. For
\texttt{dz = 1} and \texttt{dz = -1}, we find the off-diagonal blocks
$H_{i,p;i\pm1,q}$ that contain terms involving $z$ derivatives, see, e.g.,
Eqs.~\eqref{eqn_discrete_first_derivative1} and \eqref{eqn_discrete_second_derivative1}.

The full Hamiltonian matrix, a sparse matrix of dimension $8n_z\times 8n_z$, is
constructed as follows.

\begin{itemize}
 \item Initialize lists \texttt{allrows}, \texttt{allcols}, \texttt{allvals}.
 These will be used later to construct a sparse matrix in \texttt{COO} (coordinate)
 format, where nonzero matrix elements are represented as triplets $(I, J, v)$
 of row index, column index, and value.
 \item Iterate over the diagonal blocks: For $z=0,\ldots,n_z-1$, call
 \texttt{hz(z, 0, k, b, params, **kwds)}. Flatten the resulting $8\times 8$
 matrix \texttt{m} and append the values to \texttt{allvals},
\begin{verbatim}
allvals.append(m.flatten())
\end{verbatim}
 The corresponding row and column indices are $I=8z + p$, $J=8z + q$ where
 $p,q$ are the orbital indices. This is implemented as
\begin{verbatim}
allrows.append(z * norb + rows0)
allcols.append(z * norb + cols0)
\end{verbatim}
 where \texttt{rows0} and \texttt{cols0} are (precalculated) arrays of length
 $64$ with the row and column indices ($0, \ldots, 7$) corresponding to the
 flattened matrix.
 \item Iterate over the off-diagonal blocks. For $z=0,\ldots,n_z-2$, calculate
\begin{verbatim}
mm = 0.5 * (hz(z + 1, -1, k, b, params, **kwds) \
    + hz(z, 1, k, b, params, **kwds).conjugate().transpose())
mp = mm.conjugate().transpose()
\end{verbatim}
 In principle, we could have used \texttt{hz(z + 1, -1, k, b, params, **kwds)}
 and \texttt{hz(z, 1, k, b, params, **kwds)} as the two blocks above and below
 the diagonal. By construction they are each other's hermitian conjugate, but
 we symmetrize them in order to eliminate numerical errors and to make the
 Hamiltonian exactly hermitian.
 We proceed by filling the value, row, and column lists as
\begin{verbatim}
allvals.append(mp.flatten())
allrows.append(z * norb + rows0)
allcols.append((z + 1) * norb + cols0)
allvals.append(mm.flatten())
allrows.append((z + 1) * norb + rows0)
allcols.append(z * norb + cols0)
\end{verbatim} 
 \item Turn \texttt{allvals}, \texttt{allrows}, and \texttt{allcols} into 
 one-dimensional arrays and construct the sparse matrix from all nonzero values,
\begin{verbatim}
non0 = (allvals != 0)
s = coo_matrix(
    (allvals[non0], (allrows[non0], allcols[non0])),
    shape = (norb * nz, norb * nz), dtype = complex
)
\end{verbatim} 
 using \verb+coo_matrix+ from \texttt{scipy.sparse}.
\end{itemize}

The \texttt{COO} format is suitable for construction of a sparse matrix, but not
so much for other operations. Diagonalization is done most efficiently in the 
\texttt{CSC} (compressed sparse column) format. For obtaining a sparse matrix
in \texttt{CSC} format,
constructing the matrix in \texttt{COO} format and converting it to \texttt{CSC}
is more efficient than constructing it directly as a \texttt{CSC} matrix
\footnote{See the documentation of \texttt{scipy.sparse} at
\url{https://docs.scipy.org/doc/scipy/reference/sparse.html}}.

\subsubsection{One-dimensional geometry}

The construction of the Hamiltonian for 1D geometries follows the same line of
reasoning. With the additional discretization of the $y$ coordinates, the
Hilbert space has dimension $8n_yn_z$, where $n_y$ and $n_z$ are the number of
coordinate values in $y$ and $z$ direction, respectively. The Hamiltonian is
thus an $8n_yn_z \times 8n_yn_z$ matrix $H_{l,i,p;m,j,q}(k_x,k_y)$ where the
indices $l,m$ encode the $y$ coordinates, $i,j$ the $z$ coordinates and $p,q$
the orbitals. The construction is by filling a sparse matrix with $8\times 8$
blocks; the appropriate function for a 1D geometry (in absence of magnetic fields)
is \verb+hzy()+, defined with the argument signature
\begin{verbatim}
hzy(z, dz, y, dy, kx, params, **kwds)
\end{verbatim}
where the indices $l,m$ are represented by \texttt{y} and \texttt{y + dy}, and
$i,j$ by \texttt{z} and \texttt{z + dz} as before. The sparse matrix constructor
iterates over $z$ and $y$. For example, for the diagonal terms (\texttt{dz = 0}
and \texttt{dy = 0}),
\begin{verbatim}
allvals.append(m.flatten())
allrows.append(y * norb * nz + z * norb + rows0)
allcols.append(y * norb * nz + z * norb + cols0)
\end{verbatim}
where the row and column indices are obtained as $I=8n_zy+8z + p$ and
$J=8n_zy + 8z + q$, respectively. The constructor also does a similar iteration
for three types of off-diagonal blocks,
$(\mathtt{dz}, \mathtt{dy}) = (0,\pm 1)$, $(\pm 1,0)$, and $(\pm1, \pm 1)$.
Like the 2D geometry, symmetrization is applied to the off-diagonal blocks to
make the Hamiltonian matrix exactly hermitian.
The constructor function also has an option \texttt{periodicy} by means of which
periodic boundary conditions can be applied in the $y$ direction. This is done by
adding the blocks corresponding to $(l,m)=(0,n_y-1)$ and $(n_y-1,0)$.
In the presence of magnetic fields, the blocks are obtained by using \verb+hzy_magn()+,
instead of \verb+hzy()+, but the sparse construction itself is fully analogous.

\subsubsection{Landau levels in axial approximation}

In the Landau level formalism, the in-plane momentum operators $\hat{k}_x$ and
$\hat{k}_y$ are substituted by the ladder operators $a$ and $a^\dagger$ acting
on the Landau level states, see Sec.~\ref{sec_ll_formalism}.
In the axial approximation, the Landau level index $n$ is a
conserved quantum number, so that the Hamiltonian may be split in several
independent blocks $H^{\mathrm{ax},(n)}$. The diagonalization is then simply
done for each $n=-2,-1,0,\ldots,n_\mathrm{max}$, where $n_\mathrm{max}$ is the
maximum index set by the command line argument \texttt{nll}.

For the axial model, we use the \emph{symbolic} Landau level mode. (The label
\emph{symbolic} distinguishes it from the \emph{legacy} mode, which was
implemented earlier, but is no longer used.) In this mode, a symbolic form of the
2D Hamiltonian $H(k_x,k_y,z)$ is constructed. Subsequently, the operators
$\hat{k}_x$ and $\hat{k}_y$ are substituted by $a$ and $a^\dagger$ following
Eq.~\eqref{eqn_ll_ladder_operators}. The block $H^{\mathrm{ax},(n)}$ is obtained
by applying the ladder operators which yields factors involving $n$. In detail,
these steps are performed as follows:
\begin{itemize}
 \item The Taylor expansion of the Hamiltonian in $k_x$ and $k_y$ up to quadratic
 order is
 \begin{align}
   H(k_x,k_y,z)
   &= H(0,0,z) + k_x(\partial_{k_x}H)(0,0,z) + k_y(\partial_{k_y}H)(0,0,z)\nonumber\\
   &\qquad{}+ \tfrac{1}{2} k_x^2(\partial^2_{k_x}H)(0,0,z)
   + \tfrac{1}{2} k_y^2(\partial^2_{k_y}H)(0,0,z)
   + k_xk_y(\partial_{k_x}\partial_{k_y}H)(0,0,z).
   \label{eqn_hsym_taylor}
 \end{align}
 We evaluate $H_0(z)\equiv H(0,0,z)$ and its first and second degree derivatives at
 $(k_x,k_y) = (0, 0)$.
 We calculate the derivatives as
 \begin{align}
   H_x(z) &= (\partial_{k_x}H)(0,0,z) = \frac{H(\Delta k_x,0,z) - H(-\Delta k_x,0,z)}{2\Delta k_x}\nonumber \\
   H_y(z) &= (\partial_{k_y}H)(0,0,z) = \frac{H(0,\Delta k_y,z) - H(0,-\Delta k_y,z)}{2\Delta k_y}\nonumber \\
   H_{xx}(z) &= (\partial_{k_x}^2H)(0,0,z) =
     \frac{H(\Delta k_x,0,z) -2H(0,0,z) + H(-\Delta k_x,0,z)}{2(\Delta k_x)^2}\\
   H_{yy}(z) &= (\partial_{k_y}^2H)(0,0,z) =
     \frac{H(0,\Delta k_y,z) -2H(0,0,z) + H(0,-\Delta k_y,z)}{2(\Delta k_y)^2}\nonumber\\
   H_{xy}(z) &= (\partial_{k_x}\partial_{k_y}H)(0,0,z)\nonumber\\
   &=
   \frac{H(\Delta k_x,\Delta k_y,z) - H(\Delta k_x,-\Delta k_y,z)
   - H(-\Delta k_x,\Delta k_y,z)+H(-\Delta k_x,-\Delta k_y,z)}{\Delta k_x\Delta k_y}.\nonumber
 \end{align}
 Here, the values of $\Delta k_x$ and $\Delta k_y$ can be set arbitrarily,
 because the Hamiltonian is quadratic and does not contain higher-order terms.
 \item The symbolic Hamiltonian can be formed by simply substituting
 $k_x\to \hat{k}_x$ and $k_y\to \hat{k}_y$ in Eq.~\ref{eqn_hsym_taylor},
 \begin{equation}
   H(\hat{k}_x,\hat{k}_y,z)
   = H_0(z) + \hat{k}_x H_x(z) + \hat{k}_yH_y(z)
   + \tfrac{1}{2} \hat{k}_x^2H_{xx}(z) + \tfrac{1}{2} \hat{k}_y^2H_{yy}(z)
   + \tfrac{1}{2}\{\hat{k}_x,\hat{k}_y\}H_{xy}(z).
 \end{equation}
 Note that the order of the operators is important.
 To simplify the substitution of ladder operators, we first transform to
 $\hat{k}_\pm = \hat{k}_x\pm\ii\hat{k}_y$. We thus obtain the symbolic
 Hamiltonian
 \begin{align}
   H(\hat{k}_+,\hat{k}_-,z)
   &= H_0(z) + \hat{k}_+ H_+(z) + \hat{k}_-H_-(z)\nonumber\\
   &\qquad{}+ \tfrac{1}{2} \hat{k}_+^2H_{++}(z) + \tfrac{1}{2} \hat{k}_-^2H_{--}(z)
   + \tfrac{1}{2} \hat{k}_+\hat{k}_-H_{+-}(z) + \tfrac{1}{2} \hat{k}_-\hat{k}_+H_{-+}(z).\label{eqn_hsym}
 \end{align}
 with $H_\pm = \frac{1}{2}(H_x\mp H_y)$,
 $H_{\pm\pm} = \frac{1}{2}(H_{xx} - H_{yy} \mp 2\ii H_{xy})$,
 and $H_{+-} = H_{-+} = \frac{1}{2}(H_{xx}+H_{yy})$.
 \item The symbolic Hamiltonian of Eq.~\eqref{eqn_hsym} is implemented as a
 \texttt{SymbolicMatrix} object, which is essentially a wrapper around an
 operator sum represented by the \texttt{dict}
\begin{verbatim}
self.opsum = {
    "": h0, "+": hkp, "-": hkm,
    "++": hkpkp, "--": hkmkm, "+-": hkpkm, "-+": hkpkm
}
\end{verbatim}
 where \texttt{self} refers to the \texttt{SymbolicMatrix} instance. The keys of
 the operator sum \texttt{dict} are strings that represent combinations of
 operators $\hat{k}_\pm$; the values are the $8n_z \times 8n_z$ matrices $H_0$,
 $H_+$, $H_-$, etc.
 \item The \texttt{SymbolicMatrix} class has a member function
 \verb+SymbolicMatrix.ll_evaluate()+ that effectively evaluates matrix elements
 of the form $\bramidket{\Phi^{(n')}}{H(a,a^\dagger,z)}{\Psi^{(n)}}$,
 cf.\ Eq.~\eqref{eqn_ll_eigenstate}. (For the axial model, $n'=n$.)
 It acts with the ladder operators on the Landau level states $\ket{n+\delta n}$
 ($\delta n=-1,0,1,2$) and replaces them by the appropriate values (that depend
 on $n$ and the magnetic field $B_z$).
 The matrix elements are calculated for each term in the operator sum separately.
 The result is the sum over these terms. It represents $H^{\mathrm{ax},(n)}$ as
 an $8 n_z\times 8n_z$ matrix. (Except for $n=-2,-1,0$, where the number of
 orbitals is $1,4,7$, respectively, instead of $8$.)
\end{itemize}
We note that the construction of the \texttt{SymbolicMatrix} object needs to be
done only once. This object can then be evaluated repeatedly by calling
\verb+ll_evaluate()+ for each $n=-2,-1,0,\ldots,n_\mathrm{max}$ and for multiple
values of $B_z$.

\subsubsection{Landau levels in `full' mode}

In absence of axial symmetry, the Landau level index $n$ fails to be a conserved
quantum number. As argued in Sec.~\ref{sec_ll_formalism}, the nonaxial terms
$R^\mathrm{nonax}$ in the \kdotp{} Hamiltonian couple Landau levels $n'$ and $n$
with $n'-n=\pm 4$. (That is, $\bramidket{\Phi^{(n')}}{H}{\Psi^{(n)}}$ is nonzero
for $n'-n=-4,0,4$.) For lower symmetry, there may be other nonzero matrix elements
as well for other combinations of $n'$ and $n$ with $\abs{n'-n}\leq 4$.

The `full' Landau level Hamiltonian contains the Landau-level degrees of freedom
in the Hilbert space. The matrix structure of the Hamiltonian can thus be
expressed as $H_{n',i,p;n,j,q}(B)$, with $n',n$ being the Landau level indices,
$i,j$ the $z$ coordinates, and $p,q$ the orbitals. This matrix is constructed
as a sparse matrix as follows:
\begin{itemize}
 \item Initialize lists \texttt{allrows}, \texttt{allcols}, \texttt{allvals}.
 \item Iterate over the Landau indices $n=-2,-1,0,\ldots,n_\mathrm{max}$.
 For each index $n$, iterate over $n'=n,\ldots,n+4$ with $n'\leq n_\mathrm{max}$.
 Evaluate the $(n',n)$ block of the symbolic Hamiltonian as
\begin{verbatim}
ham = hsym.ll_evaluate((nprime, n), magn, ...)
\end{verbatim}
 This yields a (sparse) matrix of dimension
 $n_\mathrm{orb}(n')n_z\times n_\mathrm{orb}(n)n_z$,
 where $n_\mathrm{orb}(n)$ is the number of orbitals for Landau level index $n$,
 i.e., $1$, $4$, $7$, or $8$ for $n=-2$, $-1$, $0$, and $n\geq1$, respectively.
 \item If the block \texttt{ham} is nonzero (i.e., \texttt{ham.nnz > 0}), 
 convert it to a sparse matrix \texttt{hamcoo} in \texttt{COO} format and extract
 its values, row indices, and column indices.
 The block is added the full matrix by copying its values and taking the indices
 shifted by the correct index offsets
\begin{verbatim}
allvals.append(hamcoo.data)
allrows.append(index_offsets[nprime + 2] + hamcoo.row)
allcols.append(index_offsets[n + 2] + hamcoo.col)
\end{verbatim}
 where \verb+index_offsets+ is a (precalculated) array of index offsets, defined
 as the cumulative number of degrees of freedom $n_\mathrm{orb}(n')n_z$ for all
 Landau levels $n'<n$. The values of\linebreak[4] \verb/index_offsets[n + 2]/ are equal to
 $0, n_z, 5n_z, 12n_z$ for $n = -2,-1,0,1$ and $(4+8n)n_z$ for $n\geq 1$.
 \item Turn \texttt{allvals}, \texttt{allrows}, \texttt{allcols} into one-dimensional
 arrays and create the sparse matrix
\begin{verbatim}
s = coo_matrix(
    (allvals, (allrows, allcols)), shape = (dim, dim), dtype = complex
)
\end{verbatim}
 where $\mathtt{dim} = (12+8n_\mathrm{max})n_z$. Convert it to \texttt{CSC} format
 and return the result.
\end{itemize}
This algorithm is implemented as the function \verb+hz_sparse_ll_full()+ in the
submodule\linebreak[4] \verb+hamiltonian.hamiltonian+.

\subsubsection{Transformable Hamiltonian}
\label{sec_transf_ham}


The default way of constructing the $8\times 8$ blocks is by direct definition
of the matrices by element. These built-in definitions are appropriate for crystals
grown along the (001) direction, the most common direction used in crystal growth.
Unfortunately, constructing the Hamiltonian for a generic growth direction from
the Hamiltonian for (001) is not straightforward \cite{PfeufferJeschke2000_thesis}.

In \kdotpy, we follow a different approach and construct the Hamiltonian bottom up.
The idea is that the terms in the Hamiltonian are written as products of momenta
$k_i$ and angular momentum matrices $\sigma_j$, $J_j$, $T_j$, $T_{lm}$ ($i,j,l,m=x,y,z$), similar to Table C.5 in the book by Winkler \cite{Winkler2003_book}.
Each quantity has known transformation rules given by a representation of the
group $\mathrm{SO}(3)$ of the proper rotations of three-dimensional space.
\begin{itemize}
 \item Trivial representation: Some terms, like $k_x\sigma_x + k_y\sigma_y+k_z\sigma_z$
 in the $H_{66}$ block, are invariant under all $\mathrm{SO}(3)$ rotations.
 \item Vector representation: The irreducible three-dimensional representation
 of $\mathrm{SO}(3)$ describes the transformations of vector-like quantities. Examples are
 momentum $(k_x,k_y,k_z)$, and angular momentum $(\sigma_x,\sigma_y,\sigma_z)$ and
 $(J_x,J_y,J_z)$.
 \item Five-dimensional representation: The five-dimensional irreducible
 representation of $\mathrm{SO}(3)$ is a term in the symmetric product of two vector
 representations. The quintuplet
 \begin{equation}
 (2T_{yz}, 2T_{zx}, 2T_{xy}, T_{xx}-T_{yy}, (2T_{zz} - T_{xx} - T_{yy})/\sqrt{3})
 \end{equation}
 (in the $H_{68}$ block) transforms in this representation.
\end{itemize}
Inversion does not exist in $\mathrm{SO}(3)$, hence we do not distinguish axial vectors
from regular vectors.

We use this information define the transformation between the
\emph{lattice coordinates}, related to the crystal lattice,
and the \emph{device coordinates}, attached to the geometry of the device.
In this section, we denote the device (or sample) coordinates as $(x,y,z)$. The
$z$ direction is by definition the growth direction. For a strip geometry,
the longitudinal direction (in which momentum is defined) is $x$ and the
transversal direction (confined) is $y$. For lattice coordinates, we use the primed
symbols $(x',y',z')$. The `primed' axes are aligned with the axes of the crystal
lattice. The two coordinate systems are related by a pure rotation, $(x',y',z')=R(x,y,z)$
where $R$ is a matrix in $\mathrm{SO}(3)$, i.e., an orthogonal $3\times 3$ matrix
with determinant $+1$. Note that for the growth direction (001), the two coordinate
systems coincide; in other words, $R$ is the identity transformation.

The definition of the Hamiltonian is in terms of the lattice coordinates $(x',y',z')$,
because the electrons are subject to the (local) crystal environment. In order
to calculate the band structure for a non-trivial lattice orientation, the Hamiltonian
defined in terms of momenta $k_i$ and angular momenta $J_j$ is transformed into
device coordinates $(x,y,z)$ using the transformation matrix $R$. The following
example illustrates this principle: Take the $\gamma_3$ term in the $\Gamma_8$
block [cf.\ Eq.~\eqref{eqn_ham_kp}], which can be written as
\begin{equation}\label{eqn_hamtransform_example}
 -\gamma_3[\{J_{x'},J_{y'}\}\{k_{x'},k_{y'}\} + \{J_{y'},J_{z'}\}\{k_{y'},k_{z'}\} + \{J_{z'},J_{x'}\}\{k_{z'},k_{x'}\} ]
\end{equation}
where ${A,B}=AB+BA$ and we have explicitly indicated that it is defined in the
primed (crystal) coordinate system. The unprimed version of this term is then
found by setting
\begin{equation}
(J_{x'},J_{y'},J_{z'}) = R (J_{x},J_{y},J_{z})
\qquad\text{and}\qquad
(k_{x'},k_{y'},k_{z'}) = R (k_{x},k_{y},k_{z}).
\end{equation}
The Hamiltonian is stored in tensor form, where each term is stored as a separate
tensor. The term in this example is encoded as tensor $T_{ij;kl}$; the
relation Hamiltonian term is then found by setting $T_{ij;kl}k_i k_j J_k J_l$. 
(Einstein summation convention assumed.) The transformation to device coordinates
is then given by
\begin{equation}
T_{ij;kl} \to T'_{ij;kl} = R_{ia} R_{jb} R_{kc} R_{ld} T_{abcd},
\end{equation}
This transformation is done for all terms in the Hamiltonian separately.

For the implementation in \kdotpy, each term in the Hamiltonian is represented
by instances of the \texttt{KJTensor} class. These are then transformed and 
evaluated by substituting the angular momentum matrices $J_j$. The result is
a \texttt{KTermsDict} instance that encodes sum of terms $M_{ij} k_i k_j$ where
$M$ is a matrix. In detail, the algorithm for constructing and
transforming the Hamiltonian is as follows:

\begin{itemize}
 \item For each term, a \texttt{KJTensor} is initialized through a \texttt{dict}
 which defines the tensor components. The example term of
 Eq.~\eqref{eqn_hamtransform_example} is defined as
\begin{verbatim}
g3_tens = KJTensor(
    {'yzyz':1, 'xzxz':1, 'yxyx':1}, nk = 2
).symmetrize((0,1), fill = True).symmetrize((2,3), fill = True)
\end{verbatim}
 \item The transformation is then done by calling \texttt{KJTensor.transform(R)}
 for each term in the Hamiltonian. This yields the transformed version of the
 \texttt{KJTensor} for each term. Terms that are known to be invariant are
 left alone:
\begin{verbatim}
for tens, tens_name in zip(all_tens, tens_names):
    invariant = tens.is_invariant_under_transform(params.lattice_trans)
    if not invariant:
        tens.transform(params.lattice_trans, in_place = True).chop() 
\end{verbatim}
 \item Subsequently, calling \verb+KJTensor.apply_jmat()+ on the transformed
 \texttt{KJTensor} instances turns them into matrix form by substituting the
 spin matrices $J_j$ for their matrix representations. This returns a
 \texttt{KTerms} instance, which encodes the sums of the
 form $M_{ij} k_i k_j$ where $M$ is a matrix. Multiple \texttt{KTerms} instances
 are collected in the \texttt{KTermsDict} instance \texttt{kterms}. For the
 example term [Eq.~\eqref{eqn_hamtransform_example}
\begin{verbatim}
kterms = KTermsdict()
...
kterms['g3_88'] = \
    g3_tens.apply_jmat(spin.j3basis, symmetrize_k = True).chop()
\end{verbatim}
Here, the term labelled \verb+{'g3_88'+ in \texttt{kterms} is a \texttt{KTerms}
instance defined by substituting the `$J$ basis' (simply
the set of matrices $J_x, J_y, J_z$) on the already transformed \texttt{KJTensor}
labelled \verb+g3_tens+. We enforce symmetrization in the momentum components
and use \texttt{.chop()} to chop off almost-zero values.

\end{itemize}

A non-trivial lattice orientation can be requested from the command-line by
using the argument \texttt{orientation} with one to three further parameters that
define the transformation matrix $R$ (see Appendix~\ref{app_orientation} for
details). If this is done, the \texttt{KTermsDict} object is constructed as
described above and passed to the Hamiltonian block construction function as
argument \texttt{kterms}. These functions are then responsible for
composing the Hamiltonian from all different terms, and for substituting the
momentum components $k_i$ as usual: By discretized derivatives in the confined
directions and by momentum values in the unconfined directions.

The orientation may be set as \texttt{orientation - 001} for the default growth
direction. This applies the formalism described here with the rotation
matrix $R$ being the identity. In absence of an \texttt{orientation} (or an equivalent)
argument, the regular construction method of the Hamiltonian is used. Comparison
between the two is useful for verification that the two representations of the
Hamiltonian yield the same result. We do not use the transformable representation
of the Hamiltonian by default, because the regular construction method is
significantly faster.

\subsection{Diagonalization}
\label{sec_diagonalization}

\subsubsection{Default eigensolver: Arnoldi with shift-and-invert}
\label{sec_default_eigensolver}

Diagonalization is the process of finding the eigenvalues and eigenvectors of
a square matrix $M$. This is an ubiquitous numerical problem, for which several
algorithms have been developed over the years. Perhaps the most well-known
algorithm adapted to large sparse matrices is the Lanczos algorithm, that finds
a subset of the eigenvalues (typically the ones largest in magnitude) and the 
corresponding eigenvectors for a hermitian matrix $M$ \cite{Lanczos1950}.
The matrix $M$ is applied iteratively
to a set of independent vectors $v_j$ until sufficient convergence is reached.
The idea is that matrix-vector multiplication is a relatively cheap operation if
the matrix $M$ is sparse. The naive implementation of this method is
numerically unstable, but it can be adapted to yield reliable results \cite{OjalvoNewman1970}.
The \emph{Arnoldi iteration} method extends the Lanczos method to generic
complex matrices. The \emph{implicitly restarted Arnoldi method} is a commonly
applied method thanks to its implementation in the popular ARPACK package
\cite{ARPACK}, that has been linked to by many software packages, including the 
SciPy library for Python.

In order for these methods to be usable for \kdotpy, we need to make an additional
intermediate step. The Lanczos and Arnoldi methods find the eigenvalues with the
largest magnitude, yet for band structures, we are typically most interested in
the eigenvalues close to the charge neutrality point. Instead of finding the
eigenvalues of the Hamiltonian $H$ itself, we apply the \emph{shift-and-invert}
algorithm and apply Arnoldi/Lanczos to the matrix
\begin{equation}\label{eqn_shift_invert}
  M = (H - \sigma I)^{-1}
\end{equation}
where $I$ is the identity matrix and $\sigma$ is the \emph{target energy}.
Indeed, finding the $n_\mathrm{eig}$ largest eigenvalues of $M$ is equivalent to
finding the $n_\mathrm{eig}$ eigenvalues of $H$ with smallest distance to $\sigma$.
Implementations of the shift-and-invert algorithm generally do not store the
matrix $(H - \sigma I)^{-1}$ in memory, as it is usually not sparse even if $H$ is.
Instead, as it only needs to calculate matrix-vector products of the form
$A^{-1} \vec{b} = \vec{x}$, it solves the equivalent system of equations
$A \vec{x} = \vec{b}$. This can be solved efficiently with a sparse version of
LU factorization.

In \kdotpy, the default \emph{eigensolver} is the function \texttt{eigsh()} from
the SciPy module\linebreak[4] \texttt{scipy.sparse.linalg}. In essence, this SciPy function
is just an interface to the ARPACK implementation. For the shift-and-invert step,
it uses the SuperLU library.
The function \texttt{eigsh()} is applied to the Hamiltonian matrix
in \texttt{CSC} sparse format. The number of eigenvalues $n_\mathrm{eig}$ is set
by the command line using argument \texttt{neig}.
The \texttt{eigsh()} argument \texttt{sigma} is set to the target energy,
given by \texttt{targetenergy} on the command line.

The diagonalization step is usually the computationally heaviest part of \kdotpy{}
calculations, and as such the overall performance of the program strongly depends
on this step. In order to maximize performance, \kdotpy{} applies parallelization
to reduce calculation time by making use of all available CPU resources. In
addition, custom eigensolvers can increase performance by providing optimized
methods for the available hardware, for example using the Intel MKL framework
for optimization on supported CPUs or using CUDA for GPU acceleration; see  Appendix~\ref{app_eigensolvers} for a detailed discussion.

\subsubsection{Parallelization of the main loop}
\label{sec_parallelization_main}

In \kdotpy{}, calculation of the eigenvalues at each value of momentum $k$
and/or magnetic field $B$ constitutes a separate eigenvalue problem, and can be
treated as an independent task. For each type of Hamiltonian (1D, 2D, LL, etc.),
the \texttt{diagonalization} module provides a diagonalization function that
performs the following steps for a single value of $k$ and/or $B$:
\begin{itemize}
 \item Construct the Hamiltonian;
 \item Diagonalize the Hamiltonian (apply eigensolver);
 \item Process the eigenstates (e.g., calculate observables and Berry curvature).
\end{itemize}
This function returns a single \texttt{DiagDataPoint} instance. 
For a dispersion, the complete process of finding a band structure can be
summarized as
\begin{verbatim}
datapoints = [diag_function(k, *args, **kwds) for k in kvalues]
data = DiagData(datapoints, grid=kvalues)
\end{verbatim}
where \texttt{kvalues} is the \texttt{VectorGrid} instance containing the $k$ values.
The iterative application of \verb+diag_function+ is very suitable for
parallelization, because each call to \verb+diag_function+ requires a similar
amount of resources.

Unfortunately, parallelization in Python is challenging as a result of the
global interpreter lock (GIL). However, Python's \texttt{multiprocessing} module
avoids this obstacle by running Python code in parallel subprocesses,
each of which is affected only by its own GIL.

The basic form of parallelization in \kdotpy{} is facilitated by the
\verb+parallel_apply()+ function in the \texttt{parallel} submodule, that is
based on the \verb+Pool.apply_async()+ method of Python's \texttt{multiprocessing}
module. The iteration in the example above is replaced by 
\begin{verbatim}
datapoints = parallel_apply(
    diag_function, kvalues, args, kwds, num_processes=num_cpus
)
data = DiagData(datapoints, grid=grid)
\end{verbatim}
where \verb+num_cpus+ is the number of CPU cores in the system, or a custom value
set by the command line argument \texttt{cpu}. The function \verb+parallel_apply()+
also implements a custom \emph{signal handler} for proper handling of keyboard
interrupts (Ctrl-C by the user) and terminate and abort signals. For details,
refer to Appendix~\ref{app_parallelization}.

The \verb+parallel_apply()+ method is adequate for the default \texttt{eigsh}
solver from SciPy, where the three steps of each diagonalization task run on the
same CPU. This approach is no longer adequate when a GPU-accelerated eigensolver
is used: While the application of the eigensolver is done by the GPU, the other
steps are CPU bound. The need for additional flexibility has motivated us to
implement a parallelization framework with a higher level of abstraction.
This \emph{Task-Model framework} is built around the \texttt{Model\emph{X}},
\texttt{Task}, and \texttt{TaskManager} classes. Each \texttt{Model\emph{X}} class
(where \texttt{\emph{X}} can be \texttt{1D}, \texttt{2D}, \texttt{LL}, etc.)
implements the recipe analogous to the diagonalization function. It defines
the steps as separate functions so that they can be run as separate tasks.
The \texttt{Task} class is a wrapper around each such task. The 
\texttt{TaskManager} initializes, manages, and closes the process and/or
thread pools, and thus plays a similar role as \verb+parallel_apply()+.
Like the parallelization framework with \verb+parallel_apply()+, the Task-Model
framework is built around the \verb+Pool.apply_async()+ method of Python's
\texttt{multiprocessing} module. The benefit of using the Task-Model framework
is the higher degree of flexibility and configurability. We discuss the
implementation details in Appendix~\ref{app_parallelization}.

\subsection{Post-diagonalization}

The eigenvectors obtained from the diagonalization have as many components as the
size of the Hamiltonian matrix, and thus they require a substantial space to
store them into memory or on disk. The default approach of \kdotpy{} is to extract
information from the eigenvectors immediately after each point is diagonalized,
and to discard the eigenvectors themselves upon creating a \texttt{DiagDataPoint}
instance, i.e., usually the attribute \texttt{DiagDataPoint.eivec} remains \texttt{None}.
 
\subsubsection{Observables}
The \texttt{DiagDataPoint} class has a separate attribute \texttt{DiagDataPoint.obsvals}
which stores a two-dimensional array with the observable values for each eigenstate
in the \texttt{DiagDataPoint} instance. The attribute \verb+DiagDataPoint._obsids+
contains the labels for the observables stored in\linebreak[4] \texttt{DiagDataPoint.obsvals}.
The \texttt{DiagDataPoint} class also defines several functions for getting the
values for one or more observables for all eigenstates in the instance.

The observables are defined in \texttt{observables.py}. Each one is an instance
of the \texttt{Observable} class. The key attributes in this object are
\begin{itemize}
 \item \texttt{obsid}: The observable label.
 \item \texttt{obsfun}: A function that calculates the values from the eigenvectors.
 \item \verb+obsfun_type+: Determines the argument pattern passed to this function.
\end{itemize}
Further properties determine the textual representation of the observable, its
unit, as well as the colour map used in plots.

Each observable has two variants, namely a dimensionless and a dimensionful one.
For example, the expectation value along the growth axis $\langle z \rangle$ is
a dimensionful value, while its dimensionless partner is $\langle z \rangle/d$
where $d$ is the total thickness of the layer stack. The \texttt{obsfun} function
typically calculated the dimensionless variant. The \texttt{Observable} instance
also contains attributes for the conversion between the two, and both the
dimensionless and dimensionful variants have their own textual representations
for quantity and unit.

The source file \texttt{observables.py} also defines an \texttt{ObservablesList}
class with \verb+all_observables+ as its single instance. This instance
contains all the \texttt{Observable} objects with the observable definitions.
The output functions import \verb+all_observables+ in order to access the relevant
properties. An overview of all observables defined in \texttt{observables.py}
can be found in Appendix~\ref{app_observables_reference}.

\begin{figure}
 \includegraphics[width=95mm]{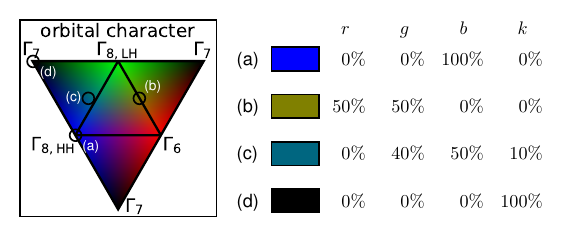}
 \caption{The legend of figures with \texttt{obs orbitalrgb} as command-line input.
 The red, green, and blue colour channels $(r,g,b)$ indicate the expectation
 values of the \texttt{gamma6}, \texttt{gamma8l}, and \texttt{gamma8h} observables,
 respectively. For example, blue (a) represents a pure heavy-hole state and
 dark yellow (b) an equal mixture of $\ket{\Gamma_6,\pm\frac{1}{2}}$ and
 $\ket{\Gamma_8,\pm\frac{1}{2}}$ states. The blackness value $k = 1-r-g-b$ is
 implied, and represents the \texttt{gamma7} observable. We show examples with
 $k=10\%$ (c) and pure black $k=100\%$ (d), where the latter represents a pure
 $\ket{\Gamma_7,\pm\frac{1}{2}}$ state.
 }
 \label{fig_rgblegend}
\end{figure}

All observable values saved in \texttt{DiagDataPoint.obsvals} are written to the
output files. If the command line argument \texttt{obs \emph{obsid}} is provided,
then that observable is used to generate extra output files, e.g., csv files
with just the values of that observable as function of momentum, and to colourize
the curves in a dispersion or magnetic-field dependence plot. The appropriate
colour scale is determined by the definitions in the appropriate \texttt{Observable}.
Depending on the type of observable, a discrete or continuous set of colours is
used. It is also possible to combine observables into a dual colour scale or an
red-green-blue (RGB) colour map. In particular, \texttt{obs orbitalrgb} visualizes
the orbital content of the states by using colours where the red, green, and blue
channels ($(r,g,b)$ with normalized values on the interval $[0,1]$) represent the
\texttt{gamma6}, \texttt{gamma8l}, and \texttt{gamma8h} observables, respectively;
the \texttt{gamma7} observable is implicitly represented as the `blackness'
$1 - r - g - b$ of the colour. Figure~\ref{fig_rgblegend} illustrates the legend
of figures with \texttt{obs orbitalrgb} with a few example colours.

\subsubsection{Overlaps}

The overlap between eigenvectors at different momenta or magnetic-field values
is often useful to identify states in complicated dispersions or Landau fans.

For dispersions in the two-dimensional geometry (\texttt{kdotpy 2d}), the
eigenvectors $\ket{\psi_i(0)}$ of the subbands at $\vec{k}=0$ are stored when
the extra command line argument \texttt{overlaps} is provided. Then,
for each momentum $\vec{k}$, \kdotpy{} calculates the subband overlaps
$|\braket{\psi_i(0)}{\psi_j(\vec{k})}|^2$ for all eigenstates $\ket{\psi_j(\vec{k})}$.
This quantity can be interpreted as the expectation values of the projection
operator
\begin{equation}
  O^{(i)} = \ket{\psi_i(0)}\bra{\psi_i(0)}
\end{equation}
on the eigenstates $\ket{\psi_j(\vec{k})}$.
The same also works for Landau level calculations (\texttt{kdotpy ll}), where
the overlaps are calculated with the subbands $i$ at zero magnetic field.

The values of the subband overlaps are stored in \texttt{DiagDataPoint.obsvals}
like any other observable and thus also written to the output files.
If one provides \texttt{obs subbandrgb} on the command line, the plot colour of
the dispersions is defined as the RGB triplet
\begin{equation}
  (r,g,b) = (\avg{O^{E1+}}+\avg{O^{E1-}},\avg{O^{H1+}}+\avg{O^{H1-}},\avg{O^{H2+}}+\avg{O^{H2-}}),
\end{equation}
where the red, green, and blue channels are normalized values on the interval $[0,1]$.
Likewise, \texttt{obs subbande1h1l1} does the same with the H2$\pm$ subbands
replaced by L1$\pm$. A larger number of subband labels may also been given, e.g.,
\texttt{obs subbande1h1h2l1}, where each label refers to a pair of subbands with
opposite spin states, like $\avg{O^{E1+}}+\avg{O^{E1-}}$. For four labels, the
colours are mixed from red, yellow, green, and blue. For $n_l>4$ labels,
the colours are mixed from $n_l$ equidistant hues (maximally saturated colours).

In Landau level mode, the command line argument \texttt{lloverlaps} can be used
in order to calculate the probability density $\braket{n}{\psi}$ in the Landau
levels $n=-2,-1,0, \ldots, n_\mathrm{max}$. These quantities can also be viewed
as expectation values of the projection operators $\ket{n}\bra{n}$. This option
is available in the full LL mode only; in the symbolic LL mode, the index $n$ is
a conserved quantum number, so that the expectation value is always $1$ for one
index and $0$ for all others.

\subsubsection{Berry curvature, Chern numbers, Hall conductance}
\label{sec_berrychern_implementation}

The calculation of the Berry curvature in dispersion mode is based on
Eq.~\eqref{eqn_berrycurvature3} (see Sec.~\ref{sec_berry}), with the slight
modification that the summation
over $\ket{\psi}$ includes all eigenstates within the energy window that has been
calculated, which may be a subset of the full spectrum. Due to the denominator
in Eq.~\eqref{eqn_berrycurvature3}, states $\ket{\psi'}$ that are far in energy
from $\ket{\psi}$ only contribute weakly to the Berry curvature. Nevertheless,
one should be aware that the omission of remote states can lead to deviations of
the calculated Chern numbers from integer values.

The implementation in \texttt{berry.py} involves numerical evaluation of the 
derivatives $\partial_{k_x}H$ and $\partial_{k_y}H$ of the Hamiltonian. These
are used to evaluate the matrices $V_i$ ($i=x,y,z$) with
\begin{equation}\label{eqn_berrycurvature_v}
 (V_i)_{pq} = \frac{\bramidket{\psi_p}{\partial_{k_i}H}{\psi_q}}{E_p-E_q}
\end{equation}
if $p\not=q$ and $V_{pp}=0$. Here, $p$ and $q$ label the eigenstates and they
run over all eigenstates obtained from the diagonalization. In terms of $V_i$,
Eq.~\eqref{eqn_berrycurvature3} may be evaluated as
\begin{equation}
 \Omega^{q}_i(\vec{k})
 = -\Im \sum_{p}  \sum_{l,m} \epsilon_{ilm} (V_l)^*_{qp}(V_m)_{pq}
 = -\Im \sum_{l,m} \epsilon_{ilm} (V_l^\dagger V_m)_{qq}
\end{equation}
The summation over $p$ is a matrix multiplication that can be evaluated efficiently
with NumPy functions. Because we only need the diagonal elements of $V_l^\dagger V_m$,
we extract the relevant columns and rows from $V_l^\dagger$ and $V_m$ and
evaluate vector-vector dot products as
\begin{verbatim}
bcurv = [
    -np.imag(np.dot(vxd[q, :], vy[:, q]) - np.dot(vyd[q, :], vx[:, q]))
    for q in range(0, neig1)
]
\end{verbatim}
for evaluating $\Omega^{q}_z(\vec{k})$ for all eigenstates $q$. In view of
efficiency, we can choose to restrict the set of eigenstates $q$ for which we
evaluate the Berry curvature. It is important to note that the set of eigenstates
$p$ in the summation (or vector-vector product) must be a large as possible as to 
not lose significant contributions from remote eigenstates.

In two-dimensional dispersion mode (\texttt{kdotpy 2d}), only $\Omega^{q}_z(\vec{k})$
is evaluated, as the Berry curvature is essentially scalar in two dimensions. The 
Berry curvature observable may be accessed with the argument \texttt{obs berry}
or \texttt{obs berryz} (equivalent).
For three dimensions (\texttt{kdotpy bulk}), the Berry curvature is a vector quantity
$\Omega^{q}_i(\vec{k})$ with $i=x,y,z$. The relevant observable labels are
\texttt{berryx}, \texttt{berryy}, and \texttt{berryz}.

For Landau levels, Berry curvature $\Omega^q$ is calculated in the same manner as for
dispersions, with the exception of how the Hamiltonian is treated: In the symbolic
and full LL modes, the derivatives $\partial_{k_i}H$ of the Hamiltonian are calculated
symbolically. The remaining factors $k_x$ and $k_y$ are substituted by the
appropriate ladder operators $a$ and $a^\dagger$ before $\partial_{k_i}H$ is
inserted into into Eq.~\eqref{eqn_berrycurvature_v}.

The Chern number $C^q$ of each eigenstate $q$ is obtained by multiplying the Berry
curvature by the Landau level degeneracy,
\begin{equation}\label{eqn_chern_ll}
  C^q = \frac{e B_z}{\hbar}\Omega^q
\end{equation}
in terms of the perpendicular magnetic field $B_z$, cf.\ Sec.~\ref{sec_chern_hall}.
The degeneracy factor is
related to the magnetic length $\lB$ as $e B_z/\hbar = 1/\lB^2$. The Chern number
from Eq.~\eqref{eqn_chern_ll} is stored as observable labelled \texttt{chern}.
These values can be used to find the Hall conductance $\sigmaH$ by summing over
all occupied states, see Eq.~\eqref{eqn_bulk_boundary}.

Due to the spectrum being incomplete, the Chern numbers calculated with
Eq.~\eqref{eqn_chern_ll} may deviate from integer values. In addition to
observable \texttt{chern}, \kdotpy{} also provides the observable \texttt{chernsim}
which `simulates' exactly integer Chern numbers. The value is simply $1$ for all
states if $B_z\not=0$ and $0$ if $B_z=0$. We note that this simple assumption
violates the property that the sum of all Chern numbers must vanish, hence the
resulting Hall conductance [from Eq.~\eqref{eqn_bulk_boundary}] is valid only in
the part of the spectrum close to the charge neutrality point.

\subsection{Band alignment}
\label{sec_bandalign}


\subsubsection{Motivation}

For all calculation modes except for the bulk mode, \kdotpy{} uses sparse
diagonalization which yields only a subset of all eigenvalues and eigenvectors.
Thus, it is not possible to label the eigenvectors from $1$ to $N$ (the matrix size)
from smallest to largest eigenvalue. Given two sets of eigenvalues at two
points in momentum space, it is not known a priori how to connect the
eigenvalues as to form a (sub)band.

\begin{figure}[h]
 \includegraphics{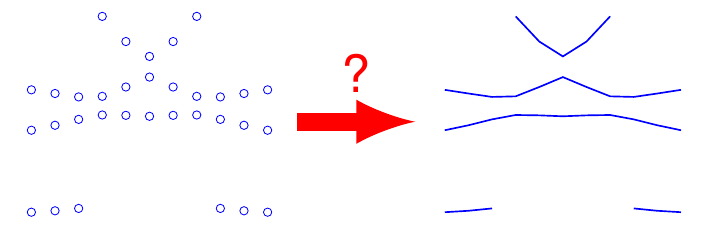}
 \caption{Illustration of the problem that is solved by the band alignment algorithm.}
\end{figure}

The band alignment algorithm \emph{connects the dots}: Considering two adjacent
points in momentum space (or in magnetic-field value), assign a `band index' to
each eigenvalue. States with identical band indices across momentum point are
considered to be one connected band. This band index is used by many other
functions of \kdotpy, for example, plot and CSV output, calculation of density
of states, dispersion derivatives, etc.

\subsubsection{Band indices definition}

Before we look at the actual recipe, let us discuss in detail the properties and
the role of the band indices. They satisfy the following basic properties:

\begin{itemize}
\item The band indices are nonzero integers.

\item The band indices are always monotonic in the
energy (eigenvalues). As a consequence, dispersions never cross (e.g., in the
output plots). Whereas this is not always the most natural choice from the perspective
of the eigenstates, the assumption of monotonicity greatly simplifies the band
alignment algorithm. The advantage of this method is
that it does not need the eigenvectors, so that the RAM usage can remain limited.
Earlier attempts with use of the eigenvectors have proved to be much less
reliable than the present implementation.

\item Additionally, \kdotpy{} tries to determine the charge neutrality point from
the band characters at zero momentum and zero magnetic-field. At this point,
states above this energy get positive band indices ($1, 2, 3, \ldots$) and those
below get negative indices ($-1, -2, -3, \ldots$). The index $0$ is never assigned.
Thus, the sign of the band index is used to determine `electron-like' or
`hole-like' states. (Here, the notions `electron' and `hole' are defined with
respect to the band energy with respect to charge neutrality.)

\item For Landau-level calculations in the axial approximation, the band indices are
calculated for each Landau level index separately. In this mode, states are
labelled by a pair of indices (LL index, band index). This is not true for full
LL mode, where \kdotpy{} uses a single band index similar to dispersion mode.

\item The band indices are quite essential for many functions to do their job.
For example, in order to calculate the derivative of the dispersion, one needs
two points of the same band at different points in momentum space, as to be
able to calculate the differential quotient. For the integrated density of
states (IDOS), the assignment of positive and negative band indices determines
at which energy the $\mathrm{IDOS} = 0$. It also increases the accuracy of the (integrated)
DOS, because it is based on interpolation of the dispersion between the
available data points.
\end{itemize}

\subsubsection{Band alignment algorithm}

The band alignment algorithm is given by the following `recipe'.

\begin{figure}
 \includegraphics[width=\textwidth]{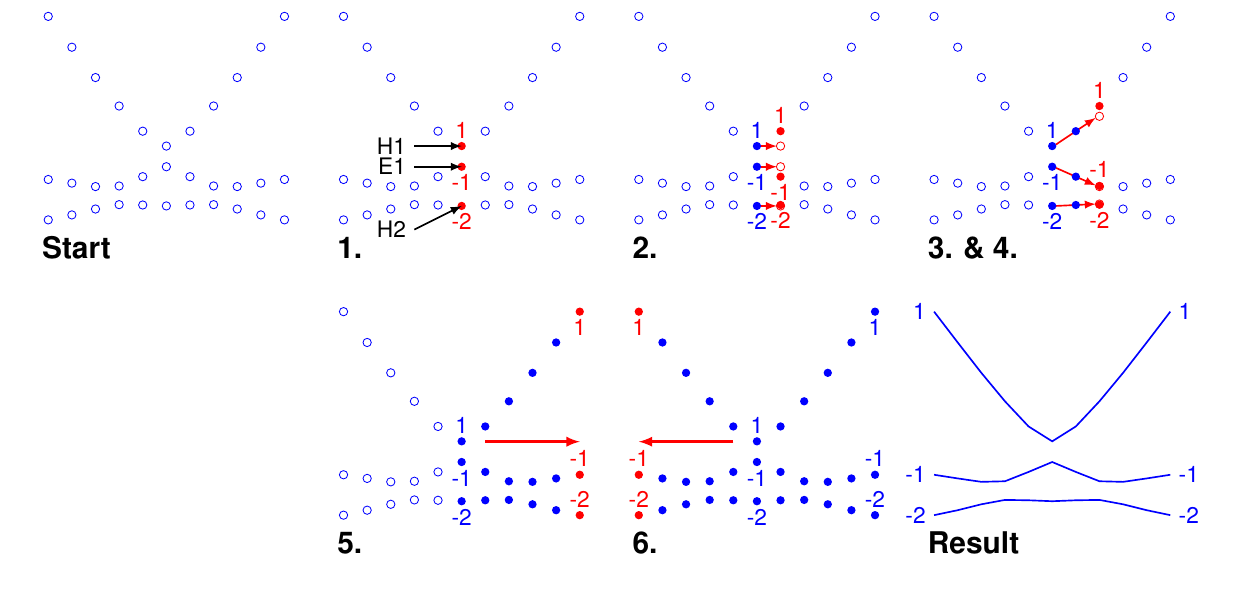}
 \caption{Band alignment recipe. 1. Determine band characters and assign band indices at zero momentum. 2. Align energies at adjacent point. 3--4. Extrapolate and align at third point. 5. Repeat previous steps forwards till the end of the domain. 6. Repeat previous steps backwards till the other end of the domain.}
\end{figure}

\begin{enumerate}
 \item Calculate band characters at zero ($\vec{k}=0$ and $\vec{B}=0$). Determine
   the charge neutrality point and assign positive band indices above, and
   negative band indices below this value.

 \item Take a data point adjacent to zero. Align the two sets of eigenvalues (see
   below). States aligned with each other get the same band index, i.e., the
   band indices for the `new' data point get them from the data point at zero.
   
 \item For all band indices that the two points have in common, extrapolate the
   energies to the next momentum or magnetic field value. That is, given
   the energies $\{E^{(0)}_i\}$ and $\{E^{(1)}_{i'}\}$, calculate
   \begin{equation}
      \tilde{E}^{(2)}_j = E^{(0)}_j + \frac{k_2 - k_0}{k_1 - k_0}(E^{(1)}_j - E^{(0)}_j)
   \end{equation}
   for all $j$ for which $E^{(0)}_j$ and $E^{(1)}_j$ are both defined.
   
 \item Align the actual eigenvalues at the new data point to the extrapolated values
   $\tilde{E}^{(2)}_j$. Thus, set the band indices for the new data point from
   the extrapolated energies.
   
 \item Repeat steps 3. and 4. until one reaches the end of the calculated domain.

 \item Repeat steps 2. for the data point on the other side of the zero point.
   Perform steps 3. through 5. towards the other end of the domain.

\end{enumerate}  

This algorithm is essentially one-dimensional. For two dimensional
grids, first perform the algorithm along the $k_x$ axis, then in the perpendicular
direction starting at $(k_x, 0)$, in a `fish bone' pattern. Analogously, in
polar coordinates, the algorithm is performed along the radial direction first,
then in the angular direction.

\begin{figure}
  \includegraphics{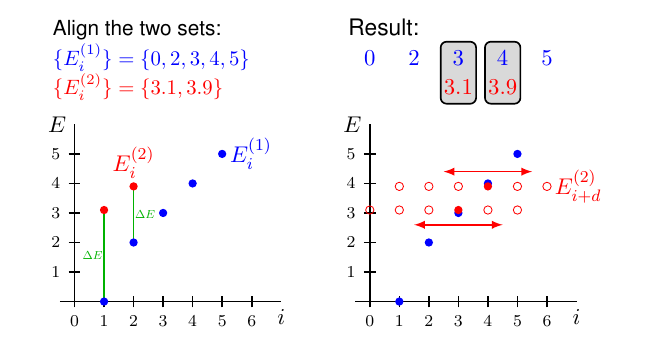}
  \caption{The alignment algorithm. (Left) Given two lists of energy values, we
  calculate the energy differences $\Delta E = |E^{(1)}_i - E^{(2)}_i|$ for each
  pair of values. The second set is shifted uniformly by index relative to the
  first one, $E^{(2)}_i \to E^{(2)}_{i+d}$. The sum of energy differences
  $\Delta(d)$ [see Eq.~\eqref{eqn_bandalign_delta}] is calculated for each shift
  $d$ and minimized over $d$.}
\end{figure}

The alignment of two sets of energies as described above, is essentially a
minimization of the energy differences. Assume two ordered sets of energies
$\{E^{(1)}_i\}$ and $\{E^{(2)}_j\}$ (i.e., the energies are monotonic).
Then vary the two indices relative to each other and find the
\emph{minimum average energy difference}. We also add a `penalty' for non-matching
energies, i.e., values in one set that do not have a partner in the other set.

Formalizing this idea, we find the value $d$ for which
\begin{equation}\label{eqn_bandalign_delta}
\Delta(d) = \left(\sum_{i} |E^{(1)}_{i} - E^{(2)}_{i+d}|^e\right) / n(d) + W / n(d).
\end{equation}
is minimal. The sum runs over all $i$ for which both values $E^{(1)}_{i}$
and $E^{(2)}_{i+d}$ are defined. The value $n_i(d)$ is the number of
terms in the sum. The coefficient $e$ is an exponent and $W$ is the `bonus
weight' for each term in the sum. These values are set by the configuration
settings \verb+band_align_exp+ (default is \texttt{4}) and
\verb+band_align_ndelta_weight+ (default is \texttt{20}).
(Setting \verb+band_align_exp=max+ replaces the sum by
$\max_i|E^{(1)}_{i} - E^{(2)}_{i+d}|$.)

Given the value $d$ for which $\Delta(d)$ is minimal, we say that the values
$E^{(1)}_{i}$ and $E^{(2)}_{i+d}$ are aligned. The result is that they will
receive the same band indices.

Note that the algorithm is based on the energies \emph{collectively}, not
\emph{individually}. This property makes the algorithm quite reliable.

\subsubsection{Manually assisted band alignment}
\label{sec_bandalign_manual}

Should the result of the band alignment not be satisfactory (e.g., incorrect),
\texttt{kdotpy merge} includes two options to manually `guide' the band alignment
to a correct result.

\begin{itemize}
\item Using the \texttt{bandalign \# [\#]} command line argument. The first value is an energy
  and the second one a gap index (omitted value means $0$). This pins the gap with
  that index to the given energy at zero momentum or magnetic field. For example,
  if one uses \texttt{bandalign -10 2}, then the first eigenstate below $-10\meV$ gets the
  band index $2$, the first one above band index $3$. The given energy need not be
  very precise; it must just lie in between two bands. For convenience, one could
  choose a large gap.

  \textsc{Note:} The gap with index $0$ lies between the bands with
  indices $-1$ and $1$. The gap with index $g > 0$ between bands $g$ and
  $g + 1$, the gap with index $g < 0$ between bands $g - 1$ and $g$.
  Zero gaps (i.e., between degenerate states) also count.
  
  This option replaces step 1. in the above recipe only. The algorithm then uses
  the same alignment strategy as described in steps 2. through 6.

\item Using an input file in CSV format with band indices and energies as function of
  momentum or magnetic field. The file format is the same as the \emph{output} files
  \texttt{dispersion.byband.csv}. This output file is essentially the result of the band
  alignment, i.e., the assignment of band indices to energies. This file may be
  edited in a spreadsheet editor by moving around values using
  cut-and-paste, for example. Then it can be saved as a CSV file and loaded into
  \kdotpy{} by using the command-line option \texttt{reconnect filename.csv}.
  
  The input file may be incomplete: Missing energy values at the left- or
  right-hand side of each line (not in the middle) will be filled in
  automatically, by extrapolation of band indices. Momentum (or magnetic field)
  values for which there is no data are filled in by interpolation or
  extrapolation from adjacent points, i.e., steps 3. and 4. of the above recipe.
  The energies only need to be approximately exact. This tolerant behaviour of
  the input is made possible by the implementation: The file input is processed
  by aligning the given energies to the actual eigenvalues. The band alignment
  algorithm does not need the energies to be exact or complete in order to be
  reliable.
  
  In this manner, it is also possible to get some hybrid result between manual
  input and the automatic algorithm.
  
\end{itemize}

\begin{figure}[h]
 \includegraphics[width=120mm]{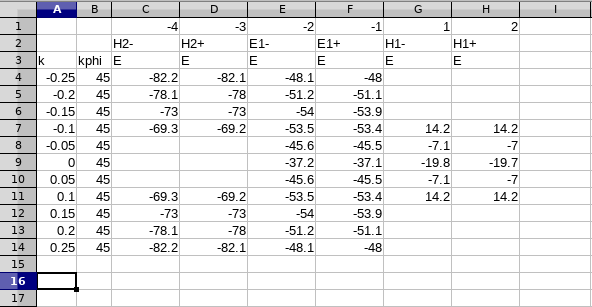}
 \caption{Manually assisted band alignment. Example of the manual input of band indices in a CSV file, as displayed in a spreadsheet editor.}
\end{figure}

\subsubsection{Challenges}

The band alignment algorithm is quite reliable, but there are scenarios that
commonly lead to incorrect results, for example:

\begin{itemize}
\item The alignment algorithm may misalign states if the two sets consist of
  almost equidistant subbands. Such patterns typically occur in both the
  conduction and the valence band individually. In order to avoid this
  condition, choose the diagonalization parameters such that at least the
  highest valence subband \emph{and} the lowest conduction subband are included at
  every point in momentum space. The alignment algorithm is then stabilized by
  the large gap.

\item The sets of eigenvalues must be connected and complete, i.e., between
  the lowest and highest eigenvalues, there must be no missing ones. For a
  single calculation, this is always true, but if one uses \texttt{kdotpy merge} for
  merging multiple data sets at the same momentum values, make sure that no
  eigenstates are missing. If the two sets overlap (share at least one
  eigenvalue, preferably more), then one is on the safe side.

\item The band alignment algorithm does not admit real crossings. Even when
  real crossings are expected in the spectrum, e.g., between states that have
  different values of a conserved quantum number, they are rendered as anticrossings.
  This also affects derived quantities: For example, density of states may be
  inaccurate around the absent crossing points. This can partially be mitigated
  by increasing the momentum or magnetic-field resolution.

\end{itemize}

\subsection{Density of states}
\label{sec_dos}

\subsubsection{Motivation}

The density of states plays an essential role in many physical quantities
observed in experiment. The position of the Fermi level determines whether the
material behaves as an insulator or conductor in transport physics. In
spectroscopy, the occupation of each state determines which optical transitions
are visible. The total carrier density is often known from experiments, is used
to determine the Fermi energy. The tuning of the Fermi level with one or more
gate electrodes is understood most straightforwardly in terms of carrier density,
as the latter typically depends linearly on the gate voltage.

The density functions of \kdotpy{} calculate the relation between carrier
density $n$ and energy $E$, and places the Fermi level at the correct position
if the carrier density is provided with the command line argument \texttt{cardens}.
The relation also allows \kdotpy{} to convert $\vec{k}$- or $\vec{B}$-dependence
plots by putting density $n$ on the vertical axis instead of energy $E$. This is
particularly useful in the Landau-level mode, where Landau fans with $n$ on the
vertical axis may be compared directly with Landau fan plots from
magnetotransport experiments.

\subsubsection{Integrated density of states (IDOS)}

The key quantity in the density calculations in \kdotpy{} is the integrated
density of states (IDOS; equivalent to carrier density), defined as the number
of occupied states lying between
the charge neutrality point and the Fermi level. The counting of occupied states
is based on the principle that each `mode' (given by its momentum) contributes
equally. An isolated, fully occupied band would correspond to one state in a
volume of $a^d$, assuming a (hyper)cubic lattice in $d$ dimensions with lattice
constant $a$. In momentum space, this state occupies the full Brillouin zone
with a volume of $(2\pi/a)^d$. A partially occupied band, which occupies a
volume of $V_\mathrm{occ}$ in momentum space, thus corresponds to a density of
$V_\mathrm{occ}/(2\pi)^d$.

In order to calculate the IDOS, we thus need to calculate the occupied volume in
momentum space for all bands and sum over them. In order to get density $n(E)$ as
function of energy, we choose a discrete array of energy values, which we call
the `energy range'. We sketch the algorithm for the IDOS calculation of a
dispersion in $d=1$ dimension. For each band $b$:

\begin{figure}
 \includegraphics[width=56mm]{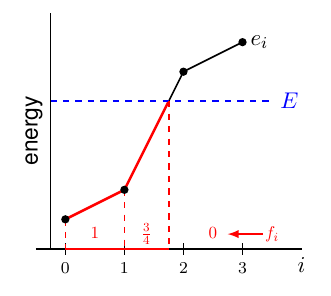}
 \caption{Calculation of the integrated density of states (IDOS) for a single band.
 The band energies at momenta $k_i$ are extracted from \texttt{DiagData}
 as the values $e_i$. Interpolate linearly to find a continuous function $e(k)$.
 For each interval $(i,i+1)$, find the fraction $f_i(E)$ of the interval for
 which $e(k) < E$. Iterate this procedure over many values of $E$. For holes,
 subtract $1$ from each $f_i$; in this example, this would yield
 $f_i-1 = (0,-\frac{1}{4},-1)$.}
 \label{fig_idos_onedim}
\end{figure}

\begin{itemize}
 \item Extract $e_i = E^{(b)}(k_i)$, the energy dispersion of the band
 with band index $b$ from \texttt{DiagData}. The index $i=1,\ldots,n_k$ runs over
 $n_k$ data points.
 \item Let $E$ be the Fermi energy. Do a piecewise linear
 interpolation $e(k)$ of the $e_i$, i.e., drawing straight lines between
 $(k_i,e_i)$ and $(k_{i+1}, e_{i+1})$. For all intervals $(k_{i},k_{i+1})$,
 determine the fraction of the interval where $e(k)\leq E$. This is given by
 \begin{equation}\label{eqn_density_fraction_onedim}
    f_i(E) = \left\{\begin{array}{ll}
    1 &\text{if } e_i < e_{i+1} \leq E\\
    \tfrac{E-e_{i}}{e_{i+1} - e_{i}} &\text{if }e_i \leq E < e_{i+1}\\
    0 &\text{if } E < e_i < e_{i+1}
   \end{array}\right.
 \end{equation}
 where we have assumed (without loss of generality) that $e_i < e_{i+1}$. See
 Fig.~\ref{fig_idos_onedim} for an illustration.
 This step is done by \verb+linear_idos_element()+, which does this for all 
 energies $E$ in the energy range simultaneously, by clever use of NumPy arrays.
 \item If the band is hole-like (band index $b < 0$), subtract $1$ from all $f_i$.
 Thus, the volume below the Fermi energy is counted as $0$ and the volume above
 it as $-1$. In this way, hole-like densities are automatically counted as being
 negative.
 \item Integrate over momentum. For all intervals $(k_{i},k_{i+1})$, calculate
 the integration volume element $dk_i$. In one dimension, the volume element
 would be $dk_i = k_{i+1} - k_{i}$ in cartesian coordinates
 and $dk_i = \pi (k^2_{i+1} - k^2_{i})$ if $k$ is a radial coordinate in polar
 coordinates. The IDOS contribution for band $b$ is thus
 \begin{equation}\label{eqn_idos_momentum_integration}
   n^{(b)}(E) = \frac{1}{(2\pi)^d}\sum_{i=1}^{n_k-1} f_i(E) dk_i,
 \end{equation}
 where we note that the $f_i(E)$ are still functions of the Fermi energy $E$,
 implemented as one-dimensional NumPy array. The result is thus also a function
 of $E$.

\end{itemize}
To find the total density as function of $E$, we sum over all bands. If
desired, the sum may also be restricted to either electron-like or hole-like bands.

The function \verb+loc_int_dos_by_band()+ in \texttt{density/base.py} takes care
of obtaining the $f_i$, where Eq.~\eqref{eqn_density_fraction_onedim} is
implemented in \texttt{density/elements.py}. The integration over momentum is
done by \verb+int_dos_by_band()+ in \texttt{density/base.py}. The result is a
one-dimensional array, where the elements encode the IDOS values $n(E_i)$, where
$E_i$ are the values in the energy range. The summation over bands may also be
skipped in order to find the contributions by band; this result is represented
by a \texttt{dict} instance where the keys are the band labels $b$ and the values
are the one-dimensional arrays $n^{(b)}(E_i)$.

\begin{figure}
 \includegraphics[width=110mm]{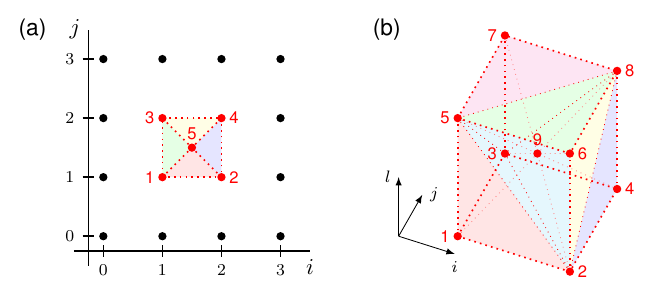}
 \caption{Division of momentum space for two and three dimensions.
 (a) In two dimensions, the momentum space is triangulated into triangular
 simplices. To this end, a fifth point is inserted in each elementary square
 (red) by setting $e_{5,i,j}$ as the average over the four corner points. The
 four simplices for each square are $125$, $135$, $245$, and $345$.
 (b) In three dimensions, the space is divided into tetrahedral simplices. A
 ninth point is inserted in the centre of each elementary cube, by setting
 $e_{9,i,j,l}$ as the average over the eight corner points. The illustrated
 division is the set of twelve tetrahedra $1239$, $2349$, $1259$, $2569$, $2489$,
 $2689$, $1359$, $3579$, $5689$, $5789$, $3489$, and $3789$. Another choice can
 be obtained by spatial inversion.
 }
 \label{fig_idos_higherdim}
\end{figure}

For higher dimensions, an additional intermediate step is required. In order to
calculate $f_i$ in two or three dimensions, we need to divide the space into
triangular or tetrahedral simplices. In two dimensions, for an elementary square
$[k_{x,i},k_{x,i+1}] \times [k_{y,j},k_{y,j+1}]$, take the centre point
$k_{5,i,j} = (\frac{1}{2}(k_{x,i}+k_{x,i+1}), \frac{1}{2}(k_{y,j}+k_{y,j+1}))$
and divide the plaquette into four triangles with $k_{5,i,j}$ as one of its
vertices, see Fig.~\ref{fig_idos_higherdim}(a). The band energy at $k_{5,i,j}$
is determined by interpolation,
\begin{equation}
  e_{5,i,j} = \tfrac{1}{4}\left( E^{(b)}(k_{x,i},k_{y,j}) + E^{(b)}(k_{x,i+1},k_{y,j})
   + E^{(b)}(k_{x,i},k_{y,j+1}) + E^{(b)}(k_{x,i+1},k_{y,j+1})\right).
\end{equation}
Labelling each simplex shape as $t=1,2,3,4$ (corresponding to the four colours
in Fig.~\ref{fig_idos_higherdim}(a)), we find all $f_{t,i,j}(E)$ with a
function analogous to Eq.~\eqref{eqn_density_fraction_onedim}, given as
Eq.~\eqref{eqn_density_fraction_twodim} in Appendix \ref{app_idos_elements} 
and implemented in \verb+triangular_idos_element()+. The integral
over momentum involves volume elements $dk_{t,i,j}$ for all simplices,
\begin{equation}\label{eqn_idos_momentum_integration_twodim}
  n^{(b)}(E) = \frac{1}{(2\pi)^d}\sum_{t=1}^4\sum_{i=1}^{n_{k_x}-1}\sum_{j=1}^{n_{k_y}-1}
  f_{t,i,j}(E) dk_{t,i,j}.
\end{equation}
For cartesian coordinates, $dk_{t,i,j} = \frac{1}{4} \Delta k_x \Delta k_y$. For
polar coordinates, we take the approximate expression
$dk_{t,i,j} = \tfrac{1}{4} \overline{k_r} \Delta k_r \Delta k_\phi$,
where $\overline{k_r}$ is the average of the radial coordinates of the three vertices
of the triangle defined by $t,i,j$ and $\Delta k_\phi$ is measured in radians.

For three dimensions, an analogous method is used, with each elementary cube
$[k_{x,i},k_{x,i+1}] \times [k_{y,j},k_{y,j+1}] \times [k_{z,l},k_{z,l+1}]$ being
divided into $12$ tetrahedra with the body-centred point
$k_{9,i,j,l} = (\frac{1}{2}(k_{x,i}+k_{x,i+1}), \frac{1}{2}(k_{y,j}+k_{y,j+1}),
\frac{1}{2}(k_{z,l}+k_{z,l+1}))$ as one of its vertices, see
Fig.~\ref{fig_idos_higherdim}(b). The energy value
$e_{9,i,j,l}$ is calculated as the average over all eight vertices of
the elementary cube. The method for finding value $f_{t,i,j,l}(E)$ with a
function analogous to Eq.~\eqref{eqn_density_fraction_onedim} is given as 
Eq.~\eqref{eqn_density_fraction_threedim} in Appendix \ref{app_idos_elements} and
is implemented in \verb+tetrahedral_idos_element()+. The volume elements are
$dk_{t,i,j,l} = \frac{1}{12}\Delta k_x\Delta k_y\Delta k_z$ for cartesian coordinates, $dk_{t,i,j,l} = \frac{1}{12}\overline{k_r}\Delta k_r\Delta k_\phi\Delta k_z$ for
cylindrical coordinates, and
$dk_{t,i,j,l} = \frac{1}{12}\overline{k^2_r}\,\overline{\sin k_\theta}\,
 \Delta k_r\Delta k_\theta\Delta k_\phi$
for spherical coordinates, where the overlines denote averages over all four
vertices of each tetrahedron.

\subsubsection{Density in Landau level mode}

In Landau level mode, the calculation of IDOS is much simpler due to the absence
of the momentum coordinates. For each magnetic field value $B=B_z$, apply the
following recipe.
\begin{itemize}
 \item For each band labelled $b$ (full LL mode) or $(n,b)$ (symbolic LL mode),
 extract the energy values $E^{(b)}(B)$ from \texttt{DiagData}.
 \item Determine the charge neutrality point $E_\mathrm{CNP}(B)$. For full LL
 mode, this is simply the centre of the gap between bands $-1$ and $1$. For
 symbolic LL mode, the band labels $(n,b)$ are first converted to a single
 \emph{universal band index} $u$; $E_\mathrm{CNP}(B)$ then lies between the states
 with $u=-1$ and $u=1$. (Details on the universal band index in
 Appendix~\ref{app_ubindex}.)
 \item Let the Fermi energy be $E$. Count the number of states with energies
 between $E_\mathrm{CNP}(B)$ and $E$.
 \item Multiply by the LL degeneracy factor $eB/2\pi\hbar$.
\end{itemize}
This algorithm is applied for each magnetic field value independently, and yields
a function of energy $E$ (represented by an array) for each $B$. The result
can be interpreted as `local' integrated density of states. The
function that takes care of this calculation is \verb+loc_int_dos()+, with
the determination of the charge neutrality points being handled by the
member function \verb+DiagData.get_e_neutral()+ of the \texttt{DiagData} class.

\subsubsection{Data structure}

The IDOS data is stored in an instance of the class \texttt{DensityData}.
This class has three important array-like attributes:
\begin{itemize}
 \item \texttt{densdata}: An array of at least one dimension, holding the IDOS
 values.
 \item \texttt{ee}: A one-dimensional array of the energy values.
 \item \texttt{xval}: Optionally, an array or \texttt{VectorGrid} object that
 holds the momentum or magnetic field values for local IDOS data.
\end{itemize}
The shape of \texttt{densdata} must match those of \texttt{xval} and \texttt{ee},
i.e., the condition
\begin{verbatim}
densdata.shape == (*xval.shape, *ee.shape)
\end{verbatim}
must evaluate to \texttt{True}, with \texttt{xval.shape} being replaced by
\texttt{()} if \texttt{xval} is \texttt{None}. The class also remembers the
dimensionality (as \texttt{kdim}) and whether or not Landau levels are considered
(as the boolean value \texttt{ll}).

The class has several member functions for common operations on the integrated
density of states. The most important ones are:
\begin{itemize}
 \item \verb+integrate_x()+: Integrate over the momentum coordinates.
 \textsc{Note}: Technically, magnetic field values can also be integrated over,
 but there is no physically relevant reason to do so.
 \item \verb+get_idos()+: Return the IDOS data, scaled to the desired units.
 \item \verb+get_dos()+: Return the density of states, defined as the energy
 derivative of the IDOS. Scale the values to the desired units, like with
 \verb+get_idos()+.
 \item \verb+get_validity_range()+: Estimate the energy range where the IDOS
 is valid, i.e., in which energy range we can be reasonably sure that no states
 have been overlooked due to the finite momentum range. This is done by considering
 the dispersion energies and their derivatives at the edge of the momentum range.
 For example, if the momentum range is $[k_\mathrm{min},k_\mathrm{max}]$ and the
 dispersion $E(k)$ approaches $E(k_\mathrm{max})$ from below (with $dE/dk>0$),
 then we can infer that there are states at $E>E(k_\mathrm{max})$ for
 $k>k_\mathrm{max}$, which are not considered for the DOS calculation. Thus,
 the upper bound of the validity range must be $\leq E(k_\mathrm{max})$. 
 The upper and lower limits of the validity range are determined by iterating
 this algorithm over all bands. The validity range may be empty or `negative'
 (when the upper bound lies below the lower bound) in some cases. The
 result is printed on the terminal and is visualized in the DOS and IDOS
 plots by shading of the regions outside the validity range.
 \item \verb+idos_at_energy()+: Get the IDOS value at a given energy. This is
 done by interpolation along the energy axis.
 \item \verb+energy_at_idos()+: Get the energy value at a given IDOS $n_\mathrm{target}$
 (carrier density). This is done by solving $E$ from the equation
 $n(E)=n_\mathrm{target}$, where $n(E)$ is a linearly interpolated function
 from the IDOS data (\texttt{densdata} versus \texttt{ee}) stored in the
 \texttt{DensityData} instance.
\end{itemize}
There are several additional attributes and member functions related to
\emph{special energies}, for example the Fermi level at zero density and at the
desired density, and the charge neutrality point.

\subsubsection{Units and scaling}

The internal density units for \texttt{DensityData.densdata} are always
$\mathrm{nm}^{-d}$, where $d$ is the dimensionality. In order to be able to
extract quantities proportional to DOS and IDOS values expressed in a
user-preferred set of units, the \texttt{DensityData.scale} may be set. If
this attribute is set as an instance of
the \texttt{DensityScale} class, functions like \verb+DensityData.get_idos()+
and \verb+DensityData.get_dos()+ automatically return the values in the
appropriate units.

The role of the \texttt{DensityScale} class is to store
the desired quantity and unit, and to scale the exponent $p$ automatically,
such that the values can be written as $x \times 10^p$ with $x$ having a
`convenient' magnitude. The default quantity is simply the IDOS as
defined before. The area or volume in momentum space is related by
a factor of $(2\pi)^d$. Charge density has the same value as IDOS, but with
units of $e\, \mathrm{nm}^{-d}$ instead of $\mathrm{nm}^{-d}$. For the units,
one may choose $\mathrm{nm}$, $\mathrm{cm}$, or $\mathrm{m}$ as the units for
length. For the extracted IDOS, the values are expressed in units of
$10^p\,\mathrm{nm}^{-d}$, $10^p\,\mathrm{cm}^{-d}$, or $10^p\,\mathrm{m}^{-d}$,
with $p$ set to a reasonable value for the chosen units. For DOS, the appropriate
unit is obtained by multiplying with $\mathrm{meV}^{-1}$.
The \texttt{DensityScale} class provides the functions
\texttt{DensityScale.qstr()} and \texttt{DensityScale.unitstr()} for formatting
the quantity and unit as a string for use in output files (text and graphics).
For convenience, \texttt{DensityData} provides these member functions as well.

\subsubsection{Integrated observables}

A special application of the density functions is the notion of integrated
observable. In the summation over the bands [similar to
Eq.~\eqref{eqn_idos_momentum_integration}], we include the value of an
observable $O$ as an additional factor. The integrated observable is defined
by the sum $O(E)=\sum_b O^{(b)}(E)$ over the contributions from all bands,
\begin{equation}\label{eqn_intobs_momentum_integration}
  O^{(b)}(E) = \frac{1}{(2\pi)^d}\sum_{i=1}^{n_k} f_i(E) O^{(b)}(k_i) dk_i,
\end{equation}
where $k_i$ are the momentum values of the grid and $O^{(b)}(k_i)$ is the
observable value of band $b$ at $k_i$. Note that unlike the IDOS calculation,
we sum over the momentum values themselves, not the intervals between them. The
implementation (function \verb+integrated_observable()+) returns an
\texttt{IntegratedObservable} class derived from the \texttt{DensityData} class.

The integrated observable $O(E)$ as function of energy can be transformed into
the corresponding function $\tilde{O}(n)$ as function of density by means of
a \emph{pushforward} transformation over the function defining density $N(E)$
as function of energy. The result is $\tilde{O} = O \circ N^{-1}$, where
$\circ$ denotes composition of functions and $N^{-1}$ is the inverse function of
$N$. The function $N^{-1}$ takes a density value $n$ and calculates the
corresponding energy $E$. Then, $O$ acting on $E$ yields $O(E)$. Thus, one may
think of $\tilde{O}$ as the function $n\mapsto O(E(n))$, with $E(n) = N^{-1}(n)$.

This transformation is implemented as \verb+IntegratedObservable.pushforward()+,
which determines $\tilde{O}$ by inverse interpolation using $O(n)$ defined from
the class instance itself and the function $N(E)$ (a \texttt{DensityData} instance)
as the first argument. The second argument is the array of density values $n$ at
which $\tilde{O}(n)=O(N^{-1}(n))$ is evaluated. In the most simplified form, the
essence of the implementation is\footnote{This function is currently a member of the parent class \texttt{DensityData}.}
\begin{verbatim}
def pushforward(self, other, values):
    return np.interp(values, other.densdata, self.densdata)
\end{verbatim}
where \texttt{self} represents $O(E)$ and \texttt{other} represents $N(E)$. In
this example, we assume that \texttt{self} and \texttt{other} are defined on the 
same array of energies and that the attribute \texttt{xval} is \texttt{None} for
both.

\subsubsection{Broadening}
\label{sec_broadening}


In essence, the density of states (DOS) is defined as the size (length, area,
volume) of the level sets of the dispersion $E(\vec{k})$. From the principles of
minimal energy and Pauli exclusion, electronic states fill up to the Fermi energy,
such that the number of carriers equals the integral over the DOS up to the Fermi
energy. This picture assumes zero temperature, so that the occupation function
$F(E)$ has a hard cutoff at the Fermi energy.

For finite temperature $T$, the occupation function smoothly goes from $1$ to $0$
in an energy window of width proportional to $\kB T$. The effective density of
carriers in the product of the DOS $g(E)$ and the smooth occupation function
$F(E)$. Due to the smoothness of this function, an electron in quantum state
$E$ can also be found at energies nearby. The probability distribution for
finding the particle at energy $E$ is $f(E) = dF(E)/dE$, which is centred
around the eigenenergy of the state. The fact that this distribution is somewhat
spread out explains the term \emph{broadening}.
Other probabilistic phenomena such as disorder can broaden also the density of
states. These can be treated on the same footing as the broadening due to finite
temperature.

Broadening is applied essentially as the convolution of the density of states
$g(E)$ with the broadening function $f(E)$.
\begin{equation}
d(E) = (f * g)(E) = \int_{E_0}^E f(E - E') g(E') dE'
\end{equation}
In \kdotpy, the central quantity is the integrated density of states $G(E)=n(E)$.
The broadened version $D(E)$ is similarly obtained as the convolution
\begin{equation}
D(E) = (f * G)(E) = \int_{E_0}^E f(E - E') G(E') dE' = (F * g)(E),
\end{equation}
where the latter equality is a property of convolution in general.

Since the energy values are defined on a grid, we need to take special care of
numerical inaccuracies. In particular, we need to make sure that the integral
over the broadening function equals 1 exactly. Note that if we simply take some
values $f(E_j)$ ($E_j=j \delta E$ with $j$ being integers) the (Riemann) sum
$\sum_{j} f(E_j) \delta E$ that approximates the integral $\int f(E) dE$ may
deviate from 1, especially if the broadening parameter is comparable in size to
$\delta E$. In order to prevent this from happening we rather use the discrete
derivative of $F(E)=1-\int^E f(E')dE$.
The reason for `1 minus' is to make the interpretation of $F(E)$ to be the
\emph{occupation functions} associated with the \emph{broadening kernel} $f(E)$.

The broadening functions and their occupation functions implemented in \kdotpy{}
are listed in Appendix~\ref{app_broadening_reference}. In general, the command
line argument is of the form \texttt{broadening \emph{w0} \emph{type} \emph{dep}},
where $w_0$ is the width parameter and \texttt{\emph{type}} determines the type
(shape of the broadening function), i.e.,
\texttt{fermi}, \texttt{thermal}, \texttt{gauss}, or \texttt{lorentz}).
The dependency argument \texttt{dep} determines the
dependence of the width parameter on $k$ or $B$; for example, when \texttt{sqrt}
is used with a magnetic-field dependence, the broadening width applied to the
IDOS is $w(B)=w_0 \sqrt{B}$ with $B$ in $\mathrm{T}$.

Generally, some of the arguments after \texttt{broadening} may be omitted, in
which case the following defaults are used:
\begin{itemize}
\item If \texttt{broadening} is not given at all, do not apply broadening.
\textsc{Note}: \texttt{dostemp \emph{T}} (where $T$ is a temperature in K)
qualifies as broadening argument and is 
equivalent to \texttt{broadening \emph{T} thermal}.
\item If \texttt{broadening} is given without type, then use the default type:
\texttt{thermal} in dispersion mode, \texttt{gauss} in LL mode.
\item If \texttt{broadening thermal} is given without width parameter
(temperature), use the temperature given by \texttt{temp \emph{T}}.
This defaults to $T=0$ if \texttt{temp} is absent.
\item If the dependence argument is omitted, broadening type \texttt{gauss}
assumes \texttt{sqrt} dependence, all others use \texttt{const}.
\end{itemize}

Finally, compound broadening might be achieved with multiple broadening arguments.
The broadening arguments are iteratively applied to the density of states.
For example, if one gives \texttt{broadening gauss 1.0 broadening thermal 0.5},
first the Gaussian broadening is applied to the DOS, then the thermal broadening.
If the broadening kernels are denoted $f_1$ and $f_2$, then the result is
\begin{equation}
d(E) = (f_2 * (f_1 * g))(E) = ((f_2 * f_1) * g)(E)
\end{equation}
By virtue of associativity of convolution, the combined broadening kernel is
$f_2 * f_1$, which generalizes to $f_n * \cdots * f_1$ if one combines $n$
broadening kernels.

Since convolution is commutative, $f_2 * f_1 = f_1 * f_2$, the order of applying
the broadening kernels is irrelevant \emph{in principle}. Due to limitations of
the numerics---convolution is calculated by numerical integration on a finite
interval with finite resolution---changing the order might lead to small
numerical differences. Some combinations of broadening may be simplified analytically,
for example the convolution of two Gaussian distribution functions with standard
deviations $\sigma_1$ and $\sigma_2$ is a Gaussian distribution function
with standard deviation $\sigma = \sqrt{\sigma_1^2 +\sigma_2^2}$. (For Lorentzians,
the widths add up, $\gamma=\gamma_1+\gamma_2$, and a combination of Fermi
distributions cannot be simplified analytically.) These simplifications may be
useful to perform manually prior to input, as to avoid unnecessary numerical errors.

\subsubsection{Density as function of $z$}
\label{sec_densityz}

Knowledge of the spatial distribution of charge $\rho(z)$ given a certain
carrier density can aid significantly in understanding the observable behaviour
of a device. It is also essential for the self-consistent Hartree algorithm 
(see Sec.~\ref{sec_selfcon}) in order to find the electric potential induced by
the charges in the material.

The idea of calculating $\rho(z)$ is analogous to the methods for the IDOS and
the integrated observable. We insert the probability density appropriate for the
interval $[k_i,k_{i+1}]$ into Eq.~\eqref{eqn_idos_momentum_integration}: for
each band, we calculate
\begin{equation}\label{eqn_densityz_momentum_integration}
  \rho^{(b)}(z,E) = \frac{1}{(2\pi)^d}\sum_{i=1}^{n_k-1}
  \frac{1}{2}\left(\abs{\psi^{(b)}_{k_i}(z)}^2 + \abs{\psi^{(b)}_{k_{i+1}}(z)}^2\right)
  f_i(E) dk_i.
\end{equation}
and we sum over all bands, $\rho(z,E) = \sum_b\rho^{(b)}(z,E)$ (or if desired,
over either electrons or holes only). In the implementation of
\verb+density_energy()+, the probability densities $\abs{\psi^{(b)}_{k_i}(z)}^2$
are extracted from the eigenvectors, summed over the orbital degree of freedom,
and supplied as an argument to \verb+int_dos_by_band()+. The result is a
two-dimensional array that encodes $\rho(z,E)$ over an array of $z$ values and
an array of energy values $E$. The function \verb+densityz()+ evaluates $\rho(z,E)$
for a specific energy $E$ and returns a one-dimensional array. The latter
function is used by the self-consistent Hartree method, discussed in
Sec.~\ref{sec_selfcon}.

In two dimensions, the algorithm is analogous, but the interpolated probability
density in \eqref{eqn_densityz_momentum_integration} is replaced by a weighted
sum over the four corner points of each momentum space plaquette,
\begin{equation}
 \sum_{v=1}^4 c_v\abs{\psi^{(b)}_{\vec{k}_{i_v}}(z)}^2,
\end{equation}
and the sum over momenta is replaced by the simplices over triangles in the
triangulated lattice [cf.\ Eq.~\eqref{eqn_idos_momentum_integration_twodim}]
One can choose between equal weights
$c_v=(\frac{1}{4}, \frac{1}{4}, \frac{1}{4}, \frac{1}{4})$ for all triangles
or adjusted weights taking into account the shape of the triangle, i.e.,
$c_v=(\frac{5}{12}, \frac{5}{12}, \frac{1}{12}, \frac{1}{12})$ for the triangle
with vertices $125$, etc.

In Landau level mode, the summation over momentum is absent, and we sum all
contributions $\abs{\psi^{(b)}_{B}(z)}^2$ for all states with eigenvalues
$E^{(b)}$ between the charge neutrality point and the Fermi level. Like the IDOS
calculation, the Landau level degeneracy factor $eB/2\pi\hbar$ is taken into
account as well.

The $z$-dependent density $\rho(z)$ is always calculated such, that
$\int \rho(z)dz$ is equal to the IDOS $n$ at the same energy. We note that until
here, we have assumed that $\rho(z)$ is uniformly zero at the charge neutrality
point, but this is not necessarily the case: In fact, at zero total density
($n=0$), the system need only be neutral on average ($\int \rho(z)dz=0$), but
the local charge density $\rho(z)$ may vary. In \kdotpy, one can define a
density offset $\rho_\mathrm{offset}(z)$ at zero density, or use a different
reference point. An in-depth analysis of physically sensible choices of the
reference density is outside the scope of this work and will be presented
elsewhere.

\subsection{Optical transitions}
\label{sec_optical_transitions}

\subsubsection{Background}
Optical spectroscopy is a powerful experimental technique for characterization and investigation of band structures of semiconductors. By analysing the optical transitions between two states, conclusions about e.g. the band gap, the band order, etc. can be drawn. Simulating these transitions and their dependence on e.g. magnetic field in \kdotpy{} and, thus, being able to easily compare them to experimental data, provides a great tool to improve the modelling capabilities of \kdotpy{} and to validate which approximations/symmetries suffice for specific system configurations. Ultimately, this refines the predictions that can be made with \kdotpy{} with regards to band structure engineering, etc.

We achieve this by including the contributions of the electromagnetic (EM) field, that drives optical transitions, as a separate vector potential $\vec{A}_\mathrm{EM}$ in the Hamiltonian
\begin{equation}
	H_\mathrm{pert} = \frac{\left(\vec{p}+e\vec{A}_0+e\vec{A}_\mathrm{EM}\right)^2}{2m} + V
	\equiv H+H_\mathrm{EM}
\end{equation}
with
\begin{equation}
	H = \frac{\left(\vec{p}+e\vec{A}_0\right)^2}{2m}+V
	\qquad\text{and}\qquad
	H_\mathrm{EM}= \frac{e\vec{A}_\mathrm{EM}\cdot\vec{p}}{m}.
\end{equation}
The identity $\vec{p}\cdot\vec{A} = \vec{A}\cdot\vec{p}$, which is always valid for divergence free fields of EM waves, was used and the energy offset terms $\vec{A}^2_\mathrm{EM}+2\vec{A}_0\cdot\vec{A}_\mathrm{EM}$ were neglected to get to the second line.
Using the dipole approximation $\vec{q}\to 0$, neglecting any spatial contribution to the phase factor, the vector potential of EM waves can be expressed as
\begin{equation}
	\vec{A}_\mathrm{EM} = \frac{\vec{E}}{2\omega}\left(\ee^{\ii\omega t}+\text{c.c.}\right)
\end{equation}
For the sake of simplicity, we drop the time varying phase factors (these will become relevant in Fermi's Golden Rule) and we assume an EM wave travelling in $z$ direction. The operator product can then be written as
\begin{equation}
	\vec{A}_\mathrm{EM}\cdot\vec{p} \propto \vec{E}\cdot\vec{p} = E_xp_x+E_yp_y
	= \frac{1}{\sqrt{2}}\left(E_+p_++E_-p_-\right)
\end{equation}
where $E_\pm=\frac{1}{\sqrt{2}}\left(E_x\mp\ii E_y\right)$ and $p_\pm=p_x\pm\ii p_y$ as circular polarization basis.

The optical transition rate $\Gamma_{i\to f}$ from initial state $\ket{\psi_i}$ to final state $\ket{\psi_f}$ can be calculated using Fermi's Golden Rule
\begin{equation}
	\Gamma_{i\to f} = \frac{2\pi}{\hbar} \left|M^{(fi)}\right|^2 \delta\left(\hbar\omega_{fi}-\hbar\omega\right)\label{eqn_fgr}
\end{equation}
with transition matrix element
\begin{equation}
	M^{(fi)} = \bra{\psi_f}H_\mathrm{EM}\ket{\psi_i}\\
	= \frac{e\vec{E}}{2\omega} \bra{\psi_f}\vec{v}\ket{\psi_i}\label{eqn_transition_matrix_element}
\end{equation}
The omitted time dependence of the electric fields is responsible for the $\delta$ distribution, assuring energy conservation for absorption and emission of photons.

\subsubsection{Evaluation of transition matrix elements}

The transition matrix elements given by \eqref{eqn_transition_matrix_element} are calculated by \verb+get_transitions()+ (symbolic mode) and \verb+get_transitions_full()+ (full mode) in \texttt{transitions.py}. This calculation requires the eigenvectors, and therefore this is done immediately after diagonalization of the Hamiltonian, before the eigenvectors are discarded from memory.

The velocity operator is evaluated by using the Ehrenfest theorem
\begin{equation}
	v_{x} = -\frac{\ii}{\hbar}\left[x, H_\mathrm{pert}\right]
	\approx -\frac{\ii}{\hbar}\left[x, H\right]
	= \ddx{H}{p_x}
	= \frac{1}{\hbar}\ddx{H}{k_x}
\end{equation}
and similar for $y$.
Let us relabel this operator $O_{x/y}=v_{x/y}$ and define the corresponding operators in circular basis
\begin{equation}
	O_{\pm} = O_x\pm\ii O_y
	=\frac{2}{\hbar}\ddx{H}{k_\mp}
\end{equation}
using $k_\pm=k_x+\ii k_y$ and
$\ddx{}{k_\pm}=\frac{1}{2}\left(\ddx{}{k_x}\mp\ii\ddx{}{k_y}\right)$.
Independent of the used LL mode, the derivation of the basic LL Hamiltonian $H$ in $k_\mp$ is performed in its symbolic representation. After derivation, the numerical values of the operator are evaluated for LLs $n$ and $n+1$. Then, the matrix product
\begin{align}
	\abs{O_\pm^{fi}}^2=\left|\bra{\psi_f}O_\pm\ket{\psi_i}\right|^2 \label{eqn_opm_squd}
\end{align} 
is evaluated for all combinations of $\bra{\psi_f}$ and $\ket{\psi_i}$. This evaluation is
done efficiently as a sequence of two matrix multiplications on NumPy matrices
\begin{verbatim}
opeivec1T = op @ eivec1T
ov = eivec2T_H @ opeivec1T
ov2 = np.real(np.abs(ov)**2)
\end{verbatim}
where \texttt{op} represents $O_\pm$, \verb+eivec1T+ holds the eigenvectors $\ket{\psi_i}$ as columns, and \verb+eivec2T_H+ holds
the conjugate eigenvectors $\bra{\psi_f}$ as rows. The result is filtered, by discarding
all transitions with an amplitude below a threshold \texttt{ampmin}, schematically
\begin{verbatim}
sel = (ov2 >= ampmin)
ov2_filtered = ov2[sel]
\end{verbatim}
which yields the filtered transitions as the one-dimensional array \verb+ov2_filtered+.
In the same manner,
one-dimensional arrays containing the energies and Landau level index of initial
and final state are constructed. These arrays are stored together in a
\texttt{TransitionsData} instance (class definition in \texttt{transitions.py}) inside \texttt{DiagDataPoint.transitions}, see \verb+ModelLL._post_solve()+ in \texttt{models.py}, for further analysis later on.

Note that we only calculate $\abs{O_+^{fi}}^2$ in \kdotpy{}, to reduce calculation time. Because of the relation $\bra{\psi_f}O_+\ket{\psi_i} = \bra{\psi_i}O_-\ket{\psi_f}^\dagger$ and the fact that we only keep the absolute square of these matrix elements, we can reinterpret the emission matrix elements of the $\abs{O_+^{fi}}^2$ matrix (negative energy difference from final to initial state) as absorption matrix elements of $\abs{O_-^{fi}}^2$.

\subsubsection{Postprocessing and filtering}

In \texttt{postprocess.py} the function \texttt{transitions()} is responsible for any further analysis of optical transitions and corresponding spectra. Depending on input parameters (e.g. using \texttt{cardens}) the calculated transitions are filtered by the method \verb+filter_transitions()+ of a \texttt{DiagData} instance. If no filtering is performed, all transitions are plotted and written into a table. For filtered data, plotting can be suppressed to speed up code execution, but the data will always be written into a table. Further, also only for filtered transitions, other quantities related to optical transitions can be optionally calculated, e.g., rotation and ellipticity. This heavily impacts calculation time.

To filter transitions the \texttt{TransitionsData} instance method \verb+at_energy()+ is called, which only returns transitions that cross the given energy level, i.e., the initial state is (partially) occupied while the final state is (partially) unoccupied, as a new \texttt{TransitionsData} instance.
If no broadening is used (delta-peak shaped energy states), every transition is checked if and only if one of the two involved states is above while the other is below the given energy.
Otherwise, the implemented occupation function for the type of broadening is used to filter transitions (e.g., Fermi distribution for thermal broadening). The occupation factor is 
\begin{equation}
	P_{i\to f}= \left|f(E_i) - f(E_f)\right|  \label{eqn_occupation_factor}
\end{equation}
where $f(E_{i/f})$ is the occupation function for the given broadening type evaluated at the energy of the initial/final state, respectively. Only transitions for which $\left|\bra{\psi_f}O_\pm\ket{\psi_i}\right|^2\cdot P_{i\to f}\geq A_\mathrm{min}$ will be kept as filtered transition, where $A_\mathrm{min}$ can be adjusted by the command line argument \texttt{transitions} followed by a floating point value or by the configuration option \verb+transitions_min_amplitude+ (default value is $0.01$).

\subsubsection{Output (tables and plots)}
\label{sec_optical_transitions_output}

Plotting and saving of transition data is performed with the same routines for both unfiltered and filtered transition data, using different file names (\texttt{transitions-all...} vs. \texttt{transitions-filtered...}). These functions can be found in \texttt{ploto/auxil.py} and \texttt{tableo/auxil.py} and are both called\linebreak[4] \texttt{transitions()}.

Only a single quantity is plotted as transition plot. By using the configuration option\linebreak[4] \verb+plot_transitions_quantity+ (valid choices see below) the user can choose which quantity is plotted, the default is \texttt{rate}. Independent of choice, the amplitude of the chosen quantity is colour-coded and plotted onto a $B$ vs. $\Delta E$ grid.\\
Contrary to the plot output, as many quantities as possible are saved into the csv file. Depending on input and material parameters, some may not be possible to calculate. Additional quantities that will always be saved to the csv, but are not available for plotting, are magnetic field values, initial/final LL indices and initial/final band indices. The remaining quantities mostly overlap with the plot quantities and are given in the following list:

\begin{itemize}
	\item \texttt{deltae}:
	The energy difference $\Delta E$ between initial and final state in meV.
	
	\item \texttt{freq}:
	The corresponding frequency for the energy difference $\Delta E$ in THz.
	
	\item \texttt{lambda}:
	The corresponding wavelength for the energy difference $\Delta E$ in \microm.
	
	\item \texttt{amplitude}:
	The absolute-squared transition matrix element $\abs{O_\pm^{fi}}^2\cdot\delta(\hbar\omega_{fi}-\hbar\omega)$ [see Eq.~\eqref{eqn_opm_squd}; including delta distribution from Fermi's Golden Rule] in nm$^2$\,ns$^{-2}$\,meV$^{-1}$.
	
	\item \texttt{occupancy}:
	The occupation factor $P_{i\to f}$ [see Eq.~\eqref{eqn_occupation_factor}], dimensionless. (Only included if occupancies can be calculated. In this case, also the LL degeneracy factor $G=eB/2\pi\hbar$ is included.)
		
	\item \texttt{rate}:
	Calculates a transition rate density per electric field intensity,
	\begin{equation}
		R = \frac{\pi}{4\hbar\omega_{fi}^2}\cdot\abs{O_\pm^{fi}}^2\cdot G\cdot P_{i\to f} \label{eqn_transition_rate_density}
	\end{equation}
	in ns$^{-1}$ mV$^{-2}$, taking LL degeneracy $G$ and occupancy into account.
	
	\item \texttt{absorption}:
	Local 2D absorption coefficient $\alpha$ as in $I(d)=\ee^{-\alpha}\cdot I(0)$
	\begin{equation}
		\alpha = \frac{1}{\varepsilon_0}\frac{2}{cn}\hbar\omega_{fi}R
	\end{equation}
	in \textperthousand, where $R$ is the transition rate per electric field intensity, with velocity of light $c$ and refractive index $n$. (Only included if the refractive index of the QW layer is known. In this case, also a signed absorption coefficient is included, where the sign indicates which circular polarization basis is absorbed.)
\end{itemize}

\subsubsection{Outlook on polarimetry spectra}

Expanding on optical absorption spectroscopy, polarimetry experiments investigate the polarization state of the EM wave after interaction with a sample. The generally elliptic polarization can be described in terms of a complex angle. For the above mentioned applications, the imaginary part of this complex angle is of great interest, which can be translated into an ellipticity $\epsilon_F$. The ellipticity describes the ratio of minor $b$ to major axis $a$ by $\tan\epsilon_F=b/a$ and is a measure for the shape of the ellipse, where the sign indicates the pseudo-spin of the EM wave. Therefore, it not only carries information on the absorption amplitude but also on the relative absorption difference between both circular polarization modes, which allows observing a change in orbital composition of states and thus, achieving further insight into the band structure.

Knowledge of the relative absorption $A_\pm$ allows direct access to the ellipticity $\epsilon_F$, by $\tan(\epsilon_F)=\frac{E_+-E_-}{E_++E_-}$, where $E_\pm\propto\sqrt{I_\pm(d)}$. Assuming $I_+(0)=I_-(0)$ and using $A_\pm = \frac{I_\pm(d)}{I_\pm(0)}=\exp(-\alpha_\pm)$ we get
\begin{align}
	\epsilon_F = \arctan\left(\frac{\sqrt{A_+} - \sqrt{A_-}}{\sqrt{A_+} + \sqrt{A_-}}\right) = \arctan\left(\frac{\exp(-\alpha_+/2) - \exp(-\alpha_-/2)}{\exp(-\alpha_+/2) + \exp(-\alpha_-/2)}\right)
\end{align}

At the time of publication of this article, only the absorption coefficients are calculated by \kdotpy{}, ellipticity may be added at a later stage in development. Until then, users may use the csv output to calculated the ellipticity on their own.
Note that these calculated ellipticity spectra are delta peak transitions. In experimental data these transitions always have finite linewidth. This can be accounted for by broadening all transitions by Cauchy-Lorentz distributions, after calculating them with \kdotpy{}.

\subsection{Other postprocessing functions}

\subsubsection{Extrema}

Local extrema in the band dispersions are important features for characterizing
the band structure. In \kdotpy, extrema analysis is performed if the
command-line argument \texttt{extrema} is given. The algorithm relies on the
notion of bands, hence the band indices are essential. The extrema analysis does
not just find the local extrema in the calculated eigenvalues, but also does a
quadratic interpolation to localize each extremum more precisely, and to
determine the effective mass as the inverse of the second derivative. Each band
extremum is stored in a light-weight class \texttt{BandExtremum} that contains
the momentum value, energy value, whether minimum or maximum, and the effective
mass. The results are displayed as standard output and written to a csv file.
The extrema are also shown in two-dimensional dispersion plots.

\begin{figure}
  \includegraphics[width=150mm]{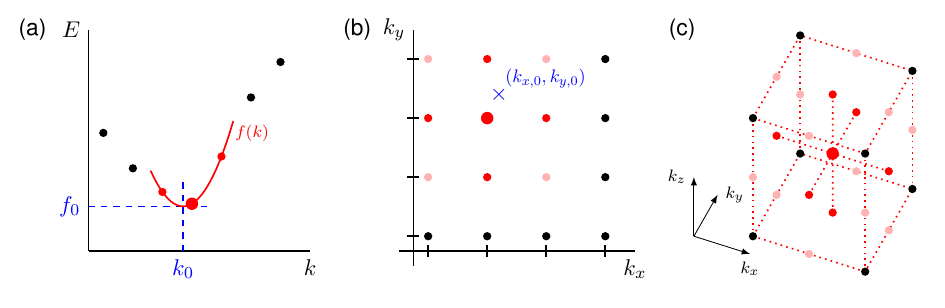}
  \caption{Extrema solvers in 1, 2, and 3 dimensions.
  (a) The three-point extrema solver for one dimension fits the function
  $f(k) = f_0 + c (k-k_0)^2$ to three consecutive data points. The location of
  the minimum in momentum space is $k_0$, which may be between the
  grid points. The local minimum has energy $f_0$.
  (b) In two dimensions, a nine-point extrema solver is used. The fit function
  [Eq.~\eqref{eqn_nine_point_extremum_function}] considers the energy values at
  the five red points [central point $(k_{x,i},k_{y,j})$ highlighted with larger
  size]. The four pink points are considered partially, i.e., only some linear
  combinations of the dispersion values, see Appendix~\ref{app_extrema_solvers}.
  The location $(k_{x,0},k_{y,0})$ of the minimum is generally between the grid
  points.
  (c) In three dimensions a nineteen-point extrema solver is used. Seven points
  (red) are considered fully, twelve more points (pink) are treated partially.
  The corner points (black) are not considered at all.
  }
  \label{fig_extrema_solvers}
\end{figure}

In one dimension, the following algorithm is applied to each band $b$:
\begin{itemize}
 \item Extract the dispersion $e_i=E^{(b)}(k_i)$ as function of the momenta
 $k_i$ on the grid.
 \item Scan over all $i=2, \ldots, n_k-1$. If a triplet $(e_{i-1},e_i,e_{i+1})$
 satisfies $e_{i-1}>e_i$ and $e_{i+1}>e_i$, label it as a minimum; if
 $e_{i-1}<e_i$ and $e_{i+1}<e_i$, label it as a maximum.
 \item For each minimum and maximum, apply \verb+three_point_extremum_solver()+.
 This function takes the points $(k_{i-1},e_{i-1})$, $(k_{i},e_{i})$, and
 $(k_{i+1},e_{i+1})$, and calculates the coefficients $f_0$, $k_0$, and $c$ of
 the quadratic function $f(k) = f_0 + c (k-k_0)^2$ that passes through these
 points. Assuming that the momentum values are spaced evenly,
 $k_{i+1} - k_i = k_i - k_{i-1}$, we have
 \begin{align}
  c   &= \frac{e_{i+1}-2e_i+e_{i-1}}{2(k_{i}-k_{i-1})^2},\nonumber\\
  k_0 &= k_i - \frac{e_{i+1}-e_{i-1}}{2 (k_{i+1}-k_{i-1}) c},
    \label{eqn_three_point_extremum_solver}\\
  f_0 &= e_{i} - c(k_i-k_0)^2\nonumber.
 \end{align}
 The coefficient $k_0$ is the momentum position of the extremum and $f_0$ is
 the energy value, see Fig.~\ref{fig_extrema_solvers}(a) for an illustration.
 The second derivative is equal to $2c$, from which we determine the
 effective mass $m_\mathrm{eff} = -\hbar/m_0 c$, where $m_0$ is the bare electron
 mass.
\end{itemize}
If the momentum range starts or ends at $k=0$, it is automatically extended by
reflection, so that extrema at zero are found as well.

In two dimensions, we use a similar method to detect extrema at the grid points
$(k_{x,i},k_{y,j})$. The location, energy, and band mass of the extremum are
determined with the\linebreak[4] \verb+nine_point_extremum_solver()+,
that tries to fit the quadratic equation
\begin{equation}\label{eqn_nine_point_extremum_function}
 f(k_x,k_y) = f_0 + a (k_x-k_{x,0})^2 + b (k_y-k_{y,0})^2 + c (k_x-k_{x,0}) (k_y-k_{y,0})
\end{equation}
to the energies at the nine points defined by indices $\{i-1,i,i+1\}$ and
$\{j-1,j,j+1\}$ in the $k_x$ and $k_y$ direction, respectively. There are only
six unknowns, namely $f_0$, $(k_{x,0}, k_{y,0})$, and $(a, b, c)$, for nine
input variables. Of the four corner point values $e_{i\pm 1,j \pm 1}$, only the
linear combination $e_{i+1,j+1}-e_{i-1,j+1}-e_{i+1,j-1}+e_{i-1,j-1}$ is considered
see Fig.~\ref{fig_extrema_solvers}(b) and Appendix~\ref{app_extrema_solvers}.
The extremum energy is $f_0$, its location
$(k_{x,0}, k_{y,0})$, and the band mass follows from the eigenvalues
$\lambda_{1,2}$ of the Hessian matrix
\begin{equation}
 \begin{pmatrix}2a & c \\ c & 2b\end{pmatrix}
\end{equation}
as $m_\mathrm{eff,\alpha} = -\hbar/m_0\lambda_\alpha$. Momentum space extension
is also used if the range of $k_{x}$ and/or $k_{y}$ values ends at zero.
For polar coordinates, we use the same algorithm as for cartesian coordinates,
except for $k=0$, where we use a special version of the one-dimensional solver.
The band masses are coordinate-system independent in principle: prior to obtaining
the eigenvalues, the Hessian matrix is transformed to cartesian coordinates
(see Appendix~\ref{app_extrema_solvers}).
If one compares the results from dispersion calculations in cartesian and polar
coordinates, they are equal up to some numerical differences from the coordinate
conversion and from the grid point interpolation.

For three dimensions, we use a \verb+nineteen_point_extremum_solver()+. The
inputs are arranged on a $3\times 3\times 3$ grid, but the eight corner points
are ignored. The fitted function is
\begin{equation}\label{eqn_nineteen_point_extremum_function}
  f(k_x,k_y,k_z) = f_0 +
  \begin{pmatrix}k'_x & k'_y & k'_z\end{pmatrix}
  \begin{pmatrix}2a & d & e\\d & 2b & f\\e & f & 2c\end{pmatrix}
  \begin{pmatrix}k'_x \\ k'_y \\ k'_z\end{pmatrix},
\end{equation}
where $(k'_x,k'_y,k'_z) = (k_x,k_y,k_z) - (k_{x,0},k_{y,0}, k_{z,0})$.
The number of unknowns is ten, i.e., the energy $f_0$, the location
$(k_{x,0},k_{y,0}, k_{z,0})$ and the six independent entries $(a,b,c,d,e,f)$ of
the Hessian matrix. Like with the nine-point extrema solver, some points are
considered partially, see Fig.~\ref{fig_extrema_solvers}(c) and
Appendix~\ref{app_extrema_solvers}. The band masses are derived from the 
eigenvalues of the Hessian matrix, with coordinate transformations applied for 
cylindrical and spherical coordinate systems.

\subsubsection{Wave functions}
\label{sec_wavefunctions}


The band structure scripts (\texttt{kdotpy 1d}, \texttt{kdotpy 2d}, etc.) have 
the option to plot the wave functions. This is done with the command line
argument \texttt{plotwf} and further arguments that determine the plot style and the 
locations (momenta or magnetic fields) at which the plots are made. 

In most cases, \texttt{plotwf} will output data files in csv format together 
with the plots. The csv files typically contain the same data as the plot in the 
plot PDF. For styles where multiple plots are collected in a single PDF, there 
will be a separate csv file for each page in the PDF (i.e., for each state),
where the file name labels the state by band index, LL index, character, and/or
energy. If the configuration option \verb+table_wf_files+ is set to \texttt{tar}, 
\texttt{targz}, \texttt{zip}, or \texttt{zipnozip} the files are collected into 
a single archive file.

\begin{figure}
\includegraphics[width=140mm]{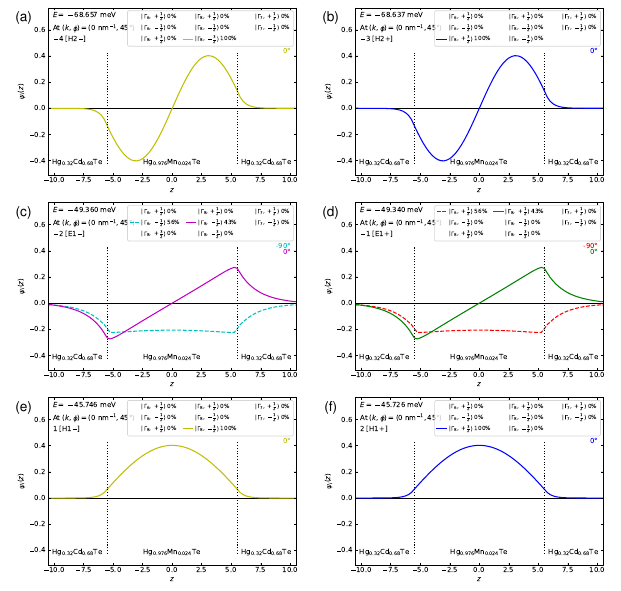}
\caption{Examples of wave function plots from \texttt{kdotpy 2d} with
\texttt{plotwf separate}. The output is a multipage PDF; here, six pages
are shown as panels (a)--(f), each representing a different eigenstate.
}
\label{fig_wf_2d_separate}
\end{figure}

\begin{figure}
\includegraphics[width=90mm]{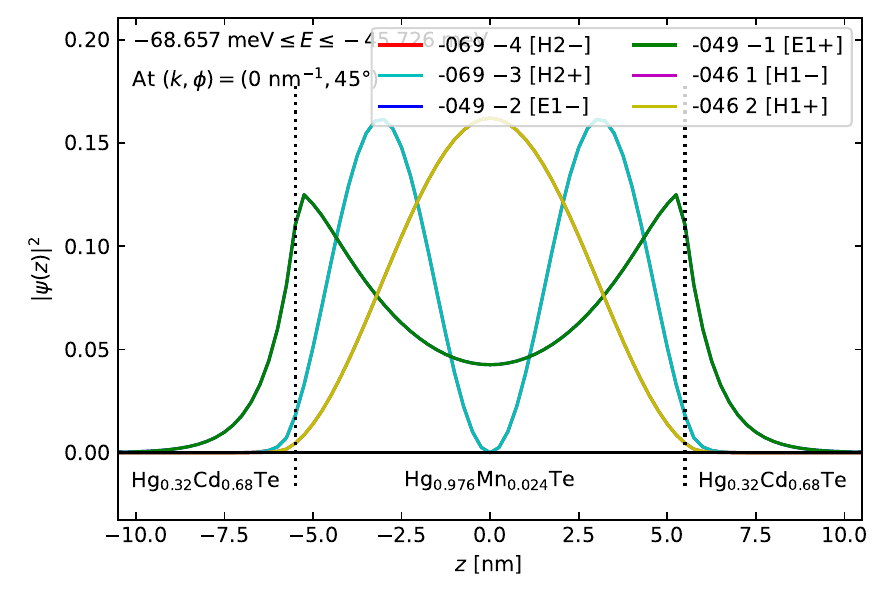}
\caption{Example of a wave function plot from \texttt{kdotpy 1d} with
\texttt{plotwf together}. The output is a single page PDF.
}
\label{fig_wf_2d_together}
\end{figure}

For the two-dimensional geometry, there are two main plot styles:
\begin{itemize}
\item \texttt{separate}: As function of z, with different curves for each orbital.
 For each state, the wave function is expanded into its orbital components
 $\psi_i(z)$. Each orbital is represented by a separate colour and solid and
 dashed lines indicate the real and imaginary parts, respectively.
 Orbitals for which the amplitude is small, are not drawn. The plot is saved as a
 multi-page PDF, with each eigenstate on a separate page,
 see Fig.~\ref{fig_wf_2d_separate}.
\item \texttt{together}: Absolute-squared, as function of z, together in one plot,
 see Fig.~\ref{fig_wf_2d_together}.
\end{itemize}

For plot style \texttt{separate}, the phase for each orbital wave function is
normalized to the orbital with largest amplitude (over z, not integrated): The
maximum value is set to a positive real number, with relative phases being
preserved. If the orbital have a definite complex phase, i.e.,
$\psi_i(z_) = f_i(z)\ee^{\ii\phi_i}$ for some real function $f_i(z)$ and phase value
$\phi_i$, the phases $\phi_i$ are listed as angles in degrees as additional
inset on the right side of the plot. The behaviour can be adjusted with the 
configuration option \verb+plot_wf_orbitals_realshift+.

In LL mode \texttt{kdotpy ll}, one can use the same plot styles 
\texttt{separate} and \texttt{together}. In symbolic LL mode, the wave function 
$\psi(z)$ is plotted for the corresponding LL index $n$. In full LL mode, plot 
only the largest contribution, i.e., $\psi_n(z)$ for which $\int|\psi_n(z)|^2 
dz$ is maximal over $n$; in this case, the probability density 
$\int|\psi_n(z)|^2 dz$ may be smaller than~$1$. For plot style \texttt{separate}, 
this LL index $n$ and the probability density are displayed as $\mathrm{LL} = n$ 
and $|\psi_\mathrm{LL}|^2=\mathrm{value}$, respectively, for each state.

\begin{figure}
\includegraphics[width=140mm]{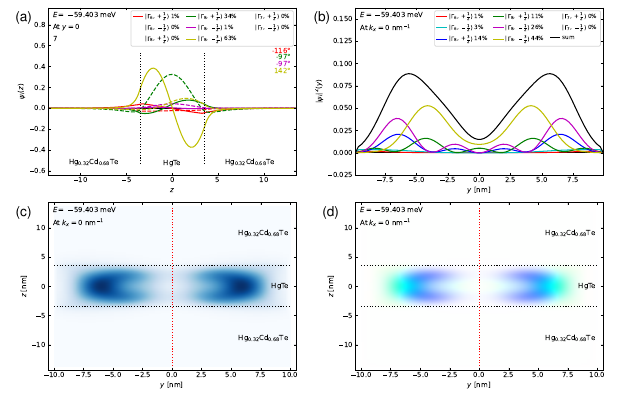}
\caption{Examples of wave function plots from \texttt{kdotpy 1d} with
(a) \texttt{plotwf z},
(b) \texttt{plotwf y},
(c) \texttt{plotwf zy}, and
(d) \texttt{plotwf color}. These examples are based on the same eigenstate. The
actual output contains the wave functions for multiple eigenstates.
}
\label{fig_wf_1d}
\end{figure}

For the one-dimensional geometry, the wave functions $\psi_i(z,y)$ depend on
two spatial coordinates. The following plot styles can be used:
\begin{itemize}
 \item \texttt{z}: As function of $z$, at $y = 0$ (at the middle of the sample),
 i.e., $\psi(z, 0)$. Like \texttt{separate} for 2D, the wave function is decomposed
 into orbitals.
 \item \texttt{y}: Absolute-squared, as function of $y$, integrated over the $z$
 coordinate, i.e.,\linebreak[4] $|\psi|^2(y) = \int |\psi(z, y)|^2 dz$. The wave function is
 decomposed into orbitals or subbands.
 \item \texttt{zy}: Absolute-squared, as function of $(z, y)$, i.e.,
 $|\psi(z, y)|^2$, summed over all orbitals. The values $|\psi(z, y)|^2$ are
 represented as colours from a colour map.
 \item \texttt{color}: Absolute-squared, with different colours for the orbitals,
 as function of $(z, y)$.
\end{itemize}
Examples are shown in Fig.~\ref{fig_wf_1d}. With \texttt{kdotpy 1d}, wave
function output is possible only if the grid contains a single point.
This restriction has been imposed to avoid problems
with the large memory footprint needed given two spatial coordinates.

\subsubsection{Symmetry analysis and symmetrization}

The point group symmetry of the crystal structure imposes symmetry constraints
on the Hamiltonian. The band structure exhibits the symmetries of the Hamiltonian,
and \kdotpy{} provides a way to verify the symmetry properties, known as
\emph{symmetry analysis}.

In \kdotpy, symmetry analysis is implemented as the member function
\verb+DiagData.symmetry_test()+. This function applies the following algorithm:
\begin{itemize}
 \item Define a symmetry transformation $T$, being a reflection, rotation, or
 roto-reflection of the point group. This object is implemented as instance of 
 the \verb+VectorTransformation+ class.
 \item For each $\vec{k}\not=0$ in the grid, verify if its image $T\vec{k}$ is
 also in the grid. If $T\vec{k}$ is not in the grid or if $T\vec{k}=\vec{k}$,
 skip this point.
 \item Compare the set of eigenvalues at $\vec{k}$ and the set of eigenvalues
 at $T\vec{k}$ and verify whether they are identical up to small numerical errors.
 If not, there is no symmetry between $\vec{k}$ and $T\vec{k}$.
 \item If the eigenvalues are the same, compare the observables at $\vec{k}$ and
 $T\vec{k}$. For scalar observables $O$, verify if $O_\vec{k}=\pm O_{T\vec{k}}$
 for all eigenstates. (Special consideration is given to degenerate states.)
 For vector observables $\vec{O}=(O_x,O_y,O_z)$, for example
 $(\mathtt{jx},\mathtt{jy},\mathtt{jz})$, verify if it transforms as one of the
 vector representations from the point group $O_h$.
 \item Collect the results by iterating over the grid points $\vec{k}$. For each
 transformation $T$ and for each observable $O$, print the $O_h$ representations
 compatible with the transformation properties of the observable.
 \item As a final step, analyze the symmetries over all transformations $T$. Try
 to deduce the relevant point group (subgroup of $O_h$) and the possible
 representations for each observable.
\end{itemize}
In version 1.0 of \kdotpy, the implementation is experimental, and may be
improved in a future version.

The \emph{symmetrization} feature of \kdotpy{} works in the opposite direction:
Knowing that the dispersion satisfies a set of symmetries, the dispersion can
be calculated on some range of momentum space and be extended to a larger
range of momenta, thus saving time compared to doing a calculation on the full
range. For the momenta $T\vec{k}$ obtained by symmetrization from the original
grid points $\vec{k}$, the eigenvalues are copied from $\vec{k}$, and the 
observable values are calculated from those at $\vec{k}$ with the appropriate
transformation applied, given by a representation of the point group.

\begin{figure}
 \includegraphics[width=140mm]{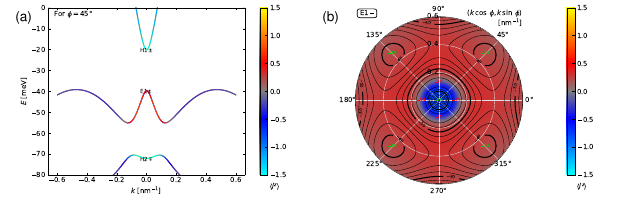}
 \caption{(a) Symmetrization of a one-dimensional momentum scan. The data for
 $k<0$ are obtained by symmetrization from those at $k>0$.
 (b) Symmetrization of a two-dimensional dispersion in polar coordinates. The
 data has been calculated explicitly for angles $k_\phi\in[0^\circ,90^\circ]$.
 Symmetrization extends it to the full circle.}
 \label{fig_symmetrization}
\end{figure}

A common use case is the reflection around zero for a one-dimensional momentum
scan in a two-dimensional geometry (\texttt{kdotpy 2d}),
\begin{verbatim}
kdotpy 2d 8o noax msubst CdZnTe 4% mlayer HgCdTe 68% HgTe HgCdTe 68%
llayer 10 7 10 zres 0.25 k 0 0.6 / 60 kphi 45 erange -80 0 split 0.01
splittype isopz obs jz legend char out -jz outdir data-qw localminmax
symmetrize
\end{verbatim}
where the eigenvalues and observable values for negative $k$ are obtained from
those at $-k$, see Fig.~\ref{fig_symmetrization}(a).
(The observable $O=\avg{J_z}$ is symmetric under this transformation.)
A similar construct works for two-dimensional dispersions: The dispersion is
calculated in the first quadrant ($k_x>0$, $k_y>0$ in cartesian coordinates or
$k_\phi\in[0^\circ,90^\circ]$ for polar coordinates) and extended to all four
quadrants, see Fig.~\ref{fig_symmetrization}(b).


\subsubsection{L\"owdin perturbation theory (BHZ-like models)}
\label{sec_bhz}

A substantial number of works in the field relies on the simplified model made
famous by Bernevig, Hughes, and Zhang (BHZ) \cite{BernevigEA2006}. This 
effective model takes the four subband degrees of freedom
$\ket{\mathrm{E}1\pm},\ket{\mathrm{H}1\pm}$ as its basis. This model is adequate
for describing the topological phase transition in quantum wells as function of
the well thickness $d$: At the critical thickness $d_c$, the energetic positions
of the electron-like subbands $\ket{\mathrm{E}1\pm}$ and the heavy-hole-like
$\ket{\mathrm{H}1\pm}$ change. For $d<d_c$, the device is trivially insulating
and for $d>d_c$, the device is a two-dimensional topological insulator that
hosts the quantum spin Hall effect \cite{BernevigEA2006} that can be measured in
a Hall bar geometry \cite{KonigEA2007}.

Outside of the context of HgTe quantum wells with $d\approx d_c$, the four-band
BHZ model is often inadequate to describe the essential physics. However, similar
models with a modified or extended basis may be useful in order to gain useful
insight, in some cases. Although we do not advocate the use of these simplified
models in subband basis, we have included the functionality to derive them so
that they can be compared against the more accurate \kdotp{} models.

The method of projecting the \kdotp{} Hamiltonian onto a basis of subband states
is known as L\"owdin perturbation theory or quasi-degenerate perturbation theory
\cite{Lowdin1951,Winkler2003_book}. The basis is a set of subband states at
$\vec{k}=0$. The contributions from the other subband states (outside of the basis)
and from $\vec{k}\not=0$ are treated by perturbation theory. The result is an
effective Hamiltonian in the chosen subband basis that is valid near $\vec{k}=0$.

In \kdotpy, L\"owdin perturbation theory is implemented up to second order in
momentum. We use the framework of symbolic Hamiltonians
(class \texttt{SymbolicHamiltonian}) that we also use for deriving Landau level
Hamiltonians. The recipe is as follows:
\begin{itemize}
 \item Take the symbolic Hamiltonian $H$ as well the set of eigenvalues and
 eigenstates at $\vec{k}=0$ as input.
 \item Select the basis for the effective model based on the command-line
 argument \texttt{bhz}. Usually, this is an even number $n_\mathrm{A}$ of
 subband states near the charge neutrality point. We label the subband states in
 the basis the `A states'.
 \item Select the subband states that are treated perturbatively. The amount can
 optionally be specified from the command line. If it is not specified, all
 states confined in the quantum well are taken by default. (Deconfined states
 are not considered, because including them leads to unpredictable and unphysical
 results.) We label these states the `B states'. For the selection procedure,
 they are separated into `lower' and `upper' B states, i.e., with energies below
 and above those of the A states.
 \item For the perturbative expansion, we use the expressions listed in the
 textbook by Winkler \cite{Winkler2003_book}. The zeroth term is simply the
 diagonal matrix of the eigenvalues $E_m$ of the A states,
 \begin{equation}
  H^{(0)} = \diag(E_m).
 \end{equation}
 The matrix size is $n_\mathrm{A}\times n_\mathrm{A}$. 
 \item The construction of the first order perturbation term $H^{(1)}$ is
 implemented in\linebreak[4] \verb+SymbolicHamiltonian.hper1()+. It takes the matrix of the
 eigenvectors $\ket{\psi_m}$ (where $m$ label the A states) as its only
 argument. It is defined as \cite{Winkler2003_book}
 \begin{equation}
  H^{(1)}_{mm'} = \bramidket{\psi_m}{H}{\psi_{m'}},
 \end{equation}
 where $H$ is the symbolic Hamiltonian. The resulting matrix $H^{(1)}$ is
 an $n_\mathrm{A}\times n_\mathrm{A}$ matrix containing the operators
 $\hat{k}_\pm$ up to second order. The function \verb+SymbolicHamiltonian.hper1()+
 returns a list of lists of \verb+SymbolicObject+ instances, representing the
 matrix elements $H^{(1)}_{mm'}$.
 \item The construction of the second order perturbation term $H^{(2)}$ is
 implemented in\linebreak[4] \verb+SymbolicHamiltonian.hper2()+. As arguments, it takes the
 eigenvalues $\{E_m\}$ and $\{E_l\}$ of the A and B states, respectively, as
 well as the associated eigenvectors $\ket{\psi_m}$ and $\ket{\phi_l}$. The
 second-order term is \cite{Winkler2003_book}
 \begin{equation}
  H^{(2)}_{mm'} = \frac{1}{2}\sum_l \bramidket{\psi_m}{H}{\phi_l}\bramidket{\phi_l}{H}{\psi_{m'}}\left(\frac{1}{E_m-E_l}+\frac{1}{E_{m'}-E_l}\right),
 \end{equation}
 where the index $l$ of the sum runs over all B states. The result $H^{(2)}$ is
 once more an $n_\mathrm{A}\times n_\mathrm{A}$ matrix involving the operators
 $\hat{k}_\pm$. The function \verb+SymbolicHamiltonian.hper2()+ returns the
 matrix elements $H^{(2)}_{mm'}$ as a list of lists of \verb+SymbolicObject+
 instances.
 \item The results are summed together, $H^\mathrm{L} = H^{(0)} + H^{(1)} + H^{(2)}$.
 Negligible coefficients (absolute value $<10^{-7}$) and terms of
 order $>2$ in $\hat{k}_\pm$ are discarded.
 \item The basis vectors may be multiplied by complex phases in an attempt to
 make the matrix elements $H^\mathrm{L}_{mm'}$ purely real or purely imaginary.
 \item If the basis can be separated into two uncoupled sectors, it is
 reordered, such that the matrix $H^\mathrm{L}$ separates in two blocks on the
 diagonal.
\end{itemize}
The result of this method is a \texttt{SymbolicHamiltonian} object, that encodes
the operator sum
\begin{equation}
  H^\mathrm{L}
  = H^\mathrm{L}_0 + H^\mathrm{L}_+ \hat{k}_+ + H^\mathrm{L}_- \hat{k}_-
  + H^\mathrm{L}_{++} \hat{k}_+^2 + H^\mathrm{L}_{--} \hat{k}_-^2
  + H^\mathrm{L}_{+-} \hat{k}_+\hat{k}_- + H^\mathrm{L}_{-+} \hat{k}_-\hat{k}_+.,
\end{equation}
where the factors $H^\mathrm{L}_0$, $H^\mathrm{L}_0\pm$, etc. are
$n_\mathrm{A}\times n_\mathrm{A}$ matrices.

\begin{figure}
 \includegraphics[width=120mm]{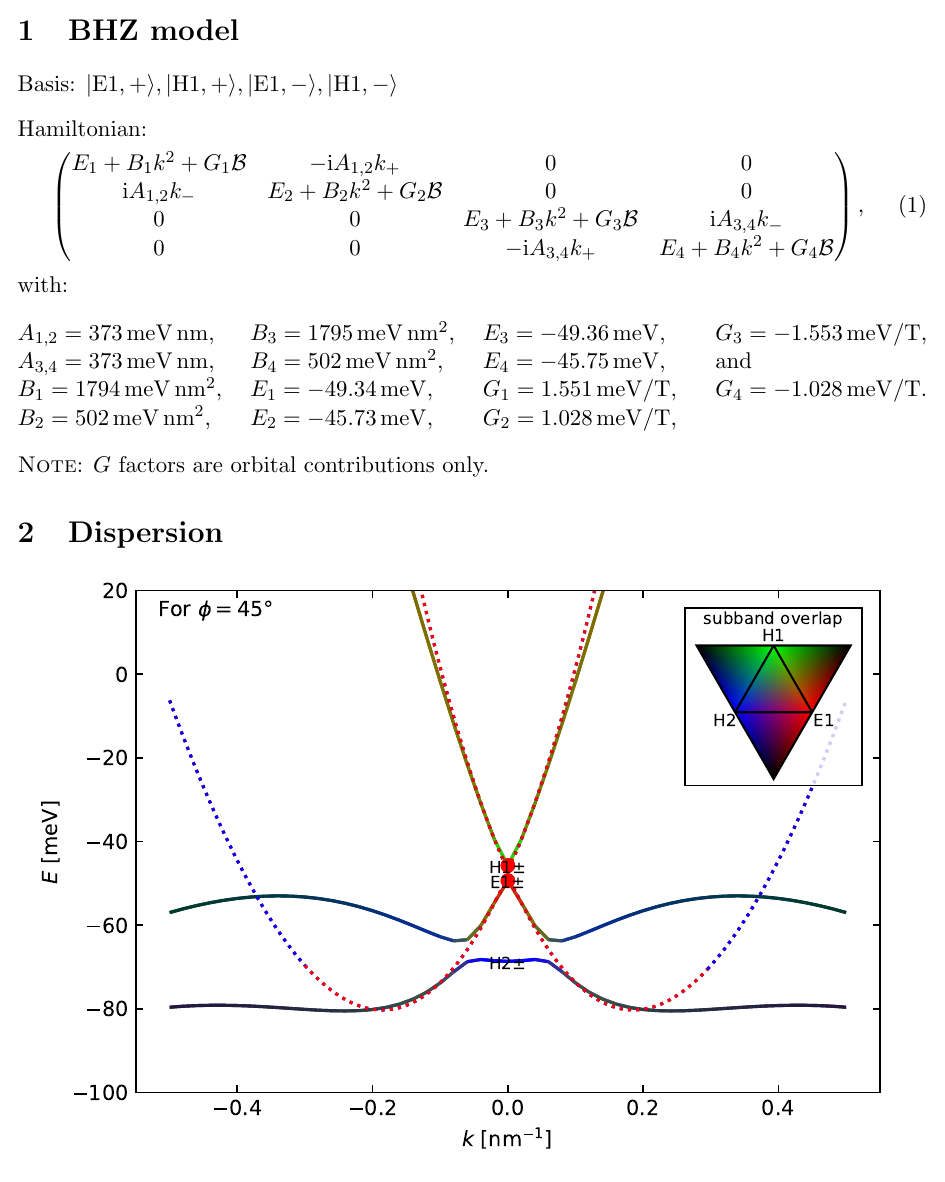}
 \caption{Example output of \kdotpy{} with the \texttt{bhz} command line
   argument. This calculation is done for a quantum well of $11\nm$ thick
   Hg$_{1-x}$Mn$_x$Te ($x=0.024$). This L\"owdin perturbation uses the same
   basis as the canonical BHZ model of Ref.~\cite{BernevigEA2006}. In the plot,
   the solid curves indicate the \kdotp{} dispersion. The dotted curves are the
   bands calculated from the L\"owdin Hamiltonian $H^\mathrm{L}$ (here,
   equivalent to the BHZ Hamiltonian). The red and blue colours of the dotted
   curves separate the two blocks; here, they are degenerate. (The layout
   has been adjusted for reasons of illustration. The actual \kdotpy{} output
   appears on two landscape A4 pages.)}
  \label{fig_bhz}
\end{figure}

The submodule \texttt{bhzprint} outputs the result as a matrix in \LaTeX{}
notation and compiles it if PDF\LaTeX{} is available. An example result is
shown in Fig.~\ref{fig_bhz}. The result is expressed in a notation that
generalizes the notation of Ref.~\cite{BernevigEA2006}. For a L\"owdin
perturbation similar to the BHZ model, i.e., two blocks of two subbands each,
the configuration setting \verb+bhz_abcdm=true+ may be used in order to express
the result in the canonical BHZ notation with coefficients $A,B,C,D$ and $M$.
In \texttt{verbose} mode, \kdotpy{} also writes the matrix elements
$H^\mathrm{L}_{mm'}$ and several intermediate results to standard
output.

\subsection{Data output}

\subsubsection{Types of data}

Output is produced by \kdotpy{} in various formats that serve various (in part
complementary) purposes.
\begin{itemize}
 \item Long-time storage: Scientific data must be maximally reproducible and is
 expected to be kept for a longer period of time. For this purpose, we use an
 XML-based data format, which contains both data and metadata.
 \item Immediate further processing: Data must be in a format that many other
 applications can read or import with as little effort as possible. We rely
 on CSV files (comma-separated values) for this purpose.
 \item Human interaction: Immediate feedback aids users in judging the
 calculation results even without further processing. Examples are graphical
 output (PDF and PNG) and console output.
\end{itemize}
Plots and CSV files are typically produced in pairs, with the CSV file containing
exactly the data that is shown in the figure.

\subsubsection{XML files}
\label{sec_xml}

The XML-format produced by \kdotpy{} is a hierarchical data format originally
designed for communication between different sessions of \kdotpy, in particular
for combining data sets with \texttt{kdotpy merge}. An important design principle
to make this possible is maximum compatibility between different versions of \kdotpy.
The hierarchical nature of the XML format is well suited to achieve this goal.
Reading the XML data is done by means of extracting the relevant tags.
When a new type of data is added to the XML file (e.g., by a newer version of
\kdotpy), the new tag is simply ignored by older versions, and the XML file
remains compatible.

The format includes a large number of metadata attributes, which is an important
asset for reproducibility. The data file contains the following tags with metadata:
\begin{itemize}
 \item \texttt{<info>}: Information on the program. This tag contains the command
 line, information on the host (computer) and its operating system, and the
 version numbers of \kdotpy{} (including the Git hash if available), Python, and
 the installed modules. 
 \item \texttt{<configuration>}: Configuration values.
 \item \texttt{<parameters>}: All information in the
 \texttt{PhysParams} object, including the evaluated material parameters.
 \item \texttt{<options>}: Options set on the command line. These are the
 values stored in the \texttt{dict} instance \texttt{opts}, which affects
 the calculations.
\end{itemize}
Especially the command line arguments and version numbers are essential for
repeating a calculation. The metadata can also be used for filtering a large
collection of data files by a certain attribute. The metadata are placed at the
top of the file to allow for human inspection.

The actual data follows the metadata. The data in the \texttt{DiagData} object
is serialized into the tag \texttt{<dispersion>} or \verb+<dependence variable="b">+,
in case of a dispersion and a magnetic-field dependence, respectively.
In both cases, it contains the \texttt{<vectorgrid>} tag for the
\texttt{VectorGrid} object. All data points (\texttt{DiagDataPoint} instances)
are serialized as \texttt{<momentum>} (for dispersion) or \texttt{variabledata}
(for magnetic-field dependence); they contain lists of the eigenenergies, band
indices, Landau level indices, and observables.

Finally, the \texttt{<extrema>} tag is included for serialization of the
\texttt{BandExtremum}. The serialization of further data container objects 
(e.g., \texttt{DensityData}) is scheduled for future versions of \kdotpy.

\subsubsection{CSV files}
\label{sec_tables}

Data for further processing is saved in CSV files. The CSV (comma separated value)
format is a universally understood lightweight format for storing one- or
two-dimensional arrays of data. The workflow in \kdotpy{} separates the
composition (or preparation) of the data and writing it to a file. In the composition
stage, \kdotpy{} prepares a list or a \texttt{dict} instance, which contains a
set of one-dimensional arrays of values, representing the columns. This data is
passed to the writer function, together with formatting information and labels
for each column. In some cases, a post-write function adds extra data to the file,
e.g., extra rows with band labels.

\begin{figure}
 \includegraphics[width=150mm]{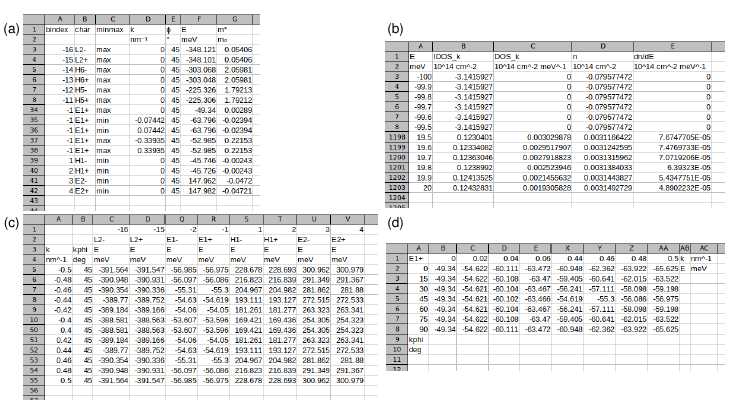}
 \caption{Examples of CSV output, as imported by a spreadsheet program.
 Some rows and columns have been hidden for clarity.
 (a) \texttt{extrema.csv}: Column-wise data, where each row lists the properties
 of a band extremum. The first and second row label the quantities and units.
 (b) \texttt{dos.csv}: The DOS and the IDOS as function of energy $E$.
 (c) \texttt{dispersion-byband.csv}: Dispersions $E_i(\vec{k})$, i.e., band energies
 as function of momentum. Each column represents a band. The bands are labelled
 by band index and character in the first and second row.
 (d) \texttt{dispersion.b\emph{i}.csv}: Two-dimensional dispersion $E(k,\phi)$ of
 a single subband.
 }
 \label{fig_csv}
\end{figure}

We use a small number of common layouts of the data. The uniformity is a
boon to both users and developers. Fewer layouts means less effort for external
processing of the data. On the development side, there is less code to maintain.
The following layouts are the most common ones:
\begin{itemize}
  \item Column-wise data: Each column in the table represents a property or
  quantity and each row represents another item. For example, in 
  \texttt{extrema.csv}, there are columns for band index, band character, 
  minimum or maximum, momentum, energy, and band mass; see Fig.~\ref{fig_csv}(a)
  for an illustration. The column headers on the first row indicate the
  quantity, and units are included (optionally) on the second row.
  \item Functions of one dimension: A set of functions $f_i(x)$ is represented
  column-wise. The $x$ values appear in the first column, and the subsequent
  columns represent $f_i(x)$. Examples: \texttt{dos.csv} [Fig.~\ref{fig_csv}(b)],
  with DOS and IDOS as function of energy $E$ (first column);
  \texttt{dispersion-byband.csv} [Fig.~\ref{fig_csv}(c)], with the dispersion
  $E_i(\vec{k})$ of many bands $i$ as function of momentum
  $\vec{k}$. The latter uses two columns, here with $k$ and $\phi$ in the
  polar coordinate system. The band indices and characters appear in two extra
  rows at the top; the position is configurable.
  \item Two-dimensional array: The data represents a function $f(x,y)$. The
  first row contains the values $x$ and the first column contains the values
  $y$. The data $f(x,y)$ starts at the second row and second column.
  The quantities and units for $x$ are printed at the end of the first row and
  those for $y$ at the end of the first column. The quantity and unit for $f$
  are at the end of the second row, i.e., the first data row. The first row and
  first column provides an extra label for the data. Examples:
  \texttt{dispersion.b\emph{i}.csv} [Fig.~\ref{fig_csv}(d)], which provides the
  dispersion $E_i(k, \phi)$ for a single subband. The file is labelled with
  the band index $i$.
  This layout is also used frequently in LL mode for functions $f(B,E)$ or $f(B,n)$,
  i.e., quantities as function of magnetic field $B$ and energy or carrier density.
  \item Three-dimensional array: Used for bulk dispersions $E(k_x,k_y,k_z)$
  (as well as for cylindrical and spherical coordinates). The third coordinate
  is inserted as the first column. The result is arranged as a stack of
  two-dimensional arrays.
\end{itemize}
At the time of composition, the data is arranged in columns. The composition
functions also determines the formatting for the data, i.e., how numerical values
are converted to strings. For several quantities \texttt{\emph{X}}, the number
of digits for numerical values can be set by configuration values
\texttt{table\uscore\emph{X}\uscore{}precision}.
The composition function also takes care of the formatting of
quantities and units in the headers. The result depends on the configuration
settings \texttt{table\uscore\emph{X}\uscore{}style}, with the following choices:
\begin{itemize}
 \item \texttt{raw}: `Without' formatting; use the raw labels (internal representation)
 for quantities and units.
 \textsc{Note}: The output is not guaranteed to be ASCII only. For example, the 
 `micro' prefix $\mu$ may be used with some units.
 \item \texttt{plain}: Plain-text formatting using common symbols, e.g.,
 square is \verb+^2+ and Greek letters are spelled out.
 \item \texttt{unicode}: Formatting using `fancy' Unicode symbols, e.g.,
 square is the superscript-2 symbol and Greek letters use their
 corresponding Unicode symbol.
 \item \texttt{tex}: \LaTeX{} formatting.
\end{itemize}

The composition functions call the general writer function \texttt{tableo.write.write()}.
This function delegates the actual work to one of the following functions,
depending on the value of the configuration value \verb+csv_style+:
\begin{itemize}
 \item \texttt{csvwrite()}: Uses the \texttt{csv} module
 provided with Python. The default \emph{dialect} is to use \texttt{,} as
 delimiter (separator between values), and \texttt{"} as quoting character.
 This writer is used if \verb+csv_style=csvinternal+ or if \verb+csv_style=csv+
 and the Pandas package is not installed.
 \item \texttt{alignwrite()}: Aligns the formatted strings in columns, using
 spaces as column separators and fill characters. The result is well suited for
 direct viewing in a text editor (with a monospace font). Importing it into a
 spreadsheet program is more tedious than with the other writers. This writer is
 slightly slower than \texttt{csvwrite()} due to the extra step needed for
 determination of the column widths.
 To use this writer, set \verb+csv_style=align+.
 \item \texttt{pdwrite()}: Uses the Pandas package to produce a CSV file. The
 data is first converted into a Pandas \texttt{DataFrame} object. The file is
 then produced by \verb+DataFrame.to_csv()+. The result is very similar to
 \texttt{csvwrite()}, but slightly less flexible in terms of setting the
 numerical precision of floating-point values. This writer is used if
 \verb+csv_style=csvpandas+ or \verb+csv_style=csv+, when Pandas is available.
\end{itemize}
More writers may be added in the future, depending on user request. After the
work done by one of these three writers, \texttt{tableo.write.write()} may call
the post-write functions \verb+write_axislabels()+, for the quantities and units
in the two- and three-dimensional array layout, and/or \verb+write_extraheader()+,
for the band labels (see Fig.~\ref{fig_csv}(c), for example).

\subsubsection{Graphics}
\label{sec_plots}

\begin{figure}[!p]
 \includegraphics[width=150mm]{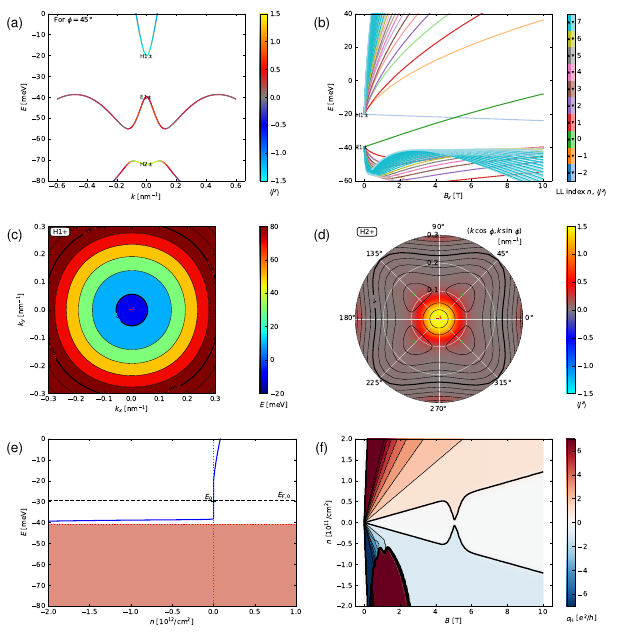}
 \caption{Examples of graphical output of \kdotpy.
 (a) A one-dimensional dispersion plot for a quantum well device with $7\nm$ HgTe
 as the active layer. The plot
 visualizes the dispersions of many bands $E^{(i)}(k)$ as function of momentum
 $k$. The colours encode the values of the observable $\avg{J_z}$.
 (b) Magnetic-field dependence of Landau levels, for the same system as in (a).
 (c) A two-dimensional dispersion plot $E^{(i)}(k_x,k_y)$ for one of the surface
 states of a 3D topological insulator device ($70\nm$ HgTe). The colour scale
 refers to energy $E$ and the red value $-2$ at $\vec{k}=0$ indicates a local
 minimum at that energy.
 (d) A dispersion plot in polar coordinates for the highest valence band state
 of the same system as (c). The colour scale refers to the observable $\avg{J_z}$.
 (e) Integrated density of states (also known as carrier density $n$) corresponding
 to (a). The vertical axis is energy and aligns with the vertical axis of (a).
 (f) Landau fan with carrier density on the vertical axis, corresponding to the
 magnetic field dependence in (b). The colour scale is given by the Hall
 conductance $\sigmaH$ in units of $e^2/h$.
 }
 \label{fig_plots_example}
\end{figure}

For graphics, \kdotpy{} uses the Matplotlib package with the \texttt{pdf} backend.
This backend produces vector graphics, which delivers the highest quality with
reasonable file sizes. (Rasterization is used with some two-dimensional colour
maps, which would lead to large files and long rendering times as pure vector
graphics.)

The results are highly configurable. First of all, command-line arguments can be
used to enable or disable plot elements, for example, legends, plot titles, and
band character labels. Secondly, configuration options determine the plot geometry
(sizes and margins), the colour scale being used, and the style of the axis labels,
for example. Finally, Matplotlib provides a variety of customization options,
named \texttt{rcParams}, that can be manipulated with style sheets. Matplotlib
style sheets are supported by \kdotpy, and can be loaded with the configuration
value \verb+fig_matplotlib_style+. The default style file \verb+kdotpy.mplstyle+
is provided with \kdotpy{} and is copied to the \texttt{\HOME/.kdotpy} directory.
The user may edit the default style file or provide additional style files.

We provide several representative examples of plots generated by \kdotpy. Like
with the CSV output, we use a small number of generic layouts for uniformity of
the output and for maintainability. Commonly used plots are:

\begin{itemize}
 \item One-dimensional dispersion or magnetic-field-dependence plot:
 The function \verb+ploto.bands_1d()+ produces the typical visualization of a
 band structure or Landau level fan, as illustrated by
 Figs.~\ref{fig_plots_example}(a) and (b) in the Introduction.
 The horizontal axis is
 a momentum or magnetic field component, and the vertical axis is energy $E$.
 When an observable is specified with the command line argument \texttt{obs},
 it is used to colour the bands; depending on whether the observable is constant
 or variable, the band is plotted as a single line object with a flat colour or
 as a collection of line segments with variable colours. Optionally, the
 energy axis may be transformed to density $n$.
 \item Two-dimensional dispersion: The function \verb+ploto.bands_2d()+
 implements the visualization of dispersions $E(k_x,k_y)$ or $E(k,\phi)$ in
 cartesian or polar coordinates, respectively;
 see Figs.~\ref{fig_plots_example}(c) and (d).
 The dispersions are represented
 as contour plots in a cartesian or polar coordinate system. The result is a
 multi-page PDF file, with one page for each band within the energy range.
 The colouring is done based on the observable given by the command-line
 argument \texttt{obs}; if this argument is omitted, the colour is based on
 energy.
 \item Total (integrated) density of states: The functions \verb+ploto.dos()+
 and \verb+ploto.integrated_dos()+ plot the DOS and IDOS, respectively, as
 stored in a \texttt{DensityData} instance. By default, the energy axis is
 vertical, and aligns with the vertical axis of dispersion plots.
 See Fig.~\ref{fig_plots_example}(e) for an example.
 \item Generic two-dimensional colour maps: The function \verb+ploto.density2d()+
 provides the framework for generic two-dimensional density plots, where a
 function $f(x,y)$ is represented by a colour map. The horizontal
 axis is typically momentum or magnetic field and the vertical axis is energy
 or density. The function \verb+ploto.density2d()+ is used for many quantities
 in the \texttt{postprocess} module, e.g., optical transitions and Hall plots,
 see Fig.~\ref{fig_plots_example}(f).
\end{itemize}

\subsection{Self-consistent Hartree method}
\label{sec_selfcon}

The \emph{self-consistent Hartree method} is an iterative solver of the
Schr\"odinger equation and the one-dimensional Poisson equation\footnote{This version of
the Poisson equation describes one-dimensional electrostatics in the $z$ direction only.}.
A simultaneous solution of the Schr\"odinger and Poisson equation cannot be found
analytically. The iterative method aims at finding a self-consistent solution,
solving the Schr\"odinger equation and the Poisson equation alternatingly.
The physical picture is that the occupied eigenstates of the Hamiltonian induce
a carrier density, leading to an electrostatic potential that in turn affects
the Hamiltonian. We emphasize that this method applies to the 2D geometry only.
The self-consistent Hartree method can be invoked by using \texttt{selfcon} on the
command line for \texttt{kdotpy 2d} and \texttt{kdotpy ll}.

\subsubsection{Program Flow}
\label{sec_selfcon_program_flow}

\begin{figure}
	\centering
	\includegraphics[width=0.7\linewidth]{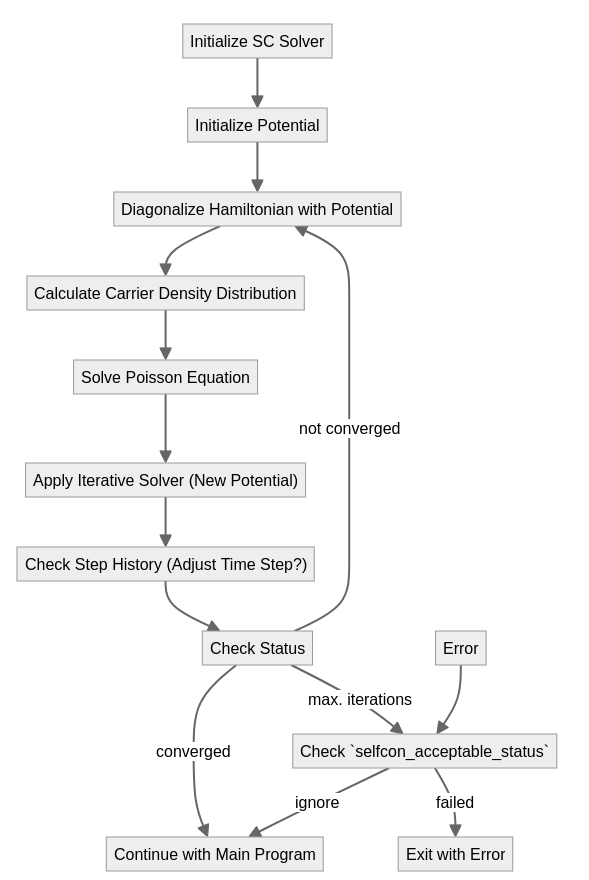}
	\caption{Flowchart for the self-consistent Hartree part of \kdotpy{}.}
	\label{fig_selfcon_flowchart}
\end{figure}

The program flow of the iterative \texttt{SelfConSolver} is sketched in Fig.~\ref{fig_selfcon_flowchart}.
After initializing the starting conditions of the solver, a starting potential is initialized.
If no initial potential is provided and the user did not request to initialize the potential based on a uniform charge density or fixed background density, the first iteration starts with a flat potential.
Next, the Hamiltonian including the Hartree potential is diagonalized and the resulting charge carrier density $\rho(z)$ along the growth direction $z$ is evaluated (see Section \ref{sec_densityz}).
The Hartree potential $\VH(z)$ is then solved from the Poisson equation,
\begin{equation}
	\partial_z \left[\varepsilon(z)\,\partial_z \VH(z)\right] = \frac{e}{\varepsilon_0}\rho(z), \label{eqn_poisson}
\end{equation}
where $\varepsilon(z)$ is the relative dielectric constant of the layers and $\varepsilon_0$
is the permittivity of the vacuum.
The electric-field or potential boundary conditions can be specified for solving Eq.~\ref{eqn_poisson}, see Appendix~\ref{app_solution_poisson} for details.
It is also possible to specify a fixed background density which is added to $\rho(z)$ to model for example ionic cores from modulating doping.

In order to ensure better convergence, the solution of Eq.~\eqref{eqn_poisson} is not directly used as a new potential in the next iteration. Instead, the potential $\VH^{(i+1)}(z)$ of iteration $i+1$ is calculated from the old potential $\VH^{(i)}(z)$ as
\begin{equation}\label{eqn_selfcon_iteration}
  \VH^{(i+1)}(z) = \VH^{(i)}(z) + \tau \left(\mathcal{S}[\VH^{(i)}](z) - \VH^{(i)}(z)\right),
\end{equation}
where $\mathcal{S}[\VH^{(i)}]$ is the potential obtained from solving Eq.~\eqref{eqn_poisson}
with the eigenstates of $H+\VH^{(i)}(z)$, and $\tau$ is a number between $0$ and $1$.
Viewing Eq.~\eqref{eqn_selfcon_iteration} as a discrete step in solving a differential equation, we can interpret the value $\tau$ as a `virtual time step'. The initial time step is $0.9$ by default, and can be adjusted by an additional numerical argument after \texttt{selfcon}.

Next, the history of potential differences $\VH^{(i+1)}(z) - \VH^{(i)}(z)$ between iterations is checked for periodic orbits or chaotic oscillations.
If either are detected, \kdotpy{} can automatically reduce the time step in an attempt to `escape' the periodic orbit or chaos and to reach convergence.
An large initial time step leads to faster convergences but makes the solver more vulnerable to developing periodic orbits or chaotic oscillations.

Finally, the convergence criterion is checked by comparing the last potential difference to a predefined convergence threshold.
If a satisfactory convergence has not yet been achieved, the above process loops again.
Once convergence is accomplished, \kdotpy{} continues with the rest of the program as if the \texttt{selfcon} options was not given but using the self-consistently calculated Hartree potential
In case the maximum number of specified iterations is reached or an error occurs in the above process, \kdotpy{} either exits with an error (by default) or continues with the rest of the program ignoring the error (if configured to do so).

The source file \texttt{selfcon.py} defines two separate classes \texttt{SelfConSolver}
and \texttt{selfConSolverLL} for momentum and Landau level mode, respectively. The latter
is derived from the former, where the differences lie in how the charge carrier densities are calculated (cf.\ Section~\ref{sec_dos}) and the fact that in the Landau level mode, a separate Hartree potential is calculated for each magnetic field value.
Furthermore, as it is not possible to calculate the density up to arbitrarily low magnetic fields in a Landau level picture with a finite number of Landau levels, the potentials at the smallest fields are set equal to the first potential that can be calculated.

\subsubsection{Two alternatives for calculating the density as function of $z$}

\kdotpy{} implements two different algorithms for calculating the charge carrier distribution $\rho(z)$ along the growth axis $z$.
The first method is the one sketched in Sec.~\ref{sec_densityz}, with the `naive' assumption
that $\rho(z)$ is identically zero at the charge neutrality point.
All states in a given band are either counted as pure holes or as pure electrons based on the position of the band relative to the charge neutrality point, i.e., whether the band index is
negative or positive.
This method was applied successfully to model modulation doping in thin HgTe quantum wells \cite{NovikEA2005}.
For larger thicknesses or strong Hartree potentials this method can fail, as it becomes difficult to clearly separate electron-like and hole-like states.

The second method avoids this problem: One applies full diagonalization, where one calculates all eigenstates of the Hamiltonian, and takes the top or bottom end of the energy spectrum as a reference for $\rho(z)$. In other words, all states are treated as a single carrier type and a suitable background density is subtracted (full-band envelope approach) \cite{Andlauer2009}.
It suffices to consider only the top of the spectrum and to calculate all conduction band eigenstates down to the charge neutrality point, plus a few valence band states in addition.
(Taking the conduction band is computationally more efficient than the valence band, in view of the smaller number of states.)
The subtracted density offset is
\begin{equation}
  n_\mathrm{offset} = 2n_z \frac{\Omega_\mathrm{grid}}{(2\pi)^2 \Delta z},
\end{equation}
where $n_z$ is the total number of $z$ points and $\Omega_\mathrm{grid}$ is the volume of the grid in $k$-space over which the calculation is performed.
For the Landau-level mode, the volumetric offset density is instead given by
\begin{equation}
  n_\mathrm{offset} = n_z [(n_\mathrm{max}+1)+(n_\mathrm{max}+2)] \frac{e B}{h} \frac{1}{n_z \Delta z},
\end{equation}
where $n_\mathrm{max}$ is the maximum Landau level index (set by the command line argument \texttt{llmax}) and $B$ is the magnetic field.

\begin{figure}
\includegraphics[width=140mm]{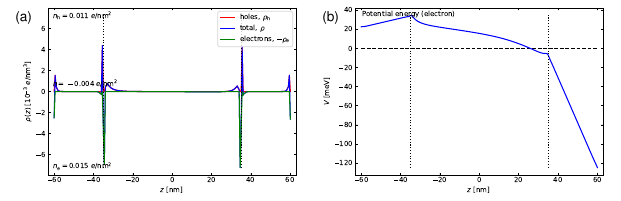}
\caption{(a) Self-consistently calculated Hartree potential for a structure with $70\,\mathrm{nm}$ thick HgTe layer and (Hg,Cd)Te barriers. The total carrier density has been set as $0.004\,\mathrm{nm}^{-2}$. The boundary conditions are determined by imposing an electric field strength of $0.48\,\mathrm{mV}\,\mathrm{nm}^{-1}$ inside the bottom barrier. This calculation has been performed with the default setting \texttt{selfcon\uscore{}full\uscore{}diag=true}.}
\label{fig_selfcon_70nm}
\end{figure}

The full-band envelope approach not only considers the active carriers in the quantum well but also results in interface dipoles and side bumps at the model edges (see Fig.~\ref{fig_selfcon_70nm}).
Note that the algorithm described in \ref{sec_selfcon_program_flow} keeps the total carrier density in the system constant, which includes these features.
Hence the carrier density inside the quantum well will be different from the one specified by the argument \texttt{cardens}.

The implementation of the full-band approach is done by the \texttt{SelfConSolver} derived classes \texttt{SelfConSolverFullDiag} and \texttt{SelfConSolverLLFullDiag} 
for dispersion and Landau-level mode, respectively. By default, the full-band
approached is used for the self-consistent Hartree method. To use the naive method,
for gaining speed at the expense of becoming less physically sound, one may set
the configuration value \verb+selfcon_full_diag=false+.

\subsubsection{Interpretation of the results}

Due to the combination of diagonalization and integration (in the steps solving the
Schr\"odinger and the Poisson equation, respectively), the iterative
dynamics of $\VH^{(i)}$ as function of iteration step $i$ is difficult to
characterize and to predict. The convergence behaviour can depend sensitively
on input parameters such as boundary conditions and the initial time step. If
the solver converges, the result only represents a single solution, out of possibly
many solutions. Whether the result is the most natural solution can only be determined
by critical scrutiny.

When one relies on the self-consistent method, it is advisable to run the calculation
multiple times with slightly different settings, in order to see if the algorithm
is stable against such variations. The terminal output from \kdotpy{} can also be a
useful information source, especially for tracking convergence behaviour. To allow for
systematic analysis of the iterative dynamics, the self-consistent solvers can
provide debug output, with the internal state (carrier density $\rho(z)$ and Hartree
potential $\VH(z)$) being written to a CSV file at each iteration. A thorough
systematic analysis of the self-consistent Hartree method is beyond the scope of
this work and will be addressed elsewhere.

\clearpage
\section{Installation}
\label{sec_installation}

\subsection{Prerequisites}

\kdotpy{} is packaged as a Python package that is intended to be
installed using PIP. For installing \kdotpy{}, you must have a working
Python installation (version 3.9.0 or above) and an up-to-date version of
PIP. When PIP installs \kdotpy{}, it automatically checks if the packages
\texttt{numpy}, \texttt{scipy}, and \texttt{matplotlib} are installed and have
compatible versions. Packages that \kdotpy{} depends on optionally may be
installed manually with PIP.

If your Python installation already contains a variety of packages, the dependencies
of the installed packages could potentially conflict with those of \kdotpy{}.
This problem may be avoided by installing \kdotpy{} in a virtual environment,
which is like a `sandbox' where \kdotpy{} and its dependencies do not interfere
with other packages. The virtual environment can be created with
\texttt{python3 -m venv \emph{directoryname}}. After activating the
virtual environment, you may now install \kdotpy{} following any of the installation
methods below. Consult the documentation of the Python \texttt{venv} module for
more information\footnote{See \url{https://docs.python.org/3/library/venv.html}.}.

Installation methods 2 and 3 below rely on git, which you must have
installed in order for these methods to work. If you do not wish to use git,
you may wish to use one of the other methods.

\subsection{Download and installation}

We list four methods for installing \kdotpy{}. You can choose any one of them,
but combining them is not recommended.

\begin{enumerate}
\item \emph{Using PIP, from the Python Package Index (PyPI):}
\begin{verbatim}
python3 -m pip install kdotpy
\end{verbatim}
\item \emph{Using PIP, from our Gitlab repository:}
\begin{verbatim}
python3 -m pip install git+ssh://git@git.physik.uni-wuerzburg.de/kdotpy/kdotpy.git
\end{verbatim}
\item \emph{Using \texttt{git clone} and installing it from your local copy:}
\begin{verbatim}
git clone https://git.physik.uni-wuerzburg.de/kdotpy/kdotpy.git
python3 -m pip install ./kdotpy 
\end{verbatim}
Note that this clones the files into the subdirectory \texttt{kdotpy} of your current
working directory.
\item \emph{Using a manual download and installing it from your local copy:}\\
If you prefer to avoid the PyPI and git, then you can download a \texttt{.zip} or \texttt{.tar.gz} file from our Gitlab repository at \url{https://git.physik.uni-wuerzburg.de/kdotpy/kdotpy} \cite{kdotpy_repo}. Download the preferred file format from the
dropdown menu under the `Code' button. Unpack the files into an empty directory. Then use
\begin{verbatim}
python3 -m pip install ./directoryname
\end{verbatim}
to install kdotpy, replacing \texttt{./directoryname} by the appropriate directory name
where you unpacked the files.
\end{enumerate}

You may obtain version 1.0.0, the exact version discussed in this article, also
from the Gitlab repository of this journal.

\subsection{Installation for active developers}

If you wish to make modifications to the \kdotpy{} code, then it is recommended
to use PIP's \texttt{-e} option (short for \texttt{--editable}). This option
dynamically links the package installation directory to the source files. If you
change the source, you thus do not have to re-install the package. To use an
editable install, use
\begin{verbatim}
git clone https://git.physik.uni-wuerzburg.de/kdotpy/kdotpy.git
python3 -m pip install -e ./kdotpy 
\end{verbatim}
If you intend to actively contribute to the project, you must change to a
non-protected branch in order to be able to push your changes to the repository.
(Where you are allowed to push depends on your membership role in our Gitlab
project. The \texttt{master} branch is protected for all except the project owner.
The journal repository shall not be used for development purposes.)

\subsection{Testing the installation}

If the install was successful, then entering
\begin{verbatim}
kdotpy version
\end{verbatim}
on the command line should return the version number. If not, then something went
wrong during the installation and you may wish to try one of the other methods.
(This assumes that if you use a virtual environment, it has been set up and activated
correctly.)

For testing the functionality of \kdotpy{}, you can use
\begin{verbatim}
kdotpy test
\end{verbatim}
for running the standardized tests. Note that the tests may take a few minutes to
complete. They will generate output in the subdirectory \texttt{test} relative to
where you start \texttt{kdotpy test}.

\clearpage
\section{Usage}
\label{sec_usage}

\subsection{General remarks}

\kdotpy{} is designed as a standalone application. If you have followed the
installation instructions above, you can simply run \texttt{kdotpy} from the command
line, followed by the `subprogram' label and further arguments. The first argument
is always the subprogram, but the order of the following arguments is usually
unimportant. (For an overview of the subprograms, see Sec.~\ref{sec_package_structure}.)
You can run \texttt{kdotpy} from any folder.

Alternatively, you can also use \texttt{python3 -m kdotpy} followed by the subprogram
and further arguments. You can also import specific functions or submodules of
\kdotpy{} from other Python scripts or the interactive interpreter using 
\texttt{from kdotpy import \emph{function}}. As we have designed \kdotpy{} primarily
as a command-line tool, such imports are not recommended for normal use.

Below, we provide detailed tutorials that describe how to set up a calculation
in \kdotpy{} step by step, and how to interpret the output. The curious user may
also find the tests defined by \texttt{kdotpy test} useful; the command lines
for these standardized tests can be extracted using \texttt{kdotpy test showcmd}.

\subsection{Example calculation: Basic dispersion}
\label{sec_tutorial_basic}

\subsubsection{Introduction}

This basic tutorial describes a systematic manner to compose the command line
from scratch, adding the necessary arguments step by step.
As an example, we will calculate the dispersion of a $7\,\mathrm{nm}$ HgTe quantum well:
We assume the substrate is $\mathrm{Cd}_{0.96}\mathrm{Zn}_{0.04}\mathrm{Te}$
and the barriers are $\mathrm{Hg}_{0.32}\mathrm{Cd}_{0.68}\mathrm{Te}$.
We will calculate the dispersion along the diagonal axis (110) and try to answer
the following questions about the maxima of the dispersion at finite momentum
in the valence band (also known as \emph{camel back}):
\begin{itemize}
 \item Where is the finite-momentum maximum in momentum and energy?
 \item Do we have a direct or an indirect gap?
 \item What is the orbital character at the finite-momentum maximum?
\end{itemize}

\subsubsection{Setting it up step by step}

\begin{enumerate}
 \item We first have to determine the geometry and/or mode, defined by the
 number of translationally invariant dimensions. In this example, we have
 a `2D' geometry, with momentum coordinates $(k_x,k_y)$. Thus, the appropriate
 subprogram to use here is \texttt{kdotpy 2d}.
 
\begin{verbatim}
kdotpy 2d
\end{verbatim}
 
 \item Next, we determine the number of orbitals in the model. This is usually
 eight orbitals, for which we use the option \texttt{8o}. Alternatively \texttt{6o}
 for six orbitals may be used. Let us also specify that we do not want to use the
 axial approximation by entering \texttt{noax}.
\begin{verbatim}
kdotpy 2d 8o noax
\end{verbatim}

 \item Let us now enter the substrate material and `layer stack'. We use
 \texttt{msubst} for the substrate material, \texttt{mlayer} for the layer
 materials and \texttt{llayer} for the layer thicknesses. We also enter the
 resolution of the discretization in the $z$ direction with \texttt{zres};
 $0.25\,\mathrm{nm}$ is a good value to start with.
\begin{verbatim}
kdotpy 2d 8o noax msubst CdZnTe 4% mlayer HgCdTe 68% HgTe HgCdTe 68%
llayer 10 7 10 zres 0.25
\end{verbatim}
 Note that we have taken the barriers to be 10 nm thick. Usually, even if the
 barriers are much larger in reality, setting the thickness to $10\,\mathrm{nm}$
 gives an accurate representation of the dispersion while reducing calculation
 time. Note also how the materials are entered.

 \item We define the parametrization of the momentum. Here, we take 60 points
 along the chosen momentum axis, symmetric around zero, with
 \verb+k -0.6 0.6 / 120+. The (110) axis has a $45$ degree angle with the $k_x$ axis,
 hence we enter \verb+kphi 45+.
\begin{verbatim}
kdotpy 2d 8o noax msubst CdZnTe 4% mlayer HgCdTe 68% HgTe HgCdTe 68%
llayer 10 7 10 zres 0.25 k -0.6 0.6 / 120 kphi 45
\end{verbatim}
 If we would run this command, we already get a result, \verb+dispersion.pdf+. It
 looks promising, but \kdotpy{} indicates that there are a few problems:
\begin{verbatim}
Calculating bands (k=0)...
Warning (band_type): Unable to determine band character and/or number of
nodes for 50 eigenstates.
Possible causes: spin degeneracy not broken, nonzero potential,
one-dimensional geometry, etc.
1 / 1
ERROR (estimate_charge_neutrality_point): Failed, because E+ and/or E-
bands are missing
\end{verbatim}

 \item The assignment of band characters has failed because the spin degeneracy
 has not being broken. The band characters are used to determine the 
 position of the charge neutrality point and the band indices, and if this fails
 this can lead to subsequent problems, especially with post-processing functions
 that rely on the band indices, like density of states. We avoid these problems
 by splitting the degeneracy with \texttt{split 0.01}. The value
 $0.01$ (in meV) hardly ever needs to be changed.
\begin{verbatim}
kdotpy 2d 8o noax msubst CdZnTe 4% mlayer HgCdTe 68% HgTe HgCdTe 68%
llayer 10 7 10 zres 0.25 k -0.6 0.6 / 120 kphi 45 split 0.01
\end{verbatim}

 \item With \texttt{erange -80 0} we zoom in on the energy range of our interest.
 Moreover, let us choose a colour scale representing the orbital degree of freedom of the 
 eigenstates. We set this by \texttt{obs orbitalrgb}. We also include the figure
 legend with \texttt{legend}. We add subband character labels with \texttt{char}.
\begin{verbatim}
kdotpy 2d 8o noax msubst CdZnTe 4% mlayer HgCdTe 68% HgTe HgCdTe 68%
llayer 10 7 10 zres 0.25 k -0.6 0.6 / 120 kphi 45 split 0.01 erange -80 0
obs orbitalrgb legend char
\end{verbatim}

 \item We choose a label for the filenames with \texttt{out -7nm} and the target
 folder with \texttt{outdir data-qw}.
\begin{verbatim}
kdotpy 2d 8o noax msubst CdZnTe 4% mlayer HgCdTe 68% HgTe HgCdTe 68%
llayer 10 7 10 zres 0.25 k -0.6 0.6 / 120 kphi 45 split 0.01 erange -80 0
obs orbitalrgb legend char out -7nm outdir data-qw
\end{verbatim}
 Note that if \texttt{outdir} is omitted, the files will end up in the subfolder
 \texttt{data} if it exists, and in the current folder otherwise.

 \item Finally, we add the post-processing option \texttt{extrema}, which will
 gives us useful information about the extrema.
\begin{verbatim}
kdotpy 2d 8o noax msubst CdZnTe 4% mlayer HgCdTe 68% HgTe HgCdTe 68%
llayer 10 7 10 zres 0.25 k -0.6 0.6 / 120 kphi 45 split 0.01 erange -80 0
obs orbitalrgb legend char out -7nm outdir data-qw extrema
\end{verbatim}
\end{enumerate}

Other than the subprogram \texttt{kdotpy 2d}, which must be at the beginning of
the command line, the other command-line arguments may be added in any order. 

\subsubsection{Results and interpretation}

\begin{figure}
 \includegraphics[width=150mm]{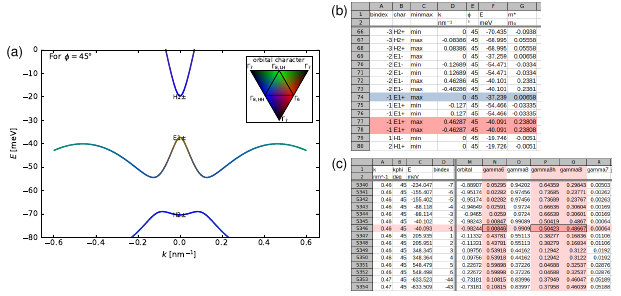}
 \caption{
   (a) Dispersion of a $7\,\mathrm{nm}$ HgTe quantum well as shown by
   \texttt{dispersion-7nm.pdf}. The colours indicate
   orbital character, see the legend. At $k=0$, the band character labels are shown,
   where `$\pm$' indicates that the states are approximately two-fold degenerate.
   (b) The properties of the extrema in \texttt{extrema-7nm.csv}. The maxima of
   the band with \texttt{bindex = -1} are highlighted (highlighting added manually,
   for illustration).
   (c) In \texttt{dispersion-7nm.csv}, we find the observable values for all
   eigenstates that \kdotpy{} has calculated. The state closest to the maximum
   at finite momentum has been highlighted. We read off the expectation values
   for the observables \texttt{gamma6}, \texttt{gamma8h}, and \texttt{gamma8l}
   to gain information on the orbital character at this point.
 }
 \label{fig_tutorial_basic}
\end{figure}

After running this command, we will find the following files in the target
output folder \texttt{data-qw}:

\begin{itemize}
 \item \verb+dispersion-7nm.pdf+:
 The dispersions $E(k)$ for all calculated eigenstates, see
 Fig.~\ref{fig_tutorial_basic}(a). The colours visualize
 the values of the observable(s) chosen with the option \texttt{obs}.
 \item \verb+dispersion-7nm.csv+: A csv file where all data points (momentum, energy)
 are given with the values of the observables. The points are ordered by momentum.
 \item \verb+dispersion-7nm.byband.csv+: A csv file with the band energies for
 each curve of\linebreak[4] \verb+dispersion-7nm.pdf+, i.e., ordered by band.
 \item \verb+extrema-7nm.csv+: A csv file listing all the extremal values.
 \item \verb+output-7nm.xml+: The XML data file (see Sec.~\ref{sec_xml}) that can be
 read again by the subprograms \verb+kdotpy merge+ and \verb+kdotpy compare+
 for the purpose of replotting and comparing data sets and that contains a large
 amount of metadata that can be used for diagnostics and reproducing the results
 later. Discarding this file is strongly discouraged.
\end{itemize}
 
With these results, we can answer the questions above:

\begin{itemize}
\item \emph{Where is the finite-momentum maximum in momentum and energy?}---
 In the plot, we see that these maxima appear in the band with
 character $\mathrm{E}1\pm$ at $k = 0$. We find the following data in
 \texttt{extrema.csv}, see Fig.~\ref{fig_tutorial_basic}(b).
 We find three maxima of the $\mathrm{E}1+$ band, one at $k = 0$ (highlighted in
 blue) and two at finite $k$ (highlighted in red), namely, at
 $k = \pm 0.463\ \mathrm{nm}^{-1}$ and $E = -40.1\,\mathrm{meV}$.
 Since we have calculated along the (110) axis, with \texttt{kphi 45}, these
 points are located at
 $\vec{k} = \pm 0.463\,\mathrm{nm}^{-1} \times  (\cos 45^\circ, \sin 45^\circ) = \pm (0.327, 0.327)\,\mathrm{nm}^{-1}$.

\item \emph{Do we have a direct or an indirect gap?}---
 The maximum of the $\mathrm{E}1+$ band at $k = 0$ lies at $E = -37.2\,\mathrm{meV}$.
 The maxima at finite $k$ are lower, so the global maximum is at $k = 0$. The value
 \texttt{bindex = -1} confirms that this is the highest valence band state.
 The $\mathrm{H}1-$ state with \texttt{bindex = 1} is the lowest conduction band state.
 It only has one minimum, at $k = 0$ and $E = -19.7\,\mathrm{meV}$. We find that
 the gap is direct (at $k = 0$) and its size is $\abs{\Delta} = 17.5\, \mathrm{meV}$.

\item \emph{What is the orbital character at the finite-momentum maximum?}---
 We extract this information from \texttt{dispersion-7nm.csv} that contains the
 values of the observables.
 The maximum is located at $k = \pm 0.463\,\mathrm{nm}^{-1}$, so we check the states
 at the nearest value $k = 0.46\,\mathrm{nm}^{-1}$ (as a first approximation).
 The band with \texttt{bindex = -1} is the correct one. We could also identify
 it by energy. The orbital character is determined by the probability densities
 in the $\ket{\Gamma_6,\pm\frac{1}{2}}$, $\ket{\Gamma_8,\pm\frac{3}{2}}$, and
 $\ket{\Gamma_8,\pm\frac{1}{2}}$ states, which is encoded by the observables
 \texttt{gamma6}, \texttt{gamma8h} and \texttt{gamma8l}, respectively. We thus
 find $0.8\%$ $\Gamma_6$ (`electron or s orbital'), $50.4\%$ $\Gamma_{8\mathrm{H}}$
 (`heavy hole'), and $48.7\%$ $\Gamma_{8\mathrm{L}}$ (`light hole'). We confirm
 in the dispersion plot that the colour at this position is approximately cyan,
 a mixture of $50\%$ green and $50\%$ blue.

\end{itemize}

\subsection{Example calculation: Landau levels}
\label{sec_tutorial_ll}
\subsubsection{Introduction}

We set up a Landau-level calculation for the same structure as in
Sec.~\ref{sec_tutorial_basic}. We aim to calculate the Landau levels up to
$B=10\,\mathrm{T}$ and to answer the following questions, centred around the
properties of the `lowest Landau levels', i.e., those that border the band gap:

\begin{itemize}
 \item Which Landau levels are the `lowest' ones?
 \item What is the critical field $B_c$ where the inversion of the lowest LLs is undone?
 \item What orbital character do the states have at this crossing?
\end{itemize}

We set up the calculation step by step, reusing some of the steps of
Sec.~\ref{sec_tutorial_basic}.

\subsubsection{Setting it up step by step}
\begin{enumerate}

\item First, we determine the subprogram. The geometry is 2D, and we use the
 Landau level formalism. We thus use
\begin{verbatim}
kdotpy ll
\end{verbatim}

\item As before, we use the eight-orbital model. In order to speed up the
 calculation, let us use the axial approximation. For \texttt{kdotpy ll}, this
 is achieved by omitting \texttt{noax}.
\begin{verbatim}
kdotpy ll 8o
\end{verbatim}

\item The layer stack parameters are the same as in Sec.~\ref{sec_tutorial_basic}. 
\begin{verbatim}
kdotpy ll 8o msubst CdZnTe 4% mlayer HgCdTe 68% HgTe HgCdTe 68%\
llayer 10 7 10 zres 0.25
\end{verbatim}

\item Next, we specify the grid for the magnetic field values. Let is choose 100
 points with the upper limit $10\,\mathrm{T}$. We could choose evenly spaced
 points (every $0.1\,\mathrm{T}$ with \texttt{b 0 10 / 100}, but for LL calculations,
 quadratic stepping is recommended. This is achieved by the `double slash' notation
 \texttt{b 0 10 // 100}
\begin{verbatim}
kdotpy ll 8o msubst CdZnTe 4% mlayer HgCdTe 68% HgTe HgCdTe 68%
llayer 10 7 10 zres 0.25 b 0 10 // 100
\end{verbatim}

\item Like in Sec.~\ref{sec_tutorial_basic}, we zoom in at the energy range
 \texttt{erange -80 0} and we use \texttt{split 0.01}.
 This time, we also enter the diagonalization parameters explicitly.
 We choose the maximum LL index to be 20 using \texttt{nll 20}.
 For each LL index, we aim to get approximately 12 (subband) states. For this we
 need about 240 eigenvalues: \texttt{neig 240}. We also put the target energy
 for the diagonalization somewhere near the gap, which we know is at
 approximately $-30\,\mathrm{meV}$. Note, however, that if we put the target energy
 at that value, the $\mathrm{E}2\pm$ subbands might be out of range. These subbands
 are required for the determination of the charge neutrality point to work properly,
 so we put the target energy higher up, i.e., we put \texttt{targetenergy 0}.
 (For the same reason, we aim for 12 subband states instead of the minimally
 required number of 8.)
\begin{verbatim}
kdotpy ll 8o msubst CdZnTe 4% mlayer HgCdTe 68% HgTe HgCdTe 68%
llayer 10 7 10 zres 0.25 b 0 10 // 100 split 0.01 erange -80 0 nll 20
neig 240 targetenergy 0
\end{verbatim}

\item For colouring the states, let us choose a combination of Landau level index
 and total angular momentum. We set this by \texttt{obs llindex.jz}. Again, we
 include the figure legend with \texttt{legend} and the band characters with \texttt{char}.
\begin{verbatim}
kdotpy ll 8o msubst CdZnTe 4% mlayer HgCdTe 68% HgTe HgCdTe 68%
llayer 10 7 10 zres 0.25 b 0 10 // 100 split 0.01 erange -80 0 nll 20
neig 240 targetenergy 0 obs llindex.jz legend char
\end{verbatim}

\item We choose a label for the filenames \verb+out -7nm-landau+ and the folder
 where the files should go \verb+outdir data-landau+.
\begin{verbatim}
kdotpy ll 8o msubst CdZnTe 4% mlayer HgCdTe 68% HgTe HgCdTe 68%
llayer 10 7 10 zres 0.25 b 0 10 // 100 split 0.01 erange -80 0 nll 20
neig 240 targetenergy 0 obs llindex.jz legend char out -7nm-landau
outdir data-landau
\end{verbatim}

\end{enumerate}
    
\subsubsection{Results and interpretation}

\begin{figure}
 \includegraphics[width=150mm]{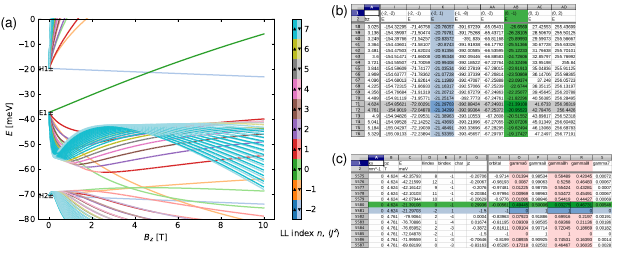}
 \caption{
 (a) The Landau level spectrum, \texttt{bdependence-7nm-landau.pdf}.
 The colours indicate the Landau level index $n$ and the shade (brighter/darker)
 indicates the sign of $\avg{J_z}$, see the colour legend.
 (b) Screenshot of \texttt{bdependence-7nm-landau.byband.csv} which provides the
 energies of the `lowest' Landau levels $\ket{n,b}$ with $(n,b)=(-2,1)$ and $(0,-1)$
 as a function of the magnetic field $B$ (column A, labelled \texttt{bz}).
 (c) Screenshot of \texttt{bdependence-7nm-landau.csv} which contains the
 expectation values for all relevant observables for all states.
 In (b) and (c), some columns were hidden and the coloured highlights were added
 manually to lay emphasis on the lowest Landau level states near the crossing.
 }
 \label{fig_tutorial_ll}
\end{figure}

After running this command, we get the following files in the folder \verb+data-landau+:
\begin{itemize}
\item \verb+bdependence-7nm-landau.pdf+:
 A plot of the Landau level spectrum, i.e., the eigenenergies of the states
 $\ket{n, b}$ as function of the magnetic field, see Fig.~\ref{fig_tutorial_ll}(a).
\item \verb+bdependence-7nm-landau.csv+: A csv file where all data points
 (magnetic field, energy) are given with the values of the observables. The points
 are ordered by magnetic field value.
\item \verb+bdependence-7nm-landau.byband.csv+: A csv file with the band energies
 for each curve, i.e., ordered by LL band.
\item \verb+output-7nm-landau.xml+: The XML file with all parameters and data.
\end{itemize}

We can extract sufficient information to answer the questions above:

\begin{itemize}
\item  \emph{Which Landau levels are the `lowest' ones?}---
    Let us examine the plot of Fig.~\ref{fig_tutorial_ll}(a). The lowest LL index
    formally is $-2$, but the actual lowest index differs by subband character.
    Let us focus on the $\mathrm{H}1\pm$ and $\mathrm{E}1\pm$ subbands, as these
    are the most relevant for the physics. The `$\pm$' indicates the expectation
    value $\avg{J_z}$ at
    zero momentum and zero magnetic field. This degree of freedom is encoded in
    the colour shading, see the legend: dark colours represent the $+$ states
    ($\avg{J_z} > 0$) and bright colours the $-$ states ($\avg{J_z} < 0$).
    \begin{itemize}
    \item  For $\mathrm{H}1+$, dark colours emanating from `$\mathrm{H}1\pm$', we
    find that the lowest index is $1$ (red colour); $0$, $-1$, and $-2$ are missing.
    \item  For $\mathrm{H}1-$, bright colours emanating from `$\mathrm{H}1\pm$',
    the baby-blue LL can be identified as index $-2$, so all indices are present
    for $\mathrm{H}1-$.
    \item  For $\mathrm{E}1+$, dark colours emanating from `$\mathrm{E}1\pm$',
    the lowest index is $0$ (dark green).
    \item  For $\mathrm{E}1-$, bright colours emanating from `$\mathrm{E}1\pm$', the lowest index is $-1$ (bright orange green).
    \end{itemize}
    These observations can be confirmed from the data in the csv files. The
    lowest LLs near the gap are thus ($\mathrm{E}1+, n = 0$) and ($\mathrm{H}1-, n = -2$).

\item \emph{What is the critical field $B_c$ where the inversion of the lowest LLs
    is undone?}---
    In principle we could estimate this from Fig.~\ref{fig_tutorial_ll}(a), but
    we use one of the csv files in order to answer this question more precisely:
    In \verb+bdependence-7nm-landau.byband.csv+, see Fig.~\ref{fig_tutorial_ll}(b),
    the bands are
    ordered in columns labelled by $(n, b)$ in the first row, where $n$
    is the LL index and $b$ is the band index. The band indices are counted for
    each LL separately, which makes intuition a bit tricky for the lowest LLs
    ($n = -2, -1, 0$), but it is guaranteed that the LLs at the crossing have
    $(n, b) = (0, -1)$ and $(-2, 1)$ respectively. The respective columns are
    highlighted. We find the crossing between $4.624\,\mathrm{T}$ and $4.761\,\mathrm{T}$.
    Linear interpolation may be used as to obtain a more accurate value.

\item \emph{What orbital character do the states have at this crossing?}---
    From \verb+bdependence-7nm-landau.csv+, depicted in Fig.~\ref{fig_tutorial_ll}(c),
    we extract the orbital character from the respective states at $B=4.624\,\mathrm{T}$
    (which lies closer to the actual crossing than $B=4.761\,\mathrm{T}$.
    The desired information is encoded in the observables \texttt{gamma6},
    \texttt{gamma8h}, and \texttt{gamma8l} for the two respective states there.
    The lowest LL coming from the $\mathrm{E}1+$ band, $(n, b) = (0, -1)$, is a
    mixture of $49.4\%$ $\Gamma_6$, $46.7\%$ $\Gamma_{8\mathrm{L}}$, and $3.3\%$
    $\Gamma_{8\mathrm{H}}$. We thus find that it is predominantly a $J_z = \pm\frac{1}{2}$ state. The one from the $\mathrm{H}1-$ band is purely $\Gamma_{8\mathrm{H}}$,
    and has $\avg{J_z} = -\frac{3}{2}$ exactly.
\end{itemize}

\subsection{Example calculation: Magneto-transport}
\label{sec_tutorial_ll_advanced}
\subsubsection{Introduction}

In the previous example calculation, we have set up a basic LL calculation. With
some postprocessing in \kdotpy{}, we are able to extract much more useful
information and simulate magneto-transport experiments. The key ingredients here
are density of states (DOS) and Chern numbers (or Berry curvature). We will
set up an example simulation, where we will address the following questions:

\begin{itemize}
\item Which Landau levels are filled at the constant carrier density $2 \times 10^{11} \,\mathrm{cm}^{-2}$?

\item Do we see plateaus in the Hall resistance $R_{xy}$?

\item Are there Shubnikov-de Haas (SdH) oscillations at low magnetic fields?

\item Can we simulate a Landau fan such that it resembles experimental data?
\end{itemize}

\subsubsection{Setting up the simulation step by step}

We take the the command line from Sec.~\ref{sec_tutorial_ll} as a starting point:

\begin{verbatim}
kdotpy ll 8o msubst CdZnTe 4% mlayer HgCdTe 68% HgTe HgCdTe 68%
llayer 10 7 10 zres 0.25 b 0 10 // 100 split 0.01 erange -80 0 nll 20
neig 240 targetenergy 0 obs llindex.jz legend char out -7nm-landau
outdir data-landau
\end{verbatim}

We modify and add command line arguments as follows:

\begin{enumerate}
\item We change some parameters slightly for this calculation. We choose
   \texttt{erange -60 40} instead of \texttt{erange -80 0}, in order to give a
   better view of the data. We increase the number of eigenvalues from
   \texttt{neig 240} to \texttt{neig 300}, so that we obtain a sufficient number
   of states to view this energy range. We also change the output filenames and
   output folder.
\begin{verbatim}
kdotpy ll 8o msubst CdZnTe 4% mlayer HgCdTe 68% HgTe HgCdTe 68%
llayer 10 7 10 zres 0.25 b 0 10 // 100 split 0.01 erange -60 40 nll 20
neig 300 targetenergy 0 obs llindex.jz legend char out -7nm-hall
outdir data-hall
\end{verbatim}

\item In order to calculate the density of states, we add the options \texttt{dos}
   and \texttt{localdos} for calculating the total and local density of states,
   respectively. For the DOS, we use Gaussian broadening as in Ref.~\cite{NovikEA2005},
   with a broadening width (standard deviation) $\sigma(B)=\sigma_1\sqrt{B\,[\mathrm{T}]}$
   with $\sigma_1=0.5\,\mathrm{meV}$ (see also Appendix~\ref{app_broadening_reference}).
   The appropriate command is \texttt{broadening gauss 0.5 sqrt}, but this may
   be shortened to \texttt{broadening 0.5}; for \texttt{kdotpy ll}, \texttt{gauss}
   and \texttt{sqrt} are default settings for the broadening. 
\begin{verbatim}
kdotpy ll 8o msubst CdZnTe 4% mlayer HgCdTe 68% HgTe HgCdTe 68%
llayer 10 7 10 zres 0.25 b 0 10 // 100 split 0.01 erange -60 40 nll 20
neig 300 targetenergy 0 obs llindex.jz legend char out -7nm-hall
outdir data-hall dos localdos broadening 0.5
\end{verbatim}
   
\item For calculating the Hall conductance, we need to calculate the Chern numbers
   and set its broadening correctly. We add \texttt{chern} to
   calculate the Chern numbers.
\begin{verbatim}
kdotpy ll 8o msubst CdZnTe 4% mlayer HgCdTe 68% HgTe HgCdTe 68%
llayer 10 7 10 zres 0.25 b 0 10 // 100 split 0.01 erange -60 40 nll 20
neig 300 targetenergy 0 obs llindex.jz legend char out -7nm-hall
outdir data-hall chern dos localdos broadening 0.5
\end{verbatim}

\item We replace \verb+broadening 0.5+ by \verb+broadening 0.5 10%+ to set the
   broadening of the Hall conductance as function of energy to $10\%$ of that of
   the carrier density (integrated density of states).
\begin{verbatim}
kdotpy ll 8o msubst CdZnTe 4% mlayer HgCdTe 68% HgTe HgCdTe 68%
llayer 10 7 10 zres 0.25 b 0 10 // 100 split 0.01 erange -60 40 nll 20
neig 300 targetenergy 0 obs llindex.jz legend char out -7nm-hall
outdir data-hall chern dos localdos broadening 0.5 10%
\end{verbatim}

\item For convenience, one may also use the command shortcut \texttt{hall},
   which is equivalent to \texttt{chern dos localdos broadening 0.5 10\%}.
\begin{verbatim}
kdotpy ll 8o msubst CdZnTe 4% mlayer HgCdTe 68% HgTe HgCdTe 68%
llayer 10 7 10 zres 0.25 b 0 10 // 100 split 0.01 erange -60 40 nll 20
neig 300 targetenergy 0 obs llindex.jz legend char out -7nm-hall
outdir data-hall hall
\end{verbatim}
   This command line is equivalent to the one in step 4.
\end{enumerate}

For this calculation, the following configuration options are relevant:

\begin{itemize}

\item  \verb+dos_unit+:
   The plots in Fig.~\ref{fig_tutorial_lladvanced} have been produced with
   \verb+dos_unit=cm+, so that densities are expressed in units of
   $\mathrm{cm}^{-2}$ with an appropriate power of ten.

\item \verb+dos_energy_points+:
   The default energy resolution used for DOS calculations (etc.) is\linebreak[4]
   \verb+dos_energy_points=1000+, which is usually is a good compromise between
   accuracy and computation time. If the value is insufficiently large, one may
   see visible artifacts, like jagged equal-density lines where one would expect
   smooth curves. If that is the case, one should increase this value.

\item  \verb+berry_ll_simulate+:
   By default (\verb+berry_ll_simulate=false+), the plots related to Hall conductance
   are extracted from the calculated Chern numbers (Berry curvature), stored as
   the observable \texttt{chern}. When the number of states, set by command-line argument
   \texttt{neig}, is not sufficient, some values are inaccurate or incorrect; see
   Sec.~\ref{sec_berrychern_implementation} for a detailed discussion.
   In bad cases, this renders the data to be partially or completely unusable.
   One can mitigate this problem by setting \verb+berry_ll_simulate=true+.
   In this case, the output functions use a simulated Chern number, which is
   exactly $1$ for all Landau-level eigenstates and is stored as the observable \texttt{chernsim}. This value coincides with the
   calculated Chern number (up to numerical error) for almost all states, except
   for states the edges if the spectrum, which should usually be mistrusted
   anyway. If \verb+berry_ll_simulate=true+, the output file names will be changed
   by insertion of \texttt{sim} or \texttt{simul} in order to be able to easily
   spot whether the calculated or simulated Chern numbers have been used.
   It is recommended to run \kdotpy{} with both settings and to compare the results
   at least once in order to get an intuition for the
   implications of this setting and for the physics behind it.
\end{itemize}

The configuration values may be set permanently with
\begin{verbatim}
kdotpy config 'dos_unit=cm;dos_energy_points=1000;berry_ll_simulate=true'
\end{verbatim}
or used once by appending
\begin{verbatim}
config 'dos_unit=cm;dos_energy_points=1000;berry_ll_simulate=true'
\end{verbatim}
to the command line that starts \texttt{kdotpy ll}.

\subsubsection{Results and interpretation}

The above commands produce many pdf and csv files, and here we shall only discuss
the ones relevant for answering the questions above. In most cases, the pdf is
accompanied by a csv file which contains exactly the data shown in the figure.

\begin{figure}
 \includegraphics[width=150mm]{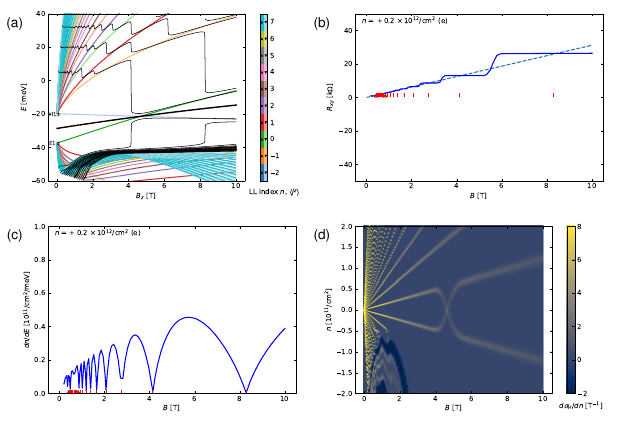}
 \caption{
 (a) In the file \texttt{bdependence-density.pdf}, equal-density contours are
 overlaid on top of the magnetic field depdendence plot, cf.\ Fig.~\ref{fig_tutorial_ll}(a).
 The thick line near $-20\,\mathrm{meV}$ is the charge neutral point $n = 0$.
 Contours are drawn for each multiple of $1 \times 10^{11} \,\mathrm{cm}^{-2}$ for
 $\abs{n}\leq 15\times 10^{11} \,\mathrm{cm}^{-2}$.
 The medium-thick lines are at odd multiples of $5 \times 10^{11} \,\mathrm{cm}^{-2}$
 and the very thick lines are at multiples of $10 \times 10^{11} \,\mathrm{cm}^{-2}$.
 (b) The plot in \texttt{rxy-constdens-7nm-hall.pdf} shows the simulated
 Hall resistance $R_{xy} \equiv 1/\sigmaH$ as function of magnetic field $B$
 for the constant density $2 \times 10^{11} \,\mathrm{cm}^{-2}$. The dashed line
 indicates the classical Hall slope given by $R_{xy} = B / ne$.
 (c) The file \texttt{dos-constdens-7nm-hall.pdf} visualizes the
 density of states as function of magnetic field for a constant carrier density.
 (d) In \texttt{dsigmah-dn-7nm-hall.pdf}, we plot the derivative $d\sigmaH/dn$
 of the Hall conductance with respect to the carrier
 density $n$. This type of plot can easily be compared to plots obtained from
 experiments which measure Hall conductance as function of magnetic field and
 gate voltage.
 The figures (b) and (c) are single pages in a multipage PDF unless \texttt{cardens}
 has been used in order to specify a single density.
 For this figure, we have used the configuration settings \texttt{dos\uscore{}unit=cm}, \texttt{dos\uscore{}energy\uscore{}points=1000}, and
 \texttt{berry\uscore{}ll\uscore{}simulate=false}.
 }  
 \label{fig_tutorial_lladvanced}
\end{figure}

\begin{itemize}
 \item \emph{Which Landau levels are filled at the constant carrier density
 $2 \times 10^{11} \,\mathrm{cm}^{-2}$?}---
  The answer to this question of course depends on the magnetic field. It is
  encoded in the file\linebreak[4] \texttt{bdependence-density.pdf}, see
  Fig.~\ref{fig_tutorial_lladvanced}(a), which
  is generated when \texttt{kdotpy ll} is used with the \texttt{dos} option.
  Equal-density contours are overlaid onto the magnetic-field dependence plot,
  (LL energies as function of magnetic field $B$).
  In Fig.~\ref{fig_tutorial_lladvanced}(a), we find the equal-density curve for
  $n=2 \times 10^{11} \,\mathrm{cm}^{-2}$ to be at approximately $20\,\mathrm{meV}$
  for low magnetic fields. A different visualization may be obtained by
  adding \texttt{cardens 0.002} to the command line, in which case only the
  contour line for the given carrier density
  ($0.002\,\mathrm{nm}^{-2}=2\times 10^{11} \,\mathrm{cm}^{-2}$) is shown.

 \item \emph{Do we see plateaus in the Hall resistance $R_{xy}$?}---
  The Hall resistance is shown in the plot\linebreak[4] \texttt{rxy-constdens-7nm-hall.pdf},
  see Fig.~\ref{fig_tutorial_lladvanced}(b). We find plateaus corresponding to
  the filling factors
  $\nu=1,2,\ldots$. At higher filling fractions (lower magnetic fields), the
  plateau transitions are more washed out due to the broadening. Here, the curve
  tends more towards the classical value $R_{xy} = B / ne$, indicated by the
  dashed line. This type of visualization can be compared directly to experimental
  results. For completeness, \kdotpy{} also generates the plot\linebreak[4]
  \texttt{sigmah-constdens-7nm-hall.pdf}, which shows the Hall
  conductance $\sigmaH$ as function of magnetic field for constant carrier density.
  In both cases, these output files correspond to the density specified by
  \texttt{cardens}, if this argument is provided on the command-line. Otherwise,
  the output will be multi-page PDFs with multiple densities, namely integer
  multiples of $1 \times 10^{11} \,\mathrm{cm}^{-2}$.

\item \emph{Are there Shubnikov-de Haas (SdH) oscillations at low magnetic fields?}---
  The constant-carrier-density curves in Fig.~\ref{fig_tutorial_lladvanced}(c)
  clearly show oscillations with
  magnetic field. The plot shows density of states, which is thought to correlate
  with the longitudinal resistance $R_{xx}$. The red markers at the bottom
  indicate the values of $1/B$ equal to integer multiples of $e/hn$, where $n$ is
  the carrier density. The minima of the density of states thus align well with
  the multiples of $e/hn$ for $B\gtrsim 1\,\mathrm{T}$. At lower fields,
  the resolution of $B$ values is insufficient to make any conclusive statement.
  This may easily be mitigated by an additional calculation at lower fields,
  for example with \texttt{b 0 1 // 100}.

\item \emph{Can we simulate a Landau fan such that it resembles experimental data?}---
  In magneto-transport experiments, Landau fans are obtained by measuring $R_{xy}$
  while sweeping the gate voltage for many values of the magnetic field. This
  yields $R_{xy}$ as function of magnetic field $B$ and gate voltage $V_\mathrm{g}$. In
  simulations, however, the natural quantity on the `vertical axis' is energy.
  The DOS calculation in \kdotpy{} bridges this gap: The relation of carrier
  density $n$ as function of energy $E$ can be reversed as to obtain a spectrum as
  function of density. The latter often correlates approximately linearly with
  gate voltage $V$, so that comparison between the two is physically meaningful.
  In Fig.~\ref{fig_tutorial_lladvanced}(d), we visualize the plateau
  transitions by taking the derivative $d \sigmaH/dn$ in order to highlight
  the plateau transitions. This method is also customarily applied to
  experimental data.
\end{itemize}

\subsection{Example calculation: Optical transitions}
\label{sec_tutorial_optical_transitions}
\subsubsection{Introduction}

Instead of extracting information on magneto-transport experiments in postprocessing, we can also simulate magneto-optical transition spectra in \kdotpy{}. We will use an example simulation to answer the following questions:

\begin{itemize}
\item How large is the band gap of the system?
\item What are the involved LL states for a specific transition?
\item What information regarding band ordering can we extract from the transition spectrum?
\end{itemize}

\subsubsection{Setting up the simulation step by step}
Again, we take the the command line from Sec.~\ref{sec_tutorial_ll} as a starting point:

\begin{verbatim}
kdotpy ll 8o msubst CdZnTe 4% mlayer HgCdTe 68% HgTe HgCdTe 68%
llayer 10 7 10 zres 0.25 b 0 10 // 100 split 0.01 erange -80 0 nll 20
neig 240 targetenergy 0 obs llindex.jz legend char out -7nm-landau
outdir data-landau
\end{verbatim}

We modify and add command line arguments as follows:

\begin{enumerate}
\item As for the magneto-transport analysis, we slightly adjust some parameters. We choose \texttt{erange -80 50} for a better view of the data. We increase the number of eigenvalues from \texttt{neig 240} to \texttt{neig 300} to get a sufficient number of states to view in this energy range. We also change the output filenames and output folder.
\begin{verbatim}
kdotpy ll 8o msubst CdZnTe 4% mlayer HgCdTe 68% HgTe HgCdTe 68%
llayer 10 7 10 zres 0.25 b 0 10 // 100 split 0.01 erange -80 50 nll 20
neig 300 targetenergy 0 obs llindex.jz legend char
out -7nm-optical-transitions outdir data-optical-transitions
\end{verbatim}

\item In order to calculate all optical transitions, we add the option \texttt{transitions}.
\begin{verbatim}
kdotpy ll 8o msubst CdZnTe 4% mlayer HgCdTe 68% HgTe HgCdTe 68%
llayer 10 7 10 zres 0.25 b 0 10 // 100 split 0.01 erange -80 50 nll 20
neig 300 targetenergy 0 obs llindex.jz legend char
out -7nm-optical-transitions outdir data-optical-transitions transitions
\end{verbatim}

\item It is also advisable to filter transitions by state occupancy, since a transition can only be observed if the initial state is (partially) occupied while the final state is (partially) unoccupied. Thus, we add the option \texttt{cardens 0}, assuming a charge neutral sample. (There is also the option to calculate filtered transitions for multiple carrier densities at once by using the input \verb+cardens # # / #+, analogous to the range input for \texttt{b}.) Additionally, we also use the options \texttt{dos} and \texttt{broadening 0.5}. \texttt{dos} enables calculation of the (electro-)chemical potential, which is used to determine the occupation of states, while \texttt{broadening} applies a Gaussian square-root broadening to the LL states (shortened, see Sec.~\ref{sec_tutorial_ll_advanced}). Omitting broadening often leads to abrupt jumps in the (electro-)chemical potential, leading to unphysical gaps along specific transition features. Note, that broadening only is applied to DOS-related quantities, transition spectra will always be delta peaks.

Per default, these options suppress the output of all possible transitions.
\begin{verbatim}
kdotpy ll 8o msubst CdZnTe 4% mlayer HgCdTe 68% HgTe HgCdTe 68%
llayer 10 7 10 zres 0.25b 0 10 // 100 split 0.01 erange -80 50 nll 20
neig 300 targetenergy 0 obs llindex.jz legend char
out -7nm-optical-transitions outdir data-optical-transitions transitions
cardens 0 dos broadening 0.5
\end{verbatim}

\item Last, we change some configuration values using \texttt{config}. With \verb+transitions_max_deltae=80+ we set the upper limit of the energy axis in the transition plot to $80\meV$, while\linebreak[4] \verb+transitions_min_amplitude=3e11+ filters out all transitions with transition matrix elements smaller than the given threshold, keeping only the most prominent transitions.
\begin{verbatim}
kdotpy ll 8o msubst CdZnTe 4% mlayer HgCdTe 68% HgTe HgCdTe 68%
llayer 10 7 10 zres 0.25 b 0 10 // 100 split 0.01 erange -80 50 nll 20
neig 300 targetenergy 0 obs llindex.jz legend char
out -7nm-optical-transitions outdir data-optical-transitions transitions
cardens 0 dos broadening 0.5
config 'transitions_max_deltae=80;transitions_min_amplitude=3e11'
\end{verbatim}
\end{enumerate}

The quantity that will be plotted in the transitions spectrum can be changed by using the configuration option \verb+plot_transitions_quantity+. We will use the default value \texttt{rate}, plotting the rate density as defined in Eq.~\eqref{eqn_transition_rate_density}. More information on this topic can be found in Sec.~\ref{sec_optical_transitions_output}.

\subsubsection{Results and interpretation}

\begin{figure}
 \includegraphics[width=150mm]{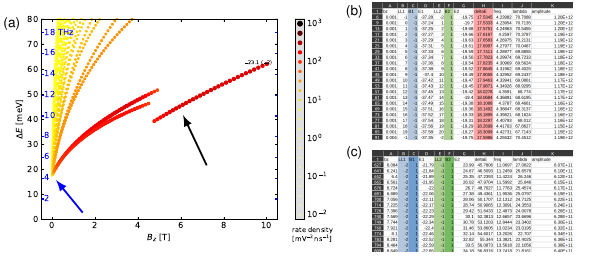}
 \caption{
 (a) The filtered transitions spectrum,
 \texttt{transitions-filtered-7nm-optical-}\linebreak[4]\texttt{transitions.pdf}.
 The strength of each individual transition is indicated by the size and colour of its respective symbol.
 The blue arrow points towards the transition feature that extrapolates to the band gap energy at small magnetic fields, while the black arrow indicates the feature for which we want to find the involved LL states. Both were added in post-production and are normally not include in the plot.
 The screenshots on the right are taken from the \texttt{transitions-filtered-7nm-optical-transitions.csv} file.
 In (b) the data is filtered by the band indices \texttt{B1} and \texttt{B2}, taking the values $-1$ and $1$, respectively.
 For (c) data was filtered for magnetic fields \texttt{bx} $\geq 6\,\mathrm{T}$.
 The Landau level spectrum corresponding to this figure is the one shown in Fig.~\ref{fig_tutorial_lladvanced}.
 }  
 \label{fig_tutorial_optical-transitions}
\end{figure}

Using the commands above produces additional output files. This includes
\texttt{bdependence-}\linebreak[4]\texttt{transitions-}\linebreak[3]\texttt{7nm-optical-transitions.pdf}, where all filtered transitions are drawn as vertical lines into the LL fan chart, and \texttt{transitions-filtered-7nm-optical-transitions.pdf}, together with its csv file, showing the filtered optical transitions spectrum. We will address the question given above by discussing the transitions spectrum in Fig.~\ref{fig_tutorial_optical-transitions}(a).

\begin{itemize}
 \item \emph{How large is the band gap of the system?}---
 The interband transition between the highest valence and the lowest conduction subband at low magnetic fields extrapolates to the energetic value of the direct band gap, see e.g. Fig.~\ref{fig_tutorial_lladvanced}(a). We suspect, that the transition feature marked with the blue arrow in Fig.~\ref{fig_tutorial_optical-transitions}(a) is this specific transition. To confirm this, we use the\linebreak[4] \texttt{transitions-filtered-7nm-optical-transitions.csv} file and filter the initial state \texttt{B1} by highest valence subband and the final state \texttt{B2} by lowest conduction subband. Per definition, these subbands, respectively, have the band index $-1$ and $1$. The blue and green columns in Fig.~\ref{fig_tutorial_optical-transitions}(b) show these indices, while the red column shows the transition energies. The lowest calculated magnetic field \texttt{bx} is $0.001\,\mathrm{T}$ and the corresponding transition energies \texttt{deltaE} are $>17.5\meV$, confirming that this feature extrapolates to the same band gap as discussed in Sec.~\ref{sec_tutorial_basic}. For zero magnetic field all LLs in a subband are degenerate, thus, all transition energies would be identical (up to the artificial offset added by \texttt{split}).

 \item \emph{What are the involved LL states for a specific transition?}---
 We analyse the transition marked with the black arrow in Fig.~\ref{fig_tutorial_optical-transitions}(a) regarding which subbands and LL indices are involved. We use the same output file \texttt{transitions-filtered-7nm-optical-transitions.csv} and filter the magnetic field values \texttt{bx} to be $>6\,\mathrm{T}$, see Fig.~\ref{fig_tutorial_optical-transitions}(c), so that we can clearly distinguish this transition from other transitions at higher energies than shown in the plot. The marked columns are the initial (blue) and the final (green) LL (light) and band indices (dark). As can be seen, this transition is an intraband transition inside the subband with band index $1$, which is the inverted subband H1. The respective LL indices of initial and final state are $-2$ and $-1$, hence the polarization of this transition corresponds to $O_+$ [see Sec.~\ref{sec_optical_transitions}].
 
 \item \emph{What information regarding band ordering can we extract from the transition spectrum?}---
 Continuing the analysis for the same transition, we discuss why it can only be observed at magnetic fields $\gtrapprox5\,\mathrm{T}$. For that, we use the LL fan chart shown in the previous tutorial section Fig.~\ref{fig_tutorial_lladvanced}, where the same sample was simulated. The relevant (electro-)chemical potential for the current case is the thick line near $-20\meV$, the charge neutral point. At magnetic fields slightly below $5\,\mathrm{T}$ the lowest LL band of H1 and E1 cross, reverting to a trivial band order, while simultaneously changing from completely unoccupied to fully occupied in the case of the H1 state (vice versa for the E1 state). This enables the corresponding intraband transition in the H1 subband, which we see in the transition spectrum. Consequently, we can indirectly identify the critical magnetic field, where the lowest LL states revert to trivial band order in the magneto-optical transition spectrum. The abruptness of the crossing is a consequence of the axial approximation, which we used here for illustrative purposes. In a more realistic picture, the non-axial and bulk-inversion asymmetric terms lead to an anticrossing instead.
\end{itemize}

\subsection{Example calculation: Dispersions and wave functions of a 3D TI}
\label{sec_tutorial_3dti}
\subsubsection{Introduction}

In thick layers of a band inverted material like HgTe, so called \emph{three-dimensional
topological insulators}, the transport properties
are dominated by topological surface states. At an interface between materials
with inverted and normally ordered bands, the band inversion induces eigenstates
with a linear (two-dimensional Dirac) dispersion, confined to the interface
\cite{VolkovPankratov1985}. The physics is markedly different from those of
narrow quantum wells (see Sec.~\ref{sec_tutorial_basic}), where instead the strong
confinement in the $z$ direction leads to the inversion of subbands.

In this usage example, we investigate the key features of a 3D topological
insulator by analyzing the dispersions and wave functions. We try to answer the
following questions:

\begin{itemize}
 \item Can we identify the Dirac point and the surface states in the dispersion?
 \item How well are the surface states confined?
 \item How much is the overlap between the surface states?
 \item What is the orbital content of the wave functions?
\end{itemize}

\subsubsection{Setting it up step by step}

\begin{enumerate}
 \item Despite the system being called `3D topological insulator', the appropriate
 geometry is still 2D, because there is confinement in the $z$ direction. We take
 the previous example, Sec.~\ref{sec_tutorial_basic}, and adjust the layer stack
 parameters \texttt{msubst} and \texttt{llayer}. We also use a different momentum
 and energy range. We raise the number of eigenvectors with \texttt{neig 100},
 because there are more subbands due to weaker confinement in the $z$ direction.
 We choose the observable \texttt{z} in order to see whether we can locate the
 surface states at the top and bottom interface.
\begin{verbatim}
kdotpy 2d 8o noax msubst CdTe mlayer HgCdTe 68% HgTe HgCdTe 68%
llayer 10 70 10 zres 0.25 k -0.3 0.3 / 120 kphi 0 split 0.01 erange -100 40
neig 100 obs z legend char out -70nm outdir data-3dti
\end{verbatim}
 
 \item We obtain the dispersion plot of Fig.~\ref{fig_tutorial_3dti}(a). At $E\approx-90\meV$,
 we observe the linear dispersions characteristic of a Dirac point. Between
 valence band and conduction bands (negative and positive energies, respectively),
 there are states bridging the gap. The gray colour indicates
 that the expectation values $\avg{z}\approx 0$ for all states, which seemingly
 suggests absence of surface character. In order to diagnose this counterintuitive
 result, let us plot plot the wave functions at $k=0.14\nm^{-1}$ using
 \texttt{plotwf separate 0.14}.
\begin{verbatim}
kdotpy 2d 8o noax msubst CdTe mlayer HgCdTe 68% HgTe HgCdTe 68%
llayer 10 70 10 zres 0.25 k -0.3 0.3 / 120 kphi 0 split 0.01 erange -100 40
neig 100 obs z legend char out -70nm outdir data-3dti plotwf separate 0.14
\end{verbatim}

 \item We obtain two files with wave functions, i.e., at $k=-0.14\nm^{-1}$ and
 at $k=0.14\nm^{-1}$. In the latter file, let us look for one of the states
 at $E\approx 13.5\meV$, see Fig.~\ref{fig_tutorial_3dti}(b). The wave function contains
 symmetric and antisymmetric components for different orbitals. In order to
 break the degeneracy due to the mirror symmetry in $z$ direction, let us add a
 weak potential difference between top and bottom surface with \texttt{vinner 0.1}.
 This argument leads to an electrostatic potential consistent with a constant
 displacement field $\vec{D}=\epsilon(z)\vec{E}(z)$, such that the potential
 difference between top and bottom surface is $0.1\meV$.
\begin{verbatim}
kdotpy 2d 8o noax msubst CdTe mlayer HgCdTe 68% HgTe HgCdTe 68%
llayer 10 70 10 zres 0.25 k -0.3 0.3 / 120 kphi 0 split 0.01 erange -100 40
neig 100 obs z legend char out -70nm outdir data-3dti plotwf separate 0.14
vinner 0.1
\end{verbatim}
 
\end{enumerate}

\subsubsection{Results and interpretation}

\begin{figure}
 \includegraphics[width=150mm]{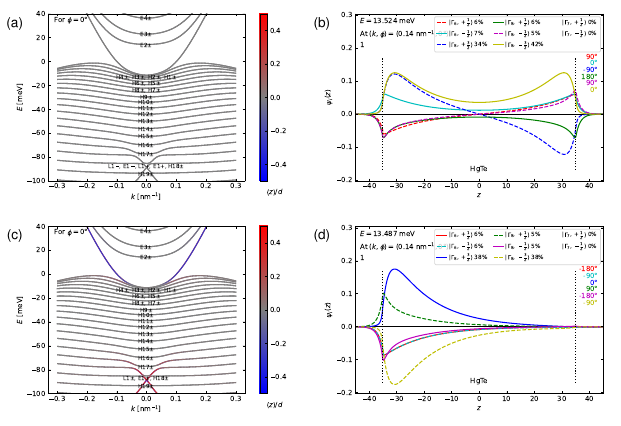}
 \caption{
   (a) Dispersion of a $70\,\mathrm{nm}$ thick HgTe layer on a CdTe substrate,
   \texttt{dispersion-70nm.pdf}. The colours indicate the expectation value
   $\avg{z}$. Due to mirror symmetry $z\to-z$, $\avg{z}=0$ for all states.
   (b) A wave function at $k=0.14\nm^{-1}$ and $E\approx 13.5\meV$. The solid
   curves indicate the real parts, the dashed curves the imaginary parts for
   each orbital component.
   (c) The dispersion with a degeneracy breaking potential (argument \texttt{vinner 0.1}).
   The surface character of the states in the gap and near the Dirac point is
   indicated by red and blue colour. States with bulk character are gray.
   (d) One of the surface states at $k=0.14\nm^{-1}$ and $E\approx 13.5\meV$.
   Clearly, the degeneracy is resolved. The angles on the right-hand side
   indicate the complex arguments in each orbital component.
 }
 \label{fig_tutorial_3dti}
\end{figure}

After applying \texttt{vinner 0.1} in order to break the mirror symmetry, we obtain
\verb+dispersion-70nm.pdf+ and \verb+wfs-70nm_0.140_0.pdf+ as shown in Fig.~\ref{fig_tutorial_3dti}(c) and (d). With these plots and the 
corresponding csv files, we can answer the questions above:

\begin{itemize}
\item \emph{Can we identify the Dirac point and the surface states in the dispersion?}---
 In Fig.~\ref{fig_tutorial_3dti}(c), we find two pairs of linearly dispersing
 states, crossing at $E\approx-89\meV$ at zero momentum. The subband labels are
 $\mathrm{E}1\pm$ and $\mathrm{L}1\pm$. The states emanating
 from the Dirac point have a distinct surface character is indicated by the
 red and blue colour (where the latter is poorly visible because the dispersions
 almost coincide).
 
 These four Dirac states can be described by a massless two-dimensional Dirac
 Hamiltonian. We note that the Dirac model for the surface states is distinct
 from the BHZ model \cite{BernevigEA2006}, which has a different set of basis
 states, namely the $\mathrm{E}1\pm$ and $\mathrm{H}1\pm$ subbands; see also
 the example in Sec.~\ref{sec_tutorial_basic}.
 
 The surface states seem to disappear into the bulk valence band for energies
 $-70\meV\leq E\leq -10\meV$ approximately, due to hybridization with the
 bulk valence band. In the bulk gap ($0\meV\leq E\leq 20\meV$ approximately),
 the surface states cross the gap from valence to conduction band.
 
\item \emph{How well are the surface states confined?}---
 The level of confinement can be extracted from the wave functions. In this
 example, we investigate one of the surface states in the bulk gap,
 at $k=0.14\nm^{-1}$ and $E\approx 13.5\meV$. From the wave functions shown in
 Fig.~\ref{fig_tutorial_3dti}(d), we estimate a characteristic `size' of
 about $10$--$20\nm$.
 
 For a more rigorous value, we can take the file
 \verb+{wfs-70nm_0.140_0.1.csv+ (depending on the configuration settings, this
 might be inside a tar or zip archive) and calculate
 $\abs{\psi}^2(z)=\sum_i\abs{\psi_i}^2(z)$, where $i$ are the orbitals.
 We determine the characteristic length $\xi$ by fitting $\psi(z)\sim\ee^{-z/\xi}$.
 From approximating $\xi^{-1}\approx\partial_z\ln(\abs{\psi}^2)$ by a discrete
 derivative around $z=0$, we find $\xi\approx 7\nm$. Note that the apparent
 discrepancy with the estimate below comes from the difference between $\psi_i(z)$
 and $\abs{\psi}^2(z)$.

\item \emph{How much is the overlap between the surface states?}---
 In order to estimate the overlap between the surface states at top and bottom surface,
 we calculate the integral of $\abs{\psi}(z)$ for $z>0$, for one of the surface states.
 We find $\int_0^{z_\mathrm{max}}\abs{\psi}^2(z) \approx 0.01$ by
 summing up the appropriate values in \verb+wfs-70nm_0.140_0.1.csv+.
 We finally point out that the overlap and the confinement length are momentum
 dependent, so that for a complete picture of the physics, this analysis has to
 be repeated for other momenta.
 
\item \emph{What is the orbital content of the wave functions?}---
 The data in Fig.~\ref{fig_tutorial_3dti}(d) visualizes the wave function as
 $\psi(z)\sum_i\psi_i(z)\ket{i}$, where $\ket{i}$ are the orbital basis
 functions (see Eq.~\ref{eqn_orbital_basis}). The curves being solid or dashed
 indicate the $\psi_i(z)$ being real or imaginary. In many cases, the functions
 $\psi_i(z)$ can be factorized as a complex constant $\ee^{\ii\phi_i}$ times a real
 function $f_i(z)$. The angles in degrees in the right hand side of
 Fig.~\ref{fig_tutorial_3dti}(d) indicate the complex arguments $\psi_i$. Thus,
 the wave function for this state can be written as
 \begin{equation}
  \psi(z) = f_{1,2}(z) (-\ketbf{1}-\ii\ketbf{2})
           + f_{3,6}(z)(\ketbf{3}-\ii\ketbf{6})
           + f_{4,5}(z)(\ii\ketbf{4}-\ketbf{5})
           + f_{7,8}(z)(\ii\ketbf{7}+\ketbf{8})
 \end{equation}
 in terms of four real functions $f_{i,i'}(z)$, where components $i$, $i'$ with
 identical shapes of the envelope function have been taken together pairwise.

\end{itemize}

\subsection{Auxiliary scripts}
\label{sec_auxiliary_scripts}

\subsubsection{kdotpy merge}

The subprogram \texttt{kdotpy merge} imports multiple XML files and
merges the result into a single data set and plots the result in a single
dispersion or magnetic-field dependence plot.
(Further postprocessing is not supported in version 1.0.0 of \kdotpy.)
The merging is done `horizontally', i.e., data sets with different values of $k$
or $B$) as well as `vertically', i.e., with data at the same values of $k$ and $B$,
but for example with respect to different target energies.
The command-line syntax may be as simple as
\begin{verbatim}
kdotpy merge data/output.1.xml data/output.2.xml
\end{verbatim}
Usually, additional options are provided, for example setting the output filenames
and directory.
\begin{verbatim}
kdotpy merge bandalign -25 2 out .merged
outdir data-merged -- data/output.1.xml data/output.2.xml
\end{verbatim}
The \verb+--+ (double hyphen) symbol separates the input file names from the
other options. It may be omitted if there is no potential confusion between
the other options and the input file names.
The command line option \texttt{bandalign} can be used to do or redo band alignment
on the combined data set. Manually assisted realignment as described in
Sec.~\ref{sec_bandalign_manual} is also supported.

For data integrity, \texttt{kdotpy merge} compares the physical parameters
associated with the source data files (stored in the \texttt{<parameter>} tag).
If values differ, it will show a warning as to inform the user that the data
sets might not fit together. There is also a check on whether the grids of the
data files are compatible, i.e., whether the combination of the grids can also
be represented as a \texttt{VectorGrid}.

One of the important use cases of \texttt{kdotpy merge}
is the ability to split up computationally demanding computations into more
manageable pieces. For example, strip calculation with \texttt{kdotpy 1d}
(which has a high RAM footprint) can be split up across
multiple nodes of a high-performance computing cluster.
Another boon of \texttt{kdotpy merge} is that incomplete data sets can be
completed with just the missing data and using the already existing data to
save calculation time. This situation may occur, for example, when one out of
many cluster job fails, or when in hindsight the value of $n_\mathrm{eig}$
was set too low to cover the desired certain energy interval.
In Refs.~\cite{ShamimEA2020SciAdv} and \cite{ShamimEA2022NatCommun}, we have used
this strategy for the strip calculations with \texttt{kdotpy 1d}, (with kdotpy
versions v0.42 and v0.64, respectively \footnote{The actual versions used were
development versions preceding v0.42 (for Ref.~\cite{ShamimEA2020SciAdv}) and
v0.64 (for Ref.~\cite{ShamimEA2022NatCommun}).}.
For the dispersion figures, $3$--$10$ data sets were merged.

\subsubsection{kdotpy compare}

The subprogram \texttt{kdotpy compare} is similar to \texttt{kdotpy merge}, but
it can take an arbitrary number of data files and merge them into multiple data
sets, which can then be compared in a single plot. The command-line syntax is
similar to \texttt{kdotpy merge}, however data sets must be separated using
\texttt{vs}, as in the following example,
\begin{verbatim}
kdotpy compare data/output.1a.xml data/output.1b.xml vs data/output.2.xml
\end{verbatim}
In this example, the first data set is constructed by merging the data from
\texttt{data/output.1a.xml} and \texttt{data/output.1b.xml}, and the second
data set is the single file \texttt{data/output.2.xml}. The two data sets are
visualized with different markers and/or colours. It is also possible to enter
more than two data sets. The syntax for extra options is similar to that of
\texttt{kdotpy merge}.

\texttt{kdotpy compare} only handles one-dimensional dispersions and magnetic
field dependence. It is primarily used for quick comparisons between two related
systems, for example to study the effect of enabling or disabling a term in the
Hamiltonian. It is also useful as a debugging tool.

\subsubsection{kdotpy batch}

The auxiliary subprogram \texttt{kdotpy batch} is a tool for running any of the
calculation subprograms in batch mode, where many calculations are run in sequence,
but with a variation in one or more parameters. The philosophy is similar to a
shell (bash) script, but \texttt{kdotpy batch} may have a slightly more intuitive
way of iterating over the parameters and it can handle runs in parallel.

The behaviour is best illustrated from an example command line,
\begin{verbatim}
kdotpy batch [opts] @x 5.0 9.0 / 4 @y [CdTe, HgTe, HgCdTe 68%] do kdotpy 2d
8o noax llayer 10 @x 10 mater HgCdTe 68% @y HgCdTe 68% out -@0of@@ ...
\end{verbatim}
This \texttt{kdotpy batch} command runs iteratively the \kdotpy{} subprogram after the 
argument \texttt{do}. Here, \texttt{kdotpy 2d} is run fifteen times, with \verb+@x+
replaced by the values \texttt{5.0}, \texttt{6.0}, \texttt{7.0}, \texttt{8.0},
and \texttt{9.0}, and \verb+@y+ replaced by \texttt{CdTe}, \texttt{HgTe}, and
\texttt{HgCdTe 68\%}. The command syntax for ranges is the same as for the grid
components, see Sec.~\ref{sec_vector_grids}. The special notations \verb+@0+ and
\verb+@@+ denote a counter and the total value respectively, i.e., \verb+@0of@@+
yields \texttt{1of15}, \texttt{2of15}, etc. All `\texttt{@} variables' defined before
\texttt{do} may appear after \texttt{do} an arbitrary number of times. Curly
brackets may be used (e.g., \verb+@{x}+) if variables appear inside a longer string.

Multiple processes may run in parallel with the \texttt{cpu} and \texttt{proc}
arguments, to be specified before \texttt{do}. These options also take into account
the number of CPUs of each process, if it is given by the argument \texttt{cpu}
after \texttt{do}. Another useful argument is \texttt{dryrun}, which prints
the command lines that result from the `\verb+@+ substitutions', but does not 
execute the tasks.

In a batch calculation, the console output of the calculation scripts is written
to the files\linebreak[4]\texttt{stdout.1.txt}, \texttt{stdout.2.txt}, etc. and
\texttt{stderr.1.txt}, \texttt{stderr.2.txt}, etc., where the extension can be
adjusted with the configuration values \verb+batch_stdout_extension+
and\linebreak[4] \verb+batch_stderr_extension+. The \texttt{kdotpy batch} subprogram itself
writes to the standard stdout and stderr streams. At the end of the batch run,
\texttt{kdotpy batch} provides a status message that summarizes how many tasks
were completed successfully.

\subsubsection{kdotpy config}
\label{sec_kdotpy_config}

Whereas the user may edit the configuration file \texttt{\HOME/.kdotpy/kdotpyrc}
directly, we also provide a command line tool \texttt{kdotpy config}, which
can do the following simple operations on the configuration:
\begin{itemize}
\item \texttt{kdotpy config list} or \texttt{kdotpy config show}:
  Show all non-default configuration values.

\item \texttt{kdotpy config all}:
  Show all configuration values.

\item \texttt{kdotpy config help \emph{key}}:
  Show information on \texttt{\emph{key}} (taken from the help file; multiple
  \texttt{\emph{key}} arguments possible).

\item \texttt{kdotpy config set \emph{key}=\emph{value}}:
  Set \texttt{\emph{key}} to a new value (multiple \texttt{\emph{key}=\emph{value}} pairs possible).

\item \texttt{kdotpy config reset \emph{key}}:
  Set \texttt{\emph{key}} to its default value (multiple \texttt{\emph{key}} arguments possible).

\item \texttt{kdotpy config file}:
  Print the full path of the configuration file.

\item \texttt{kdotpy config edit}:
  Open the configuration file in an editor. (Use environment variable \texttt{VISUAL} or
  \texttt{EDITOR} to set your command line editor.)

\item \verb+kdotpy config fig_lmargin=10 fig_rmargin=10 fig_hsize+
  (no operation specified): Set the values for \verb+fig_lmargin+ and \verb+fig_rmargin+, show
  all three values.

\end{itemize}
Multiple \texttt{\emph{key}} and/or \texttt{\emph{key}=\emph{value}} can also be
combined with a semicolon \texttt{;}, for example
\begin{verbatim}
kdotpy config 'fig_lmargin=10;fig_rmargin=10;fig_hsize'
\end{verbatim}
(In bash one must use single quotes because of the special meaning of \texttt{;}.)
The syntax is thus analogous to the \texttt{config}
argument for the calculation subprograms, for example \texttt{kdotpy 2d}.

\subsubsection{kdotpy test}

The package includes a set of standardized tests that can be run with the command
\texttt{kdotpy test}. The purpose of these tests is to catch errors during code
development and to catch performance issues. Each standardized test is defined 
in \texttt{kdotpy-test.py} by a command line. 
Success or failure is determined by the exit code. The output (by default written
into the subdirectory \texttt{test} relative to the current working directory) is
not checked for errors and has to be inspected manually, if desired. After all
tests have been attempted, a summary is printed to the terminal indicating
success or failure and the time it took to run.

Currently, in \kdotpy{} version 1.0, there are 19 standardized tests,
labelled with the following test ids:
\verb+1d+, \verb+2d_cartesian+, \verb+2d_offset+, \verb+2d_orient+,
\verb+2d_polar+, \verb+2d_qw+, \verb+2d_qw_2+, \verb+2d_qw_bia+,
\verb+2d_selfcon+, \verb+batch+, \verb+bulk+, \verb+bulk_3d+, \verb+bulk_ll+,
\verb+ll_axial+, \verb+ll_bia+, \verb+ll_legacy+, \verb+compare_2d+,
\verb+compare_ll+, and \verb+merge+.
The full test suite may be run by typing
\begin{verbatim}
kdotpy test
\end{verbatim}
To run only a selection (of one or more tests), one may add the test ids as
arguments, for example
\begin{verbatim}
kdotpy test 2d_qw ll_axial
\end{verbatim}
to run the specified ones only. The test suite may be interrupted with Ctrl-C.

The test suite supports the following extra commands:
\begin{itemize}
\item \texttt{kdotpy test list}: Shows all test ids on the terminal.

\item \texttt{kdotpy test showcmd \emph{testids}}: Shows the command lines without
running the tests. The command lines may be copied, modified, and run manually
for the purpose of testing and debugging.

\item \texttt{kdotpy test verbose \emph{testids}}: Runs the tests in verbose mode,
by adding the argument \texttt{verbose} to their command lines.
                                   
\item \texttt{kdotpy test python3.9 \emph{testids}}: Uses a specific Python command
(in this example\linebreak[4] \texttt{python3.9}) for running the test commands. It is recommended
to use a virtual environment for this purpose.
\end{itemize}
In these commands, \texttt{\emph{testids}} may be any sequence of test ids, or
it may be omitted to apply the command to all 19 tests.

The tests are defined in \texttt{kdotpy-test.py} in a way that is compatible 
with the \texttt{pytest} package for automated testing. We have included
automated testing with \texttt{pytest} into the CI/CD workflow in our Gitlab
project.

\clearpage
\section{Conclusion and outlook}

This publication marks the public release of \kdotpy{} as an open source
software project. We release it with the expectation that \kdotpy{} is
beneficial to the research community. Conversely, the open source nature 
encourages input from the community as to improve and extend the project even
further.
It is our hope that interaction with the community will lead to new
insights and new directions for the future development of \kdotpy. We
encourage our users to actively participate in the discussions in our Gitlab
repository \cite{kdotpy_repo} or via other channels. 

At the time of writing, we are already in the process of designing new features
for future releases. Improved strain handling is scheduled to be included in
an upcoming release in the near future. The treatment of electrostatics is
currently undergoing continuous improvement, where in particular we are
reconsidering the boundary conditions appropriate for several experimental
setups. We will also add the necessary infrastructure for treatment of dopants
in the material. We also expect to gain more experience with materials other than
the built-in ones. In general, the future directions of \kdotpy{} will also be
influenced by input from new experimental results from transport and spectroscopy
and by feedback and suggestions from our users.

\section*{Acknowledgements}
We thank
Domenico Di~Sante,
Giorgio Sangiovanni,
Bj{\"o}rn Trauzettel,
Florian Goth, and
Fakher Assaad
for feedback and support at various stages of the project.

\paragraph{Author contributions}
W.B. designed the software package.
W.B., F.B., C.B., and M.H. developed the source code, with W.B. acting as lead developer.
W.B., F.B., C.B., L.B., M.H., and M.S. performed code review and benchmarks.
W.B., C.F., S.S., L.-X.W., H.B., and L.W.M. provided input on the physics of quantum transport.
F.B., C.B., L.B., M.H., M.S., and T.K. provided input on the physics of optical spectroscopy.
W.B., J.B., and E.M.H. developed the theoretical model at the early stages of the project.
W.B., T.K., and L.W.M. managed the project.
The writing of the paper was led by W.B. with input from all authors.

\paragraph{Funding information}
We acknowledge financial support from the Deutsche Forschungsgemeinschaft
(DFG, German Research Foundation) in the project SFB 1170 \emph{ToCoTronics}
(Project ID 258499086) and in the W\"urzburg-Dresden Cluster of Excellence on
Complexity and Topology in Quantum Matter \emph{ct.qmat} (EXC 2147, Project ID 39085490).

\clearpage
\begin{appendix}
\numberwithin{equation}{section}

\section{Implementation details}

\subsection{Parallelization}
\label{app_parallelization}

\subsubsection{Basic method}

The Python programming language was not designed with parallelization in mind.
The global interpreter lock (GIL) is often considered (see, e.g., Ref.~\cite{PEP703})
as a major obstacle
for running parallelized Python code. However, the \texttt{multiprocessing}
module avoids this obstacle by running Python code in separate parallel subprocesses,
such that each process is affected only by its own GIL.

In \kdotpy, the \texttt{multiprocessing} module is used to parallelize
iterative tasks of the form
\begin{verbatim}
data = [f(x, *f_args, **f_kwds) for x in vals]
\end{verbatim}
where each call to the function \texttt{f} is an independent task. The main loop
in \kdotpy{} is the iterated diagonalization of the Hamiltonian over the momentum $k$
and magnetic field $B$ values of the grid. Due to the uniform nature of these tasks,
this iteration lends itself well for parallelization.

In \kdotpy, the function \verb+parallel_apply()+ in the \texttt{parallel} submodule
facilitates parallelization over the diagonalization function by implementing
the necessary boilerplate code around the \texttt{multiprocessing.Pool} object.
The following code illustrates the recipe implemented in\linebreak[4] \verb+parallel_apply()+
(simplified compared to the actual code for the sake of illustration):
\begin{verbatim}
pool = multiprocessing.Pool(processes = num_processes)
output = [
    pool.apply_async(f, args=(x,) + f_args, kwds = f_kwds) for x in vals
]
while True:
   jobsdone = sum(1 for x in output if x.ready())
   if jobsdone >= n:
       break
   print(f"{jobsdone} / {n}")
   time.sleep(poll_interval)
data = [r.get() for r in output]
pool.close()
pool.join()
\end{verbatim}
Here, the \texttt{while} loop, that runs in the parent process, is used for
process monitoring by printing the number of completed tasks to standard output
\footnote{In the actual code, an instance of the \texttt{Progress} class handles the
progress counter. In the code example here, we have replaced this by a \texttt{print}
statement for clarity.}.
This loop runs one iteration every \verb+poll_interval+
seconds in the main process until all tasks are completed. The result
\texttt{data} is collected before the pool is joined and closed. 

Unfortunately, \texttt{multiprocessing} does not handle keyboard interrupts
(pressing Ctrl-C to abort the program) and other signals properly out of the
box. Thus, \verb+parallel_apply()+ contains additional code in order to handle
these asynchronous events gracefully: Once a worker process is aborted or
terminated, the parent process of \kdotpy{} detects it and aborts the full
calculation, including the remaining worker processes. If this is not done
correctly, worker processes can become \emph{orphaned} (i.e., detached from
the parent process) and continue working even when the parent process no longer
exists. These problems are mitigated by using a custom signal handler with \verb+parallel_apply()+.

\subsubsection{Advanced method}

The Task-Model framework is also built around the process pools of the
\texttt{multiprocessing} module, but with a higher level of abstraction, that
allows a more fine-grained control of how tasks are run. The \kdotpy{} submodules
\texttt{tasks} and \texttt{models} define the following classes that constitute
this framework:
\begin{itemize}
 \item \texttt{Model\emph{X}} class: This class stores the `recipe', analogous to a
 diagonalization function. For the different `models' (1D, 2D, LL, etc.), a
 separate class is derived from the \texttt{ModelBase} class.
 It stores the steps of the recipe as a list of functions
\begin{verbatim}
self.steps = [
    self.load_ddp,
    self.construct_ham,
    self.solve_ham,
    self.post_solve
] 
\end{verbatim}
 where \verb+self.load_ddp+ enables loading of a \texttt{DiagDataPoint} object,
 and\linebreak[4] \verb+self.construct_ham+, \verb+self.solve_ham+, and \verb+self.post_solve+
 apply the construction of the Hamiltonian, application of the eigensolver, and
 the processing of eigenstates (i.e., the steps listed in
 Sec.~\ref{sec_parallelization_main}).

 \item \texttt{Task} class: This class contains the functions 
 \verb+self.worker_func+, \verb+self.callback+, and\linebreak[4] \verb+self.error_callback+
 as well as references to the process or thread pool. The \verb+Task.run()+
 method calls \verb+pool.apply_async()+, where \texttt{pool} is either a process
 or a thread pool.

 \item \texttt{TaskManager} class: The single instance of this class is
 responsible for initializing, managing, and joining the process or thread pool,
 somewhat similar to \verb+parallel_apply()+.
 The primary job of the \texttt{TaskManager} is to send the tasks in the queue to
 the available CPU or GPU workers and to join if all tasks are completed.
 The handling of asynchronous events is also done by the \texttt{TaskManager}.
 The \texttt{TaskManager} class is derived from \texttt{PriorityQueue} from the 
 \texttt{queue} module of Python.
\end{itemize}

The Task-Model framework has several benefits compared to \verb+parallel_apply()+.
Firstly, the \texttt{Model\emph{X}} classes can be derived from. The derived
class does not need to redefine all steps, but only those one where it differs
from its parent class.
Secondly, the data points are split into subtasks, so that each subtasks can be
executed where this is done most optimally (CPU or GPU).
The added flexibility of the Task-Model framework comes with a small price: As a
result of more overhead, the Task-Model framework is marginally slower than \verb+parallel_apply()+, but the difference is often barely noticeable.

\subsection{Density of states}
\label{app_idos_elements}

\subsubsection{Triangular IDOS element}

In Eq.~\eqref{eqn_density_fraction_onedim}, we have determined the fraction of
the interval $[k_{i},k_{i+1}]$ where a linear function is below a certain value $E$.
For two dimensions, the analogous problem would be considering a linearly
interpolated function on a triangle in the two dimensional plane. Let us label
the vertices $1,2,3$, and assume the function values are $e_{1,2,3}$ with
$e_1\leq e_2 \leq e_3$ (without loss of generality). The momentum coordinates of
the vertices are irrelevant. The interpolated function is given by
\begin{equation}
  e(u,v) = e_1 + (e_2-e_1)u + (e_3-e_1)v = (1-u-v)e_1+u\,e_2+v\,e_3
\end{equation}
where $(u,v)$ are scaled coordinates with $0\leq u \leq 1$, $0\leq v \leq 1$,
and $u+v\leq 1$, such that the vertices of the triangle correspond to $(u,v)=(0,0)$,
$(1,0)$, and $(0,1)$. We are interested in the fraction $f(E)$ of the triangle
where $e(u,v)<E$. This can be found by finding the intersection points of the line
$e(u,v)=E$ with the sides of the triangle and calculating the area of the resulting
polygon. The fraction $f(E)$ is this area divided by $\frac{1}{2}$ (the area of the
full triangle), and is given by
\begin{equation}\label{eqn_density_fraction_twodim}
   f(E) = \left\{\begin{array}{ll}
   1 &\text{if } E\geq e_{3}\\
   1 - \frac{(E-e_3)^2}{(e_3-e_2)(e_3-e_1)} &\text{if }e_2 \leq E < e_3\\
   \frac{(E-e_1)^2}{(e_2-e_1)(e_3-e_1)} &\text{if }e_1 \leq E < e_2\\
   0 &\text{if } E < e_1
  \end{array}\right.
\end{equation}
The conditions are such that the denominators are nonzero; for example, if
$e_2 = e_1$, then the condition $e_1 \leq E < e_2$ is never fulfilled.

The implementation in \verb+triangle_idos_element()+ takes as input the
three-dimensional array $e_{i,j,v}$, where $i,j$ label the elementary squares in
momentum space, and $v=1,2,3$ labels the three vertices. It applies the following
recipe.
\begin{itemize}
 \item The momentum space indices $(i,j)$ are flattened to $I$. The result is a
  two-dimensional array $e_{I,v}$ of shape $(n_k, 3)$, where $n_k=n_{k_x}n_{k_y}$
  is the number of elementary squares in momentum space.
 \item Sort along the last axis so that $e_{I,1}\leq e_{I,2} \leq e_{I,3}$
  is satisfied for all $I$.
 \item Evaluate $f^{(1)} =(E-e_{I,1})^2/(e_{I,2}-e_{I,1})(e_{I,3}-e_{I,1})$ and
  $f^{(2)} = 1 - (E-e_{I,3})^2/(e_{I,3}-e_{I,2})(e_{I,3}-e_{I,1})$ (see
  Eq.~\eqref{eqn_density_fraction_twodim}) for all $I$ and for all $E$ in
  the energy range. At the points where the denominator vanishes, substitute $0$.
  In both cases, the resulting arrays \texttt{f1} and \texttt{f2} are
  two-dimensional arrays of shape $(n_k, n_E)$, where $n_E$ is the number of
  energy values in the energy range.
 \item Evaluate the conditions $E<e_1$, $E<e_2$, and $E<e_3$ as the two-dimensional
  boolean arrays \texttt{cond0}, \texttt{cond1}, and \texttt{cond2}. Then evaluate
  $f_I(E)$ as
\begin{verbatim}
f = np.where(cond0, zeros,
    np.where(cond1, f1, np.where(cond2, f2, ones))
)
\end{verbatim}
  where \verb+zeros = np.zeros_like(f1)+ and \verb+ones = np.ones_like(f1)+.
  The output is an array of shape $(n_k, n_E)$, like all input arrays.
  \item Subtract $1$ if the band is hole-like.
  \item For all momenta indices $I$, where any $e_{I,v}$ ($v=1,2,3$) is
  undefined (\texttt{NaN}), substitute $f_I(E)=0$ for all $E$.
\end{itemize}

\subsubsection{Tetrahedral IDOS element}

For three dimensions, we follow similar ideas to arrive at an expression for
$f(E)$. Let the energy values at the vertices of an elementary tetrahedron be
$e_{1,2,3,4}$, where $e_1\leq e_2 \leq e_3\leq e_4$. The interpolated function
is given by $e(u,v,w) = e_1 + (e_2-e_1)u + (e_3-e_1)v + (e_4-e_1)w$, with
$0\leq u,v,w \leq 1$ and $u+v+w \leq 1$. The fraction of the tetrahedra (with
volume $\frac{1}{6}$ that satisfies $e(u,v,w)<E$ is
\begin{equation}\label{eqn_density_fraction_threedim}
   f(E) = \left\{\begin{array}{ll}
   1 &\text{if } E\geq e_{4}\\
   1 - \frac{(E-e_4)^3}{(e_4-e_3)(e_4-e_2)(e_4-e_1)} &\text{if }e_3 \leq E < e_4\\
   \frac{(E-e_1)^3}{(e_2-e_1)(e_3-e_1)(e_4-e_1)} - \frac{(E-e_2)^3}{(e_2-e_1)(e_3-e_2)(e_4-e_2)}&\text{if }e_2 \leq E < e_3\text{ and }e_1 < e_2\\
   \frac{(E-e_1)^2}{(e_3-e_1)(e_4-e_1)}
    \left(1+\frac{e_3-E}{e_3-e_1}+\frac{e_4-E}{e_4-e_1}\right) &\text{if }e_2 \leq E < e_3\text{ and }e_1=e_2\\
   \frac{(E-e_1)^3}{(e_2-e_1)(e_3-e_1)(e_4-e_1)} &\text{if }e_1 \leq E < e_2\\
   0 &\text{if } E < e_1
  \end{array}\right.
\end{equation}
In \verb+tetrahedral_idos_element()+, the arrays defining $f(E)$ are calculated
analogous to those in \verb+triangle_idos_element()+ but with some additional
intermediate steps to avoid recalculation of the products of $(e_w-e_v)$ in the
denominators. The conditions $E<e_1$, $E<e_2$, $E<e_3$, and $E<e_4$ are applied
as
\begin{verbatim}
f = np.ones_like(f1)
f[cond3] = f3[cond3]
f[cond2] = f2[cond2]
f[cond1] = f1[cond1]
f[cond0] = 0.0
\end{verbatim}
which is equivalent to a sequence of calls to \texttt{np.where()}, but with a
better performance for the larger arrays in three momentum dimensions.

\textsc{Note:}
In the expression for $f(E)$ with $e_2 \leq E < e_3$ and $e_1<e_2$, it is possible
to factor out $e_2-e_1$. Doing this yields
\begin{equation}
 f^{(2)}(E)=\frac{\left(e_{31}e_{41}+e_{41}(e_3-E)+e_{31}(e_4-E)\right)(E-e_1)^2
    + e_{21}(E-e_1)^3-3e_{21}e_{31}e_{41}(E-e_1)+e_{21}^2e_{31}e_{41}}
    {e_{32}e_{42}e_{31}e_{41}},
\end{equation}
using the short-hand notation $e_{vw}=e_v-e_w$. Unlike
Eq.~\eqref{eqn_density_fraction_threedim}, there is no factor $e_2-e_1$ in the
denominator, hence it is valid for $e_1=e_2$ as well. If we substitute $e_1=e_2$,
we retrieve the expression for $e_2 \leq E < e_3$ and $e_1=e_2$ of Eq.~\eqref{eqn_density_fraction_threedim}.

\subsection{Charge neutrality point in symbolic Landau level mode}
\label{app_ubindex}

In dispersion mode and Landau level mode \texttt{full}, the charge neutrality point
always lies between the bands with band indices $-1$ and $1$, by definition.
In the Landau level mode \texttt{sym}, each state is characterized by a pair
$(n,b)$ of Landau level index $n=-2,-1,0,\ldots$ and band index $b$, so that
finding the charge neutrality point requires an extra step: For each state, we
determine a \emph{universal band index} $u$, equivalent to ordinary band indices
in full Landau level mode, with the property that the charge neutrality point
lies between the states with $u=-1$ and $u=+=1$.

The algorithm for finding the universal band indices is implemented in\linebreak[4]
\verb+DiagDataPoint.get_ubindex()+ and proceeds as follows. The states in
\verb+DiagDataPoint+ are sorted by eigenvalue. We define an array with ones
for all states with band index $b>0$ and an array with ones where $b<0$,
\begin{verbatim}
pos = np.where(
    bindex_sort > 0,
    np.ones_like(bindex_sort),
    np.zeros_like(bindex_sort)
)
neg = 1 - pos
\end{verbatim}
The arrays are summed cumulatively from below and above, respectively,
\begin{verbatim}
npos = np.cumsum(pos)
nneg = neg.sum() - np.cumsum(neg)
\end{verbatim}
The array of universal band indices is then simply the difference between the two
cumulative sums
\begin{verbatim}
ubindex = npos - nneg
ubindex[ubindex <= 0] -= 1
\end{verbatim}
where the nonpositive values have to be decreased by one in order to make sure
that the integers $u$ are nonzero. By definition, the resulting array is an
increasing sequence as function of energy, and identical to that obtained from
regular band indices in the Landau level mode \texttt{full}.

\subsection{Solving Poisson's equation}
\label{app_solution_poisson}

\newcommand{\II}{\mathcal{I}}


As already mentioned in Sec.~\ref{sec_potentials}, a way to incorporate electrostatic potentials into the Hamiltonian is by parsing a set of boundary conditions. This set is used to solve Poisson's equation within the function \verb+solve_potential()+ in \texttt{potential.py}, for `static' as well as self-consistent potentials (see Secs.~\ref{sec_potentials} and \ref{sec_selfcon}, respectively). For the remainder of this section we will refer to this potential as Hartree potential $\VH$. It should be emphasized that $\VH$ parametrizes the potential energy of the carriers in the electrostatic potential, not the electric potential itself \footnote{These two quantities differ by a factor of $-e$, which is $-1$ in the chosen system of units.} The Hartree potential $\VH$ satisfies Poisson's equation
\begin{equation}
  \partial_{z}\left[\varepsilon(z)\,\partial_{z}\VH(z)\right] = \frac{e}{\varepsilon_0}\rho(z) \label{eqn_poisson_details}
\end{equation}
[identical to Eq.~\eqref{eqn_poisson}] with dielectric function/constant $\varepsilon(z)$, elementary charge $e$, vacuum permittivity $\varepsilon_0$ and spatial charge density $\rho(z)$. In the self-consistent case, as described in Secs.~\ref{sec_selfcon} and \ref{sec_densityz}, $\rho(z)$ is extracted from all eigenvectors, whereas in the `static' case $\rho(z)=0$.

The generalized solution can be calculated by integrating twice over $z$, with generally different lower integration limits $z_a$ for the first integration and $z_b$ for the second. The first integral over Eq.~\eqref{eqn_poisson_details} yields
\begin{equation}\label{eqn_poisson_integrated_once}
  \varepsilon(z)\,\partial_{z}\VH(z) - \varepsilon(z_a)\,\partial_{z}\VH(z_a)
  = \frac{e}{\varepsilon_0}\int_{z_a}^{z} dz' \rho(z')
  = \frac{e}{\varepsilon_0}\left[\II_\rho(z) - \II_\rho(z_a)\right],
\end{equation}
where we use the shorthand notation $\II_\rho(z) \equiv \int_0^{z}  dz' \rho(z')$.
As this equation is valid for all $z$, it follows that
\begin{equation}\label{eqn_poisson_integration_constant}
 \mathcal{C}(z)\equiv\varepsilon(z)\,\partial_{z}\VH(z) - \frac{e}{\varepsilon_0}\II_\rho(z)
\end{equation}
is constant. We evaluate this constant at $z=z_a$, divide by $\varepsilon(z)$ and integrate a second time to find the Hartree potential,
\begin{align}\label{eqn_poisson_integrated_twice}
  \VH(z) =\;& \VH(z_b) + \frac{e}{\varepsilon_0} \left(\int_0^z dz' \frac{\II_\rho(z')}{\varepsilon(z')} - \int_0^{z_b} dz' \frac{\II_\rho(z')}{\varepsilon(z')}\right)
   + \mathcal{C}(z_a) \left(\int_0^z dz' \frac{1}{\varepsilon(z')} - \int_0^{z_b} dz' \frac{1}{\varepsilon(z')}\right)
\end{align}
We have written the integrals $\int_{z_{b}}^z$ explicitly as $\int_0^z - \int_0^{z_b}$, which reflects the implementation in the code.

In order to find a unique solution, two boundary conditions need to be given, which
fix $\VH$ or $\partial_z \VH$ (which can be interpreted as electric
field) at specific $z$ coordinates. The function \verb+solve_potential()+ takes
the boundary conditions as a set of keyword arguments, namely a subset of values
\texttt{v1}, \texttt{v2}, \texttt{v3}, \texttt{dv1}, \texttt{dv2}, and \texttt{v12}
and coordinates \texttt{z1}, \texttt{z2}, and \texttt{z3}. Out of the former set,
only certain combinations of values may be set to a numerical value (the other being
\texttt{None}); any other combination will cause a \texttt{ValueError} exception to be raised.
The following combinations are valid:
\begin{enumerate}
	\item
	    \texttt{dv1} and \texttt{v1}, with \texttt{z1} as coordinate:
		$\partial_{z}\VH(z_1) = \partial V_1$ and $\VH(z_1)=V_1$.
		The $z$ coordinate for both boundary conditions is the same, thus $z_a=z_b=z_1$.    
		\begin{align}
			\VH(z) =\;& V_1 + \frac{e}{\varepsilon_0} \left(\int_0^z dz' \frac{\II_\rho(z')}{\varepsilon(z')} - \int_0^{z_1} dz' \frac{\II_\rho(z')}{\varepsilon(z')}\right)\nonumber\\
			& + \left(\varepsilon(z_1)\,\partial V_1 - \frac{e}{\varepsilon_0} \II_\rho(z_1)\right) \left(\int_0^z dz' \frac{1}{\varepsilon(z')} - \int_0^{z_1} dz' \frac{1}{\varepsilon(z')}\right).\label{eqn_poisson1}
		\end{align}
	\item \texttt{dv2} and \texttt{v2}, with \texttt{z2} as coordinate: Analogous to case 1, using $z_a=z_b=z_2$ instead.
	\item
	    \texttt{dv1} and \texttt{v2}, with \texttt{z1} and \texttt{z2} as coordinates:
		$\partial_{z}\VH(z_1) = \partial V_1$ and $\VH(z_2)=V_2$.
		The $z$ coordinates for both boundary conditions are different, thus $z_a=z_1$ and $z_b=z_2$.  
		\begin{align}
			\VH(z) =\;& V_2 + \frac{e}{\varepsilon_0} \left(\int_0^z dz' \frac{\II_\rho(z')}{\varepsilon(z')} - \int_0^{z_2} dz' \frac{\II_\rho(z')}{\varepsilon(z')}\right)\nonumber\\
			& + \left(\varepsilon(z_1)\,\partial V_1 - \frac{e}{\varepsilon_0} \II_\rho(z_1)\right) \left(\int_0^z dz' \frac{1}{\varepsilon(z')} - \int_0^{z_2} dz' \frac{1}{\varepsilon(z')}\right).\label{eqn_poisson3}
		\end{align}
	\item \texttt{v1} and \texttt{dv2}, with \texttt{z1} and \texttt{z2} as coordinates: Analogous to case 3, using $z_a=z_2$ and $z_b=z_1$ instead.
	\item
        \texttt{v1} and \texttt{v2}, with \texttt{z1} and \texttt{z2} as coordinates:
		$\VH(z_1) = V_1$ and $\VH(z_2) = V_2$.
		Only potential values are given. The solution is given by Eq.~\eqref{eqn_poisson_integrated_twice}, but we cannot evaluate the constant
		$\mathcal{C}(z_a)$ directly, because the derivative $\partial_z\VH(z_a)$ is unknown.
		By setting $z_b=z_1$ and $z=z_2$ (or vice versa), we find that
		\begin{align}\label{eqn_poisson_solution_dv}
			\mathcal{C}(z_a)
			=\;& \frac{\VH(z_2) - \VH(z_1) -  \frac{e}{\varepsilon_0} \left(\int_0^{z_1} dz' \frac{\II_\rho(z')}{\varepsilon(z')} - \int_0^{z_1} dz' \frac{\II_\rho(z')}{\varepsilon(z')}\right)}{\int_0^{z_2} dz' \frac{1}{\varepsilon(z')} - \int_0^{z_1} dz' \frac{1}{\varepsilon(z')}}.
		\end{align}
		By substitution of $\mathcal{C}(z_a)$ into Eq.~\eqref{eqn_poisson_integrated_twice},
		we thus find
		\begin{align}
			\VH(z) =\;& V_1 + \frac{e}{\varepsilon_0} \left(\int_0^z dz' \frac{\II_\rho(z')}{\varepsilon(z')} - \int_0^{z_1} dz' \frac{\II_\rho(z')}{\varepsilon(z')}\right)\nonumber\\
			& + \frac{V_2 - V_1 - \frac{e}{\varepsilon_0} \int_{z_1}^{z_2} dz' \frac{\II_\rho(z')}{\varepsilon(z')}}{\int_{z_1}^{z_2} dz' \frac{1}{\varepsilon(z')}} \left(\int_0^z dz' \frac{1}{\varepsilon(z')} - \int_0^{z_1} dz' \frac{1}{\varepsilon(z')}\right).
		\end{align}
	\item
	    \texttt{v12} and \texttt{v3}, with \texttt{z1}, \texttt{z2}, and \texttt{z3}
	    as coordinates:
		$\VH(z_2)-\VH(z_1) = V_{12}$ and $\VH(z_3) = V_3$.
		We substitute $\mathcal{C}(z_a)$ from Eq.~\eqref{eqn_poisson_solution_dv} into Eq.~\eqref{eqn_poisson_integrated_twice} with $z_b=z_3$, and obtain
		\begin{align}
			\VH(z) =\;& V_3 + \frac{e}{\varepsilon_0} \left(\int_0^z dz' \frac{\II_\rho(z')}{\varepsilon(z')} - \int_0^{z_3} dz' \frac{\II_\rho(z')}{\varepsilon(z')}\right)\nonumber\\
			& + \frac{V_{12} - \frac{e}{\varepsilon_0} \int_{z_1}^{z_2} dz' \frac{\II_\rho(z')}{\varepsilon(z')}}{\int_{z_1}^{z_2} dz' \frac{1}{\varepsilon(z')}} \left(\int_0^z dz' \frac{1}{\varepsilon(z')} - \int_0^{z_3} dz' \frac{1}{\varepsilon(z')}\right).
		\end{align}
\end{enumerate}

In \verb+solve_potential()+ the NumPy arrays \texttt{densz} [$\rho(z)$] and \texttt{epsilonz} [$\varepsilon(z)$] first are integrated incrementally, yielding the arrays
\begin{verbatim}
int_densz = integrate_arr(densz) * dz
int_invepsilonz = integrate_arr(1. / epsilonz) * dz
\end{verbatim}
which represent the integrals $\II_\rho(z)$ and $\int_0^{z} dz' 1 / \varepsilon(z')$ for each individual value of $z$. Subsequently, $\int_0^z dz' \II_\rho(z') / \varepsilon(z')$ is calculated as
\begin{verbatim}
int_dens_over_epsz = integrate_arr(int_densz / epsilonz) * dz
\end{verbatim}
The value of the integrals at the chosen $z$ coordinates ($z_1$, $z_2$, $z_3$) are evaluated by using \texttt{np.interp()}, e.g.,
\begin{verbatim}
int_dens_over_eps_z1, int_dens_over_eps_z2, int_dens_over_eps_z3 = \
    np.interp([z1, z2, z3], zval, int_dens_over_epsz)
\end{verbatim}
where \texttt{zval} is the array of $z$ coordinates where the other arrays are defined.
A similar evaluation is done for \verb+epsilonz+, \verb+int_densz+, \verb+int_invepsilon+, followed by 
\begin{verbatim}
int_dens_over_epsz_1_2 = int_dens_over_eps_z2 - int_dens_over_eps_z1
int_invepsilonz_1_2 = int_invepsilon_z2 - int_invepsilon_z1
\end{verbatim}
Depending on the combination of boundary conditions, the corresponding solution is selected from the six cases given above. Taking case 1 (\texttt{dv1} and \texttt{v1}) as example, 
the Hartree potential is then evaluated as
\begin{verbatim}
int_const = epsilon_z1 * dv1 - eovereps0 * int_dens_z1
vz = v1 + eovereps0 * (int_dens_over_epsz - int_dens_over_eps_z1) 
        + int_const * (int_invepsilonz - int_invepsilon_z1)
\end{verbatim}
where \verb+int_const+ is the integration constant $\mathcal{C}(z_a)$ from Eq.~\eqref{eqn_poisson_integration_constant}. The implementations for the other
cases are analogous.

The boundary conditions are chosen automatically by parsing command line arguments like \texttt{vinner}, \texttt{vouter}, \texttt{vsurf} or \texttt{efield}. In self-consistent calculations the automatically determined boundary conditions can be manually overwritten by using the command line argument \texttt{potentialbc}.

On a side note, one could also imagine solving the Poisson equation as a matrix
equation involving
the vector $\VH(z_i)$ on the grid of coordinates $z_i$, obtained by replacing the
derivative operator by the appropriate matrix (cf.\ Sec.~\ref{sec_dimensional_reduction}).
This results in a matrix-vector equation, which can be solved by matrix inversion.
In the continuum limit (grid resolution $\Delta z \to 0$), this method converges
to the same solution as the numerical integration method detailed above. Our
benchmarks show that the numerical integration method outperforms the matrix method
within the context of \kdotpy, with smaller numerical errors at finite resolutions
$\Delta z$. We thus decided to include the numerical integration approach only.

\subsection{Extrema}
\label{app_extrema_solvers}

\subsubsection{Nine-point extremum solver}

The function \verb+nine_point_extremum_solver()+ takes a $3\times 3$ grid of
momentum values $(k_{x,i'},k_{y,j'})$ values and a $3\times 3$ grid of energy
values $e_{i',j'}$ as inputs. The indices $i'$ and $j'$ are taken as
$i'=i-1,i,i+1$ and $j'=j-1,j,j+1$, where $i,j$ is a location where an extremum
is detected: We say there is a minimum (maximum) at $i,j$ if the values
\begin{align}
  e_{i+1,j},e_{i-1,j},e_{i,j+1},e_{i,j-1},e_{i+1,j+1},e_{i-1,j+1},e_{i+1,j-1},e_{i-1,j-1}
\end{align}
are all $> e_{i,j}$ ($< e_{i,j}$). If this is the case, the
\verb+nine_point_extremum_solver()+ locates the extremum more precisely by
fitting Eq.~\eqref{eqn_nine_point_extremum_function} to the input data. Assume
that the momenta are aligned on an equally spaced cartesian grid, and define
$\Delta x = k_{x,i+1}-k_{x,i} = k_{x,i}-k_{x,i-1}$ and 
$\Delta y = k_{y,j+1}-k_{y,j} = k_{y,j}-k_{y,j-1}$. First calculate the
coefficients of the Hessian matrix,
\begin{align}
  a&=\frac{e_{i+1,j}-2e_{i,j}+e_{i-1,j}}{2(\Delta x)^2}, &
  c&= \frac{e_{i+1,j+1} - e_{i-1,j+1} - e_{i+1,j-1} + e_{i-1,j-1}}{4\, \Delta x\,\Delta y},\\
  b&=\frac{e_{i,j+1}-2e_{i,j}+e_{i,j-1}}{2(\Delta y)^2}. \nonumber
\end{align}
We note that $c$ is the only coefficient where the $e_{i\pm1,j\pm1}$ appear;
the other linear combinations of $e_{i\pm1,j\pm1}$ are not considered.
We calculate the auxiliary variables
\begin{equation}\label{eqn_nine_point_extremum_aux}
 X = \frac{e_{i+1,j} - e_{i-1,j}}{2\Delta x},\qquad
 Y = \frac{e_{i,j+1} - e_{i,j-1}}{2\Delta y}
\end{equation}
and use them to find the location $(k_{x,0},k_{y,0})$ of the extremum in momentum
space,
\begin{equation}\label{eqn_nine_point_extremum_momentum}
  k_{x,0} = k_{x,i} + \frac{cY - 2bX}{4ab-c^2},\qquad
  k_{y,0} = k_{y,i} + \frac{cX - 2aY}{4ab-c^2}.
\end{equation}
The denominator $4ab-c^2$ is the determinant of the Hessian matrix. If the location
defined by Eq.~\eqref{eqn_nine_point_extremum_momentum} lies outside of the
rectangle $[k_{x,i-1},k_{x,i+1}]\times[k_{y,j-1},k_{y,j+1}]$, raise the warning
\emph{``Poorly defined extremum found''} and use the location $(k_{x,i}-X/2a,k_{y,j}-Y/2b)$
instead, i.e., effectively setting $c=0$.
The energy value $f_0$ of the extremum is
\begin{equation}
  f_0=e_{i,j} - a (k_{x,i} - k_{x,0})^2 - b (k_{y,j} - k_{y,0})^2 - c (k_{x,i} - k_{x,0}) (k_{y,j} - k_{y,0}).
\end{equation}
The function \verb+nine_point_extremum_solver()+ returns the result as $f_0$,
$(k_{x,0},k_{y,0})$, $(a,b,c)$.

The band masses are calculated from the eigenvalues of the Hessian matrix,
constructed from $(a,b,c)$. For cartesian coordinates, the eigenvalues are
equal to $\lambda_\pm = \frac{1}{2}(a+b \pm \sqrt{(a-b)^2+c^2})$. For polar
coordinates, $k_x$ and $k_y$ should be interpreted as $k_r=\abs{k}$ and $k_\phi$
respectively. In order to obtain band masses for cartesian coordinates, the
elements of the Hessian matrix must be rescaled as
\begin{equation}
 h_\mathrm{polar}=
 \begin{pmatrix}
   2a & c / k_r\\ c / k_r & 2b / k_r^2
 \end{pmatrix},
\end{equation}
where the factor $1/k_r$ comes from the Jacobian of the coordinate transformation
between polar and cartesian coordinates (assuming the angular coordinate is
in radians). The band masses are then obtained from the eigenvalues of
$h_\mathrm{polar}$.

\subsubsection{Nineteen-point extremum solver}

The function \verb+nineteen_point_extremum_solver()+ for three dimensions
effectively applies the method of the nine-point extremum solver in the $k_xk_y$,
$k_xk_z$, and $k_yk_z$ planes separately. The elements of the Hessian matrix $h$
[the $3\times 3$ matrix in Eq.~\eqref{eqn_nineteen_point_extremum_function}]
are calculated as
\begin{align}
  a&=\frac{e_{i+1,j,l}-2e_{i,j,l}+e_{i-1,j,l}}{2(\Delta x)^2}, &
  d&= \frac{e_{i+1,j+1,l} - e_{i-1,j+1,l} - e_{i+1,j-1,l} + e_{i-1,j-1,l}}{4\, \Delta x\,\Delta y},\nonumber\\
  b&=\frac{e_{i,j+1,l}-2e_{i,j,l}+e_{i,j-1,l}}{2(\Delta y)^2}, &
  e&= \frac{e_{i+1,j,l+1} - e_{i-1,j,l+1} - e_{i+1,j,l-1} + e_{i-1,j,l-1}}{4\, \Delta x\,\Delta z},\\
  c&=\frac{e_{i,j,l+1}-2e_{i,j,l}+e_{i,j,l-1}}{2(\Delta z)^2}, &
  f&= \frac{e_{i,j+1,l+1} - e_{i,j-1,l+1} - e_{i,j+1,l-1} + e_{i,j-1,l-1}}{4\, \Delta y\,\Delta z}.\nonumber
\end{align}
The coefficients $d$, $e$, and $f$ involve the twelve pink-coloured points in
Fig.~\ref{fig_extrema_solvers}(c), but for each group of four, only one linear
combination is considered. We note that the corner points $e_{i\pm1,j\pm1,l\pm1}$
[coloured black in Fig.~\ref{fig_extrema_solvers}(c)] are included in the input
arguments, but not considered.

If the Hessian matrix is non-singular (in \verb+nineteen_point_extremum_solver()+,
we use the condition $\abs{\det h}>10^{-6}$), we find the momentum location as
\begin{equation}
  \begin{pmatrix}
   k_{x,0} \\ k_{y,0} \\ k_{z,0}
  \end{pmatrix}
  =
  \begin{pmatrix}
   k_{x,i} \\ k_{y,j} \\ k_{z,l}
  \end{pmatrix}
  +
  h^{-1}
  \begin{pmatrix}
   -X \\ -Y \\ -Z
  \end{pmatrix}
\end{equation}
where $h^{-1}$ is the inverse of the Hessian matrix and
$X=(e_{i+1,j,l} - e_{i-1,j,l})/2\Delta x$, $Y=(e_{i,j+1,l} - e_{i,j-1,l})/2\Delta y$,
and $Z=(e_{i,j,l+1} - e_{i,j,l-1})/2\Delta z$ [analogous to
Eq.~\eqref{eqn_nine_point_extremum_aux}]. If $\abs{\det h}\leq10^{-6}$ or the
point lies outside of the box
$[k_{x,i-1},k_{x,i+1}]\times[k_{y,j-1},k_{y,j+1}]\times[k_{z,l-1},k_{z,l+1}]$,
the \emph{``Poorly defined extremum found''} warning is raised and the location
replaced by $(k_{x,i}-X/2a,k_{y,j}-Y/2b,k_{z,l}-Z/2c)$.
The energy value $f_0$ of the extremum is given by
\begin{equation}\label{eqn_nineteen_point_extremum_energy}
  f_0 = e_{i,j,l} -
  \begin{pmatrix}\tilde{k}_x & \tilde{k}_y & \tilde{k}_z\end{pmatrix}
  \begin{pmatrix}2a & d & e\\d & 2b & f\\e & f & 2c\end{pmatrix}
  \begin{pmatrix}\tilde{k}_x \\ \tilde{k}_y \\ \tilde{k}_z\end{pmatrix},
\end{equation}
with
$(\tilde{k}_x, \tilde{k}_y, \tilde{k}_z)
= (k_{x,i},k_{y,j},k_{z,l}) - (k_{x,0},k_{y,0},k_{z,0}) $.
The \verb+nineteen_point_extremum_solver()+ function returns the result as $f_0$,
$(k_{x,0},k_{y,0},k_{z,0})$, $(a,b,c,d,e,f)$. The band masses are calculated from
the eigenvalues of the Hessian matrix. If the coordinate system is cylindrical or
spherical, the appropriate coordinate transformation is applied prior to finding
the eigenvalues.

\clearpage
\section{Reference}

\subsection{Units and constants}
\label{app_units_reference}

\subsubsection{Units}
In \kdotpy, physical quantities are represented by numerical data types without
the explicit specification of units. For this reason, it is necessary to agree
on a system of units such that calculations that physical quantities can be
done intuitively, possibly without the need of conversion factors.
Whereas a system where units are treated explicitly would be more fail-safe, it
would create a lot of overhead which may reduce performance.

We choose a system of units such that physical quantities in context of
solid-state physics have reasonable values. The basic units are:
\begin{itemize}
\item Length: $\mathrm{nm}$
\item Time: $\mathrm{ns}$
\item Energy: $\mathrm{meV}$  (\emph{not} $\mathrm{eV}$)
\item Voltage: $\mathrm{mV}$   (\emph{not} $\mathrm{V}$)
\item Temperature: $\mathrm{K}$
\item Electric charge: $\mathrm{e}$
\item Magnetic field: $\mathrm{T}$
\end{itemize}

With \emph{magnetic field} we mean \emph{magnetic flux density}, more accurately speaking.
The unit for magnetic flux density in this unit system is defined as Tesla,
$\mathrm{T} = \mathrm{V}\,\mathrm{s} / \mathrm{m}^2$. This unit is `incompatible'
with the combination of units for voltage, length, and time, which would be
$\mathrm{mV}\,\mathrm{ns} / \mathrm{nm}^2$.
The conversion, given by $1\,\mathrm{T} = 10^{-6}\,\mathrm{mV}\,\mathrm{ns}/\mathrm{nm}^2$
is done internally by \kdotpy{} when it interprets magnetic field values.

From the basic units, we derive units for other physical quantities, for example,
\begin{itemize}
\item  Momentum (more appropriately wave vector) $k$: $\mathrm{nm}^{-1}$
\item  Density (such as particle/carrier density): $\mathrm{nm}^{-1}$, $\mathrm{nm}^{-2}$,
or $\mathrm{nm}^{-3}$ depending on dimensionality
\item  Density of states: $\mathrm{nm}^{-1}\,\mathrm{meV}^{-1}$, $\mathrm{nm}^{-2}\,\mathrm{meV}^{-1}$, or $\mathrm{nm}^{-3}\,\mathrm{meV}^{-1}$
\item  Charge density: $e\,\mathrm{nm}^{-1}$, $e\,\mathrm{nm}^{-2}$, or $e\,\mathrm{nm}^{-3}$ 
\item  Electric field: $\mathrm{mV} / \mathrm{nm}$
\item  Velocity: $\mathrm{nm} / \mathrm{ns} = \mathrm{m} / \mathrm{s}$
\end{itemize}
Inputs and outputs use the basic and derived units if not explicitly stated otherwise.

\subsubsection{Physical constants}

The following physical constants are defined in \texttt{physconst.py}.
\begin{itemize}
\item  \verb+m_e = 0.510998910e9+ in $\mathrm{meV}$:
   Electron mass in energy equivalents ($E = m_\mathrm{e} c^2$).
\item  \verb+e_el = 1.6021766208e-19+ in $\mathrm{C}$:
   Elementary charge $e$. We define this value to be positive.
\item  \verb+cLight = 299792458.+ in $\mathrm{nm} / \mathrm{ns} = \mathrm{m} / \mathrm{s}$:
   Speed of light $c$.
\item  \verb+hbar = 6.582119514e-4+ in $\mathrm{meV}\,\mathrm{ns}$:
   Reduced Planck constant $\hbar$.
\item  \verb+hbarm0 = hbar**2 * cLight**2 / m_e / 2+ in $\mathrm{meV}\,\mathrm{nm^2}$:\\
   $\hbar^2/2m_e = 38.09982350(23)\,\mathrm{meV}\,\mathrm{nm^2}$.
   The approximate value $38\,\mathrm{meV}\,\mathrm{nm}^2$ is useful for making
   estimates.
\item  \verb+eoverhbar = 1e-6 / hbar+ in $1 / (\mathrm{T}\,\mathrm{nm}^2)$:
   $e/\hbar$ defined such, that \texttt{eoverhbar} times magnetic field in T yields
   a density in $\mathrm{nm}^{-2}$. This takes into account the conversion factor
   $10^{-6}$ in $\mathrm{T} = 10^{-6}\,\mathrm{mV}\,\mathrm{ns}/\mathrm{nm}^2$.
   The product \texttt{eoverhbar * A}, with the vector potential $\vec{A}$ in
   $\mathrm{T}\, \mathrm{nm}$, yields a value in $\mathrm{nm}^{-1}$, as appropriate
   for a momentum quantity. This appears in the Peierls substitution, for example.
   The product \texttt{eoverhbar * B}, with the magnetic field $B$ in $\mathrm{T}$,
   yields a quantity in units of $\mathrm{nm}^{-2}$, cf. $\lB^2 = \hbar/eB$
\item  \verb+muB = 5.7883818012e-2+ in $\mathrm{meV} / \mathrm{T}$:
   Bohr magneton $\muB = e\hbar/2m_e$.
\item  \verb+kB = 8.6173303e-2+ in $\mathrm{meV} / \mathrm{K}$:
   Boltzmann constant $\kB$.
\item  \verb+eovereps0 = 1.80951280207e4+ in $\mathrm{mV}\,\mathrm{nm}$:
   $e/\varepsilon_0$, electron charge divided by permittivity constant (also called
   vacuum permittivity).
\item  \verb+gg = 2+ (dimensionless):
   Gyromagnetic ratio.
\item  \verb+r_vonklitzing = 25812.8074555+ in $\Omega$ (ohm):
   Von Klitzing constant $R_\mathrm{K}$, resistance value corresponding to one
   quantum of conductance.
\end{itemize}

These values have been extracted from the NIST Reference on Constants, Units,
and Uncertainty, revision 2014, see Ref.~\cite{CODATA2014} and online at
\url{https://physics.nist.gov/cuu/Constants/index.html}. The revised NIST values
from 2018 may deviate slightly (but negligibly) from the values listed here and
in \texttt{physconst.py}.

\subsection{Material parameters}
\label{app_matparam_reference}


\subsubsection{Chemistry}

\begin{itemize}
\item \texttt{compound}: The chemical formula of the compound, for example \texttt{HgTe}.

\item \texttt{elements}: Comma-separated list of the elements, for example \texttt{Hg, Te}.
                 If \texttt{elements} is not given, it is determined automatically
                 from \texttt{compound} (and vice versa).

\item \texttt{composition}: Numbers for each of the elements in the chemical formula, that
                 indicate their (stoichiometric) proportions in the compound. The
                 numbers must be separated by commas and may be a function of the
                 variables \texttt{x}, \texttt{y}, and/or \texttt{z}. For example,
                 for a crystal with molecular formula Hg$_{1-x}$Cd$_{x}$Te, use
                 \texttt{compound = HgCdTe} and \texttt{composition = 1-x, x, 1}.
\end{itemize}

\subsubsection{Special commands}

\begin{itemize}
\item \texttt{copy}:
  Copy all valid parameters of another material into the present one. The value
  is the id of the source material. For example
\begin{verbatim}
[mat2]
copy = mat1
\end{verbatim}
  copies all parameters from material \texttt{mat1} into material \texttt{mat2}. Each
  parameter may be subsequently overwritten manually.

\item \texttt{linearmix}:
  Define a material as a linear combination of two others, where the parameters
  are linearly interpolated. The value is a 3-tuple of the form \texttt{mat1, mat2, var},
  where \texttt{mat1} and \texttt{mat2} are the source materials and \texttt{var} is the interpolation
  variable (\texttt{x}, \texttt{y}, or \texttt{z}). For example
\begin{verbatim}
[HgCdTe]
linearmix = HgTe, CdTe, x
\end{verbatim}
  makes material \texttt{HgCdTe} a linear combination of \texttt{HgTe} and \texttt{CdTe}. Each
  parameter $p$ is interpolated as
  $p_\mathrm{HgCdTe} = (1-x)\,p_\mathrm{HgTe} + x\,p_\mathrm{CdTe}$.
  Each parameter may be subsequently overwritten manually.
\end{itemize}
  
\subsubsection{Band energies}

\begin{itemize}
\item \texttt{Ev}: `Valence' band energy in meV. This is the energy of the
                 $\Gamma_8$ orbitals at $k=0$ for the \emph{unstrained bulk}
                 material. For inverted materials, \texttt{Ev} indicates
                 the energy of the $\Gamma_8$
                 orbitals, not that of the actual valence band.

\item \texttt{Ec}: `Conduction' band energy in meV. This is the energy of the
                 $\Gamma_6$ orbitals at $k=0$ for the \emph{unstrained bulk}
                 material. For inverted materials, \texttt{Ec} indicates
                 the energy of the $\Gamma_6$ orbitals, not that of the actual
                 conductance band.

\item \verb+delta_so+:    `Split-off' energy difference $\Delta_\mathrm{SO}$
                 in meV between the $\Gamma_8$ and $\Gamma_7$ orbitals in the
                 unstrained bulk material. To put it more precisely, the energy
                 of the $\Gamma_8$ orbital at $k=0$ is\linebreak[4] \verb+Ev - delta_so+.
\end{itemize}

\subsubsection{Quadratic terms}

\begin{itemize}
\item \texttt{P}: Kane matrix element $P=(\hbar/m_0)\langle S | p_x | X \rangle$
                 in meV nm.\
                 \textsc{Note}: The definition of $P$ differs between references by
                 a sign and/or a factor of $\ii$. Here, we take a positive real
                 value for \texttt{P}.

\item \texttt{gamma1}:      Luttinger parameter $\gamma_1$, dimensionless value. This
                 Luttinger parameter describes the spherically isotropic component
                 of the (bare) band mass of the $\Gamma_8$ orbitals.

\item \texttt{gamma2}:      Luttinger parameter $\gamma_2$, dimensionless value. This
                 Luttinger parameter describes the component proportional to
                 $k_x^2 + k_y^2 - 2k_z^2$ for the $\Gamma_8$ orbitals. This
                 component is axially, but not spherically symmetric.

\item \texttt{gamma3}:      Luttinger parameter $\gamma_3$, dimensionless value. This
                 Luttinger parameter describes the component proportional to
                 $k_x^2 - k_y^2$ for the $\Gamma_8$ orbitals. This
                 component breaks axial symmetry.

\item \texttt{F}:           Band mass parameter for the $\Gamma_6$ orbitals. The bare
                 band mass term is $\frac{\hbar^2}{2m_0}(2F+1)(k_x^2 + k_y^2 + k_z^2)$.
                 In $2F+1$, the $1$ is the contribution of free electrons and $2F$
                 is the contribution from perturbative corrections from remote
                 bands in \kdotp{} theory.

\item \texttt{kappa}:       See Section \emph{Magnetic couplings} below.
\end{itemize}

\subsubsection{ Bulk-inversion asymmetry}

\begin{itemize}
\item \verb+bia_c+:   Linear bulk-inversion asymmetry coefficient $C$ in
                        meV nm$^{-1}$. 

\item \verb+bia_b8p+: Quadratic bulk-inversion asymmetry coefficient $B_{8v}^+$ in
                        meV nm$^{-2}$. 

\item \verb+bia_b8m+: Quadratic bulk-inversion asymmetry coefficient $B_{8v}^-$ in
                        meV nm$^{-2}$. 

\item \verb+bia_b7+: Quadratic bulk-inversion asymmetry coefficient $B_{7v}$ in
                        meV nm$^{-2}$. 
\end{itemize}

\subsubsection{Magnetic couplings}

\begin{itemize}
\item \texttt{ge}: Gyromagnetic factor $g_e$ for the $\Gamma_6$
                 orbitals. The magnetic coupling is also known as the Zeeman term
                 and is equal to $g_\mathrm{e} \mu_\mathrm{B} \vec{B} \cdot \vec{S}$.
                 The value of $g_e$ contains contributions from the free
                 electron (equal to $2$) and perturbative corrections from remote
                 bands.

\item \texttt{kappa}: Gyromagnetic factor $\kappa$ for the $\Gamma_8$ and $\Gamma_7$
                 orbitals. The magnetic coupling is of the form
                 $-2\mu_\mathrm{B}\kappa \vec{J}\cdot \vec{B}$ in the
                 ($\Gamma_8$, $\Gamma_8$) block of the Hamiltonian and
                 analogous couplings in the ($\Gamma_8$, $\Gamma_7$) and
                 ($\Gamma_7$, $\Gamma_7$) blocks. The coefficient \texttt{kappa}
                 also appears as a non-magnetic term containing a commutator
                 $[\kappa, k_z]$, which contributes only at material interfaces.

\item \texttt{q}: Coefficient $q$ of the $\Gamma_8$ magnetic coupling
                 $-2\mu_\mathrm{B} q \vec{\mathcal{J}}\cdot \vec{B}$, where
                 $\vec{\mathcal{J}}=(J_x^3,J_y^3,J_z^3)$.

\item \verb+exch_yNalpha+:
  Exchange energy parameter in meV for the $\Gamma_6$ orbitals. The
  paramagnetic exchange interaction is relevant for paramagnetic materials like
  (Hg,Mn)Te. The interaction strength typically depends linearly on the Mn
  concentration ($y$ in Ref.~\cite{NovikEA2005}, hence the \texttt{y} in the
  parameter name). In the materials file, this dependence must be given
  explicitly, e.g., by setting \verb+exch_yNalpha = y * 400+. The given dependence
  may be chosen non-linear if desired. See Ref.~\cite{NovikEA2005} for more
  details on the physics of this paramagnetic coupling in (Hg,Mn)Te.

\item \verb+exch_yNbeta+: Exchange energy parameter in meV for the $\Gamma_8$ and
                 $\Gamma_7$ orbitals.

\item \verb+exch_g+: The gyromagnetic factor $g_\mathrm{ex}$ in the argument of the
  Brillouin function. For (Hg,Mn)Te, this quantity is usually written as $g_\mathrm{Mn}$
  see Eq.~\eqref{eqn_magnetization_mn} and Ref.~\cite{NovikEA2005}. The default
  value is 2.

\item \verb+exch_TK0+: The offset temperature $T_{0}$ in K that appears in the
  denominator of the argument of the Brillouin function. The default value is $10^{-6}$.
  See Sec.~\ref{sec_exchange_mn}.

\end{itemize}

\subsubsection{Electrostatic properties}

\begin{itemize}
\item \verb+diel_epsilon+:
  Dielectric constant $\varepsilon_\mathrm{r}$, also known as relative permittivity. The
  value is dimensionless. The present version of kdotpy implements a constant
  value only, i.e., no frequency dependence is considered.

\item \verb+piezo_e14+:   Piezo-electric constant $e_{14}$ in $e$ nm$^{-2}$. Note
                 that this is the only independent component of the piezo-electric
                 tensor for the zincblende crystal structure. To convert from
                 units of C m$^{-2}$, multiply the value by $10^{18} e$,
                 for example \verb+piezo_e14 = 0.035 * 1e18 * e_el+.
\end{itemize}
                 
\subsubsection{Lattice constant and strain}

\begin{itemize}
\item \texttt{a}:           Lattice constant in nm of the unstrained material.

\item \verb+strain_C1+:   Strain parameter (deformation potential) $C_1$. The
                 corresponding strain term is $C_1 \mathop{\mathrm{tr}} \epsilon$
                 acting on the $\Gamma_6$ orbitals. In
                 Refs.~\cite{PfeufferJeschke2000_thesis, NovikEA2005}, the
                 notation $C$ is used.

\item \verb+strain_Dd+:   Strain parameter (deformation potential) $D_d$. The
                 corresponding strain term is $D_d \mathop{\mathrm{tr}} \epsilon$
                 acting on the $\Gamma_8$ and $\Gamma_7$ orbitals. In
                 Refs.~\cite{PfeufferJeschke2000_thesis, NovikEA2005},
                 the notation $a$ is used.

\item \verb+strain_Du+:   Strain parameter (deformation potential) $D_u$. The
                 corresponding strain term is a linear combination of the
                 diagonal entries $\epsilon_{ii}$ of the strain tensor, and
                 acts on the $\Gamma_8$ and $\Gamma_7$ orbitals. In
                 Refs.~\cite{PfeufferJeschke2000_thesis, NovikEA2005},
                 the notation $b=-\frac{2}{3}D_u$ is used.

\item \verb+strain_Duprime+:
                 Strain parameter (deformation potential) $D'_u$. The
                 corresponding strain term is a linear combination of the
                 off-diagonal entries $\epsilon_{ij}$ (shear components) of the
                 strain tensor, and acts on the $\Gamma_8$ and $\Gamma_7$
                 orbitals. In Refs.~\cite{PfeufferJeschke2000_thesis, NovikEA2005},
                 $d=-\frac{2}{\sqrt{3}}D'_u$ is used.
\end{itemize}

\subsection{Lattice orientation}
\label{app_orientation}


The $\mathrm{SO}(3)$ rotation matrix $R$ can be
parametrized in different ways, for example directly in terms of the device coordinate
axis in terms of the lattice direction, or in terms of Euler angles.
The command line option \texttt{orientation} is the most generic way to input a rotation.
By giving up to three angles or directions, the crystal lattice can be rotated to any
possible orientation with respect to the device coordinate system:

\begin{itemize}
\item \texttt{orientation \#ang}: Rotation around $z$, equivalent to \texttt{stripangle}.
\item \texttt{orientation \#ang \#ang}: Tilt the $z$ axis, then rotate around the $z$ axis.
\item \texttt{orientation \#ang \#ang \#ang}: Euler rotation $z,x,z$, i.e.,
    rotate around the $c$ crystal axis, tilt the $z$ axis, and rotate around the $z$ axis.
\item \texttt{orientation \#dir}: Longitudinal direction $x$, equivalent to \texttt{stripdir}.
\item \texttt{orientation - \#dir}: Growth direction $z$.
\item \texttt{orientation \#dir \#dir}: Longitudinal and growth direction $x$, $z$.
\item \texttt{orientation - \#dir \#dir}: Transversal and growth direction $y$, $z$.
\item \texttt{orientation \#dir \#dir \#dir}: Longitudinal, transversal, and growth direction $x$, $y$, $z$.
\item \texttt{orientation \#dir \#ang} or \texttt{orientation \#ang \#dir}: Growth direction $z$ and rotation around $z$.
\end{itemize}

Angles \texttt{\#ang} are entered an as explicit floating point number containing
a decimal sign \texttt{.} or with the degree symbol ($^\circ$ or \texttt{d}).
Directions  \texttt{\#dir} are triplets either of digits without separators and possibly
with minus signs (e.g., \texttt{100}, \texttt{111}, \texttt{11-2}, \texttt{-110})
or of numbers separated by commas without spaces
(e.g., \texttt{1,1,0} or \texttt{10,-10,3}). If multiple directions are given as
arguments, they must be pairwise orthogonal, which is tested by calculating the
inner products between them.

The command line arguments \texttt{stripangle} and \texttt{stripdir} are simplified
options for rotation of a strip geometry around the $z$ axis.

\subsection{Custom eigensolvers}
\label{app_eigensolvers}


As discussed in Sec.~\ref{sec_default_eigensolver}, the diagonalization
involves two components: The actual eigensolver and an efficient LU solver for the
shift-and-invert step. \kdotpy can be configured to use custom solvers provided
by external packages for both steps. In this section, we list them and discuss
their strengths and weaknesses.

\subsubsection{Eigensolvers}

\begin{itemize}
 \item \texttt{eigsh} from SciPy: The default implementation that ships with the
 SciPy module. It interfaces to the external ARPACK package \cite{ARPACK} that
 implements the Lanczos/Arnoldi algorithm.\\
 \textsc{Pro:} It is stable and there are no convergence problems. It comes with
 SciPy, so it does not require any steps to set up.\\
 \textsc{Contra:} A single calculation can only make use of one CPU core (multi-core
 usage possible by parallelizing over $k$ or $B$ points). The memory bandwidth may
 be a bottleneck for large problems, due to \texttt{gemm} operations (matrix-matrix
 products with general matrices in BLAS \cite{DongarraEA1990BLAS}) during eigenvector
 orthonormalization, which is the slowest operation per Lanzcos/Arnoldi iteration.
 \item \texttt{eigsh} from CuPy \cite{CUPY2017}: Also a Lanzcos/Arnoldi algorithm,
 written in pure Python. The CuPy library mimics, extends and replaces the
 NumPy/SciPy packages. Where possible and
 efficient, the computational workload is done on a CUDA capable GPU. Performance
 is best on larger problem sizes. Depending on the availability of single and 
 double precision GPU cores, using single precision floating-point arithmetic may
 be faster, at the expense of lower precision. Using CuPy
 requires a CUDA capable GPU and a compatible version of the CUDA library
 \footnote{The CUDA API, provided by Nvidia, is an extension of C/C++ that
 facilitates computation on GPUs that support it.}. \\
 \textsc{Pro:} It is much faster for large problems compared to the CPU \texttt{eigsh}
 solver, for example about 6 times in non-axial LL mode, depending on setup and
 problem parameters. The GPUs typically have higher memory bandwidth.
 TensorCores can drastically speed up the \texttt{gemm} operation.\\
 \textsc{Contra:} The RAM limits on GPU are generally tighter than those on CPU.
 The solver can fail to converge, especially if single precision floating-point
 arithmetic is used. This can be mitigated by switching to higher precision or
 falling back to the CPU solver (done automatically by default), but this leads
 to a reduction in speed. Setting up and configuring this solver requires more
 effort.
 \item FEAST solver: Either provided via Intel MKL (version 2.1) or by manual
 compilation of a shared library (version 4.0).\\
 \textsc{Pro:} Can be more efficient than default \texttt{eigsh};
 no additional LU solver is required.\\
 \textsc{Contra:} This solver can be unreliable with densely clustered eigenstates,
 which is frequently the case for \kdotpy.
 \item \texttt{eigh} from jax:
 Converts the sparse matrix to a dense matrix, then uses \texttt{eigh} from the
 Google JAX library to solve
 on accelerators like GPUs using highly optimized instructions. Scales across
 multiple accelerators. Currently implemented as fixed double precision.\\
 \textsc{Pro:} Extremely fast when a large fraction of eigenvalues are requested
 (in fact, the algorithm does a full diagonalization internally).\\
 \textsc{Contra:} Very memory intensive on GPUs. The entire Hamiltonian (including
 solutions) needs to fit into GPU memory.
\end{itemize}

\subsubsection{LU solvers}

\begin{itemize}
 \item SuperLU: The default implementation shipped with current SciPy module.\\
 \textsc{Pro:} It performs well and does not require additional effort to set up.\\
 \textsc{Contra:} Can fail with a memory allocation error for very large matrix sizes.
 \item UMFPACK: Uses external library shipped with scikit/UMFPACK. Previously,
 this was used as default by SciPy, but this was changed due to license issues.
 If UMFPACK is installed (scikit libraries and Python packages), SciPy chooses it
 automatically for use with \texttt{eigsh}.\\
 \textsc{Pro:} It is stable and efficient, and usually fast.\\
 \textsc{Contra:} In some cases, it may rather be slower than SuperLU.
 \item PARDISO: Provided by the Intel MKL library. It requires the Intel MKL libraries
 to be installed along with a suitable Python package that links to it.\\
 \textsc{Pro:} It is fast, for example about 3 to 4 times with \texttt{kdotpy 1d},
 efficient and stable.\\
 \textsc{Contra:} The official Python package is no longer maintained. It may require more
 work to set up.
\end{itemize}

\subsubsection{Performance considerations}

Optimizing solver speed is connected to the problem to be solved, as well as to
the computation environment used for the calculation. For the very different
workloads created by various \kdotpy{} problem sets and the large range of
hardware that \kdotpy{} could be run on (from small dual core, low RAM PCs over
powerful GPU accelerated workstations up to multiple HPC cluster nodes), it is
not possible to identify a single eigensolver and a set of optimal parameters
that satisfies every scenario.

Automatic settings in \kdotpy{} always try to use settings for highest stability,
maximum speed and minimal RAM requirements as a rough guideline (in order of priority).
However in most scenarios, performance data is not available and has to be found
by user experience. The following hardware parameters are important aspects to
that affect eigensolver performance:

\begin{itemize}
 \item CPU/GPU core clock speed: This parameter is mostly fixed. For some systems
 one might be able to increase the speed by overclocking without affecting other
 limiting factors, but this typically gives minor speed boosts on the order of
 few percent.

 \item CPU instruction sets: Vector and matrix BLAS operations profit much from 
 modern CPU instructions such as AVX. Intel MKL downthrottles AMD CPUs by not 
 making use of some of those features, but this behaviour can be 
 disabled. An older version of MKL might perform best in such cases, but should 
 be benchmarked against current versions.
 
 On CPU cores: While Python itself has 
 many pitfalls in terms of efficient threading and multi core usage, prominently 
 due to its global interpreter lock (GIL), we can still make good use of
 additional CPU cores by solving multiple matrices in parallel
 processes by parallelization over $k$ or $B$ points. As long as we do not run
 into other limits listed below, this yields 
 almost linear performance improvements. Some of the external libraries (all ones
 mentioned above, except ARPACK) can also use more cores through 
 multithreading, however this higher core usage is not always rewarded with speed 
 gains. A high degree of parallelization can be energy-inefficient due to
 the extra overhead. On hardware with inadequate cooling, pushing the CPU to its
 limits for an extensive time may also lead to performance degradation from
 thermal throttling.

 \item RAM size: The maximum amount of RAM in the system sets a hard limit on 
 the number of matrices that can be solved in parallel. Each independent problem 
 requires a similar amount of RAM space during calculation, that can not be 
 shared. Once the threshold is exceeded, the whole process is likely to crash.

 \item CPU cache size and RAM bandwidth: This is the amount of data the CPU can 
 request from RAM (per second). For some calculations, it is possible, that data 
 for CPU operations can not be stored efficiently enough in the CPU cache (cache 
 miss) and a lot of data has to fetched from RAM. This is a limitation for large 
 matrix-matrix-multiplications (BLAS function \texttt{gemm}), as the operation's data 
 cannot fit in cache. In non-axial LL mode, this can be observed 
 when requesting many eigenvalues (large $n_\mathrm{eig}$ for thick layers,
 as the orthogonalization of  eigenvectors (as part of the Lanczos algorithm) is
 basically done using two \texttt{gemm}
 calls. Running more simultaneous processes does not increase total solution 
 speed, as it slows down each process due to shared RAM access.

\end{itemize}

\subsection{Observables}
\label{app_observables_reference}


\subsubsection{Spatial observables}

The following observables are defined as integrals over spatial coordinates. In
the discrete coordinates used in \kdotpy, they are implemented as summations, e.g.,
$\sum_i \psi^*_i z_i \psi_i$ for the expectation value $\langle z \rangle$.

\begin{itemize}
 \item \texttt{y}: Expectation value $\langle y \rangle$, where $y$ is the coordinate
       transverse to the strip.

\item \texttt{y2}: Expectation value $\langle y^2 \rangle$.

\item \texttt{sigmay}: Standard deviation of y, $\sigma_y=\sqrt{\langle y^2 \rangle - \langle y \rangle^2}$.

\item \texttt{z}: Expectation value $\langle z \rangle$, where $z$ is the coordinate
       parallel to the growth direction.

\item \texttt{z2}: Expectation value $\langle z^2 \rangle$.

\item \texttt{sigmaz}: Standard deviation of $z$, $\sigma_y=\sqrt{\langle z^2 \rangle - \langle z \rangle^2}$.

\item \texttt{zif}: Expectation value $\langle z_\mathrm{if} \rangle$, where $z_\mathrm{if}$
         is the (signed) distance to the nearest interface.

\item \texttt{zif2}: Expectation value $\langle z_\mathrm{if}^2 \rangle$.

\item \texttt{sigmazif}: Standard deviation of $z_\mathrm{if}$, $\sigma_{z_\mathrm{if}}=\sqrt{\langle z_\mathrm{if}^2 \rangle - \langle z_\mathrm{if} \rangle^2}$.

\item \texttt{well}: Probability density inside the well.

\item \texttt{wellext}: Probability density inside the well and the adjacent 2 nm of barriers.

\item \texttt{interface}: Probability density less than 1 nm from each interface. This quantity
  cannot be used to compare samples of different size.

\item \texttt{interface10nm}: Probability density less than 10 nm from each interface. This quantity
  cannot be used to compare samples of different size.

\item \texttt{custominterface[]}: Probability density up to a length in nm, set
  by the command-line argument \texttt{custominterfacelength}.

\item \texttt{interfacechar}: `Interface character'. Probability density less than 1 nm
  from each interface divided by what this quantity would be for a uniform
  probability density, i.e., for a normalized wave function with constant
  magnitude given by $|\psi|^2 \equiv 1/V$, where $V$ is the volume (size) of the complete sample. If $\Omega$ denotes the domain near the interfaces (e.g., less than 1 nm away) and
  $V_\mathrm{if}=\mathrm{vol}(\Omega)$ its size, then the interface character is
  $(V/V_\mathrm{if})\int_\Omega |\psi(r)|^2 dr$. Regardless of sample size,
  values larger than 1 indicate strong interface character.

\item \texttt{interfacechar10nm}: Interface character, like \texttt{interfacechar},
  but with an interface region of 10 nm instead of 1 nm.

\item \texttt{custominterfacechar[]}: Interface character, like \texttt{interfacechar},
  with the size of the interface region set by the command-line argument \texttt{custominterfacelength}.
\end{itemize}

\subsubsection{Inverse participation ratio}

Generically, an inverse participation ratio (IPR) for a probability distribution
is defined in terms of the fourth moment (kurtosis) scaled by the square of the
second moment (variance). For a wave function $\psi(r)$ in terms of spatial
coordinate(s) $r$, we define
\begin{equation}
  \mathrm{IPR}
  = \frac{\left(\int |\psi(r)|^2 dr \right)^2}{\int |\psi(r)|^4 dr}
  = \left(\int |\psi(r)|^4 dr\right)^{-1},
\end{equation}
where the second equality assumes normalization of the wave function. The resulting
IPR has dimensions of length (for a one-dimensional spatial coordinate). \kdotpy{}
provides IPR observables for three choices of coordinates:

\begin{itemize}

\item \texttt{ipry} : IPR over the coordinate $y$. If necessary, we first integrate over 
  the coordinate z in order to obtain $|\psi(y)|^2$. The dimensionful variety
  has units of length. For the dimensionless variety, divide by the width $w$ in
  $y$ direction.

\item \texttt{iprz} : IPR over the coordinate $z$. If necessary, we first integrate over 
  the coordinate y in order to obtain $|\psi(z)|^2$. The dimensionful variety
  has units of length. For the dimensionless variety, divide by the thickness $d$
  in $z$ direction.

\item \texttt{ipryz} : IPR over the coordinates $(y, z)$. The dimensionful variety
  has units of area (length squared). For the dimensionless variety, divide by $w d$.

\end{itemize}

\subsubsection{Internal degrees of freedom}

The following observables are defined in terms of internal degrees of freedom,
i.e., (combinations of) orbitals. This includes spin operators.

\begin{itemize}
\item \texttt{sx}, \texttt{sy}, \texttt{sz}: Spin expectation value $\langle S_i \rangle$
  ($i = x, y, z$).
  This is proper spin in units of $\hbar$, so that the range of possible
  values is $[-\frac{1}{2}, \frac{1}{2}]$.

\item \texttt{jx}, \texttt{jy}, \texttt{jz}: Expectation values $\langle J_i \rangle$
  ($i = x, y, z$) of the total angular momentum. The range of possible values is
  $[-\frac{3}{2}, \frac{3}{2}]$.

\item \texttt{yjz}: Expectation values $\langle y J_z \rangle$, i.e., the product of the
  coordinate $y$ and the angular momentum $J_z$. This is roughly chirality for edge
  states.

\item \texttt{gamma6}, \texttt{gamma7}, \texttt{gamma8}. Expectation values $\langle P_{\Gamma_i} \rangle$ 
  ($i = 6, 7, 8$) of the projection to the $\Gamma_i$ orbitals. This is the
  probability density in these orbitals.

\item \texttt{gamma8h}: Expectation values $\langle P_{\Gamma_{8H}} \rangle$ of the
  projection to the orbital $|\Gamma_8,j_z=\pm 3/2\rangle$. `H' stands for heavy
  hole.

\item \texttt{gamma8l}: Expectation values $\langle P_{\Gamma_{8L}} \rangle$ of the
  projection to the orbital $|\Gamma_8,j_z=\pm 1/2\rangle$. `L' stands for light
  hole.

\item \texttt{orbital}: Difference of \texttt{gamma6} and \texttt{gamma8}. The value ranges between $-1$ for purely
  $\Gamma_8$ states and $1$ for purely $\Gamma_6$ states.

\item \texttt{jz6}, \texttt{jz7}, \texttt{jz8}. Expectation values $\langle J_z P_{\Gamma_i} \rangle$ 
  ($i = 6, 7, 8$) of the angular momentum in $z$ direction in the $\Gamma_i$ orbitals.

\item \texttt{split}: Expectation value of the `artificial split' Hamiltonian, which can be
  encoded as $\mathrm{sgn}(J_z)$.

\item \texttt{orbital[j]}: The squared overlaps of the eigenstates
  within orbital number $j$, where $j$ runs from 1 to $n_\mathrm{orb}$ (the number of
  orbitals, i.e., 6 or 8). Available for \texttt{kdotpy 2d} only, if the
  command line option \texttt{orbitalobs} is given.

\end{itemize}

\subsubsection{Plain parity and `isoparity' operators}

The following observables implement several plain parity operators (a purely
spatial operation) as well as `isoparity' operators, which combine a spatial
and a spin component \cite{Beugeling2021PRB}. The latter are given as actions of
the appropriate representations of the relevant point group, namely the
double group of $T_d$ or one of its subgroups.

\begin{itemize}
 \item \texttt{pz}: Expectation value of the parity operator in the $z$ coordinate, i.e.,
  $\langle \mathcal{P}_z \rangle = \int \psi^*(z) \psi(-z) dz$.

\item \texttt{px}, \texttt{py}: Expectation value of the parity operator in
  the coordinate $x$ or $y$,
  i.e., $\langle \mathcal{P}_x \rangle$, $\langle \mathcal{P}_y \rangle$

\item \texttt{pzy}: Expectation value of the combined parity operator
  $\langle \mathcal{P}_z\mathcal{P}_y \rangle$

\item \texttt{isopz}: Expectation value of isoparity $\langle \tilde{\mathcal{P}}_z \rangle$:
  this is the parity operator
  combined with the diagonal matrix $Q$ in orbital space, given by
  $Q = \mathrm{diag} (1, -1, 1, -1, 1, -1, -1, 1)$
  in the basis given by Eq.~\ref{eqn_orbital_basis}. This is a conserved quantum
  number under a few assumptions \cite{Beugeling2021PRB}.

\item \texttt{isopx}, \texttt{isopy}: Expectation value of in-plane isoparity
  $\langle \tilde{\mathcal{P}}_x \rangle$, $\langle \tilde{\mathcal{P}}_y \rangle$.
  This is the parity operator in x or y, combined with the appropriate matrix
  $Q_x$, $Q_y$ in orbital space, which can be found from the appropriate
  representations of the point group \cite{Beugeling2021PRB}.

\item \texttt{isopzy}: Expectation value of the combined isoparity operator
  $\langle \tilde{\mathcal{P}}_z\tilde{\mathcal{P}}_y \rangle$

\item \texttt{isopzw}: Expectation value of a modified isoparity operator like \texttt{isopz},
  that acts only in the quantum well layer. Formally speaking, if the well
  ranges from $z=z_\mathrm{min}$ to $z_\mathrm{max}$, then it maps $z' = z - c$
  ($c = (z_\mathrm{min} + z_\mathrm{max}) / 2$ being the center of the well)
  to $-z'$. While \texttt{isopz} is not a conserved quantity for asymmetric geometries
  (e.g., a well layer and two barriers with unequal thickness), \texttt{isopzw} can
  remain almost conserved in that case. Due to incomplete confinement in the
  well region, the eigenvalues may deviate significantly from $\pm 1$. 

\item \texttt{isopzs}: Expectation value of a modified isoparity operator like \texttt{isopz},
  that acts only in a symmetric region around the centre of the quantum well
  layer. It is like \texttt{isopzw}, but $z_\mathrm{min}$ and $z_\mathrm{max}$ are chosen
  such that the range fits inside the stack, while keeping
  $c = (z_\mathrm{min} + z_\mathrm{max}) / 2$ at the centre of the well. It
  tends to have eigenvalues closer to $\pm1$, because it generally covers the
  probability density in (a part of) the barrier layers as well.
\end{itemize}

\subsubsection{Hamiltonian terms}

The following observables are expectation values of the individual terms in the
Hamiltonian.

\begin{itemize}
 
\item \texttt{hex}: Exchange energy (expectation value of the exchange Hamiltonian).

\item \texttt{hex1t}: Exchange energy at $1\,\mathrm{T}$. Regardless of the
  actual value of the $B$ field, give the expectation value of the exchange Hamiltonian evaluated at $B_z = 1\,\mathrm{T}$.

\item \texttt{hexinf}: Exchange energy in the large-field limit.
  Regardless of the actual value of the $B$ field, give the expectation value of the
  exchange Hamiltonian evaluated for $B_z\to\infty$. In this limit, the
  Brillouin function is saturated at its maximum absolute value.

\item \texttt{hz}: Zeeman energy (expectation value of the exchange Hamiltonian).

\item \texttt{hz1t}: Zeeman energy at $1\,\mathrm{T}$. Regardless of the actual
  value of the $B$ field, give the expectation value of the Zeeman Hamiltonian
  evaluated at $B_z = 1\,\mathrm{T}$.

\item \texttt{hstrain}: Expectation value of the strain Hamiltonian.

\end{itemize}

\subsubsection{Landau levels}

The following are observables based on the Landau level index $n$.

\begin{itemize}
\item \texttt{llindex}: In the LL mode \texttt{sym} or \texttt{legacy}, this is
  the (conserved) LL index $n$.
  The minimal value is $-2$. The lowest indices $n = -2, -1, 0$ are incomplete, i.e.,
  for these indices, not all orbitals contribute.

\item \texttt{llavg}: In the LL mode \texttt{full}, the Hamiltonian is not
  diagonal in the basis of LL indexed by $n$. This observable returns the expectation value
  $\langle n \rangle$ in this basis.

\item \texttt{llbymax}: In the LL mode \texttt{full}, write the wave function in
  the basis of LL indexed by $n$, and provide the value $n$ with the highest
  probability density.

\item \texttt{llmod4}: In the LL mode \texttt{full}, analogous to \texttt{llavg},
  returns the expectation value
  $\langle n \mod 4\rangle$. While \texttt{llavg} is not a conserved quantity if
  axial symmetry is broken, this observable is usually conserved if bulk
  inversion symmetry is not broken.

\item \texttt{llmod2}: In the LL mode \texttt{full}, analogous to \texttt{llavg},
  returns the expectation value
  $\langle n \mod 2\rangle$. While \texttt{llavg} is not a conserved quantity if
  axial symmetry is broken, this observable is usually conserved even if bulk
  inversion symmetry is broken.

\item \texttt{ll[j]}: The squared overlaps of the eigenstates within
  Landau level $j$, where $j$ runs from $-2$ to $n_\mathrm{max}$, the largest LL index.
  This observable is available for \texttt{kdotpy ll} in full LL mode only, if the
  command line option \texttt{llobs} is given.
\end{itemize}

\subsection{Broadening}
\label{app_broadening_reference}

\begin{figure}
  \includegraphics{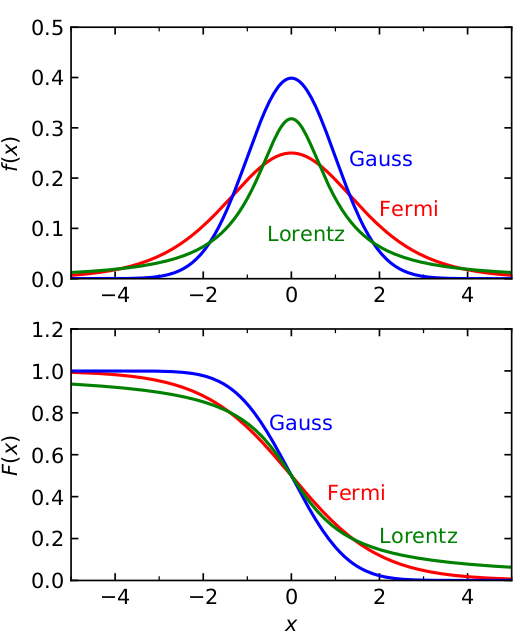}
  \caption{Broadening function kernels $f(x)$ and occupation functions $F(x)$
  for the Fermi, Gaussian, and Lorentzian shapes. The argument $x$ is scaled by
  the width parameter $\kB T$, $\sigma$, or $\gamma$, respectively.}
  \label{fig_broadening_functions}
\end{figure}

In \kdotpy, the following types of broadening are implemented. The application of
broadening is done by means of the broadening kernel $f(E)$. These are scaled
versions of the probability density functions $\mathop{\mathrm{PDF}}(x)$ of the 
corresponding distributions.
The occupation functions $F(E)$ are also implemented in the code. These are
defined as the complementary cumulative density functions
$1-\mathop{\mathrm{CDF}}(x)$ of the probability density functions
$\mathop{\mathrm{PDF}}(x)$ above. Since the PDFs are symmetric, the CDFs satisfy
$1-\mathop{\mathrm{CDF}}(x) = \mathop{\mathrm{CDF}}(-x)$. The kernels $f(x)$ and
occupation functions $F(x)$ are plotted in Fig.~\ref{fig_broadening_functions}.

\begin{itemize}
\item \emph{Thermal (Fermi function)}:\\
  The thermal broadening function models the broadening of the occupation (function) at finite temperature. The shape of the broadening is given by the Fermi function,
  \begin{equation}
  f(E) = \frac{1}{\kB T}\frac{\ee^{-x}}{(1 + \ee^{-x})^2}
  = \frac{1}{\kB T}\frac{1}{4 \cosh^2(x/2)},
  \end{equation}
  where we define $x = E / \kB T$. The characteristic width is $\kB T$, where $T$ is the temperature. The input argument is \texttt{broadening \emph{T} thermal} with the temperature value $T$. The occupation function is 
  \begin{equation}
  F(x) = \tfrac{1}{2}(1 + \tanh(-x/2)) = \frac{\ee^{-x}}{1+\ee^{-x}}.
  \end{equation}
  The corresponding distribution is also known as the \emph{logistic distribution}.
  
\item \emph{Disorder (Gaussian)}:\\
  Random disorder is modelled by smearing the density of states with a Gaussian shape,
  \begin{equation}
  f(E) = \frac{1}{\sqrt{2 \pi \sigma^2}}\ee^{-x^2/2\sigma^2},
  \end{equation}
  where $x = E / \sigma$. The broadening parameter $\sigma$ is the standard deviation of the Gaussian distribution. Note that subtly different definitions may be found in literature, for example \cite{NovikEA2005} uses the parameter $\Gamma = \sqrt{2}\sigma$.
  The input argument is \texttt{broadening $\sigma$ gauss} with the standard deviation $\sigma$.
  The occupation function is
  \begin{equation}
  F(x) = \tfrac{1}{2} \mathop{\mathrm{erfc}}(x / \sqrt{2}),
  \end{equation}
  where $\mathop{\mathrm{erfc}}(z) = 1 - \mathop{\mathrm{erf}}(z)$ is the
  complementary error function. In statistics, the distribution is usually called
  the \emph{normal distribution}.

\item \emph{Lorentzian}:\\
  Alternatively, we can use a Lorentzian broadening function,
  \begin{equation}
  f(E) = \frac{1}{\pi\gamma(1 + x^2)},
  \end{equation}
  where $x = E / \gamma$. The width parameter is $\gamma$. Lorentzian line shapes can be used to account for life-time limited broadening in optical transitions experiments. In that context, $\gamma$ is proportional to the inverse of the lifetime of excited states.
  The input argument is \texttt{broadening $\gamma$ lorentz} with the width $\gamma$.
  The occupation function is
  \begin{equation}
  F(x) = \tfrac{1}{2} + \tfrac{1}{\pi} \arctan(-x).
  \end{equation}
  The Lorentzian line shape is equivalent to the \emph{Cauchy distribution}.

\item \emph{Delta function}:\\
  In the limit of the three functions above, the occupation functions converge to
  a step function when the broadening parameter approaches $0$,
  \begin{equation}
  F(x) = \tfrac{1}{2}(1-\mathop{\mathrm{sgn}}(x))
  = \left\{\begin{array}{ll}
  1 & \text{for } x < 0\\
  0 & \text{for } x > 0
  \end{array}\right.
  \end{equation}
  where $x = E$. This function is implemented separately because the other
  functions are ill-defined in this limit. The corresponding broadening kernel
  is $f(x)=\delta(x)$, the Dirac delta function (or \emph{delta distribution}).
  
\end{itemize}

\subsection{Configuration Options}
\label{app_config}
\subsubsection{Solvers and tasks}

\begin{itemize}
\item
  \verb+diag_solver+: The implementation of the function that does
  the matrix diagonalization; see Appendix~\ref{app_eigensolvers} for
  detailed information.
  Possible values:

  \begin{itemize}
  \item
    \texttt{feast}: Use FEAST algorithm (Intel MKL). If this package is
    not available, fall back to eigsh. Can be tried as an alternative if
    eigsh fails, e.g.~for very large matrices (dim $>$ 5e6).
  \item
    \texttt{eigsh}: Use eigsh from SciPy sparse matrix library.
  \item
    \verb+superlu_eigsh+: Same as \texttt{eigsh}, but SuperLU is
    requested explicitly. Enables detailed timing statistics, as with
    other custom eigsh solvers.
  \item
    \verb+umfpack_eigsh+: Like \texttt{eigsh}, but uses umfpack
    instead of SuperLU for matrix inversion. Recommended for large
    matrices. Falls back to SuperLU if Scikit UMFPACK is not available.
    \textbf{REQUIRES} available scikit-umfpack and a suitable scipy
    version.
  \item
    \verb+pardiso_eigsh+: Like \verb+umfpack_eigsh+, but uses
    Intel MKL PARDISO instead. \textbf{REQUIRES} pyMKL package.
  \item
    \verb+cupy_eigsh+: Alternative implementation of the eigsh solver
    in python. Uses CUDA libraries for Lanczos iteration (on GPU) and
    PARDISO to SuperLU for matrix inversion, depending on availability.
    \textbf{REQUIRES} CUDA libraries and the CuPy package.
  \item
    \verb+jax_eigh+: Uses the JAX \texttt{eigh} solver. First converts sparse
    matrices to dense. Extremely memory inefficient! Fails if not enough
    VRAM can be allocated. Use the \texttt{gpus} option to reduce the
    number of workers on the GPU if this happens. This solver is best
    suited for a large number of \texttt{neig}. \textbf{REQUIRES} jax
    package.
  \item
    \texttt{auto}: Decision based on subprogram. Uses
    \verb+pardiso_eigsh+ for \texttt{kdotpy\ 1d} if available,
    otherwise uses \texttt{eigsh} for all scripts. Suggests alternative
    solvers if they could be faster. (\emph{default}; \emph{alias}:
    \texttt{automatic})
  \end{itemize}
\item
  \verb+diag_solver_worker_type+: Sets the parallelization strategy
  for solver workers. Options:

  \begin{itemize}
  \item
    \texttt{process}: Use a process pool for solve workers. Recommended
    strategy for most solvers.
  \item
    \texttt{thread}: Use a thread pool in the main process for solve
    workers. Recommended for CUDA based solver \verb+cupy_eigsh+ for
    optimal GPU workload.
  \item
    \texttt{none}: No parallel execution of the solve step. Every solve
    task is executed serially in the main thread. Recommended for
    debugging.
  \item
    \texttt{auto}: Decision based on \verb+diag_solver+.
    (\emph{default}; \emph{alias}: \texttt{automatic})
  \end{itemize}
\item
  \verb+diag_solver_cupy_dtype+: Sets the data type for the CuPy
  solver. Options:

  \begin{itemize}
  \item
    \texttt{single}: Uses complex numbers with single float precision.
    This leads to a large speed boost on GPUs with TensorCores.
    Precision of eigenvalues is worse (on the order of 10 \textmu{}eV).
  \item
    \texttt{double}: Uses complex numbers with double float precision.
    Solution speed and precision of eigenvalues comparable to other
    solvers. Medium speed boost expected on GPUs with modern
    FP64-TensorCores (e.g.~Nvidia A100). (\emph{default})
  \end{itemize}
\item
  \verb+diag_solver_cupy_iterations+: Maximum number of Lanczos
  iteration steps for both precision options. If number of iterations is
  exceeded, fall back to better precision first or CPU based solver
  next. (\emph{default}: 5)
\item
  \verb+diag_solver_cupy_gemm_dim_thr+: Maximum dimension for
  matrix matrix multiplication in single precision mode. If problem size
  exceeds this value, the solution is split into multiple smaller
  problem sets. Smaller values can lead to worse solution speeds, larger
  values can lead to more numerical problems and fallback to slower
  double precision solver. (\emph{default}: 4e-6)
\item
  \verb+task_retries+: Number of times a task is restarted after any
  exception was raised. (\emph{default}: 2)
\item
  \verb+tasks_grouped+: If set to true, all steps for a single
  \texttt{DiagDataPoint} are executed within the same worker/thread with
  the settings for the \verb+solve_ham+ step. Compared to the default
  mode, this involves less inter-worker data transfers (via pickling),
  which can give rise to issues with very large eigenvectors. As such,
  the worker communication behaves similar to kdotpy versions
  \textless{} v0.72. (\emph{default}: false)
\end{itemize}

\subsubsection{Band Alignment and Character}

\begin{itemize}
\item
  \verb+band_align_exp+: Value of the exponent in the minimization
  function of the `error' in the band
  alignment algorithm. A numerical value equal to e means that
  $\sum(|\Delta E|^e)$ is minimized. Alternatively, if the special
  value \texttt{max} is used, then the minimization function is
  $\max(|\Delta E|)$. (\emph{default}: 4)
\item
  \verb+band_align_ndelta_weight+: Coefficient of the penalty for
  reduction of the number of bands in the
  band alignment algorithm. The higher
  this value, the more the algorithm `resists' changes in the number of
  bands. The value may not be negative (however, 0 is allowed), and too
  high values should be avoided. It is recommended to use the default
  value unless the band alignment algorithm does not proceed correctly.
  (\emph{default}: 20.0)
\item
  \verb+band_char_node_threshold+: In the band character algorithm,
  this value times the resolution (\texttt{zres}) is the minimum value
  the wave function should reach such that a node (zero) is counted.
  (\emph{default}: 1e-6)
\item
  \verb+band_char_orbital_threshold+: In the band character
  algorithm, the maximum value for the probability density
  ($|\psi|^2$) in an orbital for the probability density to be
  considered zero. In that case, the orbital content of that orbital is
  ignored. (\emph{default}: 5e-3)
\item
  \verb+band_char_use_minmax+: In the band character algorithm,
  whether to use the `new' node counting method, that counts flips
  between local extrema. If set to false, use the legacy method.
  (boolean value; \emph{default}: true)
\item
  \verb+band_char_make_real+: In the band character algorithm,
  whether to divide the orbital component by the complex phase factor at
  its maximum, so that the function becomes effectively real, prior to
  counting the nodes. If set to false, consider both real and imaginary
  part as is. (boolean value; \emph{default}: false)
\item
  \verb+bandindices_adiabatic_debug+: Whether to write the
  intermediate result for adiabatic band index initialization to a csv
  file. This is useful for debugging this algorithm, for example if the
  charge neutrality point ends up at an incorrect position. (boolean
  value; \emph{default}: false)
\end{itemize}

\subsubsection{kdotpy batch}

\begin{itemize}
\item
  \verb+batch_float_format+: Format string for representation of
  float values being replaced in the command string. This is a standard
  \texttt{\%}-style conversion, with the following addition: If a
  \texttt{.} (period) is added to the end, for example \texttt{\%f.},
  apply the smart decimal option, i.e., strip superfluous zeros at the
  end, but keep the decimal point if the value is integer. Useful
  examples are, among others: \texttt{\%s}, \texttt{\%f}, \texttt{\%f.},
  \texttt{\%.3f}, \texttt{\%g}. (\emph{default}: \texttt{\%s})
\item
  \verb+batch_stderr_extension+: Extension for the file, that
  \texttt{kdotpy\ batch} writes stderr to (\emph{default}: txt)
\item
  \verb+batch_stdout_extension+: Extension for the file, that
  \texttt{kdotpy\ batch} writes stdout to (\emph{default}: txt)
\end{itemize}

\subsubsection{BHZ Calculation}

\begin{itemize}
\item
  \verb+bhz_allow_intermediate_bands+: Whether to allow a
  non-contiguous set of A bands. By default (\texttt{false}), do not
  allow B bands in between the A bands. If set to true, relax this
  restriction. This only takes effect if the input are band labels,
  e.g., \texttt{bhz\ E1\ H1\ L1}. It does not apply to numeric input
  (e.g., \texttt{bhz\ 2\ 2}), which is a contiguous set by definition.
  \textsc{Note}: Setting \texttt{true} is experimental, it may cause
  unexpected errors. (boolean value; \emph{default}: false)
\item
  \verb+bhz_points+: Number of horizontal data points for the
  BHZ dispersion plot. (\emph{default}: 200)
\item
  \verb+bhz_gfactor+: Whether to output dimensionless g factors in
  the BHZ output (tex file). If set to false (default), output
  dimensionful quantities `G' in meV / T. (boolean value;
  \emph{default}: false)
\item
  \verb+bhz_abcdm+: Whether to output (tex file) a four-band BHZ
  model in `standard form', using coefficients A, B, C, D, M. If this
  cannot be done, use the generic form instead. (boolean value;
  \emph{default}: false)
\item
  \verb+bhz_ktilde+: If BHZ is done at a nonzero momentum value
  $k_0$, whether to express the Hamiltonian in the TeX output as
  shifted momentum $\tilde{k} = k - k_0$. If set to false, express it
  in terms of unshifted momentum $k$. This option has no effect for
  BHZ done at $k_0 = 0$. (boolean value; \emph{default}: true)
\item
  \verb+bhz_plotcolor+: Colour of the BHZ dispersion in the BHZ
  output file. It may be a single matplotlib colour, a pair separated by
  a comma (separate colours for each block), or a triplet separated by
  commas (one block, other block, states without specific block).
  (\emph{default}: \texttt{red,blue,black}, legacy value: \texttt{red})
\item
  \verb+bhz_plotstyle+: Style of the BHZ dispersion in the BHZ output
  file. It may be a single matplotlib line style, a pair separated by a
  comma (separate styles for each block), or a triplet separated by
  commas (one block, other block, states without specific block).
  Examples are \texttt{solid}, \texttt{dashed}, \texttt{dotted}.
\end{itemize}

\subsubsection{Density of states}

\begin{itemize}
\item
  \verb+dos_interpolation_points+: The minimal number of points on
  the horizontal axis for some DOS and Berry curvature (Hall
  conductivity) plots. When the calculated number of (k or b) points is
  smaller, then perform interpolation to at least this number. Must be
  an integer; if equal to 0, do not interpolate. (\emph{default}: 100)
\item
  \verb+dos_energy_points+: The minimal number of points on the
  energy axis for some DOS and Berry curvature (Hall conductivity)
  plots. An energy resolution will be chosen so that the energy interval
  spans at least this many values. (\emph{default}: 1000)
\item
  \verb+dos_convolution_points+: The minimal number of points in the
  energy variable taken when applying a broadening function (a
  convolution operation) to an integrated DOS. If the energy range
  contains fewer points than this value, the integrated DOS is
  interpolated. Note that this value affects the `dummy variable' of the
  convolution integral only, i.e., the internal accuracy of the
  (numerical) integration. The broadened integrated DOS (the result)
  will always be with respect to the same energies as the input.
  (\emph{default}: 200)
\item
  \verb+dos_print_validity_range+: Print the lower and upper bound
  of the validity range for DOS and IDOS. If the lower bound (first
  value) is larger than the upper bound (second value), then the DOS and
  IDOS are invalid for all energies. See also
  \verb+plot_dos_validity_range+. (boolean value; \emph{default}:
  true)
\item
  \verb+dos_print_momentum_multiplier+: The momentum range can be
  extended by using a multiplier that takes into account the part of
  momentum space not explicitly calculated. Examples: If only positive
  momenta are calculated, simulate the negative values by multiplying by
  2; or, if in polar coordinates the calculation was done from 0 to 90
  degrees angle, multiply by 4 for a full circle. This setting
  determines whether this multiplicative factor should be printed to the
  standard output. (boolean value; \emph{default}: false)
\item
  \verb+dos_quantity+: The quantity in which to express density of
  states. Prior to version v0.95, this was done using the command line
  arguments \texttt{densitypnm}, \texttt{densityecm}, etc. Possible
  values:

  \begin{itemize}
  \item
    \texttt{k}: Occupied volume in momentum space; units
    $1/\mathrm{nm}^{d}$ (\emph{alias}: \texttt{momentum})
  \item
    \texttt{p}: Density of particles/carriers ($n$ or $dn/dE$);
    units $1/\mathrm{nm}^{d}$ (\emph{default}; \emph{alias}:
    \texttt{n}, \texttt{particles}, \texttt{carriers}, \texttt{cardens})
  \item
    \texttt{s}: Density of states (IDOS or DOS); units
    $1/\mathrm{nm}^{d}$. The only difference with \texttt{p} is the
    way the quantities are labelled. (\emph{alias}: \texttt{dos},
    \texttt{states})
  \item
    \texttt{e}: Density of charge ($\sigma$ or $d\sigma/dE$); units
    $e/\mathrm{nm}^{d}$ (\emph{alias}: \texttt{charge}) (The exponent
    d in the unit is adjusted according to the dimensionality.)
  \end{itemize}
\item
  \verb+dos_unit+: The length units used for density of states. Prior
  to version v0.95, this was done using the command line arguments
  \texttt{densitypnm}, \texttt{densityecm}, etc. Possible values:

  \begin{itemize}
  \item
    \texttt{nm}: Units of $1/\mathrm{nm}^{d}$, $e/\mathrm{nm}^{d}$
    (\emph{default})
  \item
    \texttt{cm}: Units of $1/\mathrm{cm}^{d}$, $e/\mathrm{cm}^{d}$
  \item
    \texttt{m}: Units of $1/\mathrm{m}^{d}$, $e/\mathrm{m}^{d}$ In
    the output, the density values are also scaled to a suitable `power
    of ten'. The exponent $d$ in the unit is adjusted according to the
    dimensionality.
  \end{itemize}
\item
  \verb+dos_strategy_no_e0+: The strategy to follow when trying to
  extract DOS or IDOS from the band structure, when the zero energy
  $E_0$ is not well defined. Possible values:

  \begin{itemize}
  \item
    \texttt{strict}: Neither DOS nor IDOS can be extracted.
  \item
    \texttt{dos}: DOS can be extracted, but IDOS cannot
    (\emph{default}).
  \item
    \texttt{ignore}: Both DOS and IDOS can be extracted, ignoring the
    fact that $E_0$ may lie at an arbitrary energy value. When $E_0$
    is defined (either manually or automatically), the extraction of DOS
    and IDOS is always possible, regardless of this setting.
  \end{itemize}
\end{itemize}

\subsubsection{Self-consistent Hartree}

\begin{itemize}
\item
  \verb+selfcon_acceptable_status+: Maximum status level for the
  result of the self-consistent Hartree calculation to be considered
  valid. Possible values:

  \begin{itemize}
  \item
    \texttt{0}: Successful
  \item
    \texttt{1}: Calculation skipped or aborted (\emph{default})
  \item
    \texttt{2}: Did not converge, but convergence is likely after more
    iterations
  \item
    \texttt{3}: Did not converge, convergence cannot be estimated or is
    unlikely
  \item
    \texttt{4}: Failed
  \end{itemize}
\item
  \verb+selfcon_check_chaos_steps+: Number of previous iterations
  used for the detection of chaotic behaviour. If this value is set to
  $n$, we say chaos occurs at iteration $i$ if the previous
  $V^{(j)}$ closest to $V^{(i)}$ are more than $n$ iterations ago,
  i.e., $i - j > n$. When chaos is detected, adjust the time step if
  \verb+selfcon_dynamic_time_step+ is set to true. (\emph{default}:
  4)
\item
  \verb+selfcon_check_orbit_steps+: Number of previous iterations
  used for the detection of periodic orbits. We say a periodic orbit
  occurs at iteration $i$ if the previous $V^{(j)}$ closest to
  $V^{(i)}$ show a regular pattern like $j - i = 2, 4, 6, 8$; the
  value $n$ set here is the minimum length of the regular pattern.
  When a periodic orbit is detected, adjust the time step if
  \verb+selfcon_dynamic_time_step+ is set to true. (\emph{default}:
  4)
\item
  \verb+selfcon_convergent_steps+: Number of consecutive convergent
  steps (iteration steps where the convergence condition is met)
  required for the self-consistent Hartree calculation to be considered
  successful. This prevents accidental convergence which could lead to a
  spurious solution. (\emph{default}: 5)
\item
  \verb+selfcon_debug+: Whether to enable debug mode for the
  self-consistent Hartree calculation. In debug mode, write temporary
  files and provide traceback for all exceptions (including\linebreak[4] 
  \texttt{KeyboardInterrupt}) within the iteration loop, which is useful
  for debugging. If debug mode is disabled, then do not write temporary
  files and continue on \texttt{SelfConError} and\linebreak[4]
  \texttt{KeyboardInterrupt} exceptions. (boolean value; \emph{default}:
  false)
\item
  \verb+selfcon_diff_norm+: Method that defines a measure of
  convergence for the self-consistent calculation. This method is
  essentially a function applied to the difference of potentials of the
  last two iteration steps. The result, a nonnegative value, is compared
  to the convergence criterion. Possible values:

  \begin{itemize}
  \item
    \texttt{max}: The maximum of the difference. Also known as supremum
    norm or $L^\infty$ (L-infinity) norm.
  \item
    \texttt{rms}: The root-mean-square of the difference. This is the
    $L^2$ norm. (\emph{default})
  \end{itemize}
\item
  \verb+selfcon_dynamic_time_step+: Whether the ``time'' step for
  the self-consistent calculation is adapted automatically between
  iterations. If set to false, the time step stays the same between
  iterations. (boolean value; \emph{default}: false)
\item
  \verb+selfcon_erange_from_eivals+: Whether to use the eigenvalues
  from first diagonalization result to determine the energy range used
  for calculating the density of states for the self-consistent
  calculation. If false, the energy range given in the command line is
  used instead. (boolean value; \emph{default}: false).
\item
  \verb+selfcon_full_diag+: Whether to use the full-diagonalization
  approach for the self-consistent Hartree calculation. If true, use the
  full-diagonalization approach that calculates all conduction band
  states to determine density as function of $z$. If false, use the
  standard mode that calculates bands around the charge neutrality point (CNP).
  The latter is
  significantly faster, but the results are based on an implausible
  assumption on the density at the CNP. (boolean value; \emph{default}:
  true)
\item
  \verb+selfcon_ll_use_broadening+: Whether to enable broadening
  during self-consistent calculation in LL mode. This can lead to bad
  convergence behaviour (or no convergence at all, depending on selected
  broadening), but results in more accurate Hartree potentials for the
  given broadening. This does not affect the broadening applied to the
  main diagonalization/postprocessing after the self-consistent
  calculation has finished. (boolean value; \emph{default}: false)
\item
  \verb+selfcon_energy_points+: The minimal number of points on the
  energy axis for the self-consistent calculation. An energy resolution
  will be chosen so that the energy interval spans at least this many
  values. This number may be fairly high without performance penalty.
  (\emph{default}: 1000)
\item
  \verb+selfcon_min_time_step+: The minimal value for the ``time''
  step (or ``weight'') for the self-consistent calculation. If
  \verb+selfcon_dynamic_time_step+ is set to true, the time step
  can never get lower than this value. Allowed values are between 0 and 1.
  (\emph{default}: 0.001)
\item
  \verb+selfcon_potential_average_zero+: Shift the potential such
  that its average will be zero at each iteration of the self-consistent
  calculation. Enabling this option is recommended for reasons of
  stability and for consistency of the output. (boolean value;
  \emph{default}: true)
\item
  \verb+selfcon_symmetrization_constraint+: Constraint on how the
  symmetry is checked and symmetrization performed on multiple
  quantities when solving the Poisson equation. When symmetry norm is
  below threshold the quantity is always fully symmetrized over whole
  layer stack (except for \texttt{never}). Possible values:

  \begin{itemize}
  \item
    \texttt{never}: Symmetry will not be checked. No symmetrization is
    performed.
  \item
    \texttt{strict}: Symmetry is checked over whole layer stack.
    (\emph{default})
  \item
    \texttt{loose}: Symmetry is checked over the well region only. This
    method is preferred for asymmetric layer stacks.
  \end{itemize}
\item
  \verb+selfcon_use_init_density+: Whether a uniform density
  profile (consistent with the total carrier density) is applied in the
  initialization of the self-consistent Hartree calculation. If enabled,
  calculate the potential and apply it to the Hamiltonian in the first
  iteration. If disabled, use the Hamiltonian with zero potential,
  unless an initial potential is loaded from a file. (boolean value;
  \emph{default}: false)
\end{itemize}

\subsubsection{Optical transitions}

\begin{itemize}
\item
  \verb+transitions_min_amplitude+: Minimum amplitude to consider
  for transitions. The lower this number, the larger the number of data
  points and the larger the data files and plots. (\emph{default}: 0.01)
\item
  \verb+transitions_min_deltae+: Minimum energy difference in meV to
  consider for transitions. This value is proportional to a minimal
  frequency. The smaller this number, the larger the number of data
  points and the larger the data files and plots. (\emph{default}: 0.1)
\item
  \verb+transitions_max_deltae+: Maximum energy difference in meV of
  transitions, i.e., upper limit of the vertical axis (filtered
  transitions plot only). If set to 0, determine the vertical scale
  automatically. (\emph{default}: 0)
\item
  \verb+transitions_dispersion_num+: Number of transitions to
  include in the dispersion or B dependence (LL fan) plot. If set to n,
  the transitions with n highest transitions rates will be shown. If set
  to 0, show an unlimited number of transitions. (\emph{default}: 4)
\item
  \verb+transitions_broadening_type+: Shape of the broadening
  function used for broadening the transitions in the absorption plot.
  Possible choices:

  \begin{itemize}
  \item
    \texttt{step}: A step function (\emph{alias}: \texttt{delta})
  \item
    \texttt{lorentzian}: Lorentzian function (Cauchy distribution),
    scale parameter gamma, which is the half-width at half-maximum.
    (\emph{default}; \emph{alias}: \texttt{lorentz})
  \item
    \texttt{gaussian}: Gaussian function, scale parameter sigma, which
    is the standard deviation. (\emph{alias}: \texttt{gauss},
    \texttt{normal})
  \item
    \texttt{fermi}: Fermi function (thermal distribution), scale
    parameter is energy. (\emph{alias}: \texttt{logistic},
    \texttt{sech})
  \item
    \texttt{thermal}: Fermi function (thermal distribution), scale
    parameter is temperature.
  \end{itemize}

  The broadening functions for the absorption are the probability
  density functions for all of these choices; see Appendix~\ref{app_broadening_reference}
  for further information.

\item
  \verb+transitions_broadening_scale+: Scale parameter of the
  broadening function. This may be an energy (in meV) or a temperature
  (in K) that determines the amount of broadening (i.e., its `width').
  (\emph{default}: 2.5)
\item
  \verb+transitions_spectra+: (\emph{experimental}) Output spectral
  plots and tables if a carrier density is set. If false, skip (time
  consuming) spectra calculation. (boolean value, \emph{default}: false)
\item
  \verb+transitions_plot+: Output transition plot. If false, do not
  create and save a a transitions plot. (boolean value, \emph{default}:
  true)
\end{itemize}

\subsubsection{Colours and Colormaps}

\begin{itemize}
\item
  \verb+color_bindex+ Colormap for band
  index. For band index only, the range of the observable is adjusted to
  the number of colours in the colormap. (\emph{default}:
  \texttt{tab21posneg}; \textsc{Note}: for the `old' set of colours, use
  \texttt{tab20alt})
\item
  \verb+color_dos+: Colormap for density of states (\emph{default}:
  \texttt{Blues})
\item
  \verb+color_energy+: Colormap for energy plot (2D dispersion)
  (\emph{default}: \texttt{jet})
\item
  \verb+color_idos+: Colormap for integrated density of states
  (\emph{default}: \verb+RdBu_r+)
\item
  \verb+color_indexed+: Colormap for indexed (discrete) observables
  (\emph{default}: \texttt{tab20alt,tab20})
\item
  \verb+color_indexedpm+: Colormap for indexed (discrete) observables
  using a `dual' colour scale, such as \texttt{llindex.sz}
  (\emph{default}: \texttt{tab20})
\item
  \verb+color_ipr+: Colormap for IPR observables (\emph{default}:
  \verb+inferno_r+)
\item
  \verb+color_localdos+: Colormap for local density of states
  (\emph{default}: \texttt{cividis,jet})
\item
  \verb+color_posobs+: Colormap for observables with positive values
  (\emph{default}: \texttt{grayred})
\item
  \verb+color_shadedpm+: Colormap for continuous positive observables
  using a `dual' colour scale, such as \texttt{y2.isopz}
  (\emph{default}: \texttt{bluereddual})
\item
  \verb+color_sigma+: Colormap for `sigma observables' (standard
  deviation) (\emph{default}: \verb+inferno_r+)
\item
  \verb+color_symmobs+: Colormap for observables with a symmetric
  (positive and negative) range of values (\emph{default}:
  \texttt{bluered})
\item
  \verb+color_threehalves+: Colormap for observables with range
  {[}-3/2, 3/2{]} (\emph{default}: \texttt{yrbc})
\item
  \verb+color_trans+: Colormap for transition plots (\emph{default}:
  \verb+hot_r+)
\item
  \verb+color_wf_zy+: Colormap for
  wavefunction plot
  $|\psi(z, y)|^2$ (\emph{default}: \texttt{Blues})
\end{itemize}

\subsubsection{Figures}

\begin{itemize}
\item
  \verb+fig_matplotlib_style+: Matplotlib style
  file for changing the properties of plot elements. This may be a file
  in the configuration directory \texttt{\HOME/.kdotpy} or
  in the working directory, or a built-in matplotlib style.
  (\emph{default}: \texttt{kdotpy.mplstyle})
\item
  \verb+fig_hsize+, \verb+fig_vsize+: Horizontal and vertical size
  of the figures, in mm. (\emph{default}: 150, 100, respectively)
\item
  \verb+fig_lmargin+, \verb+fig_rmargin+, \verb+fig_bmargin+,
  \verb+fig_tmargin+: Figure margins (left, right, bottom, top),
  i.e., the space in mm between the figure edge and the plot area.
  (\emph{default}: 20, 4, 12, 3, respectively)
\item
  \verb+fig_charlabel_space+: Vertical space for the character
  labels in the dispersion plot, in units of the font size. To avoid
  overlapping labels, use a value of approximately 0.8 or larger.
  (\emph{default}: 0.8)
\item
  \verb+fig_colorbar_space+: Space reserved for the colour bar
  legend in mm (\emph{default}: 30) In other words, this is the distance
  between the right-hand edges of the figure and the plot if a colour
  bar is present. It `replaces' the right margin.
\item
  \verb+fig_colorbar_margin+: Space between the right-hand edge of
  the plot and the colour bar legend, in mm. This space is taken from
  the colour bar space (set by \verb+fig_colorbar_width+), so it
  does not affect the right-hand edge of the plot. (\emph{default}: 7.5)
\item
  \verb+fig_colorbar_size+: Width of the actual colour bar in mm
  (\emph{default}: 4)
\item
  \verb+fig_colorbar_method+: Way to place the colour bar; one of
  the following options:

  \begin{itemize}
  \item
    \texttt{insert}: Take space inside the existing plot; keep the
    figure size, but decrease the plot size. (\emph{default})
  \item
    \texttt{extend}: Add extra space; keep the plot size but increase
    the figure size.
  \item
    \texttt{file}: Save into a separate file. The original figure is not
    changed.
  \end{itemize}
\item
  \verb+fig_colorbar_labelpos+: Method to determine the position of
  the label of the colour bar. One of the following options:

  \begin{itemize}
  \item
    \texttt{legacy}: The `old' method, using
    \verb+colorbar.set_label+ plus a manual shift.
  \item
    \texttt{xaxis}: As label for the `x axis', directly below the the
    colour bar.
  \item
    \texttt{yaxis}: As label for the `y axis', vertically up on the
    right-hand side.
  \item
    \texttt{center}: Centred in the whole space allocated for the
    colour bar, including margins; very similar to `legacy' for default
    settings of the colour bar size and margins. (\emph{default})
  \item
    \texttt{left}: Left aligned with the left border of the colour bar.
  \end{itemize}
\item
  \verb+fig_colorbar_abstwosided+: Whether a shaded dual colour bar,
  where the vertical value is the absolute value of the observable,
  should show the observable itself, with values running from -max to
  max (if set to true; default). Otherwise show the absolute value,
  running from 0 to max. (boolean value; \emph{default}: true)
\item
  \verb+fig_extend_xaxis+: Relative extension of the horizontal plot
  range for dispersion and magnetic field dependence. For example, a
  value of \texttt{0.05} means 5\% of the range is added left and right
  of the minimum and maximum x value (where x is k or B), respectively.
  This does not affect the range if the command line argument
  \texttt{xrange} is used. Use 0 to not extend the plot range. The value
  may not be negative. (\emph{default}: 0.05)
\item
  \verb+fig_inset_size+: Size (width and height) of the inset legend
  in mm. Values smaller than 30 are not recommended. (\emph{default}:
  30)
\item
  \verb+fig_inset_margin+: Space between inset edge and plot edge in
  mm. (\emph{default}: 3)
\item
  \verb+fig_inset_color_resolution+: Number of color gradations
  along each axis for the RGB (inset) legend. Do not change unless file
  size is an issue. (\emph{default}: 20)
\item
  \verb+fig_legend_fontsize+: Specify the font size of the legend.
  May also be set to `auto' for automatic; this yields a font size of 8
  for RGB (inset) legend, 10 for other legend or colorbars (may be
  subject to settings in matplotlibrc and/or style files).
  (\emph{default}: \texttt{auto})
\item
  \verb+fig_spin_arrow_length+: Arrow length in spin plots. The
  value is the length in mm for arrows representing spin value 0.5 or a
  direction. (\emph{default}: 5)
\item
  \verb+fig_max_arrows+: Maximum number of arrows in a vector plot
  in each dimension. The value 0 means no limit. (\emph{default}: 20)
\item
  \verb+fig_arrow_color_2d+: Color of the arrows in a 2D vector
  plot. This must be a valid matplotlib color. (\emph{default}:
  \texttt{\#c0c0c0})
\item
  \verb+fig_ticks_major+: Strategy to determine the major ticks in
  the plots. Possible choices:

  \begin{itemize}
  \item
    \texttt{none}: No major ticks
  \item
    \texttt{auto}: Determine number of ticks automatically (based on
    plot size). (\emph{default})
  \item
    \texttt{fewer}: A few ticks per axis (typically 3)
  \item
    \texttt{normal}: A moderate amount of ticks per axis (typically 6)
  \item
    \texttt{more}: Many ticks per axis (typically 12)
  \end{itemize}

  One can use different choices for the horizontal and vertical axis, as
  follows:\\
  \verb+fig_ticks_major=normal,fewer+
\item
  \verb+fig_ticks_minor+: Strategy to determine the minor ticks in
  the plots. Possible choices:

  \begin{itemize}
  \item
    \texttt{none}: No minor ticks (\emph{default})
  \item
    \texttt{auto}: Determine automatically (matplotlib's algorithm)
  \item
    \texttt{fewer}: Few minor ticks (major interval divided by 2)
  \item
    \texttt{normal}: Moderately many ticks (major interval divided by 4
    or 5).
  \item
    \texttt{more}: Many minor ticks (major interval divided by 10)
  \end{itemize}

  One can use different choices for the horizontal and vertical axis, as
  follows:\\
  \verb+fig_ticks_minor=fewer,none+
\item
  \verb+fig_unit_format+: Opening and closing bracket of the units
  in axis and legend labels. (\emph{default}: \texttt{{[}{]}})
\end{itemize}

\subsubsection{Plot output}

\begin{itemize}
\item
  \verb+plot_constdens_color+: The colour of the curves in the
  `constdens' plots. The value must be a valid matplotlib colour.
  (\emph{default}: \texttt{blue})
\item
  \verb+plot_dispersion_default_color+: The uniform colour of the
  dispersion curves, if there is no colour scale set for the given
  observable (or if no observable is set). The value must be a valid
  matplotlib colour. (\emph{default}: \texttt{blue})
\item
  \verb+plot_dispersion_energies+: Plot special energies, e.g.,
  charge-neutrality point, Fermi energy/chemical potential at zero and
  finite density in dispersion plots. (boolean value; \emph{default}:
  true)
\item
  \verb+plot_dispersion_energies_color+: The line colour for
  special energies. The value must be a valid matplotlib colour. If left
  empty, take \texttt{lines.color} from matplotlibrc or a style file.
  (\emph{default}: \texttt{black})
\item
  \verb+plot_dispersion_parameter_text+: Write an indication in the
  plot for constant parameter values, e.g., when plotting along $k_x$ for a
  nonzero ky value, write ``For $k_y$ = ''. (boolean value;
  \emph{default}: true)
\item
  \verb+plot_dispersion_stack_by_index+: If enabled, make sure the
  data with the lowest band or Landau-level index is shown on top, to
  make sure the `most interesting data' (low-index states) is not
  obscured by `less interesting data' (high-index states). Otherwise,
  the plot function uses the default plot stacking order: the data is
  then drawn simply in the order by which it is processed. (boolean
  value; \emph{default}: false)
\item
  \verb+plot_dos_color+: Colour of the curves in the (integrated)
  density of states (IDOS/DOS) plots. The value must be a valid
  matplotlib colour. (\emph{default}: \texttt{blue})
\item
  \verb+plot_dos_energies+: Plot special energies, e.g.,
  charge-neutrality point, Fermi energy/chemical potential at zero and
  finite density in density (DOS) plots. (boolean value; \emph{default}:
  true)
\item
  \verb+plot_dos_fill+: Fill the area between the curve and zero in
  the DOS plot (not integrated DOS). (boolean value; \emph{default}:
  false)
\item
  \verb+plot_idos_fill+: Fill the area between the curve and zero in
  the integrated DOS plot. (boolean value; \emph{default}: false)
\item
  \verb+plot_dos_units_negexp+: Use negative exponents in DOS units
  in density plots. If set to true, write $\mathrm{nm}^{-2}$ instead
  of $1/\mathrm{nm}^{2}$, for example. (boolean value; \emph{default}:
  false)
\item
  \verb+plot_dos_validity_range+: Shade the area in the
  (integrated) DOS plot where the value is expected to be incorrect due
  to missing data (due to momentum cutoff). (boolean value;
  \emph{default}: true)
\item
  \verb+plot_dos_vertical+: Plot the (integrated) DOS sideways, so
  that energy is plotted on the vertical axis. The vertical scale will
  match the dispersion plot, so that these figures can be put
  side-by-side with a common axis. (boolean value; \emph{default}: true)
\item
  \verb+plot_ecnp+: Plot the charge neutral energy as function of k
  or B. This is the boundary between ``electron'' and ``hole'' states
  (positive and negative band indices, respectively). (boolean value;
  \emph{default}: false)
\item
  \verb+plot_rasterize_pcolormesh+: Whether to rasterize plot
  elements created with \texttt{pcolormesh} from matplotlib. This is used
  primarily for two-dimensional color plots with \texttt{kdotpy\ ll}
  when one uses quadratic stepping for the magnetic field values.
  Rasterization leads to improved performance both in creating the plots
  as well as in rendering them with a pdf viewer. The resolution can be
  controlled with the matplotlibrc parameters \texttt{figure.dpi} and
  \texttt{savefig.dpi}. If the old behaviour is desired, i.e., that the
  data is rendered as vector graphics, set the value of
  \verb+plot_rasterize_pcolormesh+ to \texttt{false}. (boolean
  value; \emph{default}: true)
\item
  \verb+plot_rxy_hall_slope+: Plot the Hall slope
  $R_{xy} = B / (n e)$, where $B$ is magnetic field, $n$ is
  density and $e$ is electron charge, in the plots for $R_{xy}$
  (\texttt{rxy-constdens.pdf}) as a dashed line. (boolean value;
  \emph{default}: true)
\item
  \verb+plot_sdh_markers+: Whether to show markers for the period of
  the Shubnikov-de Haas (SdH) oscillations in the `constdens' plot (both `normal' and `SdH'
  versions). The markers are placed at the values for which 1 / B is a
  multiple of $e / (2 \pi \hbar n)$, where n is the density. (boolean
  value; \emph{default}: true)
\item
  \verb+plot_sdh_markers_color+: Colour of the SdH markers in the
  `constdens' plots. The value must be a valid matplotlib colour.
  (\emph{default}: \texttt{red})
\item
  \verb+plot_sdh_scale_amount+: The maximum number of SdH
  oscillations to be shown in the SdH plot. If set to a nonzero value,
  the scale on the horizontal axis is magnified to this amount of SdH
  oscillations. The scale is never shrunk, so there may be fewer SdH
  oscillations on the axis. The `constdens' plot linear in B is
  unaffected. If set to 0 (default), do not scale the axis. A typical
  useful nonzero value is 20.
\item
  \verb+plot_transitions_labels+: Show some labels in transitions
  plot. (boolean value; \emph{default}: true)
\item
  \verb+plot_transitions_quantity+: Which quantity to use for
  colouring in the transitions plot. Possible choices:

  \begin{itemize}
  \item
    \texttt{amplitude}: `Raw' amplitude gamma from Fermi's golden rule
  \item
    \texttt{rate}: Transition rate density, $n \Gamma (f_2 - f_1)$
    (\emph{default}; \emph{alias}: \verb+rate_density+)
  \item
    \texttt{occupancy}: Occupancy difference $f_2 - f_1$
  \item
    \texttt{deltae}: Energy difference $|E_2 - E_1|$ in meV
  \item
    \texttt{freq} : Corresponding frequency in THz (\emph{alias}:
    \verb+freq_thz+)
  \item
    \texttt{lambda}: Corresponding wave length in \textmu{}m (\emph{alias}:
    \texttt{lambda\uscore\textmu{}m}, \verb+lambda_um+)
  \item
    \texttt{absorption}: Absorption (relative attenuation of intensity)
    A
  \end{itemize}
\item
  \verb+plot_transitions_frequency_ticks+: Plot frequency ticks at
  the left and right energy axis for transitions plots. (boolean value;
  \emph{default}: true)
\item
  \verb+plot_transitions_max_absorption+: Upper limit of the colour
  scale in the transitions absorption plot. For the relative absorption,
  use \texttt{{[}-value,\ value{]}} as the colour range.
  (\emph{default}: 0.03)
\item
  \verb+plot_wf_orbitals_realshift+: Phase-shift the orbital
  functions to purely real values before plotting. This results in a
  single line plot per orbital with consistent amplitudes and signs. The
  actual phases are still given at the right side of the figure. Uses
  straight/dashed lines for +/- angular momentum orbitals. (boolean
  value; \emph{default}: false)
\item
  \verb+plot_wf_orbitals_order+: Order of the orbitals in the
  legend, for wave function plot style `separate'. Possible choices:

  \begin{itemize}
  \item
    \texttt{standard} (\emph{default}):
\begin{equation*}
\begin{array}{l@{\hspace{1em}}l@{\hspace{1em}}l}
\Gamma_6,+1/2  &\Gamma_8,+1/2  &\Gamma_7,+1/2\\
\Gamma_6,-1/2  &\Gamma_8,-1/2  &\Gamma_7,-1/2\\
\Gamma_8,+3/2  &\Gamma_8,-3/2
\end{array}
\end{equation*}
  \item
    \texttt{paired}:
\begin{equation*}
\begin{array}{l@{\hspace{1em}}l@{\hspace{1em}}l}
\Gamma_6,+1/2  &\Gamma_6,-1/2  &\Gamma_7,+1/2\\
\Gamma_8,+1/2  &\Gamma_8,-1/2  &\Gamma_7,-1/2\\
\Gamma_8,+3/2  &\Gamma_8,-3/2
\end{array}
\end{equation*}
  \item
    \texttt{table}:
\begin{equation*}
\begin{array}{l@{\hspace{1em}}l@{\hspace{1em}}l}
              &\Gamma_8,+3/2\\
\Gamma_6,+1/2 &\Gamma_8,+1/2 &\Gamma_7,+1/2\\
\Gamma_6,-1/2 &\Gamma_8,-1/2 &\Gamma_7,-1/2\\
              &\Gamma_8,-3/2
\end{array}
\end{equation*}
    For the six-orbital basis, the $\Gamma_7$ states are omitted.
  \end{itemize}
\item
  \verb+plot_wf_zy_format+: File format for wavefunction plots
  $|\psi(z, y)|^2$. Possible choices:

  \begin{itemize}
  \item
    \texttt{pdf}: Multi-page PDF if possible, otherwise separate PDF
    files. (\emph{default})
  \item
    \texttt{png}: Separate PNG files.
  \item
    \texttt{pngtopdf}: Separate PNG files are converted and merged into
    a multi-page PDF. Requires the `convert' command to be available.
    (\emph{alias}: \verb+png_to_pdf+)
  \end{itemize}
\item
  \verb+plot_wf_mat_label_rot+: For wave function plots, the
  rotation (in degrees) of material labels inside the layers. Can be
  used to fit long labels in thin layers. (\emph{default}: 0)
\item
  \verb+plot_wf_zy_bandcolors+: Colour model for the wavefunction
  plots $|\psi(z, y)|^2$ separated by bands. Possible choices:

  \begin{itemize}
  \item
    \texttt{hsl}: Hue-saturation-lightness. The colour (hue) is
    determined by the relative content of the bands, the saturation and
    lightness by the density.
  \item
    \texttt{hsv}: Hue-saturation-value. Like \texttt{hsl}, the colour
    (hue) is determined by the relative content of the bands, the
    saturation by the density, and the value is equal to 1.
  \item
    \texttt{rgb}: Red-green-blue. The red, green, and blue channels are
    determined by the contents of the bands.
  \end{itemize}

  \textsc{Note}: This is not a colormap! For the absolute value without
  band content, use the colormap set by \verb+color_wf_zy+.
\item
  \verb+plot_wf_zy_scale+: Scaling method (colour scale
  normalization) for wavefunction plots\linebreak[4] $|\psi(z, y)|^2$. Possible
  choices:

  \begin{itemize}
  \item
    \texttt{separate}: Normalize the colour scale for each wavefunction
    individually. (\emph{default})
  \item
    \texttt{together}: Normalize the colour scale for all wavefunctions
    collectively.
  \end{itemize}
\item
  \verb+plot_wf_y_scale+: Scaling method for the vertical axis for
  wave function plots $|\psi(y)|^2$. Possible choices:

  \begin{itemize}
  \item
    \texttt{size}: Determine scale from sample size (width in y
    direction. (\emph{default}; \emph{alias}: \texttt{width})
  \item
    \texttt{magn}: Determine scale from magnetic field. For small
    fields, use the sample size.
  \item
    \texttt{separate}: Determine scale from the maximum of each wave
    function individually.
  \item
    \texttt{together}: Determine scale from the maximum of all wave
    functions collectively.
  \end{itemize}
\item
  \verb+plot_wf_delete_png+: If the wavefunction plots are saved in
  PNG format and subsequently converted to a single multi-page PDF,
  delete the PNG files if the conversion is successful. (boolean value;
  \emph{default}: true)
\item
  \verb+plot_wf_together_num+: For the wavefunction plot in
  \texttt{together} style, plot this many wave functions. Must be a
  positive integer. (\emph{default}: 12)
\end{itemize}

\subsubsection{CSV output}

\begin{itemize}
\item
  \verb+csv_style+: Formatting for csv output. Possible values:

  \begin{itemize}
  \item
    \texttt{csvpandas}: Comma separated values using pandas module
  \item
    \texttt{csvinternal}: Comma separated values using internal function
  \item
    \texttt{csv}: Choose \texttt{csvpandas} if pandas is available,
    otherwise choose \texttt{csvinternal} (\emph{default})
  \item
    \texttt{align}: Align values in columns in the text file
  \end{itemize}
\item
  \verb+csv_multi_index+: Determines how a multi-index (LL index,
  band index) is formatted in csv output. Possible values:

  \begin{itemize}
  \item
    \texttt{tuple}: As a tuple \texttt{(\#\#,\ \#\#)} (\emph{default})
  \item
    \texttt{llindex}: Only the LL index
  \item
    \texttt{bindex}: Only the band index
  \item
    \texttt{split}: LL index on first row, band index on second row
    (\emph{alias}: \texttt{tworow})
  \item
    \texttt{short}: Short version of tuple \texttt{\#\#,\#\#} (space and
    parentheses are omitted)
  \end{itemize}
\item
  \verb+csv_bandlabel_position+: Location of the band labels in the
  `by-band' CSV output. Possible values:

  \begin{itemize}
  \item
    \texttt{top}: At the very top, above the other column headings.
    (\emph{default}; \emph{alias}: \texttt{above})
  \item
    \texttt{second}: Between the data and the other column headings.
    (\emph{alias}: \texttt{between})
  \item
    \texttt{bottom}: At the bottom, below the data. (\emph{alias}:
    \texttt{below})
  \end{itemize}
\end{itemize}

\subsubsection{Table output}

\begin{itemize}
\item
  \verb+table_berry_precision+: Precision (number of decimals) for
  floating point numbers, for the Berry curvature csv files.
  (\emph{default}: 4)
\item
  \verb+table_data_label_style+: Style for expressing data labels
  in generic two-dimensional csv output, such as density of states and
  Berry curvature. The label is positioned at the end of the first row
  with data. Possible choices: \texttt{none} (\emph{alias}:
  \texttt{false}), \texttt{raw}, \texttt{plain}, \texttt{unicode},
  \texttt{tex} (for details, see \verb+table_dispersion_unit_style+ below). If
  \texttt{none}, do not write a label. (\emph{default}: \texttt{plain})
\item
  \verb+table_data_unit_style+: Style for expressing the unit in
  generic two-dimensional csv output. Possible choices: \texttt{none}
  (\emph{alias}: \texttt{false}), \texttt{raw}, \texttt{plain},
  \texttt{unicode}, \texttt{tex} (for details, see\linebreak[4]
  \verb+table_dispersion_unit_style+ below). If \texttt{none}, do
  not write a unit. Also, if\linebreak[4] \verb+table_data_label_style+ is set
  to \texttt{none}, this option is ignored and no unit is written.
  (\emph{default}: \texttt{plain})
\item
  \verb+table_dos_precision+: Precision (number of decimals) for
  floating point numbers, for the density of states csv files.
  (\emph{default}: 8)
\item
  \verb+table_dos_scaling+: Whether to apply density scaling for csv
  output of densities. If false, use the native units
  ($\mathrm{nm}^{-2}$ in two dimensions). Otherwise, use the same
  scaling as for plots. (boolean value; \emph{default}: false)
\item
  \verb+table_dos_units_negexp+: Use negative exponents in DOS
  units for csv output. If set to true, write $\mathrm{nm}^{-2}$
  instead of $1/\mathrm{nm}^{2}$, for example. (boolean value;
  \emph{default}: false)
\item
  \verb+table_dispersion_precision+: Precision (number of decimals)
  for floating point numbers, for the dispersion csv files. Energy and
  momentum values may use a different number of decimals. (minimum: 2,
  \emph{default}: 5)
\item
  \verb+table_dispersion_data_label+: Whether to include the
  observable at the end of the first data row in a multi-dimensional
  dispersion csv table (e.g., with two or three momentum variables).
  (boolean value; \emph{default}: true)
\item
  \verb+table_dispersion_units+: Whether to include units of the
  variables and observables in dispersion csv files. For a
  one-dimensional dispersion, these are included as second header row.
  For a multi-dimensional dispersion, the unit is added at the end of
  the first data row. (boolean value; \emph{default}: true)
\item
  \verb+table_dispersion_unit_style+: Style for expressing units.
  Possible choices:

  \begin{itemize}
  \item
    \texttt{raw}: `Without' formatting
  \item
    \texttt{plain}: Plain-text formatting using common symbols (e.g.,
    square is \^{}2 and Greek letters are spelled out)
  \item
    \texttt{unicode}: Formatting using `fancy' Unicode symbols (e.g.,
    square is the superscript-2 symbol and Greek letters use their
    corresponding Unicode symbol).
  \item
    \texttt{tex}: LaTeX formatting
  \end{itemize}

  (\emph{default}: \texttt{plain})\\
  \textsc{Note}: Even with \texttt{raw} or \texttt{plain}, there may
  still be some non-ASCII symbols, for example \textmu{}.
\item
  \verb+table_dispersion_obs_style+: Style for expressing
  observables/quantities. Possible choices: \texttt{raw},
  \texttt{plain}, \texttt{unicode}, \texttt{tex} (see above).
  (\emph{default}: \texttt{raw})
\item
  \verb+table_qz_precision+: Precision (number of decimals) for
  floating point numbers, for the `Q(z)' (z-dependent quantity) csv
  files. (\emph{default}: 5)
\item
  \verb+table_extrema_precision+: Precision (number of decimals) for
  floating point numbers, for the extrema csv files. (\emph{default}: 5)
\item
  \verb+table_transitions_precision+: Precision (number of decimals)
  for floating point numbers, for the transitions csv files.
  (\emph{default}: 3)
\item
  \verb+table_absorption_precision+: Precision (number of decimals)
  for floating point numbers, for the absorption csv files (associated
  with transitions). (\emph{default}: 5)
\item
  \verb+table_transitions_ratecoeff_unit+: Unit for the rate
  coefficient for optical transitions. (\emph{default}:
  \texttt{nm\^{}2/mV/ns})
\item
  \verb+table_wf_files+: Which type of files should be written for
  the wave function data. Possible choices:

  \begin{itemize}
  \item
    \texttt{none}: No files are written.
  \item
    \texttt{csv}: Write csv files only. (\emph{default})
  \item
    \texttt{tar}: Write csv files, pack them into a tar file.
  \item
    \texttt{targz}: Write csv files, pack them into a gzipped tar file
    (compression level 6). (\emph{alias}: \texttt{gz}, \texttt{gzip},
    \texttt{tar.gz})
  \item
    \texttt{zip}: Write csv files, pack them into a zip file with
    `deflate' compression.
  \item
    \texttt{zipnozip}: Write csv files, pack them into a zip file
    without compression.
  \end{itemize}

  For the archive options (\texttt{tar}, \texttt{zip}, etc.), the csv
  files are deleted if the archive has been written successfully;
  otherwise they are kept. Some options may be unavailable depending on
  the installed Python modules.
\item
  \verb+table_wf_precision+: Precision (number of decimals) for
  floating point numbers, for the wave function csv files.
  (\emph{default}: 5)
\end{itemize}

\subsubsection{XML output}

\begin{itemize}
\item
  \verb+xml_omit_default_config_values+: If set to true, do not
  save all configuration values to the XML output file, but only the
  ones that are set to a value different than the default value.
  Otherwise save all values (\emph{default}); this is recommended for
  reproducibility. (boolean value; \emph{default}: false)
\item
  \verb+xml_shorten_command+: If set to true, replace the script
  path in the \verb+<cmdargs>+ tag by
  \texttt{kdotpy\ xx} (where \texttt{xx} = \texttt{1d}, \texttt{2d},
  etc.) if typing \texttt{kdotpy} on the command line refers to the
  kdotpy main script. For this, the main script (or a link to it) must
  be in the \texttt{PATH} environment variable; this is generally the
  case if kdotpy has been installed with pip. (boolean value;
  \emph{default}: false)
\end{itemize}

\subsubsection{Miscellaneous}

\begin{itemize}
\item
  \verb+berry_dk+: Momentum step size (in $\mathrm{nm}^{-1}$) for
  calculating the derivative of the Hamiltonian in the calculation of
  the Berry curvature as function of momentum. It does not apply to the
  Berry curvature calculation in Landau-level mode. The value must be
  positive. (\emph{default}: \texttt{1e-3})
\item
  \verb+berry_ll_simulate+: Whether to use simulated Berry curvature
  (more accurately: Chern numbers) for Berry / Hall output, for
  \texttt{kdotpy\ ll}, instead of the calculated one. The calculated
  value may sometimes show artifacts that cannot be easily resolved by
  increasing number of eigenstates for example. The simulated Berry
  curvature (observable \texttt{berrysim}) is set to exactly 1 for all
  states at nonzero magnetic field. (boolean value; \emph{default}:
  false)\\
  \textbf{Hint:} One may do a comparison by doing the calculation twice
  with settings \texttt{true} and \texttt{false}, respectively. The
  output is written to different file names as to ease the comparison.
\item
  \verb+diag_save_binary_ddp+: Whether and how to save intermediate
  binary files for each\linebreak[4] \texttt{DiagDataPoint} (diagonalization data
  point). Possible choices:

  \begin{itemize}
  \item
    \texttt{npz}: The NumPy (compressed) binary file format\footnote{See
    \url{https://numpy.org/doc/stable/reference/generated/numpy.lib.format.html}.}
    (\emph{alias}: \texttt{numpy})
  \item
    \texttt{h5}: HDF5 data format. This requires the Python module h5py
    to be installed\footnote{See \url{https://docs.h5py.org}.}. (\emph{alias}:
    \texttt{hdf5})
  \item
    \texttt{false}: Do not save intermediate files (\emph{default})
  \end{itemize}

  \textsc{Note}: This configuration value is independent from the
  command line option \texttt{tempout}. The npz and hdf5 formats are
  meant for permanent data storage, the \texttt{tempout} files are only
  safe for immediate re-use and should not be used for long-term
  storage.
\item
  \verb+job_monitor_limit+: If the number of data points is smaller
  than this value, show the full job monitor with information about the
  intermediate steps. Otherwise, show the simple in-line progress
  indicator. For the value 0, always show the simple progress indicator.
  (\emph{default}: \texttt{101})
\item
  \verb+lattice_regularization+: Enables or disables lattice
  regularization. The settings \texttt{true} and \texttt{false}
  correspond to the obsolete command-line arguments \texttt{latticereg}
  and \texttt{nolatticereg}, respectively. The recommended value and
  default value is \texttt{false}. Note that for older kdotpy versions
  (kdotpy v0.xx), the default value was \texttt{true} for compatibility
  reasons. (boolean value; \emph{default}: false)
\item
  \verb+lattice_zres_strict+: Enables or disable strict check of
  commensurability of z resolution with thickness of the layers, i.e.,
  whether the thicknesses are integer multiples of the z resolution. If
  they are incommensurate, quit with an error if strict checking is
  enabled. If disabled, change the thicknesses to match the z resolution
  and raise a warning. (boolean value; \emph{default}: true)
\item
  \verb+magn_epsilon+: Numeric value that determines whether small
  values near zero need to be inserted if the grid contains magnetic
  fields. The value zero means disabling this feature. Otherwise, +/-
  the absolute value of \verb+magn_epsilon+ is inserted at either
  side of B = 0, whichever side (positive or negative) is included in
  the range. If negative, insert the values only if the range is
  two-sided. The motivation for including this option is to reduce some
  plot artifacts for ranges that contain positive and negative magnetic
  fields. For this option to be effective, it might also be necessary to
  set the \texttt{split} parameter to a small value. (\emph{default}:
  -1e-4)
\item
  \verb+numpy_linewidth+: Sets the (approximate) line width for NumPy
  array output. (This output is used in verbose mode mostly.) The value
  is passed to \verb+numpy.set_printoptions()+.
  The value has to be an integer $\geq 0$. The output is always
  at least one column, so small values may be exceeded. (\emph{default}:
  \texttt{200})
\item
  \verb+numpy_printprecision+: The number of digits of precision for
  NumPy array floating point output. (This output is used in verbose
  mode mostly.) The value is passed to\linebreak[4]
  \verb+numpy.set_printoptions()+.
  The value has to be an integer $\geq 0$. The number of digits
  shown does not exceed the number needed to uniquely define the values,
  e.g., 17 digits for 64-bit floating point numbers. (\emph{default}:
  \texttt{6})
\item
  \verb+wf_locations_exact_match+: If set to true (\emph{default}),
  the wave function locations should match the momentum/magnetic field
  values exactly. If no exact match is found, skip the location (`old
  behaviour'). If set to false, find the nearest value to each location.
  (boolean value; \emph{default}: true)
\item
  \verb+wf_locations_filename+: Whether to label the wave function
  files using the position (momentum/magnetic field). If set to false,
  label with numbers. (boolean value; \emph{default}: true)
\end{itemize}

\textsc{Note}: Boolean configuration options may have the following
values (not case sensitive):

\begin{itemize}
 \item For ``True'': \texttt{yes}, \texttt{y},
\texttt{true}, \texttt{t}, \texttt{1}, \texttt{enabled}, \texttt{on}
 \item For ``False'': \texttt{no}, \texttt{n}, \texttt{false}, \texttt{f},
\texttt{0}, \texttt{disabled}, \texttt{off}
\end{itemize}

\clearpage
\section{Command-line arguments}
\label{app_commands}

This command line reference lists the commands for the present version,
\kdotpy{} v1.0.0. We note that commands may change (sometimes only subtly)
between versions. Thus, we recommend the user to refer to the wiki~\cite{kdotpy_wiki}
and/or the built-in help for up-to-date information if a newer version is used.

In the list below, we use the symbol \texttt{\#} to indicate additional arguments.
The bracketed \texttt{[\#]} stands for an optional argument. The list is ordered
thematically. Some commands have aliases which may be used instead of the listed
command, with identical functionality. The input of the listed commands is case
insensitive and ignores underscores, so that for example \texttt{DOS} may be used
instead of \texttt{dos} and \verb+z_res+ instead of \verb+zres+. The additional
arguments are usually case sensitive. For example, in \verb+out -HgTe+, the
distinction between uppercase and lowercase is respected.

\subsection{Options affecting computation}
\label{app_commands_calcopts}

\subsubsection{Modelling}

Determine the model, i.e., the type of Hamiltonian that needs to be
constructed.

\begin{itemize}
\item
  \texttt{norb\ \#}: Number of orbitals in the Kane model. The argument
  can be either 6 or 8, which means exclusion or inclusion,
  respectively, of the $\Gamma_7$ orbitals. (\emph{Alias}:
  \texttt{orbitals}, \texttt{orb}).\\
  Shorthand for \texttt{norb\ 6}: \texttt{6o}, \texttt{6orb},
  \texttt{6orbital}, \texttt{6band}, \texttt{sixband}\\
  Shorthand for \texttt{norb\ 8}: \texttt{8o}, \texttt{8orb},
  \texttt{8orbital}, \texttt{8band}, \texttt{eightband}\\
  \textsc{Note}: Omission of this input is not permitted.
\item
  \texttt{noren}: Do not renormalize the parameters if using anything
  else than then eight-orbital Kane model. (\emph{Alias}:
  \texttt{norenorm}, \texttt{norenormalization},
  \texttt{norenormalisation})
\item
  \texttt{lllegacy}, \texttt{llfull}: Force Landau level mode to be
  `legacy' or `full'. By default, the Landau level calculation uses
  either the symbolic mode `sym' if possible or the full mode if
  necessary. The legacy mode may not be used if the full mode were
  required. By giving \texttt{llfull}, one may also use the full mode if
  the automatically chosen `sym' mode does not give the desired results.
  Beware that full mode is much heavier on resources.
  (\texttt{kdotpy\ ll} and \texttt{kdotpy\ bulk-ll})
\item
  \texttt{llmax}: Maximum Landau level index. This has to be an integer
  $\geq 0$. If omitted, 30 is used. Larger values yield a more
  complete result, but require more computation time and memory.
  (\emph{Alias}: \texttt{nll})
\end{itemize}

\subsubsection{Regularizations, degeneracy lifting, etc.}

These options fine-tune the model.

\begin{itemize}
\item
  \texttt{noax}: Include non-axial terms, i.e., break the axial
  symmetry. (\emph{Alias}: \texttt{noaxial}, \texttt{nonaxial})
\item
  \texttt{ax}: Use axial symmetry, i.e., the axial approximation. This
  is the default for Landau level mode (\texttt{kdotpy\ ll} and
  \texttt{kdotpy\ bulk-ll}). For dispersions, it is mandatory to provide either
  \texttt{ax} or \texttt{noax}. (\emph{Alias}: \texttt{axial})
\item
  \texttt{split\ \#}: Splitting (in meV) to lift the degeneracies. It is
  recommended to keep it small, e.g., 0.01.
\item
  \texttt{splittype\ \#}: Type of degeneracy splitting. One of the
  following choices:

  \begin{itemize}
  \item
    \texttt{automatic}: Choose \texttt{sgnjz} if BIA is disabled,
    \texttt{bia} if BIA is enabled. (\emph{Alias}: \texttt{auto};
    \emph{default})
  \item
    \texttt{sgnjz}: Use the operator $\mathrm{sgn}(J_z)$, i.e., the
    sign of the total angular momentum. Despite the fact that this
    quantity is not a conserved quantum number, it works remarkably
    well. This is also the default for calculations without BIA.
  \item
    \texttt{sgnjz0}: Use the operator $\mathrm{sgn}(J_z)$ at $\mathbf{k} = 0$
    only. This type can be useful if the degeneracy is broken for
    $\mathbf{k}{}\ne 0$ as a result of some other term, for example an
    electric field.
  \item
    \texttt{isopz}: Use isoparity (`isopz') $\tilde{\mathcal{P}}_z$.
    This observable distinguishes the two blocks and is a conserved
    quantum number for symmetric geometries in many circumstances.
    Sometimes gives cleaner results than \texttt{sgnjz}. See Ref.
    \cite{Beugeling2021PRB} for more information about
    isoparity.
  \item
    \texttt{isopzw}: Use isoparity applied to the well layer only, like
    observable
    \texttt{isopzw}. While \texttt{isopz} is not a conserved quantity
    for asymmetric geometries (e.g., a well layer and two barriers with
    unequal thickness), \texttt{isopzw} can remain almost conserved in
    that case. Due to incomplete confinement in the well region, the
    eigenvalues may deviate significantly from $\pm1$.
  \item
    \texttt{isopzs}: Use isoparity applied to a region symmetric around
    the centre of the well layer, like observable
    \texttt{isopzs}. Like \texttt{isopzw}, the observable
    \texttt{isopzs} is also an almost conserved quantity for asymmetric
    geometries and tends to have eigenvalues closer to $\pm1$, because it
    generally takes into account the decaying wave function in (a part
    of) the barriers.
  \item
    \texttt{bia}: Modified form of \texttt{sgnjz}, that works better if
    bulk inversion asymmetry (BIA) is present.
  \item
    \texttt{helical}: Momentum dependent splitting, with the
    quantization axis along the momentum direction. The splitting is
    proportional to $\mathbf{k}\cdot\mathbf{S}/|k|$ for
    $\mathbf{k}{}\ne 0$. It is set to zero at $\mathbf{k} = 0$.
  \item
    \texttt{helical0}: Same as \texttt{helical}, but with
    $\mathrm{sgn}(J_z)$ at $\mathbf{k} = 0$. This option may be useful to prevent
    issues caused by degeneracies at $\mathbf{k} = 0$ for the option
    \texttt{helical}.
  \item
    \texttt{cross}: Momentum dependent splitting, with the quantization
    axis perpendicular to the in-plane momentum direction, i.e.,
    $(k_x S_y - k_y S_x)/|k|$ for $\mathbf{k}\ne 0$. It is set to
    zero at $\mathbf{k} = 0$.
  \item
    \texttt{cross0}: Same as \texttt{cross}, but with
    $\mathrm{sgn}(J_z)$ at $\mathbf{k} = 0$. This option may be useful to prevent
    issues caused by degeneracies at $\mathbf{k} = 0$ for the option
    \texttt{cross}.
  \end{itemize}
  For the relevant observables, see Appendix~\ref{app_observables_reference}.
\item
  \texttt{bia}: Include bulk inversion asymmetry. Note that combination
  of BIA with `split' may cause unwanted asymmetries, for example under
  $k_{z} \to -k_{z}$. (For \texttt{kdotpy\ 2d}, \texttt{kdotpy\ ll},
  and \texttt{kdotpy\ bulk})
\item
  \texttt{ignoremagnxy}: Ignore the in-plane components of the magnetic
  field in the gauge field (i.e., the `orbital field'). The in-plane
  components still have an effect through the Zeeman and exchange
  couplings even if this option is enabled. Enabling this option
  `simulates' the calculation before the in-plane orbital fields were
  implemented, as of version v0.58 (\texttt{kdotpy\ 1d}) or v0.74
  (\texttt{kdotpy\ 2d}), respectively. (\emph{Alias}:
  \texttt{ignoreorbxy}, \texttt{ignorebxy})
\item
  \texttt{gaugezero\ \#}: Set the $y$ position where the magnetic gauge
  potential is zero. The position coordinates are relative: -1.0 and
  +1.0 for the bottom and top edges of the sample, 0.0 for the centre
  (\emph{default}). (\emph{Alias}: \texttt{gauge0})
\item
  \texttt{yconfinement\ \#}: Set a confinement in the $y$ direction; local
  potential on the outermost sites, in meV. A large value (such as the
  \emph{default}) suppresses the wave function at the edges, which
  effectively imposes Dirichlet boundary conditions (wave functions =
  0). If the value is set to zero, the boundary conditions are
  effectively of Neumann type (derivative of wave functions = 0).
  \emph{Default}: 100000 (meV). (\emph{Alias}: \texttt{yconf},
  \texttt{confinement})
\end{itemize}

\subsubsection{Diagonalization options}\label{diagonalization-options}

Options that affect the diagonalization, i.e., which energy eigenvalues
are calculated from the Hamiltonian.

\begin{itemize}
\item
  \texttt{neig\ \#}: Number of eigenvalues and -states to be asked from
  the Lanczos method. (\emph{Alias}: \texttt{neigs})
\item
  \verb+targetenergy # [# ...]+: Energy (meV) at which the
  shift-and-invert Lanczos is targeted. If multiple values are given,
  then apply Lanczos at each of these energies (experimental feature).
  If large numbers of eigenvalues are to be calculated (e.g., 500), it
  may be faster to calculate multiple sets with a smaller number of
  eigenvalues (e.g., 5 sets of 150). Note that the values need to be
  chosen carefully. If there is no overlap between the intervals where
  eigenvalues are found, the calculation is aborted. For smaller numbers
  of eigenvalues, it is recommended to use a single value for
  \texttt{targetenergy}. (\emph{Alias}: \texttt{e0})
\item
  \texttt{energyshift\ \#}: Shift energies afterwards by this amount (in
  meV). Other energy values may still refer to the unshifted energies.
  This is an experimental feature that should be used with care. In case
  one intends to merge data (e.g., using \texttt{kdotpy\ merge}), then
  one should avoid using this option for the individual runs.
  Afterwards, this option may be used with \texttt{kdotpy\ merge}.
  (\emph{Alias}: \texttt{eshift})
\item
  \texttt{zeroenergy}: Try to align the charge-neutral gap with $E = 0\meV$.
  In combination with\linebreak[4] \texttt{energyshift}, align at that energy
  instead of $0\meV$. See also the warnings under
  \texttt{energyshift}.
\item
  \verb+bandalign [# [#]]+: Try to (re)connect the data
  points, by reassigning the band indices. The first optional argument
  determines the `anchor energy', i.e., the energy at $\vec{k} = 0$ (or $\vec{B} = 0$)
  that separates bands with positive and negative indices. If the second
  argument is given, treat the gap at the given energy as having this
  index. Omission of the first argument causes the anchor energy to be
  determined automatically. Explicit specification of this energy is
  necessary only if the automatic method appears to fail, or if the
  correct assignment of the band indices is important (e.g., for
  calculation density of states). When the second argument is omitted,
  use the default gap index 0. Alternatively,
  \texttt{bandalign\ filename.csv} may be used to use the energies in
  the csv file in order to do assign the band indices. The format is
  that of \texttt{dispersion.byband.csv}, i.e., energies of each of the
  bands in columns. This option may be used to manually `correct' an
  incorrect band connection result. If the data is not sorted (as
  function of $\vec{k}$ or $\vec{B}$), then try to sort the data automatically before
  applying the band alignment algorithm. See also
  Sec.~\ref{sec_bandalign} (\emph{Alias}:
  \texttt{reconnect}) (For \texttt{kdotpy\ merge}, \texttt{kdotpy\ 2d},
  and \texttt{kdotpy\ ll})
\end{itemize}

\subsubsection{System options}

\begin{itemize}
\item
  \texttt{cpus\ \#}: Number of parallel processes to be used. Note that the
  processes do not share memory, so it should be chosen such that the
  total memory requirement does not exceed the available memory. Can
  also be set to value \texttt{max}, \texttt{auto} or
  \texttt{automatic}, for using all available cores; this is the
  \emph{default}. For a single-core run, \texttt{cpus\ 1} must be given
  explicitly. (\emph{Alias}: \texttt{cpu}, \texttt{ncpu})
\item
  \texttt{threads\ \#}: Number of threads used per process in external
  libraries like Intel MKL (PARDISO), FEAST, LU decomposition. (Defaults
  to 1 if omitted; \emph{Alias}: \texttt{nthreads})
\item
  \texttt{gpus\ \#}: Number of parallel workers using the GPU when
  running a CUDA capable solver. (Defaults to \texttt{cpus} if omitted;
  \emph{Alias}: \texttt{gpu}, \texttt{ngpu})
\item
  \texttt{showetf}: Show estimated completion time in the progress
  monitor. When this option is omitted, the `estimated time left' (ETL)
  is shown. This option is particularly convenient for longer jobs.
  (\emph{Alias}: \texttt{monitoretf})
\item
  \texttt{verbose}: Show more information on screen (written to stdout).
  Useful for debugging purposes.
\end{itemize}

\subsubsection{Intermediate results}

Arguments that allow saving intermediate result from partially completed
diagonalization runs, for example for saving calculation time in
debugging. It saves the data container for each data point (called
\texttt{DiagDataPoint}), into a binary file that may be reloaded later.
Do this only at own risk; read the warnings below.

\begin{itemize}
\item
  \texttt{tempout}: Create a timestamped subdirectory in the output
  directory. After each step that updates a \texttt{DiagDataPoint} instance, it
  is `pickled' (using Python library `pickle' and saved to disk as temporary
  binary file. This
  output can be loaded with the \texttt{resume} argument. See also the
  notes below.
\item
  \texttt{keepeivecs}: Keep eigenvectors in memory for all
  \texttt{DiagDataPoint}s and also for temporary output files (see
  \texttt{tempout}). Warning: This can drastically increase the RAM
  usage.
\item
  \verb+resume #1 [#2]+: Path to folder (\texttt{\#1}) created
  by argument \texttt{tempout} during a previous kdotpy run. If a
  matching DiagDataPoint is found in this folder, it is restored into
  RAM and already processed calculation steps are skipped.\\
  Optionally, an integer step index (\texttt{\#2}) may be specified to
  overwrite the step from which the process is resumed. This can be
  used, e.g.~to redo the postprocessing for each DiagDataPoint, if
  eigenvectors have been saved (see \texttt{keepeivecs}).
\end{itemize}

\textsc{Note}: Some command line arguments may be changed between runs
(e.g., cpu and/or threads configuration) without affecting the validity
of older \texttt{DiagDataPoint}s for new runs. Apart from matching $\vec{k}$
and $\vec{B}$ values, there is no further automatic validation.

\textsc{Note}: The binary file format is not a suitable research data format.
Compatibility between different versions of kdotpy is \emph{not} guaranteed.

\textsc{Note}: These files should be used for temporary storage and
immediate re-use only. This is not a suitable data format for long-time
storage. Compatibility between different versions of kdotpy is \emph{not}
guaranteed. For permanent storage of eigenvectors, enable the
configuration option\linebreak[4] \verb+diag_save_binary_ddp+. See also the
security warnings for the pickle module.

\textsc{Hint}: Usage suggestions are resuming a preemptively cancelled
job (e.g., due to walltime limit, out of resources, etc), and testing or
debugging restartable from partial solutions in order to save
calculation time.

See also system options related to output, Appendix~\ref{app_commands_outputopts}.

\subsection{Density of states, electrostatics, etc.}
\label{app_commands_dos}

\subsubsection{Post-processing functions}
\label{app_commands_dos_postprocessing}

\begin{itemize}
\item
  \texttt{dos}: Plot density of states and integrated density of states.
  The Fermi energy and chemical potential are also indicated in the
  plots and printed to stdout, if their calculation has been successful.
  The range of validity is shown in red: In the shaded regions,
  additional states (typically at larger momentum $\vec{k}$) are not taken into
  account in the present calculation and may cause the actual DOS to be
  higher than indicated. The validity range typically grows upon
  increasing the momentum range (argument \texttt{k}). For \texttt{kdotpy\ ll},
  \texttt{dos} will generate equal-DOS contours and put them in the LL
  plot. This is done either at a number of predefined densities, or at
  the density given by \texttt{cardens}. For this script, also plot the
  total DOS, and the `numeric DOS' (roughly the number of filled LL).
\item
  \texttt{localdos}: For \texttt{kdotpy\ 2d}, plot the `local DOS', the
  momentum-dependent density of states. For \texttt{kdotpy\ ll}, plot
  the equivalent quantity, DOS depending on magnetic field. For
  \texttt{kdotpy\ ll}, additionally plot the `differential DOS', the
  integrated DOS differentiated in the magnetic field direction.
\item
  \texttt{banddos}: For \texttt{kdotpy\ 2d} and \texttt{kdotpy\ bulk},
  output the DOS by band. One obtains two csv files, for DOS and IDOS,
  respectively. Each column represents one band. (\emph{Alias}:
  \texttt{dosbyband})
\item
  \texttt{byblock}: For \texttt{kdotpy\ 2d} and \texttt{kdotpy\ ll}, in
  combination with \texttt{dos}. Give density of states where all states
  are separated by isoparity value ($\tilde{\mathcal{P}}_z = \pm1$). Note: By nature,
  this function does not take into account spectral asymmetry of the
  individual blocks. (\emph{Alias}: \texttt{byisopz})
\item
  \texttt{densityz}: For \texttt{kdotpy\ ll}, plot density as function
  of $z$ at the Fermi level, for all values of the magnetic field $\vec{B}$. The
  output is a multipage pdf file and a csv file with $z$ and $\vec{B}$ values over
  the rows and columns, respectively. The output is for one carrier
  density only.
\end{itemize}

\subsubsection{Self-consistent Hartree}

\begin{itemize}
\item
  \verb+selfcon [# [#]]+: Do a selfconsistent calculation
  of the electrostatic potential (``selfconsistent Hartree''). This
  method solves the Poisson equation iteratively, taking into account
  the occupied states in the well. This option also provides plots of
  the density as function of $z$ and of the potential. Two optional
  numerical arguments: maximum number of iterations (\emph{default}: 10)
  and accuracy in meV (\emph{default}: 0.01)
\item
  \texttt{selfconweight\ \#}: Use this fractional amount to calculate
  the new potential in each iteration of the self-consistent Hartree
  method. This has to be a number between 0 and 1. The \emph{default}
  value is 0.9. It may be set to a smaller value in case the iteration
  goes back and forth between two configurations, without really
  converging. A small number also slows down convergence, so the number
  of iterations may need to be increased. (\emph{Alias}:
  \texttt{scweight}, \texttt{scw})
\end{itemize}

\subsubsection{Potentials}

The following options define a background potential that affects the
calculation of the dispersion and/or the self-consistent Hartree
calculation.

\begin{itemize}
\item
  \texttt{vtotal\ \#}: Add a potential difference between top and bottom
  of the whole layer stack. The value is in meV and may be positive as
  well as negative. (\emph{Alias}: \verb+v_outer+, \texttt{vouter})
\item
  \texttt{vwell\ \#}: Add a potential difference between top and bottom
  of the `well region'. The value is in meV and may be positive as well
  as negative. (\emph{Alias}: \verb+v_inner+, \texttt{vinner})
\item
  \verb+vsurf # [# [#]]+: Add a surface/interface
  potential. The first argument is the value of the potential at the
  interfaces (barrier-well) in meV. The second parameter determines the
  distance (in nm) for which the potential decrease to 0
  (\emph{default}: 2.0). If the latter argument is \texttt{q}
  (\emph{alias}: \texttt{quadr}, \texttt{quadratic}), then the potential
  has a parabolic shape. Otherwise, the decrease to 0 is linear.
  (\emph{Alias}: \texttt{vif})
\item
  \verb+potential ## [# ...]+: Read potential from a file.
  The file must be in CSV format, i.e., with commas between the data
  values. The columns must have appropriate headings; only \texttt{z}
  and \texttt{potential} are read, whereas other columns are ignored. If
  the $z$ coordinates of the file do not align with those of the current
  calculation, then values are found by linear interpolation or
  extrapolation. If extrapolation is performed, a warning is given.\\
  The first argument must be a valid file name. The following arguments
  may be further filenames, and each filename may be followed by a
  number, interpreted as multiplier. For example,
  \texttt{potential\ v1.csv\ -0.5\ v2.csv} will yield the potential
  given by $V(z) = -0.5 V_1(z) + V_2(z)$. Multiple arguments
  \texttt{potential} are also allowed; the results are added. Thus, the sequence
  of arguments \verb+... potential v1.csv -0.5 ... potential v2.csv ...+
  is equivalent to the previous example.
\item
  \verb+cardens # [# / #]+: Carrier density in the well in
  units of $e/\mathrm{nm}^2$. This value sets the chemical potential, i.e.,
  ``filling'' of the states in the well. The sign is positive for
  electrons and negative for holes. In combination with
  \texttt{kdotpy\ ll\ ...\ dos}, specify the density at which the
  equal-density contour should be drawn. (See \texttt{dos} above.) The argument
  may also be a range (\verb+# # / #+; this affects most (but
  not all) postprocessing output functions. If omitted, the default
  range is equivalent to \verb+-0.015 0.015 / 30+ (i.e., $-1.5\times 10^{12}$
  to $1.5\times10^{12}\,e/\mathrm{cm}^2$ in steps of $10^{11}\,e/\mathrm{cm}^2$). (\emph{Alias}:
  \texttt{carrdens}, \texttt{carrierdensity}, \texttt{ncarr},
  \texttt{ncar}, \texttt{ncarrier})
\item
  \verb+ndepletion # [#]+: Density of the depletion layer(s)
  in the barrier, in units of $e/\mathrm{nm}^2$. The sign is positive for holes and
  negative for electrons. The whole sample is neutral if the arguments
  \texttt{cardens} and \texttt{ndepletion} come with the same value. If one value is
  specified, the charge is equally divided between top and bottom
  barrier. If two values are specified, they refer to bottom and top
  layer, consecutively. (\emph{Alias}: \texttt{ndepl}, \texttt{ndep})
\item
  \verb+ldepletion # [#]+: Length (thickness) of the depletion
  layers in nm. The values may be numbers $> 0$ or
  \texttt{inf} or \texttt{-} for infinite (which means zero charge
  volume density). The numbers refer to the bottom and top barrier,
  respectively. If a single value is given, use the same value for both
  bottom and top barrier. The default (if the argument is omitted) is
  infinity. (\emph{Alias}: \texttt{ldepl}, \texttt{ldep})
\item
  \texttt{efield\ \#\ \#}: Electric field at the bottom and top of the
  sample in $\mathrm{mV}/\mathrm{nm}$. Alternatively, one may enter a single value for
  either the top or the bottom electric field:

  \begin{itemize}
  \item
    \verb+efield -- #+, \verb+efield top #+,
    \verb+efield t #+, \verb+efield # top+, etc. (top)
  \item
    \verb+efield # --+, \verb+efield btm #+,
    \verb+efield b #+, \verb+efield # btm+, etc. (bottom)
  \end{itemize}

  If the variant with two values is used, the carrier density is
  calculated automatically. In that case, the explicit input of the
  carrier density (option \texttt{cardens}) is not permitted.\\
  \textsc{Note}: \verb+efield 0 #+ is not the same as
  \verb+efield -- #+.\\
  \textsc{Note}: A positive electric field at the top boundary
  corresponds to a negative gate voltage, and vice versa.
\item
  \texttt{potentialbc\ \#}: Apply custom boundary conditions for solving
  Poisson's equation in selfconsistent calculations. (\emph{Alias}:
  \texttt{potbc})\\
  The argument must be a string, which can be one of three different
  formats:

  \begin{enumerate}
  \item
    Input like a Python dict instance without any spaces:\\
    \verb+"{'v1':5,'z1':-10.,'v2':7,'z2':10.}"+\\
    All boundary names must be given explicitly, the order is irrelevant.
  \item
    Input single quantities as string separated with semicolon without
    any spaces:\\
    \texttt{"v1=5;z1=-10.;v2=7;z2=10."}\\
    All boundary names must be given explicitly, the order is irrelevant.
  \item
    Input quantity pairs as string separated with semicolon without any
    spaces:\\
    Either explicit:
    \verb+'v1[-10.]=5;v2[10.]=7'+\\
    Or implicit:
    \verb+'v[-10.]=5;v[10.]=7'+\\
    When using the explicit format, the order is irrelevant. When using the
    implicit format there is an internal counter, which applies an index
    to the quantity name, thus, the order does matter.
  \end{enumerate}

  Here, all given examples will result in the same boundary condition
  dictionary:\\
  \verb+{'v1':5,'z1':-10.,'v2':7,'z2':10.}+\\
  The $z$ values must be given as coordinate in nm, or as one of the
  following labels:

  \begin{itemize}
  \item
    \texttt{bottom}: Bottom end of the layer stack
  \item
    \verb+bottom_if+: Bottom interface of the ``well'' layer
  \item
    \texttt{mid}: Center of the ``well'' layer
  \item
    \verb+top_if+: Top interface of the ``well'' layer
  \item
    \texttt{top}: Top end of the layer stack
  \end{itemize}

  If less than 4 key-value pairs are given, only the corresponding
  values of the automatically determined boundary conditions are
  overwritten. The ones that do not appear in the automatic determined
  boundary conditions, are ignored. This also means, you can decide to
  only overwrite the $z$-coordinates but keep the automatic determined
  values for \texttt{v1}, \texttt{v2}, etc. If two full boundary
  conditions are given (4 or 5 key-value pairs), automatic boundary
  conditions are always fully overwritten.\\
  \textsc{Note}: A special case for the implicit type 3 input is
  \texttt{v12}. The input \verb+v[-10.,10.]=0;v[0.]=7+
  for example, yields
  \verb+{'v12':0,z1':-10.,'z2':10.,'v3':7,'z3':0.}+.
  Another special case for the same input type is the combination of
  \texttt{dv1} and \texttt{v1} (or \texttt{dv2} and \texttt{v2}). Here
  you can use \verb+dv[-10.]=0;v[-10.]=7+ (or
  \verb+v[-10.]=7;dv[-10.]=0+); note the same $z$ coordinates.
\end{itemize}

\subsubsection{Broadening options}\label{broadening-options}

The following options affect the broadening applied to the density of
states. A combination of options is possible, in particular also using
multiple instances of \texttt{broadening} and \texttt{berrybroadening}.
See Sec.~\ref{sec_broadening} and Appendix~\ref{app_broadening_reference} for
details and some physical background.

\begin{itemize}
\item
  \verb+broadening [#1 #2 ...]+: Broadening parameter for the
  density of the eigenstates (for\linebreak[4] \texttt{kdotpy\ ll}: Landau levels).
  The broadening is determined by a numerical width parameter $w_1$
  which may be supplemented by additional parameters, the broadening
  shape, the scaling function, and a parameter for the Berry broadening
  (the latter for \texttt{kdotpy\ ll} only). The broadening types are:

  \begin{itemize}
  \item
    \texttt{thermal}: Fermi distribution, width parameter is
    temperature; if the width parameter is omitted, use the temperature
    set by \texttt{temp}.
  \item
    \texttt{fermi}: Fermi distribution, width parameter is energy
    (\emph{alias}: \texttt{logistic}, \texttt{sech})
  \item
    \texttt{gauss}: Gaussian distribution, width parameter is energy
    (\emph{alias}: \texttt{gaussian}, \texttt{normal})
  \item
    \texttt{lorentz}: Lorentzian distribution, width parameter is energy
    (\emph{alias}: \texttt{lorentzian})
  \item
    \texttt{step}: Dirac-delta or Heaviside-step function, if width
    parameter is given, it is ignored (\emph{alias}: \texttt{delta})
  \end{itemize}

  If omitted, use the default \texttt{auto} which selects
  \texttt{thermal} for dispersion mode and \texttt{gauss} for LL mode.
  The scaling function determines how the width
  $w$ scales as function of $x$ (momentum $k$ in
  $\mathrm{nm}^{-1}$ or field $B$ in $\mathrm{T}$) ($w_1$ is the
  input width parameter):

  \begin{itemize}
  \item
    \texttt{auto}: Use \texttt{const} for dispersion mode and
    \texttt{sqrt} for LL mode (\emph{alias}: \texttt{automatic})
  \item
    \texttt{const}: Use constant width, $w = w_1$
  \item
    \texttt{lin}: The width scales as $w = w_1 x$ (\emph{alias}:
    \texttt{linear})
  \item
    \texttt{sqrt}: The width scales as $w = w_1 \sqrt{x}$
  \item
    \texttt{cbrt}: The width scales as $w = w_1 \sqrt[3]{x}$
  \item
    \texttt{\^{}n}: Where \texttt{n} is a number (integer, float, or
    fraction like \texttt{1/2}). The width scales as $w = w_1 x^n$.
  \end{itemize}

  The final optional value (\texttt{kdotpy\ ll} only) is numeric
  (floating point number or a percentage like \texttt{10\%}) that
  defines a different broadening width for the Berry curvature/Hall
  conductivity. Floating point input is interpreted as the broadening
  width itself, a percentage defines this broadening width as percentage
  of the density broadening width. The Berry/Hall broadening inherits
  the shape and scaling function from the density broadening.

  Multiple \texttt{broadening} arguments may be combined; these will
  then be iteratively applied to the (integrated) DOS, in the given
  order. See Sec.~\ref{sec_broadening} and Appendix~\ref{app_broadening_reference}
  for details.\\
  \textsc{Note}: Due to limitations of the numerical integration
  (convolution operation), combining multiple broadening functions may
  lead to larger numerical errors than a single broadening function. The
  convolution operation is commutative only up to numerical errors, so
  changing the order may lead to slight differences in the result.

  \emph{Examples}:

  \begin{itemize}
  \item
    \texttt{broadening\ 2\ thermal\ const}: A thermal broadening with
    width of $T = 2\ \mathrm{K}$, constant in momentum $k$. This is
    the default shape and scaling for dispersion mode.
  \item
    \texttt{broadening\ 2\ gauss\ sqrt}: A Gaussian broadening of width
    $2\ \mathrm{meV}$ at $1\,\mathrm{T}$ scaling proportionally to
    $\sqrt{B}$. This is the default shape and scaling for LL mode.
  \item
    \texttt{broadening\ 2}: For dispersion mode, equivalent to
    \texttt{broadening\ 2\ thermal\ const}. For LL mode, equivalent to
    \texttt{broadening\ 2\ gauss\ sqrt}.
  \item
    \texttt{broadening\ 2\ 10\%}: In LL mode, set the Berry/Hall
    broadening width to 10\% of that of the density broadening. That is,
    for the Berry/Hall broadening, the parameters are effectively
    \texttt{0.2\ gauss\ sqrt}.
  \end{itemize}
\item
  \verb+berrybroadening [#] [#] [#]+: Broadening
  parameter for the Berry curvature/Hall conductivity. The syntax is the
  same as the ordinary \texttt{broadening} parameter. Also multiple ones
  can be combined. Note that it is not permitted to combine
  \texttt{berrybroadening} with a \texttt{broadening} with argument with
  an extra numerical argument (for example
  \verb+broadening 0.5 gauss 10%+). For \texttt{kdotpy\ ll} only.
  (\emph{Alias}: \texttt{hallbroadening}, \texttt{chernbroadening})
\item
  \texttt{dostemp\ \#}: Temperature used for thermal broadening of the
  DOS. This argument is equivalent to the setting
  \texttt{broadening\ \#\ thermal\ const} (but only one of these may be
  used at a time). This temperature may be different than the
  temperature set by \texttt{temp} on the command line (which controls
  the temperature in the exchange coupling, for example). If neither
  \texttt{dostemp} nor \texttt{broadening} is given, no broadening is
  applied. If both \texttt{dostemp} and \texttt{broadening} are given, the
  setting for \texttt{broadening} takes priority. This option is
  especially useful for calculating the DOS with \texttt{kdotpy\ merge}
  and \texttt{kdotpy\ compare}: In that case, \texttt{temp} has no
  effect, because the value is read from the data files, whereas
  \texttt{dostemp} \emph{can} be used to set the thermal broadening.
\end{itemize}

\subsubsection{Additional options}

The following options affect calculations of the density of states,
self-consistent Hartree, etc.

\begin{itemize}
\item
  \texttt{ecnp\ \#}: Set charge-neutral energy (zero density) to this
  energy. The value determines the point where the density is zero. This
  affects integrated density of states in dispersion mode only. In order
  to manipulate band indices by determining the zero gap, use the
  \texttt{bandalign}
  argument. (\emph{Alias}: \texttt{cnp}, \texttt{efermi}, \texttt{ef0})
\item
  \texttt{densoffset\ \#}: Set a density offset. Basically, this number
  is added to the integrated DOS / carrier density in the selfconsistent
  calculation. The value is in units of charge density and can be
  interpreted as free carriers inside the quantum well. (\emph{Alias}:
  \texttt{noffset}, \texttt{ncnp})
\item
  \texttt{cardensbg\ \#}: Set a background density. Calculates a
  rectangular carrier distribution for this number, which is then added
  to the carrier distribution used in solving Poisson's equation. The
  value is in units of charge density and can be interpreted as immobile
  background charge.
\item
  \texttt{idosoffset\ \#}: Set an offset to the density of states, in
  appropriate DOS units. This option is identical to \texttt{densoffset}
  up to a factor of $4 \pi^2$. (\emph{Alias}: \texttt{dosoffset})
\end{itemize}

\subsubsection{Output options}

\begin{itemize}
\item
  \verb+dosrange [#1] #2+: Plot range of integrated density
  plots. If just one value $n_\mathrm{max}=\texttt{\#2}$ is given, use
  $[0, n_\mathrm{max}]$ for densities and $[-n_\mathrm{max}, n_\mathrm{max}]$
  for integrated densities. Omission means that
  the plot range is determined automatically. If a density unit is
  given, e.g., \texttt{densityenm}, the values are interpreted in the
  quantity being plotted. Here, large numbers ($>1000$) are
  interpreted as having units of $\mathrm{cm}^{-1}$, $\mathrm{cm}^{-2}$,
  or $\mathrm{cm}^{-3}$ and small numbers as $\mathrm{nm}^{-1}$, $\mathrm{nm}^{-2}$,
  or $\mathrm{cm}^{-3}$ (with the appropriate dimension).
  (\emph{Alias}: \texttt{densityrange}, \texttt{dosmax},
  \texttt{densitymax})
\end{itemize}

\subsection{Extra functions}
\label{app_commands_extra}

\textsc{Note}: Most of these functions are available only for a limited
number of kdotpy scripts.

\subsubsection{Pre-diagonalization}

\begin{itemize}
\item
  \verb+plotfz+: Plot several parameters as a function of $z$:
  \begin{itemize}
    \item $E_c$ and $E_v$ (conduction and valence band edges)
    \item $F$, $\gamma_{1,2,3}$, $\kappa$ (Luttinger and miscellaneous band parameters)
    \item $y N_0 \alpha$ and $y N_0 \beta$ (Mn exchange energies)
  \end{itemize}
  The option \texttt{legend} will include a legend in these plots (recommended).
  (\emph{Alias}: \texttt{plotqz})
\end{itemize}

\subsubsection{Using wave functions or eigenvectors}

Functions that derive extra data based on the eigenvectors.

\begin{itemize}
\item
  \texttt{overlaps}: Calculate overlaps between the eigenstates with
  those at zero momentum. By default, the overlaps are calculated with
  $\ket{\mathrm{E}1\pm}$, $\ket{\mathrm{H}1\pm}$, $\ket{\mathrm{H}2\pm}$, and
  $\ket{\mathrm{L}1\pm}$. A nice visualization can be obtained
  with \texttt{obs\ subbandrgb}, which assigns colours depending on the
  overlaps with E1, H1, and H2. A visualization with different bands can
  be obtained by using \texttt{obs\ subbandh1e1e2}, for example, where
  the observable id ends with $\geq 3$ pairs of subband
  identifiers. Each subband identifier is a band character (\texttt{e},
  \texttt{l}, or \texttt{h} followed by a number) denoting a pair of
  subbands, a single subband (the previous followed by \texttt{+} or
  \texttt{-}), or a band index (a signed integer preceded by \texttt{b}
  or parenthesized, e.g., \texttt{b+2}, \texttt{(-25)}). See also
  observables, Appendix~\ref{app_observables_reference}
  (\texttt{kdotpy\ 2d} and \texttt{kdotpy\ ll})
\item
  \verb+transitions [#1 #2] [#3]+: Calculate and plot
  transitions between levels. There can be up to 3 optional numerical
  arguments: The first pair is the energy range where transitions are
  calculated. If omitted, calculate transitions between all calculated
  states (which may be controlled with \texttt{neig} and
  \texttt{targetenergy}). The last argument is the square-amplitude
  threshold above which the transitions are taken into account. If
  omitted, the program uses the \emph{default} value 0.05.
\item
  \texttt{berry}: Calculate and plot the Berry curvature for the states
  close to the neutral gap. Also plot the integrated Berry curvature as
  function of energy. If combined with the option \texttt{dos}, then
  also plot the integrated Berry curvature as function of density. For
  LL mode (\texttt{kdotpy\ ll}), the Berry curvature is implicitly
  integrated, and the resulting output values are the Chern numbers of
  the eigenstates instead.
\item
  \texttt{hall}: Shortcut that activates all options for calculation of
  Hall conductivity with \texttt{kdotpy\ ll}. It is equivalent to the
  combination \texttt{berry\ dos\ localdos\ broadening\ 0.5\ 10\%}. The
  default value of the broadening can be overridden with the explicit
  option \verb+broadening # [#]+ combined with \texttt{hall}.
  See also options for density of states and broadening,
  Appendix~\ref{app_commands_dos}.
\item
  \verb+plotwf [# ...]+: Plot wave functions. The extra
  arguments are the plot style for the wave function plot and the
  locations (momenta) for which the plots are made. See also
  Sec.~\ref{sec_wavefunctions}.
\end{itemize}

\subsubsection{Post-diagonalization (postprocessing)}

The postprocessing functions rely predominantly on the eigenvalues (dispersion).

\begin{itemize}
\item
  \texttt{dos}: See Appendix~\ref{app_commands_dos_postprocessing}.
\item
  \texttt{localdos}: See Appendix~\ref{app_commands_dos_postprocessing}.
\item
  \texttt{banddos}: See Appendix~\ref{app_commands_dos_postprocessing}.
\item
  \texttt{minmax}: Output the minimum, maximum, and zero-momentum energy
  of each subband. (\texttt{kdotpy\ 2d} and \texttt{kdotpy\ bulk})
\item
  \texttt{extrema}: Output the local extrema of each subband. The output
  contains the type (min or max), the momentum, the energy, and an
  estimate for the effective inertial mass along the momentum direction.
  (\texttt{kdotpy\ 2d} and \texttt{kdotpy\ bulk}; \emph{Alias}:
  \texttt{localminmax}, \texttt{mimaxlocal})
\item
  \texttt{symmetrytest}: Analyze the symmetries of the eigenvalues and
  observables under various transformations in momentum space. This
  results in a list of compatible representations of the maximal point
  group $O_\mathrm{h}$, from which the program tries to determine the
  actual symmetry group (point group at the $\Gamma$ point).\\
  \textsc{Note}: For a reliable result, the momentum grid must be
  compatible with the symmetries; a cartesian grid should be used for
  cubic symmetry, a polar or cylindrical grid otherwise.\\
  For \texttt{kdotpy\ 2d}, \texttt{kdotpy\ bulk}: full analysis. For
  \texttt{kdotpy\ 1d}, \texttt{kdotpy\ merge}: partial analysis (full
  analysis to be implemented).
\item
  \texttt{symmetrize}: Extend the data in the momentum space by
  symmetrization. For example, a 1D range for positive $k$ can be extended
  to negative $k$, or a 2D range defined in the first quadrant can be
  extended to four quadrants. The extension is done by taking the known
  eigenvalues and observables and transforming them appropriately.\\
  \textsc{Note}: The algorithm relies on some pre-defined transformation
  properties of the observables, and should be used with care. A
  cross-check with a symmetric range and \texttt{symmetrytest} is
  advised.\\
  (\texttt{kdotpy\ 1d}, \texttt{kdotpy\ 2d}, and \texttt{kdotpy\ bulk},
  \texttt{kdotpy\ merge})
\item
  \verb+bhz [# ...]+: Do a L\"owdin expansion around zero
  momentum in order to derive a simplified Hamiltonian in the subband
  basis. This is the generalization of the BHZ model. For information on
  the argument pattern and further information, see Sec.~\ref{sec_bhz}.
\item
  \texttt{kbhz\ \#}: Set the reference momentum for the BHZ (L\"owdin)
  expansion. The argument refers to a momentum value on the $k_x$
  axis. This experimental option can be used for expansions around
  nonzero momentum. Please consider the results with care, as they are
  not always meaningful. See also Sec.~\ref{sec_bhz}. (\emph{Alias}:
  \texttt{bhzk}, \texttt{bhzat})
\end{itemize}

\subsection{Definition of the layer stack}
\label{app_commands_layerstack}

\begin{itemize}
\item
  \texttt{lwell\ \#}: Thickness of the active layer (e.g., quantum well) in nm. (\emph{Alias}:
  \texttt{lqw}, \texttt{qw})
\item
  \verb+lbarr #1 [#2]+: Thickness of the barrier layers in nm.
  If one thickness is given, assume this value for both barrier layers
  (bottom and top). If two thicknesses are given, the first and second
  argument refer to the bottom and top layer, respectively. The input
  \verb+lbarr #1 lbarr #2+ is equivalent to
  \verb+lbarr #1 #2+. (\emph{Alias}: \texttt{lbar},
  \texttt{lbarrier}, \texttt{bar}, \texttt{barr}, \texttt{barrier})
\item
  \verb+llayer #1 [#2 ...]+: Thicknesses of the layers in nm.
  The number of layers may be arbitrary, but the number of thicknesses
  must always be equal to the number of materials. This argument may not
  be combined with \texttt{lwell} and \texttt{lbarr}. (\emph{Alias}:
  \texttt{llayers}, \texttt{layer}, \texttt{layers}, \texttt{thickness},
  \texttt{thicknesses}, \texttt{thicknesses}, \texttt{thick})
\item
  \verb+mwell #1 [#2]+: Material for the well layer. See below
  for instructions on how to input a material. (\emph{Alias}:
  \texttt{mqw})
\item
  \verb+mbarr #1 [#2]+: Material for the barrier layers. See
  below for instructions on how to input a material. (\emph{Alias}:
  \texttt{mbarrier}, \texttt{mbar})
\item
  \verb+mlayer #1 [#2 ...]+: Material specification for an
  arbitrary number of layers. See below for instructions on how to input
  a material. The number of specified materials must match the number of
  thicknesses (\texttt{llayer}; \texttt{lwell} and \texttt{lbarr}).
  (\emph{Alias}: \texttt{mater}, \texttt{material})
\item
  \verb+msubst #1 [#2]+: Material for the substrate. This only
  sets the lattice constant which is used to calculate strain. If this
  argument is omitted, the strain is taken from the \texttt{strain} or
  the \texttt{alattice} argument. (\emph{Alias}: \texttt{msub},
  \texttt{substrate}, \texttt{msubstrate})
\item
  \texttt{ltypes\ \#1}: Define the type (purpose of each layer). The
  argument must be a string of the following letters whose length must
  match the number of layers in the stack:

  \begin{itemize}
  \item
    \texttt{b}: barrier
  \item
    \texttt{c}: cap
  \item
    \texttt{d}: doping
  \item
    \texttt{q} OR \texttt{w}: well
  \item
    \texttt{s}: spacer
  \end{itemize}

  (\emph{Alias}: \texttt{ltype}, \texttt{lstack})\\
  \textsc{Note}: Some functions will work properly only if there is
  exactly one `well' layer.
\item
  \verb+ldens #1 [#2 ...]+: For each layer, the `background
  density' of charge, for example doping. There need to be as many
  values as there are layers. The values are expressed in $e/\mathrm{nm}^2$.
  (\emph{Alias}: \texttt{layerdens}, \texttt{layerdensity})
\end{itemize}

\subsubsection{Material input}

Each material instance is a material id or compound (e.g.,
\texttt{HgMnTe}, \texttt{HgCdTe}), optionally followed by extra
numerical arguments that define the composition. The composition can
either be specified as par of the compound (chemical formula) or as
these extra arguments. Fractions and percentages are both accepted.
Thus, all of the following are equivalent: \verb+HgMnTe 2%+,
\verb+HgMnTe 0.02+, \verb+HgMn0.02Te+, \verb+Hg0.98Mn0.02Te+,
\verb+HgMn2%Te+, \verb+HgMn_{0.02}Te+, etc.\\
The chemical formulas (or material ids) are case sensitive, which eliminates
ambiguity.

\subsubsection{Material parameters}

\begin{itemize}
\item
  \texttt{matparam\ \#}: Modify the material parameters. The argument
  can either be a materials file or a sequence of
  \texttt{parameter=value} pairs. For the latter, multiple parameters
  must be separated by semicolons (\texttt{;}) and must be preceded by
  the material identifier, like so:\\
  \verb+matparam 'HgTe:gamma1=4.1;gamma2=0.7;CdTe:gamma1=1.6'+\\
  Spaces are ignored and the colon (\texttt{:}) after the material may
  be replaced by period (\texttt{.}) or underscore (\texttt{\uscore}). The
  argument must be quoted in the shell if it contains spaces. The
  material need not be repeated for subsequent parameters, so that in
  the example, \texttt{gamma2} refers to the material HgTe. The values
  may be Python expressions, but restrictions apply (see
  Appendix~\ref{app_matparam_reference} for information for material
  parameter files). Note that all expressions must resolve to numerical
  values in order for kdotpy to run successfully. Multiple
  \texttt{matparam} arguments will be processed in order of appearance
  on the command line. (\emph{Alias}: \texttt{materialparam})
\end{itemize}

\subsection{Other geometrical parameters}
\label{app_commands_othergeo}

\begin{itemize}
\item
  \texttt{zres\ \#}: Resolution in the $z$ direction in nm. (\emph{Alias}:
  \texttt{lres})
\item
  \texttt{width\ \#}: Width of the sample (in the $y$ direction. If a
  single number is given, this determines the width in nm. If the
  argument is given as \verb+#1*#2+ or \verb+#1 * #2+, where
  \verb+#1+ is an integer, then the sample has \verb+#1+ sites in
  the $y$ direction spaced by a distance of \verb+#2+ nm each. If the
  argument is given as \verb+#1/#2+ or \verb+#1 / #2+, then
  the total width is \verb+#1+ and the resolution \verb+#2+.
  (\emph{Alias}: \texttt{W})
\item
  \texttt{yres\ \#}: Resolution in the $y$ direction in nm. (\emph{Alias}:
  \texttt{wres})
\item
  \texttt{linterface\ \#}: Smoothing width of the interface in nm.
  (\emph{Alias}: \texttt{interface})
\item
  \texttt{periodicy}: Enables periodic boundary conditions in the $y$
  direction. (Only applies to 1D geometry.)
\item
  \texttt{stripangle\ \#}: Angle in degrees between the translationally
  invariant direction
  of the strip (or ribbon) and the (100) lattice vector (\texttt{kdotpy\ 1d}
  only). \emph{Default}: 0 (\emph{Alias}: \texttt{ribbonangle})
\item
  \texttt{stripdir\ \#}: Direction of the translationally invariant
  direction of the strip/ribbon in lattice coordinates. The argument may
  be a lattice vector, e.g., \texttt{130} for (1,3,0) or any of
  \texttt{x}, \texttt{y}, \texttt{xy}, and \texttt{-xy} (equivalent to
  0, 90, 45, and -45 degrees). Only one argument \texttt{stripangle} or
  \texttt{stripdir} should be given. (\emph{Alias}: \texttt{ribbondir})
\item
  \texttt{radians}: Use radians for angular coordinate values. If
  omitted, use degrees (\emph{default}).
\item
  \verb+orientation # [#] [#]+: Orientation of the lattice,
  see also Sec.~\ref{sec_transf_ham} and Appendix~\ref{app_orientation}.
  (\emph{Alias}: \texttt{orient})\\
  Possible patterns (\verb+#ang+ and \verb+#dir+ denote angles and
  direction triplets, respectively):
  \begin{itemize}
  \item
    \verb+#ang+: Rotation around $z$ (like \texttt{stripangle})
  \item
    \verb+#ang #ang+: Tilt $z$ axis, then rotate around $z$ axis
  \item
    \verb+#ang #ang #ang+: Euler rotation $z,x,z$. Rotate around $c$
    axis, tilt $z$ axis, rotate around $z$ axis.
  \item
    \verb+#dir+: Longitudinal direction $x$ (like \texttt{stripdir})
  \item
    \verb+- #dir+: Growth direction $z$
  \item
    \verb+#dir #dir+: Longitudinal and growth direction $x, z$
  \item
    \verb+- #dir #dir+: Transversal and growth direction $y, z$
  \item
    \verb+#dir #dir #dir+: Longitudinal, transversal, and growth
    direction $x, y, z$.
  \item
    \verb+#dir #ang+ \emph{or} \verb+#ang #dir+: Growth direction $z$
    and rotation around $z$.
  \end{itemize}

  Format for the inputs: For angles \verb+#ang+, enter an explicit
  floating point number containing a decimal sign (period~\texttt{.}).
  Integer values (for example 45 degrees) can be entered as
  \texttt{45.}, \texttt{45.0}, \texttt{45d}, or \texttt{45$^\circ$}. Direction
  triplets (\verb+#dir+) are a triplet of digits without separators
  and possibly with minus signs (e.g., \texttt{100}, \texttt{111},
  \texttt{11-2}, \texttt{-110}) or numbers separated by commas without
  spaces (e.g., \texttt{1,1,0} or \texttt{10,-10,3}).

  If \texttt{orientation} is combined with \texttt{stripangle} or
  \texttt{stripdir}, the latter are ignored.

  \textsc{Note}: If multiple \verb+#dir+ inputs are given, they must
  be orthogonal directions. If the inner product between any pair of
  them is nonzero, an error is raised.

  \textsc{Note}: With the option \texttt{orientation}, the program uses
  an alternative construction method for the Hamiltonian, which may
  cause the time consumption by this step to increase by a factor of
  approximately 4. There is no exception for trivial orientations, like
  \verb+orientation - 001+, which still invokes the alternative
  construction method.

\end{itemize}

\subsection{Other physical parameters}
\label{app_commands_otherphys}

\subsubsection{External parameters}

\begin{itemize}
\item
  \texttt{b\ \#}: External magnetic field in T. Ranges may be input
  using the same syntax as the momenta $\vec{k}$. For information on
  vectors and ranges, see Sec.~\ref{sec_vector_grids}.
\item
  \texttt{temp\ \#}: Temperature in K. The temperature affects the gap
  size (band edges) and the Mn exchange coupling. Optionally, it sets
  the thermal broadening of the density of states if the argument
  `broadening thermal' (without value) is given, see
  Appendix~\ref{app_commands_dos}).
  
  \textsc{Note}: Thermal broadening is \emph{not implied} by \texttt{temp}.
  In order to apply thermal broadening, specifying
  \texttt{broadening\ thermal} or \texttt{dostemp} is required, see
  Appendix~\ref{app_commands_dos} for
  more information.
\end{itemize}

\subsubsection{Specification of strain}

\begin{itemize}
\item
  \texttt{ignorestrain}: Ignore the strain terms in the Hamiltonian.
  (\emph{Alias}: \texttt{nostrain})
\item
  \verb+strain # [# #]+: Set strain value. The value may be
  set as a number or percentage (e.g., \texttt{-0.002} or
  \texttt{-0.2\%}). The value is interpreted as the `relative strain'
  $\epsilon = (a_\mathrm{strained} / a_\mathrm{unstrained}) - 1$,
  where $a_\mathrm{unstrained}$ refers to the well material. (In layer
  stacks with more than three layers, the well may not be identified,
  and then this option cannot be used. Setting \texttt{strain\ none} is
  equivalent to \texttt{ignorestrain}. It is also possible to specify
  more than one argument; then the values are interpreted as
  $\epsilon_{xx}, \epsilon_{yy}, \epsilon_{zz}$. It is possible to
  enter \texttt{-} for one or two values; then the strain values
  corresponding to these components are determined from the other
  one(s). If \texttt{strain} is used together with
  \texttt{ignorestrain}, the latter has priority, i.e., no strain is
  applied.
\item
  \texttt{alattice\ \#1}: Set the lattice constant of the strained
  materials. (\emph{Alias}: \texttt{alatt},\linebreak[4] \texttt{latticeconst})
\end{itemize}

\textsc{Note}: Exactly one of the three options \texttt{msubst},
\texttt{alattice}, and \texttt{strain} must be used at once.

\subsection{Options affecting plots}
\label{app_commands_outputopts}

\subsubsection{Observables}

\begin{itemize}
\item
  \texttt{obs\ \#}: Use observable \texttt{\#} (see Appendix~\ref{app_observables_reference}
  for the colouring of the plot. It must be one of the
  available observables in the data files. There is a special case
  \texttt{orbitalrgb}, which colours the states with RGB colours
  determined by the \texttt{gamma6,gamma8l,gamma8h} expectation values.
  In \texttt{kdotpy\ compare}, using \texttt{obs} will leave only the markers
  to distinguish the data sets; without \texttt{obs}, distinct markers and
  colours are used.
\item
  \verb+obsrange [#] #+: Minimum and maximum value of the
  observable that determines the colour scale. If one value is given, it
  is the maximum and the minimum is either 0.0 or the minus the maximum,
  which is determined by whether the standard scale is symmetric or not.
  If this option is omitted, use the predefined setting for the colour
  scale (recommended). (\emph{Alias}: \texttt{orange},
  \texttt{colorrange}, \texttt{colourrange})
\item
  \texttt{dimful}: Use dimensionful observables. Some observables, for
  example \texttt{z} and \texttt{y}, are dimensionless by default, and this option changes
  them to observables with a dimension (for example length in nm). This
  option affects output data (xml and csv) and graphics. (\emph{Alias}:
  \texttt{dimfull})
\item
  \texttt{orbitalobs}: Calculate the observables
  \verb+orbital[j]+, that is the squared overlaps of the
  eigenstates within orbital number \texttt{j}, where $j$ runs
  from 1 to $n_\mathrm{orb}$ (the number of orbitals). (For
  \texttt{kdotpy\ 2d} only.) (\emph{Alias}: \texttt{orbitaloverlaps},
  \texttt{orbobs}, \texttt{orboverlaps})
\item
  \texttt{llobs}: Calculate the observables \verb+ll[j]+, that is
  the squared overlaps of the eigenstates within Landau level
  \texttt{j}, where \texttt{j} runs from -2 to \texttt{llmax} (the
  largest LL index). This option is available for \texttt{kdotpy\ ll} in
  full LL mode only. (\emph{Alias}: \texttt{lloverlaps})
\item
  \texttt{custominterfacelengthnm\ \#}: When given, calculate additional
  `interface (character)' observables, but within a custom length
  interval given by \texttt{\#} (integer value in nm).
\end{itemize}

\subsubsection{Data and plot range}

\begin{itemize}
\item
  \texttt{erange\ \#1\ \#2}: Energy range, minimum and maximum value
  in meV. The energy range determines the vertical plot range in plots.
  It is also used as range for density of states calculations. For
  Landau level calculations, states outside the energy range are not saved in
  the B dependence data file.
\item
  \verb+xrange [#1] #2+: Horizontal range to display in the
  plot. If just one value is given, the range runs from 0 to the
  specified value. (\emph{Alias}: \texttt{krange}, \texttt{brange})
\item
  \verb+dosrange [#1] #2+: Vertical range for density of
  states plots.
\item
  \texttt{plotvar\ \#}: Plot against the given variable, instead of the
  default variable (coordinate component).
\item
  \texttt{xoffset}: Offsets the data points slightly in horizontal
  direction, so that (almost) degenerate points can be resolved. The
  direction (left or right) is determined by the sign of the requested
  observable.
\end{itemize}

\subsubsection{Plot style}

\begin{itemize}
\item
  \texttt{plotstyle\ \#}: Choose the plot style. (\emph{Alias}:
  \texttt{plotmode}). The second argument is one of the following plot
  styles:

  \begin{itemize}
  \item
    \texttt{normal}: Unconnected data points
  \item
    \texttt{curves}: Connect the data points horizontally, i.e., by band
    index. This option replaces the old \texttt{join} option.
    (\emph{Alias}: \texttt{join})
  \item
    \texttt{horizontal}: Group the data points `horizontally', but plot
    them as separate data points.
  \item
    \texttt{auto}: Use \texttt{curves} if possible; otherwise use
    \texttt{normal}. (\emph{Alias}: \texttt{automatic})
  \item
    \texttt{spin}: Use different markers based on the \texttt{jz} observable
    value. (\textsc{Note}: $J_z$ is the total angular momentum, not the
    actual `proper' spin)
  \item
    \texttt{spinxy}, \texttt{spinxz}, \texttt{spinyz}: Like the `normal'
    plot, but add arrows to indicate the spin components $(S_x,S_y)$, $S_x,S_z)$
    or $(S_y, S_z)$, respectively.
  \item
    \texttt{spinxy1}, \texttt{spinxz1}, \texttt{spinyz1}: Like \texttt{spinxy},
    \texttt{spinxz}, and \texttt{spinyz}, but rather plot directions (unit vectors) that
    indicate the spin direction in the given plane.
  \item
    \texttt{berryxy}, \texttt{berryxz}, \texttt{berryyz},
    \texttt{berryxy1}, \texttt{berryxz1}, \texttt{berryyz1} Arrows
    indicating Berry curvature, analogous to the above spin arrow modes
  \item
    \texttt{isopz}: Use different markers based on the \texttt{isopz}
    observable value.
  \end{itemize}

  Upon omission, the \emph{default} value is \texttt{auto}.
\item
  \texttt{spin}: Indicate spin expectation value (up, down) with
  different plot markers/symbols.
\end{itemize}

\subsubsection{Other plot elements, style, etc.}

\begin{itemize}
\item
  \texttt{labels}: Display band characters or band labels at $\vec{k} = 0$ ($\vec{B} = 0$
  if the horizontal axis is magnetic field) and Landau level indices,
  if applicable. 
  (\emph{Alias}: \texttt{plotlabels}, \texttt{char})
\item
  \texttt{title\ \#}: Assign plot title. One may use \verb+{var}+ to
  substitute the variable named \texttt{var}. In order to find out which
  are the available variable names (keys), use \texttt{title\ ?} to get
  a list. The format syntax follows Python's string format function,
  including the format specification ``Mini-Language'' \footnote{See the Python
  documentation at \url{https://docs.python.org/3/library/string.html\#formatspec}.}.
  Here, only named variables can be specified. Positional ones, like
  \verb+{0}+ or \verb+{1}+ are not permitted. Some special
  variable names are:

  \begin{itemize}
  \item
    \texttt{llayer(\#)}: For layer properties, append parenthesized
    integer index \texttt{(n)}, e.g.,\linebreak[4] \texttt{llayer(1)}, for the
    property of the $n$'th layer.
  \item
    \verb+b_x+, \verb+b_y+, \verb+b_z+: Cartesian vector
    components
  \item
    \verb+b_phi+, \verb+b_theta+: Angular coordinates of a vector
    in degrees
  \item
    \verb+b_len+, \verb+b_abs+: Vector length (\texttt{len} and \texttt{abs} are
    equivalent)
  \end{itemize}

  (\emph{Alias}: \texttt{plottitle})
\item
  \texttt{titlepos\ \#}: Position of the plot title. This may be any of
  the following:

  \begin{itemize}
  \item
    \texttt{l}, \texttt{r}, \texttt{t}, \texttt{b};
  \item
    \texttt{left}, \texttt{right}, \texttt{top}, \texttt{bottom},
    \texttt{center};
  \item
    \texttt{top-center}, \texttt{bottom-center};
  \item
    \texttt{tl}, \texttt{tr}, \texttt{bl}, \texttt{br};
  \item
    \texttt{top-left}, \texttt{top-right}, \texttt{bottom-left},
    \texttt{bottom-right}
  \item
    \texttt{n}, \texttt{s}, \texttt{ne}, \texttt{nw}, \texttt{se},
    \texttt{sw};
  \item
    \texttt{north}, \texttt{south}, \texttt{north-east},
    \texttt{north-west}, \texttt{south-east}, \texttt{south-west}.
  \end{itemize}

  \textsc{Note}: \texttt{left} and \texttt{right} are synonyms to
  \texttt{top\ left} and \texttt{top\ right}\\
  \textsc{Note}: Double words can be with hyphen (\texttt{top-center}),
  underscore (\verb+top_center+), space (\verb+"top center"+;
  quotes are usually needed) or be joined (\texttt{topcenter}).\\
  \textsc{Note}: \texttt{e}, \texttt{east}, \texttt{w}, \texttt{west}
  are not legal values\\
  (\emph{Alias}: \texttt{plottitlepos}, \texttt{titleposition},
  \texttt{plottitleposition})
\item
  \texttt{legend}: Include a legend in the plot. For coloured
  observables, this is a colour bar plus the indication of the
  observable. (\emph{Alias}: \texttt{filelegend})
\item
  \verb+legend label # [label # ...]+: If the argument
  \texttt{legend} is directly followed by \texttt{label} followed by the
  label text, use this text in the legend instead of the file names. The
  label text must be quoted on the command line if it contains spaces.
  (For \texttt{kdotpy\ compare} only.)
\end{itemize}

\subsubsection{System options}

\begin{itemize}
\item
  \texttt{out\ \#}: Determines the names of the output files. For
  example, if the argument is \texttt{1}, the program produces
  \texttt{output1.xml}, \texttt{plot1.pdf}, etc. This option also uses
  variable substitution using Python's string format function; see
  command \texttt{plottitle} above. (\emph{Alias}: \texttt{outfile},
  \texttt{outid}, \texttt{outputid}, \texttt{outputname})
\item
  \texttt{outdir\ \#}: Name of the output directory. If the directory
  does not exist, try to create it. If omitted, try to write to the
  subdirectory \texttt{data} if it exists, otherwise in the current
  directory. (\emph{Alias}: \texttt{dir}, \texttt{outputdir})
\end{itemize}

See also Appendix~\ref{app_commands_calcopts} for system options related to
calculations.

\end{appendix}



\end{document}